\documentclass{aa}
\usepackage{mathptmx}
\usepackage{graphicx}
\usepackage{rotating}
\usepackage{txfonts}
\usepackage{amsmath}
\usepackage[T1]{fontenc}
\usepackage{amssymb}
\usepackage{wasysym}
\usepackage{epstopdf}
\usepackage[english]{babel}
\usepackage{multirow}
\usepackage{subcaption}
\usepackage{ulem} 
\usepackage{geometry}
\usepackage{xcolor}

\usepackage{natbib,twoopt}
\usepackage[breaklinks=true]{hyperref} 
\bibpunct{(}{)}{;}{a}{}{,} 
\makeatletter
\newcommandtwoopt{\citeads}[3][][]{\href{http://adsabs.harvard.edu/abs/#3}%
{\def\hyper@linkstart##1##2{}%
\let\hyper@linkend\@empty\citealp[#1][#2]{#3}}}
\newcommandtwoopt{\citepads}[3][][]{\href{http://adsabs.harvard.edu/abs/#3}%
{\def\hyper@linkstart##1##2{}%
\let\hyper@linkend\@empty\citep[#1][#2]{#3}}}
\newcommandtwoopt{\citetads}[3][][]{\href{http://adsabs.harvard.edu/abs/#3}%
{\def\hyper@linkstart##1##2{}%
\let\hyper@linkend\@empty\citet[#1][#2]{#3}}}
\newcommandtwoopt{\citeyearads}[3][][]%
{\href{http://adsabs.harvard.edu/abs/#3}
{\def\hyper@linkstart##1##2{}%
\let\hyper@linkend\@empty\citeyear[#1][#2]{#3}}}
\makeatother

\begin{document}

 \defcitealias{pierre_xxl_2016}{XXL~Paper~I}
   \defcitealias{lieu_xxl_2016}{XXL~Paper~IV}
   \defcitealias{mantz_xxl_2014}{XXL~Paper~V}
   \defcitealias{ziparo_xxl_2016}{XXL~Paper~X}
   \defcitealias{lavoie_xxl_2016}{XXL~Paper~XV}
   \defcitealias{adami_xxl_2018}{XXL~Paper~XX}
   \defcitealias{faccioli_xxl_2018}{XXL~Paper~XXIV}
   \defcitealias{chiappetti_XXL_2018}{XXL~Paper~XXVII}
   \defcitealias{ricci_xxl_2018}{XXL~Paper~XXVIII}
%

\title{The XXL survey}
\subtitle{XLII. Detection and characterization of the galaxy population of distant galaxy clusters in the XXL-N/VIDEO field: A tale of variety
\thanks{Based on observations obtained with XMM-Newton, an ESA science mission with instruments and contributions directly funded by ESA Member States and NASA.}}

\author{A. Trudeau\inst{1}, C. Garrel\inst{2}, J. Willis\inst{1}, M. Pierre\inst{2}, F. Gastaldello\inst{3}, L. Chiappetti\inst{3}, S. Ettori\inst{4,5}, K. Umetsu\inst{6}, C. Adami\inst{7}, N. Adams\inst{8}, R. A. A. Bowler\inst{8}, L. Faccioli\inst{2}, B. H\"au\ss ler\inst{9}, M. Jarvis\inst{8, 10}, E. Koulouridis\inst{2, 11}, J.P. Le Fevre\inst{12}, F. Pacaud \inst{13}, B. Poggianti\inst{14}, \and T. Sadibekova\inst{2}
}


\institute{Department of Physics \& Astronomy, University of Victoria, 3800 Finnerty Road, Victoria, British Columbia, V8W 2Y2, Canada\\  \email{arianetrudeau@uvic.ca}
\and AIM, CEA, CNRS, Université Paris-Saclay, Université Paris Diderot, Sorbonne Paris Cité, F-91191 Gif-sur-Yvette, France
\and INAF - IASF Milan, via A. Corti 12, I-20133 Milano, Italy
\and INAF - Osservatorio di Astrofisica e Scienza dello Spazio di Bologna, via Piero Gobetti 93/3, I-40129 Bologna, Italy 
\and INFN, Sezione di Bologna, viale Berti Pichat 6/2, I-40127 Bologna, Italy 
\and Academia Sinica Institute of Astronomy and Astrophysics (ASIAA), No. 1, Section 4, Roosevelt Road, Taipei 10617, Taiwan 
\and Université Aix-Marseille, CNRS, LAM (Laboratoire d’Astrophysique de Marseille) UMR 7326, 13388 Marseille, France
\and Sub-department of Astrophysics, University of Oxford, Denys Wilkinson Building, Keble Road, Oxford, OX1 2DL, UK
\and European Southern Observatory, Alonso de Cordova 3107, Vitacura, Santiago 19001, Chile
\and Department of Physics \& Astronomy, University of the Western Cape, Private Bag X17, Bellville, Cape Town, 7535, South Africa
\and Institute for Astronomy \& Astrophysics, Space Applications \& Remote Sensing, National Observatory of Athens, GR-15236 Palaia Penteli, Greece
\and CEA Saclay, DRF/Irfu/DEDIP/LILAS, 91191 Gif-sur-Yvette Cedex, France 
\and Argelander Institut f\"{u}r Astronomie, Universit\"{a}t Bonn, Auf dem Huegel 71, DE-53121 Bonn, Germany
\and INAF-Astronomical Observatory of Padova, vicolo dell'Osservatorio 5, 35122 Padova, Italy 
}
 
\authorrunning{A. Trudeau, C. Garrel, J. Willis et al.}
\date{Received ; accepted }

\abstract
{Distant galaxy clusters provide an effective laboratory in which to study galaxy evolution in dense environments and at early cosmic times.}
{We aim to identify distant galaxy clusters as extended X-ray sources coincident with overdensities of characteristically bright galaxies.}
{We use optical and near-infrared (NIR) data from the Hyper Suprime-Cam (HSC) and VISTA Deep Extragalactic Observations (VIDEO) surveys to identify distant galaxy clusters as overdensities of bright, $z_{phot}\geq 0.8$ galaxies associated with extended X-ray sources detected in the ultimate XMM extragalactic survey (XXL).}
{We identify a sample of 35 candidate clusters at $0.80\leq z\leq 1.93$ from an approximately 4.5 deg$^2$ sky area. This sample includes 15 newly discovered candidate clusters, ten previously detected but unconfirmed clusters, and ten spectroscopically confirmed clusters. Although these clusters host galaxy populations that display a wide variety of quenching levels, they exhibit well-defined relations between quenching, cluster-centric distance, and galaxy luminosity. The brightest cluster galaxies (BCGs) within our sample display colours consistent with a bimodal population composed of an old and red subsample together with a bluer, more diverse subsample.}
{The relation between galaxy masses and quenching seem to be already in place at $z\sim 1$, although there is no significant variation of the quenching fraction with the cluster-centric radius. The BCG bimodality might be explained by the presence of a younger stellar component in some BCGs but additional data are needed to confirm this scenario.} 


   \maketitle


\section{Introduction}\label{sec_intro}




Galaxy clusters are the most massive gravitationally bound structures
at any epoch. Clusters are dark matter dominated ($\sim 85$\% of the
total mass), while a hot X-ray emitting intracluster medium (ICM)
accounts for most of the baryonic mass of the cluster
\citep{plionis_2008}. Stars and galaxies correspond to less than 5\%
of the total mass \citep{plionis_2008}. Clusters provide one of the
most extreme environments in universe: infalling
galaxies are stripped of their gas by the intracluster medium ram
pressure \citep[e.g.][]{poggianti_comparison_2004,poggianti_relation_2008,poggianti_jellyfish_2016,poggianti_gasp_2019,jaffe_gasp_2018,tonnesen_journey_2019} while the centre is 
one of the densest environments found in space.

The formation and evolution of the most massive giant elliptical
galaxies, the Brightest Cluster Galaxies (BCGs), is intimately related
to the cluster environment. Located near the gravitational centre of
the galaxy cluster, they exhibit unique properties such as distinct
luminosity and surface brightness profiles, and/or supersolar
metallicities
\citep[e.g.][]{oemler_structure_1976,tremaine_test_1977,dressler_comprehensive_1978,von_der_linden_how_2007,loubser_stellar_2009}. The
classical formation scenario of these galaxies, proposed by
\citet{de_lucia_hierarchical_2007}, is one of early star formation
(mostly before $z\sim 3$), quickly suppressed by AGN feedback
\citep[e.g][]{croton_many_2006}, and of progressive, late assembly via
gas-poor mergers.


At low redshifts, BCG properties are generally consistent with this
picture
\citep[e.g.][]{stott_near-infrared_2008,stott_little_2011,lidman_evidence_2012,bellstedt_evolution_2016,edwards_clocking_2020},
although several examples of low to moderately star-forming BCGs have
been reported in individual, X-ray bright clusters
\citep[e.g.][]{egami_spitzer_2006,bildfell_resurrecting_2008,stott_near-infrared_2008,pipino_evidence_2009,loubser_stellar_2009,loubser_regulation_2016,rawle_relation_2012,green_multiwavelength_2016}. However,
there is gathering evidence against the classical scenario at
$z\gtrsim 1$: \citet{webb_star_2015} and
\citet{mcdonald_star-forming_2016} report evidence of significant
in-situ star formation in respectively $\sim 20$\% and $\sim 90$\% of
their $z>1$ samples. The triggering mechanism of this star formation remains unknown, although
\cite{mcdonald_star-forming_2016} suggested galaxy interactions, a
possibility supported by recent simulations \citep{rennehan_rapid_2020}.



The cessation of star formation activity, referred to as quenching,
plays an important role in the evolution of galaxies \---\ both for
the BCG and within the cluster environment as a whole. Indeed,
galaxies appear to evolve at an accelerated rate in clusters than in
the field at all redshifts
\citep[e.g.][]{alberts_evolution_2014,nantais_evidence_2017,foltz_evolution_2018,jian_first_2018,pintos-castro_evolution_2019,strazzullo_galaxy_2019},
although it is unclear at which redshift the passive fraction in
clusters becomes greater than in the field
\citep[e.g.][]{strazzullo_galaxy_2013,strazzullo_galaxy_2019,brodwin_era_2013,alberts_evolution_2014,nantais_evidence_2017}. Quenching
also depends on galaxy mass in the sense that higher mass galaxies are more quenched than those of 
lower mass
\citep[e.g.][]{muzzin_gemini_2012,balogh_evidence_2016,kawinwanichakij_effect_2017,jian_first_2018,pintos-castro_evolution_2019}. Because
the most massive galaxies typically reside in the cluster core, mass and
environmental effects are difficult to disentangle
\citep[e.g.][]{balogh_implications_2010,muzzin_gemini_2012,kawinwanichakij_effect_2017,jian_first_2018,pintos-castro_evolution_2019}
and require large samples of well-characterised galaxy clusters.

%


Galaxy clusters may be identified employing a range of techniques.
Optical and infrared (IR) imaging surveys identify clusters as
overdensities of galaxies \citep[e.g.][]{postman_palomar_1996,gladders_new_2000,gladders_red-sequence_2005,euclid_collaboration_euclid_2019}. In the case of the red-sequence algorithm \citep{gladders_new_2000,gladders_red-sequence_2005}, as used in the recent Spitzer Adaptation of the Red-sequence Survey \citep[SpARCS; e.g.][]{wilson_clusters_2006,wilson_spectroscopic_2009}, identified overdensities display colours consistent with the
red-sequence at a given redshift. 
However, this red-sequence selection may introduce 
a bias toward clusters with enhanced red galaxy populations
\citep[e.g.][]{donahue_distant_2002,willis_x-ray_2018}. An
alternative approach is to identify clusters using the properties of
the intra-cluster medium, either indirectly via the Suyaev-Zel'dovich
effect
\citep[e.g.][]{zeldovich_interaction_1969,sunyaev_spectrum_1970,sunyaev_observations_1972,sunyaev_microwave_1980,sunyaev_velocity_1980,carlstrom_cosmology_2002, bleem_galaxy_2015}
or directly via X-ray bremsstrahlung emission
\citep[e.g.][hereafter \citetalias{pierre_xxl_2016}; \citealt{willis_x-ray_2018}]{gursky_x-ray_1972,sarazin_x-ray_1986,pierre_xmm-lss_2004,pierre_xxl_2016}. 
X-ray selection has been successfully used in the past to find clusters of galaxies either alone \citep[e.g.][]{vikhlinin_catalog_1998,clerc_cosmological_2012} or with the aid of optical data \citep[e.g.][]{gioia_einstein_1990,bohringer_rosat-eso_2001,willis_distant_2013}. There is tentative evidence that such clusters sometimes display smaller red-sequence galaxy populations than optically selected clusters \citep{donahue_distant_2002,willis_x-ray_2018}, but a drawback is that X-ray selected samples can exhibit a bias toward relaxed, cool core clusters \citep[e.g.][]{eckert_cool-core_2011,rossetti_cool-core_2017,willis_x-ray_2018} and lower BCG-X-ray peak distances \citep[e.g.][hereafter \citetalias{lavoie_xxl_2016}; \citealt{rossetti_measuring_2016}]{lavoie_xxl_2016}. 
Hence the need for cluster studies with various, complementary selected samples \citep[e.g.][]{donahue_distant_2002,sadibekova_x-class-redmapper_2014,bleem_galaxy_2015,willis_x-ray_2018}.


\begin{figure}
\includegraphics[width=8cm]{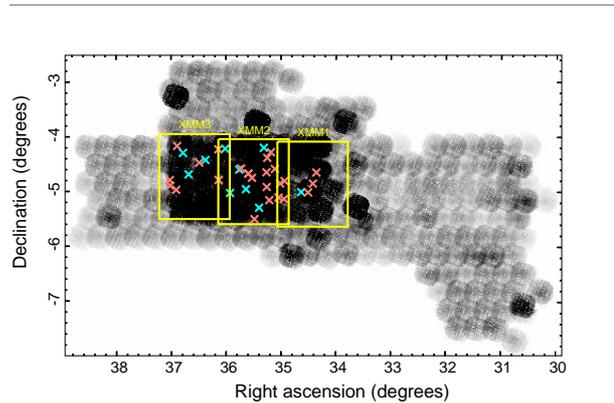}
\caption{VIDEO footprints overlaid on XXL-N exposure map. VIDEO covers eight VISTA footprints, three of them within the XXL-N field: XMM1, XMM2, and XMM3. The darker part of the exposure map correspond to the 46 ks exposure of the XMM-SERVS field. The cyan and salmon crosses correspond to the respective locations of the confirmed clusters and candidate clusters in our sample.}
\label{fig_VIDEO_footprint}
\end{figure}


In this paper we employ a multi-wavelength data set constructed as
part of the XMM-XXL 
survey to identify distant galaxy clusters and study their galaxy
populations. The XMM-XXL survey covers 50 $\mathrm{deg^2}$ divided
into two equal fields: XXL-North and XXL-South \citepalias[XXL-N and XXL-S;][]{pierre_xxl_2016}. Each field is constructed from a mosaic of 10 ks
XMM pointings. The present paper focuses on a contiguous 
sub-area of the XXL-N field covering 5.3 $\mathrm{deg^2}$, 
XMM-SERVS, which has been observed with an
exposure time of 46 ks per pointing \citep{chen_xmm-servs_2018}. This deeper sub-area of XMM data is accompanied by a range of
multi-wavelength optical and IR data \citepalias[see][]{pierre_xxl_2016},
including a high-quality data set generated by the \textit{Visible and Infrared Survey Telescope for Astronomy} (VISTA) \textit{Deep Extragalatic Observations} (VIDEO) survey \citep{jarvis_vista_2013}. We refer to the 4.5 deg$^2$ field with overlapping deep XMM and VIDEO data as the XXL-N/VIDEO field (see Fig. \ref{fig_VIDEO_footprint}).

%


This paper presents the identification and characterisation of a
sample of distant galaxy clusters selected from the XXL-N/VIDEO
field. In Sects. \ref{sec_method} and \ref{sec_sample} we describe
the identification and composition of the cluster sample. In Sect.
\ref{sec_quenching_RS}, we compute the fraction of quenched galaxies
within the cluster sample and as a function of salient properties such
as cluster-centric distance and galaxy luminosity. In Sect.
\ref{sec_BCG_ssp} we identify a sample of BCGs from the cluster sample,
and investigate the properties of their stellar populations and their
star formation histories. In Sect. \ref{sec_discuss} we discuss the
possible causes of the variety of observed cluster quenching
fractions and of the BCG colour bimodality before summarizing our main
conclusions in Sect. \ref{sec_conclu}. We employ a WMAP9 cosmology
characterised by $H_0=69.32 ~\mathrm{km~s^{-1}~Mpc^{-1}}$,
$\Omega_m=0.2865$, and $\Omega_\Lambda=0.7135$
\citep{hinshaw_nine-year_2013}. At redshifts of 1 and 1.5, an angular
scale of one arcminute corresponds to 489 and 518 kpc
respectively. All photometry is quoted in the AB magnitude
system. 

The present paper relies on the new version of XXL-XMM pipeline (V4), still in development, and on the related X-ray parameters and images. Compared to the V3 pipeline dealing with individual XMM observations (on which were based all previous XXL publications), the V4 pipeline processes co­‐added observations assembled into $1\times 1 ~\mathrm{deg^2}$ mosaics. By dealing with pointing overlaps, V4 ensures reaching the ultimate sensitivity at any position \citep[hereafter \citetalias{faccioli_xxl_2018}]{faccioli_xxl_2018}. This is especially important for the VIDEO region, characterized by a high level of redundancies.


Throughout this paper, we consider that a cluster is confirmed if at least 3 galaxies within the X-ray emission have matching spectroscopic redshifts, or if an obvious BCG has a spectroscopic redshift \citep[][hereafter \citetalias{adami_xxl_2018}]{adami_xxl_2018}. The expression \lq unconfirmed clusters\rq~will be used to refer to candidate clusters with insufficient information to be spectroscopically confirmed. Cluster names with the prefix \lq XLSSC\rq~pertain to spectroscopically confirmed clusters only and may be found in \citetalias{adami_xxl_2018}. Prefix \lq 3XLSS\rq~ refers to X-ray sources part of \citet[][XXL Paper XXVII]{chiappetti_XXL_2018} catalogue. New V4 detections are labelled by the prefix \lq XLSSU\rq .


\section{Observations and cluster detection}\label{sec_method}

In this paper, we attempt to identify significant galaxy overdensities observed in optical-IR imaging data associated with extended X-ray sources. 
We employ the galaxy photometric redshift (from the VIDEO catalogue) distribution of positive matches to select candidate distant clusters at $z_{phot}\geq 0.8$.

\subsection{X-ray data}\label{ssec_X-ray}


\bibpunct[; ]{(}{)}{;}{a}{}{,} 
In short, the \textsc{Xamin} pipeline tests four models to characterise the detected sources, generating likelihood estimates for point, extended, double point sources, and extended plus point source; this latter model, denoted AC, is intended to flag extended sources significantly contaminated by a central AGN. The coordinates of the X-ray source presented in Sect. \ref{sec_sample} are based on the centre of the best-fit model. Cluster sources are further classified into C1 and C2 on the basis of pipeline parameters {\tt extent} and {\tt extent\_likelihood}. The C1 sample corresponds to an almost pure sample of bright clusters, while the C2 sample, fainter, allows for 50\% of misclassified point sources \citep[i.e. up to 50\% of the sample might be point sources; see][\citetalias{faccioli_xxl_2018}]{pacaud_xmm_2006}. False C2 are routinely excluded by the examination of X­-ray/optical overlays for cluster candidates below z=1.
\bibpunct[, ]{(}{)}{;}{a}{}{,} 


We mention that the choice of the numerical pipeline parameter values used to define the C1 and C2 criteria are still those based on detections performed with the V3 pipeline from simulated individual XMM observations. These criteria will be revised when the final V4 is fully validated and applied to mosaic simulations. However, we do not expect drastic changes in the class parameters, since they are based on likelihoods. This situation does not impact the current study, because it does not explicitly involve the cluster selection function at any stage.

\subsection{Optical and near infrared photometry}\label{ssec_NIR}

The VIDEO observations consist of IR imaging undertaken with the VISTA
telescope in the $\mathrm{YJHK_s}$ photometric bands. 
In the XMM-SERVS field, these observations 
reach 5$\sigma$ depths of at least 25.1, 24.7, 24.2, and 23.8 mag within 2
arcsecond circular apertures for $\mathrm{YJHK_s}$ respectively
\citep{adams_rest-frame_2020}. The VIDEO catalogue also contains 
additional imaging data consisting of the \textit{Canada-France-Hawaii 
Telescope Legacy Survey Deep-1} field (CFHTLS-D1) and of the deep
\lq layer\rq~of the \textit{Hyper Suprime-Camera} (HSC) \textit{Subaru 
Strategic Program} \citep[HSC-SSP,][]{aihara_hyper_2018,aihara_first_2018}. The ultra-deep \lq layer\rq~of HSC-SSP overlaps with the XMM1 field. 
The photometric redshift analysis included in the catalogue employs an i-band selected source list 
where photometry in additional bands is obtained by applying {\tt
 SExtractor} to astrometrically-matched pixel data at
other wavelengths. In addition, we employ HSC-SSP $\mathrm{iz}$ and VIDEO
$\mathrm{JK_s}$ photometry from this catalogue to study the properties of
candidate cluster member galaxies directly. The HSC-SSP deep data have
5$\sigma$ limiting magnitudes of 25.4 and 24.6 respectively in the $i$
and $z$ bands, and the ultra-deep data have limiting magnitudes of 26.4 and 26.3 \citep{adams_rest-frame_2020}. 

Photometric redshifts for sources in the VIDEO catalogue are computed
using the \textsc{LePhare} photometric redshift code
\citep{ilbert_accurate_2006}. The code employs the COSMO template set \citep{ilbert_cosmos_2009}, including 32 templates from \citet{polletta_spectral_2007} and from \citet{bruzual_stellar_2003}. Dust attenuation follows a \citet{calzetti_dust_2000} law, and intergalactic medium absorption treatment is based on \citet{madau_radiative_1995}. Further details are provided by
\citet{adams_rest-frame_2020}.
%



\subsection{Identification of galaxy clusters}\label{ssec_finding_clusters}

We perform four steps to select candidate clusters from our X-ray
sample, following a similar procedure to that of \cite{willis_distant_2013}. Each step is summarised below:

\begin{enumerate}

\item Convolve the photometric redshift histogram for bright galaxies 
  with a matched Gaussian filter chosen to match the properties of
  redshift peaks of spectroscopically confirmed clusters. Galaxies are
  selected to be brighter than the characteristic luminosity,
  $\mathrm{L^\ast}$, along the line-of-sight of each X-ray source (i.e. a one arcmin radius aperture centred on the X-ray best-fit model centre).

\item Identify overdensities corresponding to a bright galaxy excess of $\gtrsim 4$, and display the position of potential members on
 colour images together with X-ray emission contours.

\item Examine the $i-z$ and $z-J$ colour-magnitude diagrams of each
 candidate cluster.


\item Employ a Gaussian model of the photometric redshift distribution
 of selected overdensities to estimate a refined mean cluster
 redshift and its standard deviation.




\end{enumerate}

\begin{figure*}
\centering
\begin{subfigure}{0.45\textwidth}
\includegraphics[width=8cm]{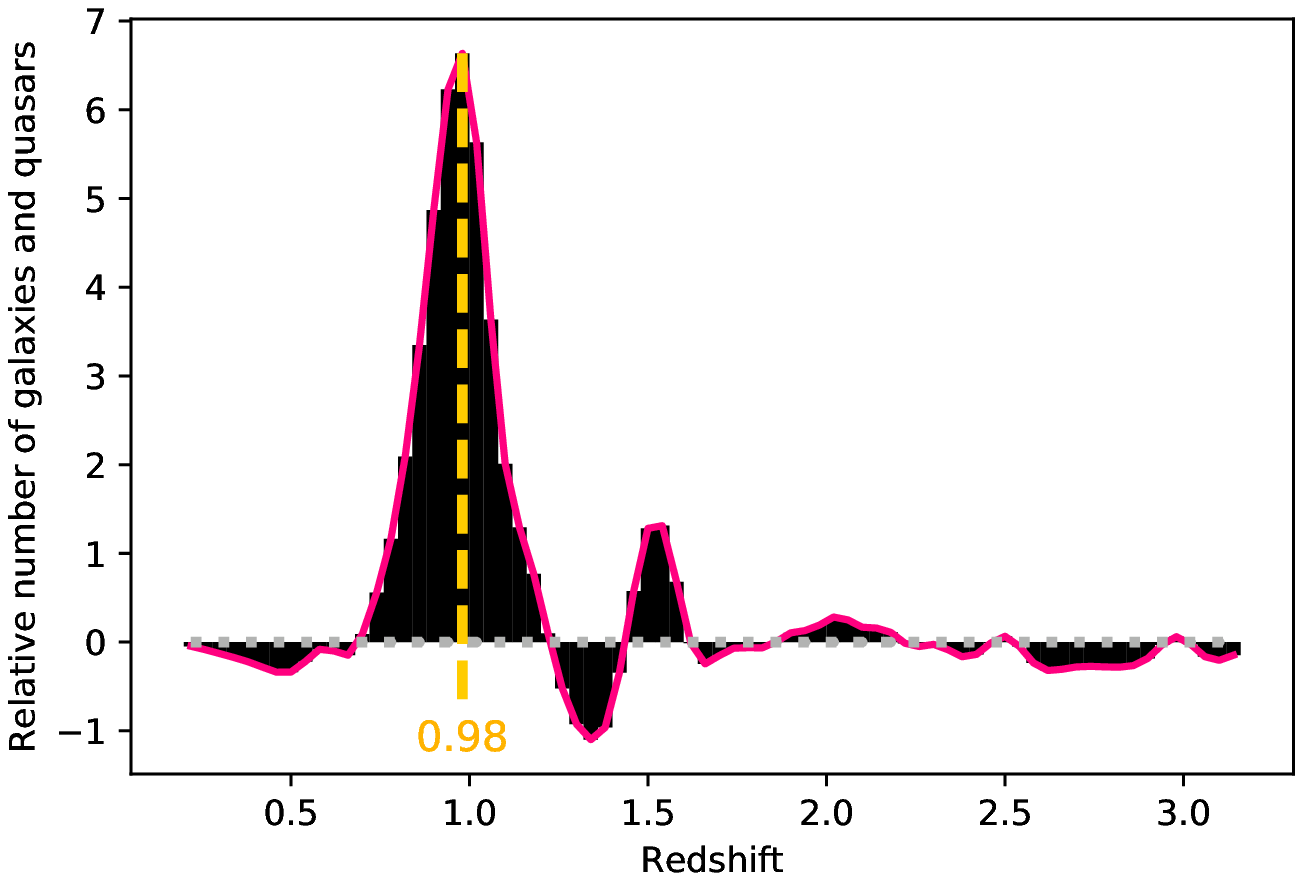}
\end{subfigure}
~\qquad
\begin{subfigure}{0.45\textwidth}
\includegraphics[width=8cm]{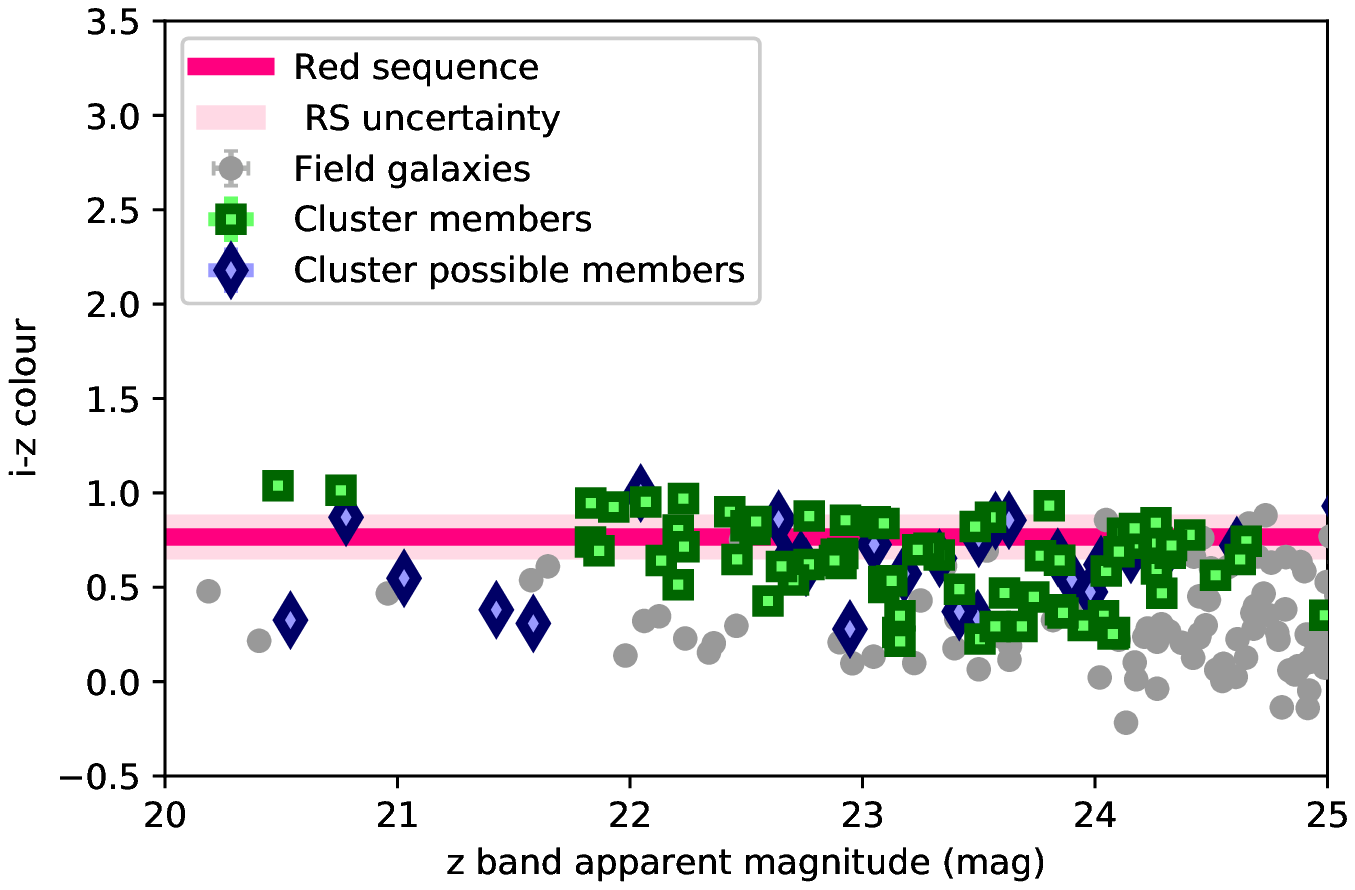}
\end{subfigure}
\\

\centering
\begin{subfigure}{0.5\textwidth}
\includegraphics[width=8cm]{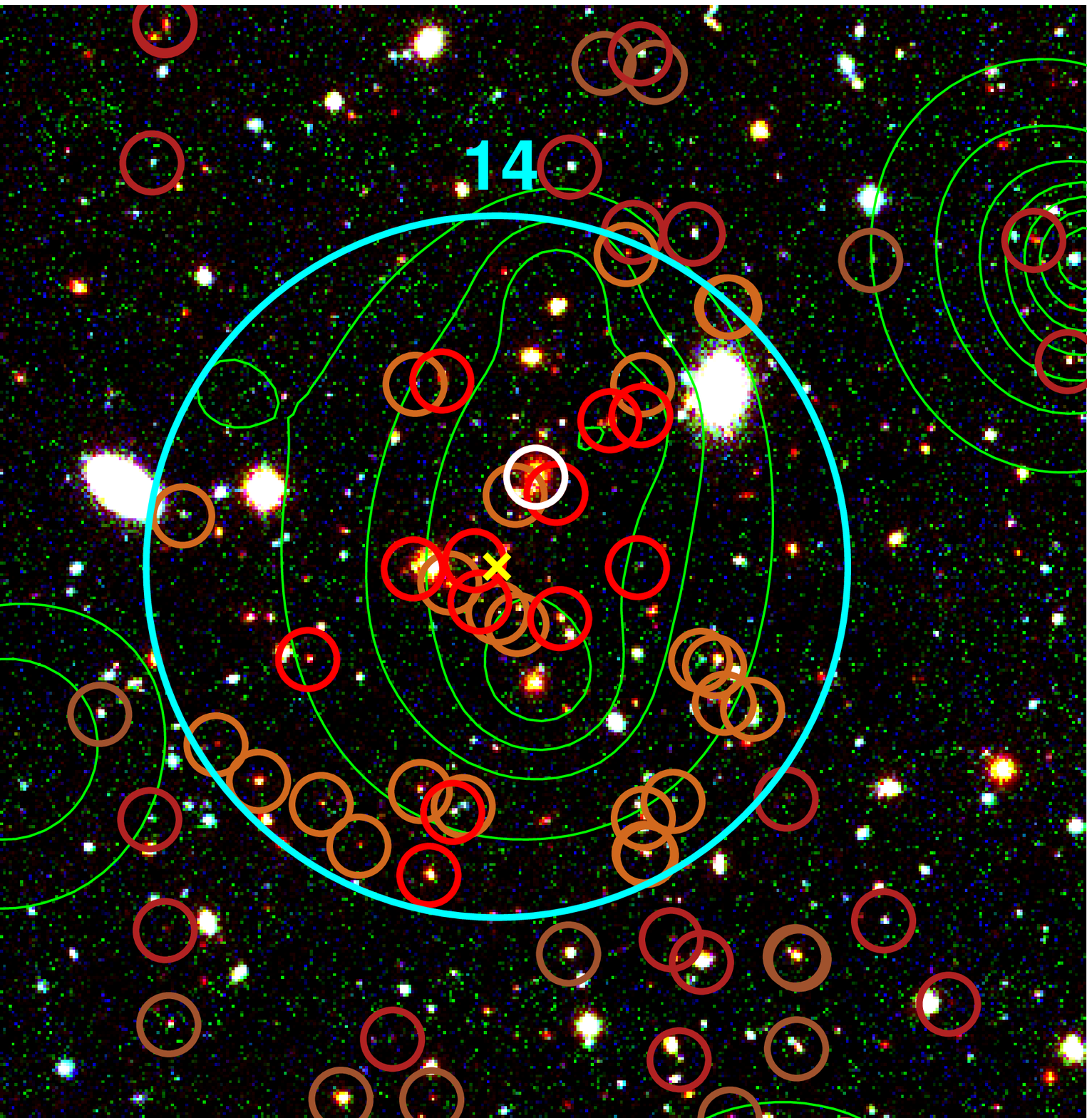}
\end{subfigure}
\caption{Visual summary of the cluster identification process. Top left: Background subtracted and Gaussian filtered
 photometric redshift distribution of the bright galaxies within the central
 arcmin of candidate 14. The dashed line indicates the highest bin in
 the redshift spike. Top right: i-z CMD plot of the galaxies above VIDEO 5$\sigma$ limit within 1
 arcmin of the centre. The green squares indicate the galaxies
 with photometric redshifts consistent with the mean redshift plus or
 minus 1.5 times the standard deviation of the most accurate Gaussian
 modelling of the redshift spike. The blue lozenges indicate galaxies
 with redshifts consistent with the sidewings of the most accurate Gaussian
 model, up to three times the standard deviation. The deep pink line
 indicates where the red sequence should be at this redshift, based on the best fit calculated in Sect. \ref{ssec_first_method}. The light pink region indicates the uncertainty on this red-sequence model, also calculated in Sect. \ref{ssec_first_method}. Bottom panel: Example of a Megacam r and i filter, and
 VIDEO H filter image for candidate 14, one of our candidate
 clusters. The cyan circle delimits the region within one arcmin of the X-ray best fit model centre, which is marked by a yellow cross. The red and brown circles highlight the bright galaxies with a redshift corresponding
 respectively to the cluster photometric peak redshift $\pm 0.02$ and
 to the cluster redshift $\pm 0.06$. Darker circles indicate galaxies outside the central region. The BCG is circled in white. 
The X-ray contours in green are logarithmically distributed in 10 levels between the maximum and minimum emission observed in a $7\times 7 ~\mathrm{arcmin^2}$ box around the X-ray source.}

\label{fig_candidates}
\end{figure*}

Figure \ref{fig_candidates} presents a visual summary of these four
steps. Similar images of the other candidate clusters are presented
in Appendix \ref{sec_images}.

%


The first step of the identification process is to select galaxies and
active galactic nuclei (AGN) that might be associated with each X-ray
sources.
We select the galaxies and AGN sources employing their goodness-of-fit
when compared to stellar, galactic, and AGN templates
\citep[see][]{jarvis_vista_2013}. Because galaxies in clusters are more
likely to host radio-loud AGN than field galaxies
\citep{best_prevalence_2007}, we keep AGN and discard only the
star-like objects. Then, we select objects within one arcmin of the
considered X-ray detection. We refer to these objects as the
\textit{field-of-view} galaxies.

We select bright galaxies in the
field-of-view employing luminosity function arguments: we compute the
apparent magnitude of $\mathrm{M^\ast}$, assuming $\mathrm{M^\ast}=-22.26$ in
the $\mathrm{K_s}$ band using \citet{cirasuolo_new_2010} with no evolution
from $z\sim 3$ to $z\sim 0$ 
and a $k$-correction described by the bandwidth term. Galaxies
brighter than the expected apparent $\mathrm{m^\ast}$ magnitude at their photometric redshift are
retained. We apply a further photometric cut based on the 5$\sigma$
depth of VIDEO $\mathrm{K_s}$ band, discarding any galaxy fainter than
23.8. The latter cut is important beyond $z\sim 1.4$, where $\mathrm{m^\ast}$ is
fainter than the VIDEO 5$\sigma$ limit.

Selected field-of-view galaxies are then binned in photometric
redshift space, over the interval $0.2<z_{phot}<3.2$. Galaxies are
sampled in bins of 0.04 in photometric redshift since this is the
redshift iteration step employed in the photometric redshift analysis. 
We then use the catalogue distribution to perform a background
subtraction, applying the same selection steps described above and
scaling the number distribution by the relative size of our
field-of-view compared to that of the full the catalogue, i.e.
\begin{equation}
N_{excess}=N_{FOV}-N_{cat}\left(\frac{\pi r_{FOV}^2}{A_{cat}} \right)
\label{bck_subtraction}
\end{equation}
where $N_{FOV}$ is the number of galaxies in a particular redshift bin
in our field-of-view, $N_{cat}$ is the number of galaxies listed in
the catalogue in the corresponding redshift bin after applying the
same cuts, $r_{FOV}$ is the one arcminute radius we use to select
galaxies associated with an X-ray detection, and $A_{cat}$ is the
catalogue area.

To identify structures in the photometric redshift histogram along the
line-of-sight to each extended X-ray source we employ a matched
Gaussian filter with a FWHM equal to 0.12 in redshift (i.e. 3
bins). The properties of the Gaussian profile are based upon the
unfiltered redshift peaks associated with spectroscopically confirmed
clusters.


\subsection{Overdensity assessment}\label{ssec_overdensities_confirmation}



We perform a visual inspection of all C1, C2, and AC X-ray sources that display a
signal consistent with $>4$ galaxies at a single photometric redshift.
We typically employ $riH$ images from CFHTLS and VIDEO with candidate
members indicated in addition to X-ray emission contours (see Fig.
\ref{fig_candidates}). We further generate $i-z$ and $z-J$ colour
magnitude diagrams of each candidate cluster to determine if a red
sequence is present.


The overdensity finding method provides a first estimate of the
candidate cluster redshift based on the median redshift of the highest
bins (see Fig. \ref{fig_candidates}, bottom left panel). To refine
this estimate, we model each candidate redshift signal as a Gaussian
employing the 13 central bins of the non-filtered redshift signal. We
then use the Gaussian mean as the cluster redshift and the standard
deviations as a estimate of the uncertainty.



 \section{The cluster sample}\label{sec_sample}

\begin{table*}[htb!]
\caption{List of detections above $z\sim 0.8$. Every detection is presented with its official designation, X-ray characterization model, X-ray coordinates, photometric redshift from VIDEO, and redshift from the literature ($z_{lit}$) when available. The sixth Col. corresponds to the significance of the detection, in terms of the numbers of galaxies in the highest bin. The ninth Col. displays the [0.5-2] keV band X-ray luminosities in the central 300 kpc of the candidate clusters, while the tenth Col. provides an X-ray luminosity based estimate of the cluster mass.} 
\label{table_detection}
\centering
\setlength{\tabcolsep}{1.5pt}
\begin{tabular}{c l c c c c c c c c c}
\hline
\# & Nearest object$^{a}$ & Flags$^{b}$ & RA & Dec & Sign. & $z_{phot}^{c}$ & $z_{lit}^{de}$ & $\mathrm{L_{X}}^{f}$ & $\mathrm{M_{500}}$ & Notes\\ 
 & & & (degrees) & (degrees) & & & & ($\mathrm{10^{43} ~erg ~s^{-1}}$) & ($\mathrm{10^{13} ~M_\odot}$) & \\ 
\hline
1 & 3XLSS J022222.9-044043 & US & 35.595 & -4.679 & 3.9 & 0.80 & 0.77 & - & - & $^{g}$ \\ 
2 & XLSSC 184 & C & 35.312 & -4.207 & 7.4 & 0.80 & 0.81 & $1.9\pm 0.2$ & $7\pm 2$ & - \\ 
3 & XLSSC 071 & C & 35.639 & -4.966 & 6.4 & 0.83 & 0.83 & $2.8\pm 0.2$ & $8\pm 3$ & - \\ 
4 & 3XLSS J022432.9-044742 & U & 36.137 & -4.796 & 5.1 & 0.83 & 0.90 & $4.3\pm 0.3$ & $9\pm 3$ & -\\ 
5 & 3XLSS J022135.2-051811 & C & 35.399 & -5.305 & 8.4 & 0.85 & 0.84 & $2.9\pm 0.2$ & $7\pm 3$ & - \\ 
6 & XLSSU J021947.4-050841 & U & 34.948 & -5.145 & 5.0 & 0.86 & 0.89 & $1.2\pm 0.3$ & $6\pm 2$ & -\\ 
7 & XLSSC 015 & C & 35.928 & -5.034 & 8.2 & 0.87 & 0.86 & $1.3\pm 0.2$ & $6\pm 2$ & - \\ 
8 & XLSSC 064 & C & 34.632 & -5.018 & 9.2 & 0.89 & 0.87 & $6.9\pm 0.3$ & $10\pm 4$ & - \\ 
9 & 3XLSS J022557.1-042845 & U & 36.489 & -4.480 & 4.0 & 0.90 & 1.05 & $1.9\pm 0.3$ & $6\pm 2$ & -\\ 
10 & 3XLSS J022156.1-053049 & U & 35.483 & -5.513 & 3.8 & 0.91 & 0.95 & $1.4\pm 0.3$ & $6\pm 2$ & - \\ 
11 & 3XLSS J021945.5-044831 & U & 34.935 & -4.814 & 4.9 & 0.91 & 0.92 & $1.1\pm 0.2$ & $5\pm 2$ & - \\ 
12 & XLSSU J022530.3-042544 & C & 36.376 & -4.429 & 5.9 & 0.87 & 0.92 & $0.8\pm 0.2$ & $5\pm 2$ & - \\ 
13 & 3XLSS J022804.6-045351 & U & 37.020 & -4.898 & 4.3 & 0.93 & 0.86 & $2.5\pm 0.4$ & $7\pm 3$ & -\\ 
14 & XLSSU J022051.0-050958 & N & 35.213 & -5.166 & 6.6 & 0.97 & - & $2.5\pm 0.5$ & $7\pm 3$ & $^{h}$ \\ 
15 & 3XLSS J022103.0-045524 & U & 35.260 & -4.924 & 5.1 & 0.97 & 1.10 & $4.7\pm 0.3$ & $8\pm 4$ & -\\ 
16 & 3XLSS J022739.0-045830 & N & 36.909 & -4.976 & 3.9 & 0.99 & - & $10.0\pm 0.6$ & $10\pm 5$ & - \\ 
17 & 3XLSS J022044.7-041713 & N & 35.185 & -4.287 & 4.1 & 1.00 & - & $6.2\pm 0.4$ & $9\pm 4$ & - \\ 
18 & XLSSC 044 ($z_f=0.27$) & US & 36.141 & -4.235 & 4.9 & 1.00 & 1.13 & - & - & $^{h}$\\ 
19 & XLSSC 124 ($z_f=0.52$) & NS & 34.419 & -4.862 & 3.9 & 1.00 & - & - & - & $^{h}$ \\ 
20 & XLSSC 029 & C & 36.016 & -4.225 & 8.3 & 1.06 & 1.05 & $13.1\pm 0.3$ & $11\pm 5$ & - \\ 
21 & XLSSC 005 & C & 36.785 & -4.300 & 5.6 & 1.04 & 1.06 & $4.5\pm 0.4$ & $7\pm 4$ & - \\ 
22 & XLSSC 192 ($z_f=0.35$) & NS & 34.507 & -5.023 & 4.6 & 1.08 & - & - & - & $^{h}$ \\ 
23 & 3XLSS J022027.0-043538 & N & 35.111 & -4.595 & 4.7 & 1.09 & - & $13.4\pm 0.8$ & $11\pm 6$ & - \\ 
24 & 3XLSS J022222.9-044043 & US & 35.595 & -4.679 & 4.0 & 1.12 & - & - & - & $^{g}$ \\ 
25 & XLSSC 141 ($z_f=0.20$) & NS & 34.356 & -4.659 & 4.9 & 1.21 & - & - & - & -\\ 
26 & XLSSC 046 & C & 35.762 & -4.605 & 9.0 & 1.18 & 1.21 & $3.5\pm 0.5$ & $6\pm 4$ & - \\ 
27 & 3XLSS J022003.6-045142 & N & 35.016 & -4.861 & 5.1 & 1.44 & - & $1.9\pm 0.5$ & $4\pm 3$ & - \\ 
28 & 3XLSS J022255.1-043508 & N & 35.726 & -4.587 & 4.5 & 1.45 & - & $16.4\pm 0.9$ & $9\pm 6$ & - \\ 
29 & 3XLSS J022100.4-042327 & N & 35.250 & -4.392 & 4.2 & 1.48 & - & $9\pm 1$ & $7\pm 5$ & -\\ 
30 & 3XLSS J022207.4-044532 & N & 35.529 & -4.758 & 3.9 & 1.49 & - & $9\pm 1$ & $7\pm 5$ & -\\ 
31 & XLSSU J022105.6-043935 & N & 35.265 & -4.656 & 4.1 & 1.54 & - & $13.2\pm 0.9$ & $8\pm 6$ & - \\ 
32 & 3XLSS J022010.3-050701 & N & 35.043 & -5.117 & 5.5 & 1.57 & - & $4.0\pm 0.8$ & $5\pm 4$ & - \\ 
33 & 3XLSS J022806.4-044803 & N & 37.025 & -4.797 & 4.9 & 1.79 & - & $8\pm 2$ & $6\pm 5$ & - \\ 
34 & JKCS 041 & C & 36.683 & -4.694 & 8.2 & 1.63 & 1.80 & $16\pm 1$ & $7\pm 6$ & - \\ 
35 & 3XLSS J022734.1-041021 & N & 36.891 & -4.174 & 4.0 & 1.93 & - & $10\pm 1$ & $6\pm 5$ & - \\ 
\hline

\multicolumn{11}{p{16cm}}{$^a$ Nearest confirmed cluster or X-ray source. If the nearest object is a confirmed foreground cluster, its spectroscopic redshift, $z_f$, is given.}\\
\multicolumn{11}{p{16cm}}{$^b$ C: spectroscopically confirmed cluster; U (for unconfirmed): not spectroscopically confirmed, but listed as a candidate cluster in the literature; N: New candidate cluster; S: superposition with a low redshift confirmed cluster or with another candidate}\\
\multicolumn{11}{l}{$^{c}$ The uncertainties on the photometric redshifts are estimated to 0.02 at $z<1.4$ and 0.14 at $z\gtrsim 1.4$}\\
\multicolumn{11}{p{16cm}}{$^{d}$ Spectroscopic redshifts (flag C) reported from \citet{pierre_xmm_2006,willis_distant_2013,andreon_jkcs_2014}; \citetalias{adami_xxl_2018}}\\
\multicolumn{11}{p{16cm}}{$^{e}$ Tentative or photometric redshifts (flag U) reported from \citet{finoguenov_x-ray_2010,durret_galaxy_2011,wen_galaxy_2011,licitra_redgold_2016}; \citetalias{adami_xxl_2018}}\\

\multicolumn{11}{p{16cm}}{$^{f}$ We do not compute a X-ray luminosity for clusters marked S (superposition), because their X-ray emission might be a blend of foreground and background emission.}\\
\multicolumn{11}{l}{$^{g}$ Archival spectroscopic observations available.}\\
\multicolumn{11}{l}{$^{h}$ Gemini GMOS observations under proprietary time.}\\

\end{tabular}
\end{table*}

\subsection{Sample selection}\label{ssec_sample_selection}

\begin{figure}
\includegraphics[width=8cm]{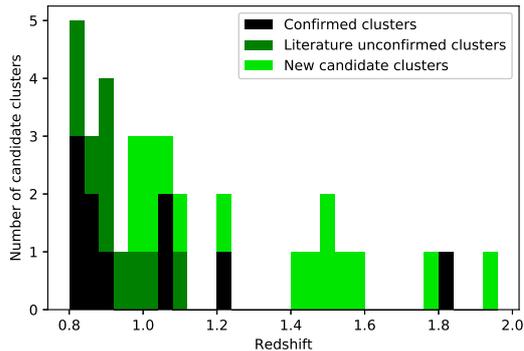}
\caption{Histogram of the candidate cluster redshifts. The black bars correspond to the spectroscopic redshifts of the confirmed clusters and the green bars to the photometric redshifts of the candidate clusters, either previously observed (dark green) or newly detected (lighter green).}
\label{fig_cluster_redshift}
\end{figure}

\bibpunct[; ]{(}{)}{;}{a}{}{,} 
We processed a total of 284 extended X-ray detections 
within the XXL-N/VIDEO region. This parent sample generated a sample of 35 candidate distant galaxy clusters. Table \ref{table_detection} and Fig. \ref{fig_cluster_redshift} present these clusters, each of which
represents a detection with a significance of approximately four galaxies
or more in a photometric redshift bin satisfying $z_{phot}\geq 0.8$ (see also Fig. \ref{fig_VIDEO_footprint}). Of these 35 candidate clusters, ten are  
spectroscopically confirmed while 15 are presented here for the first time. Ten additional candidates have been previously identified as distant clusters but, to our knowledge, never spectroscopically confirmed \citep[see][\citetalias{adami_xxl_2018}]{olsen_galaxy_2007,finoguenov_x-ray_2010,durret_galaxy_2011,wen_galaxy_2011,licitra_redgold_2016}. 

Nine of the confirmed cluster detections were confirmed by prior spectroscopy \citep[e.g.][\citetalias{adami_xxl_2018}]{pierre_xmm_2006,willis_distant_2013}, including one at $z=1.803$ \citep{andreon_jkcs_2014}. With the spectroscopic redshifts listed in the CESAM database\footnote{\url{http://cesam.lam.fr/xmm-lss/}} \citepalias{adami_xxl_2018}, we were able to confirm one additional cluster (candidate 3), bringing the total number of confirmed cluster to ten. All of these X-ray detections meet our candidate clusters criteria.
\bibpunct[, ]{(}{)}{;}{a}{}{,}

\bibpunct[; ]{(}{)}{;}{a}{}{,} 
Four confirmed distant clusters in this area \citep[][\citetalias{adami_xxl_2018}]{pierre_xmm_2006,papovich_spitzer-selected_2010}, are not part of our sample because their coordinates do not correspond to a V4 C1, C2, or AC detection. This is the case for a $z=1.62$ cluster \citep{papovich_spitzer-selected_2010}. 
Although IRC-0218A and our detections at similar redshifts possess comparable masses (IRC-0218A $M_{200}$ is $7.7\pm 3.9\times 10^{13}~\mathrm{M_{\odot}}$), IRC-0218A X-ray emission is completely dominated by a point source \citep{pierre_chandra_2012}. 

We nevertheless applied our optical/IR detection criteria to those four clusters. Two clusters satisfied these criteria, including IRC-0218A, and were included in our photometric redshift accuracy assessment (see the following section). The clusters XLSS J022609.9-043120 and XLSSC 203 \citep[see][\citetalias{adami_xxl_2018}]{pierre_xmm_2006} located respectively at $z=0.82$ and $z=1.077$, did not satisfy them. Because V4 works on co-­added and thus deeper images, we expect its source characterisation to be more reliable.
\bibpunct[, ]{(}{)}{;}{a}{}{,} 

%
%

%

\subsection{Photometric redshift accuracy}\label{ssec_test_redshift}

\begin{figure}
\includegraphics[width=8cm]{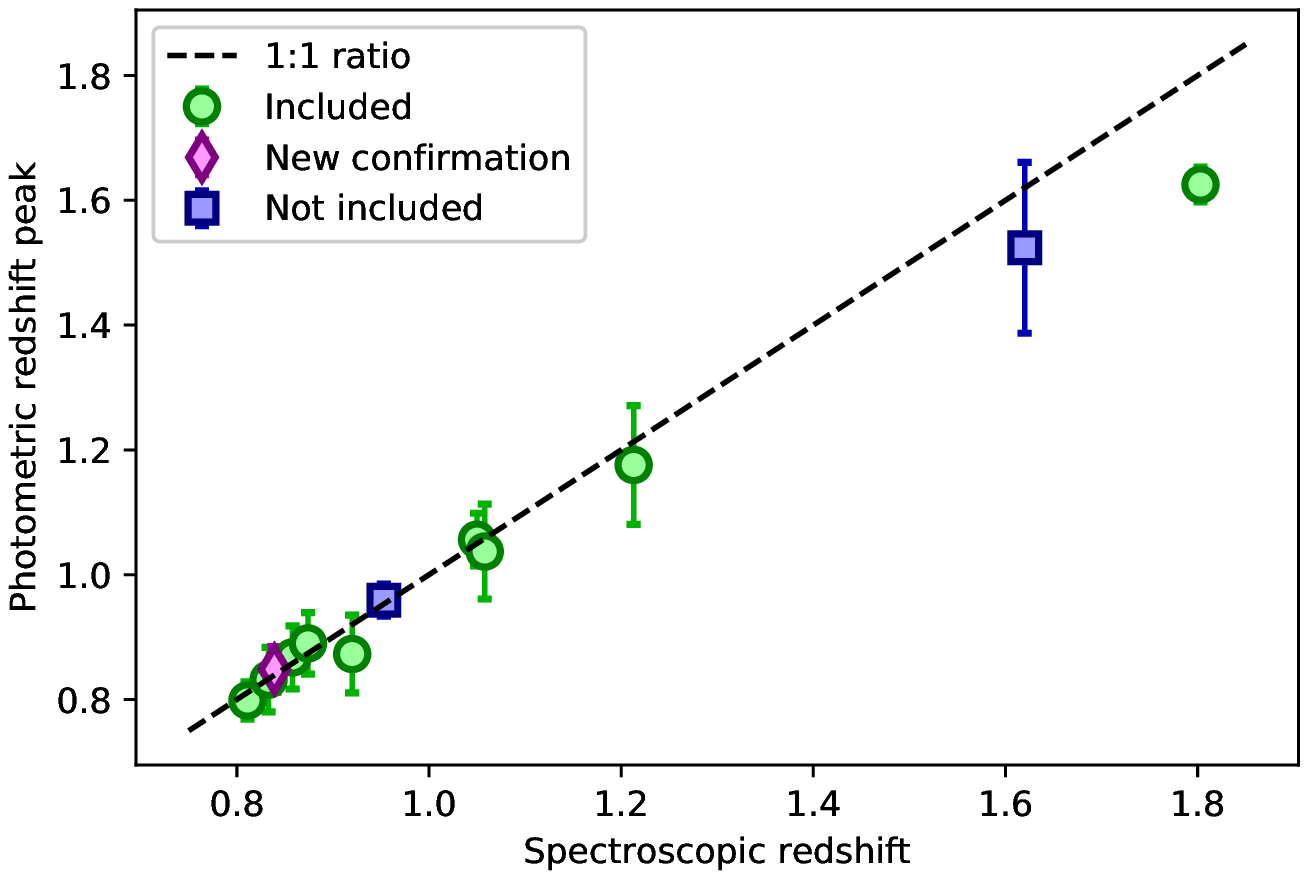}
\caption{Comparison between the spectroscopic redshifts of confirmed
 clusters within the XXL-N/VIDEO overlap and the corresponding
 photometric overdensities in VIDEO catalogue. The error bars
 correspond to the standard deviation of the Gaussian model best
 fitting the photometric spike. All clusters except two (not included), 
 are detected in optical. The green dots are the
 clusters that are part of our sample, while the blue square
 are not, since they do not meet our X-ray selection criteria. The magenta diamond is the newly confirmed candidate 3. 
The dashed line represents the ideal case, where $z_{phot}=z_{spec}$.}
\label{fig_photo_vs_spectro}
\end{figure}

To estimate the reliability of our photometric redshift estimates for
the candidate clusters, we compared the spectroscopic redshift of
12 confirmed clusters in the field \citepalias[6 clusters listed in][four from \citealt{pierre_xmm_2006,papovich_spitzer-selected_2010,willis_distant_2013,andreon_jkcs_2014}, and 2 others confirmed clusters in the XXL-N/VIDEO area]{adami_xxl_2018}
to the
photometric redshift generated by the cluster finding
procedure. 
Figure \ref{fig_photo_vs_spectro} shows the result of
this comparison. At $z\sim 1$, the differences between the photometric
and spectroscopic redshifts fall well within the photometric redshift
error estimates. These error bars represent the standard deviation of
the best-fitting Gaussian model. For the two high-redshift clusters,
the photometric redshifts seem to underestimate the spectroscopic
values. Therefore, we calculate the root mean square (RMS) of the
$z\sim 1$ and $z\gtrsim 1.5$ clusters separately. We obtained 0.02
and 0.14 respectively. For now on, we will use these RMS values as
the uncertainties on the photometric redshifts.

\subsection{Clusters estimated masses}\label{ssec_clusters_masses}

Following a similar methodology than the second data release of XXL,
we used scaling relations to provide mean parameters estimate for clusters for which the data quality is not enough to perform a direct spectral fit. A detailed description is provided in Sect. 4.3 of \citetalias{adami_xxl_2018}, 
and we provide a brief overview here. We estimated count-rates in the pn data in the [0.5-2] keV band within 300 kpc of the cluster centre using the Bayesian approach to fixed aperture photometry measurement outlined in \citet{willis_x-ray_2018}. We then converted this count-rate to the corresponding X-ray luminosity by adopting an initial gas temperature, a metallicity fixed to 0.3 times the solar value \citep[as tabulated in][]{anders_abundances_1989}, and the cluster spectroscopic (when available) or photometric redshift. With the same initial guess of the temperature, we estimate $r_{500,scal}$ from the mass-temperature relation constrained from a subset of 105 XXL clusters that have both measured HSC lensing masses and X-ray temperatures \citep[see][]{umetsu_weak-lensing_2020}. 
We stress that here we use a $M_{500}~|~T_{X}$ relation, obtained using the Bayesian regression scheme implemented in the LIRA package \citep{sereno_bayesian_2016,sereno_comparison_2016}, and not the $T_{X}-M_{500}$ relation reported in \citet{umetsu_weak-lensing_2020}. The luminosity is then extrapolated from 300 kpc to $r_{500,scal}$ assuming a $\beta$-model for the cluster emissivity with parameters $(r_c,\beta) = (0.15 r_{500,scal},2/3)$. Then a new temperature is evaluated using the best-fit result for the luminosity-temperature relation quoted in Table 6 of \citetalias{adami_xxl_2018} (XXL fit). The iteration on the gas temperature is stopped when the input and output values agree within 5\%, and in general the process converges after 2-3 steps. The uncertainties on the derived parameters and in particular the masses, are obtained by propagation of errors on the scaling parameters, including the measured correlation among them.



\subsection{Other clusters in XXL-N/VIDEO}\label{ssec_other_clusters_in_the_field}


There are 54 previously confirmed clusters either within or overlapping
with the XXL-N/VIDEO field \citepalias{adami_xxl_2018}. Of these, 47 are
located at $z<0.8$ with the remaining seven clusters located at
$z\geq 0.8$. These seven clusters correspond to a surface density of 1.6
distant clusters per square degree. This number represents a lower limit because some known clusters associated with an extended X-ray detection \citep[e.g. JKCS 041, a z=1.803 confirmed cluster][]{andreon_jkcs_2014} are excluded of \citepalias{adami_xxl_2018} compilation. 

%

Adding all our detections would bring up this number to approximately 8.2 
clusters per square degree, with a flux limit of $1.7 \times 10^{-15} ~\mathrm{erg~cm{-2}~s^{-1}}$ in the [0.5-2] keV band \citep{chen_xmm-servs_2018}. This represent 3.6 times and 0.51 times the surface densities reached by \citet{willis_distant_2013} and \cite{finoguenov_x-ray_2010} respectively, with depths of $1 \times 10^{-14}~\mathrm{erg~cm{-2}~s^{-1}}$ and $2 \times 10^{-15} ~\mathrm{erg~cm{-2}~s^{-1}}$ in the [0.5-2] keV band.

\section{Quenching and star formation in clusters} \label{sec_quenching_RS}

The fraction of quenched galaxies in a sample of distant, X-ray selected
galaxy clusters has the potential to provide an unbiased view of the
star formation conditions in massive, virialised structures. In this
section, we compute the fraction of quenched galaxies within the
XXL-N/VIDEO distant cluster sample, focussing on the clusters at $z<1.4$, because the catalogue 5$\sigma$ magnitude limit restricts the number of selected galaxies beyond that redshift. We achieve this employing four
related analysis techniques, intending to investigate
whether each provides consistent results. Only the galaxies brighter than the 5$\sigma$ limit in J and $\mathrm{K_s}$ are used in these computations.

\begin{figure}
\includegraphics[width=8cm]{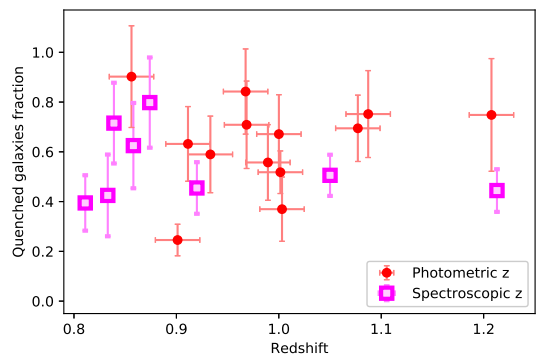}
\caption{Fraction of quenched galaxies as a function of the redshift according to method 4, for each VIDEO candidate, excluding candidates 1, 4, 10, 21, and 24 and the candidate clusters above z=1.4. Spectroscopically confirmed clusters are indicated by squares, and the other one by circles. The error bars are the propagation on the Poissonian uncertainties on the integrals of the red sequence and the blue cloud models.} 
\label{fig_quench_z}
\end{figure}

\begin{enumerate}


\item The first employs the $i-z$ colour histogram of background
 corrected galaxies in the field of each candidate cluster. This
 colour space distribution is then modelled
 using two Gaussian functions to represent the red sequence and the blue
 cloud (see Fig. \ref{fig_quenching_bck}). 

\item The second method employs the same background corrected colour
 distribution as above but employs a single colour cut to divide the
 distribution into quenched and star-forming galaxies.

\item The third method selects cluster galaxies using photometric
 redshift and then applies the boundary method used in method (2)
 above.

\item The fourth method is similar to method (2) with the additional constraint that each galaxy (including the background) must be brighter than $\mathrm{L^\ast}$ at the candidate cluster redshift. The $\mathrm{L^\ast}$ evolutionary k-correction is computed using method 1 results (see Fig. \ref{fig_quench_z}).

\end{enumerate}

\subsection{First quenching method}\label{ssec_first_method}


\begin{figure*}
\centering
\begin{subfigure}{0.3\textwidth}
\includegraphics[width=6cm]{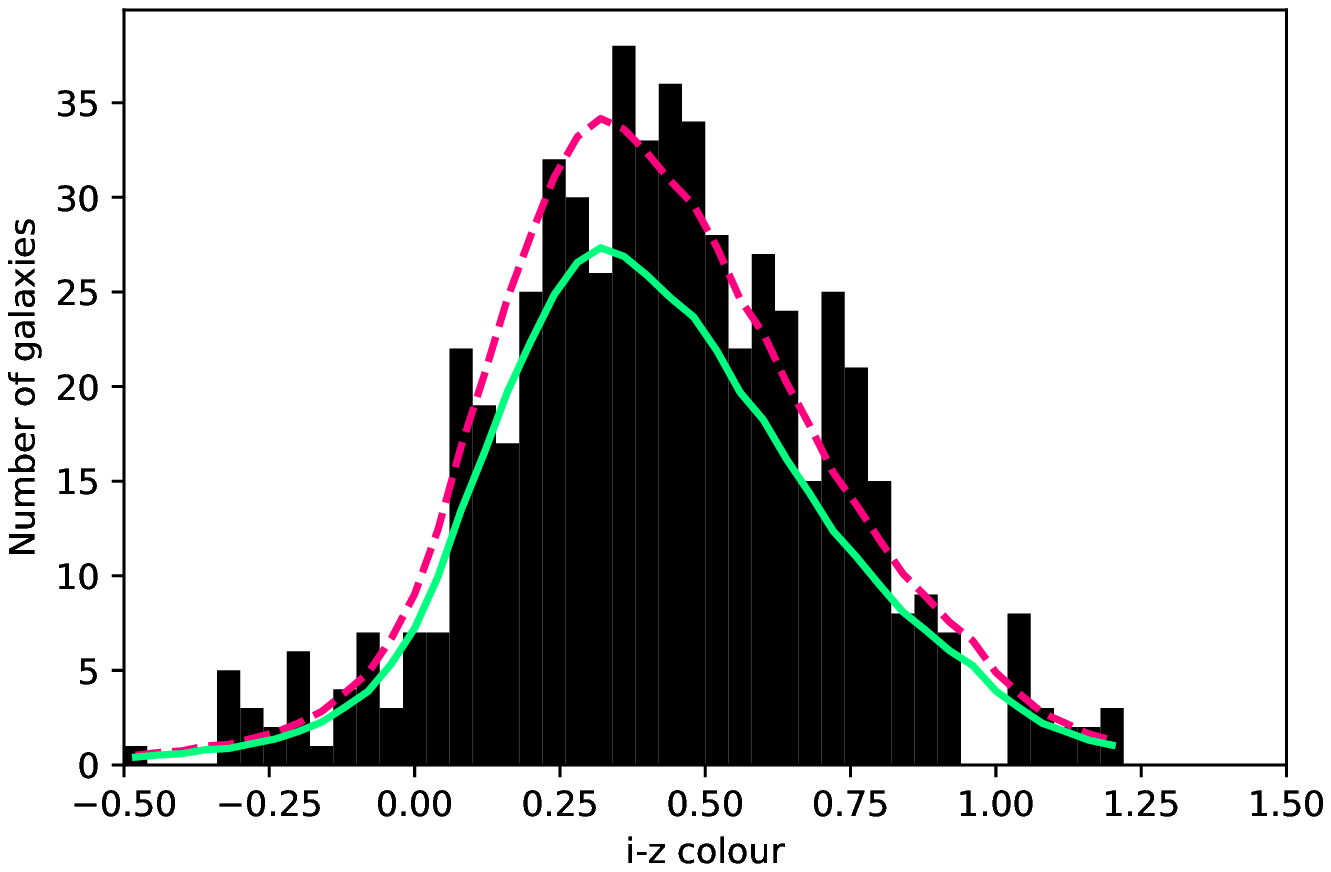}
\end{subfigure}
\hfill
\begin{subfigure}{0.3\textwidth}
\includegraphics[width=6cm]{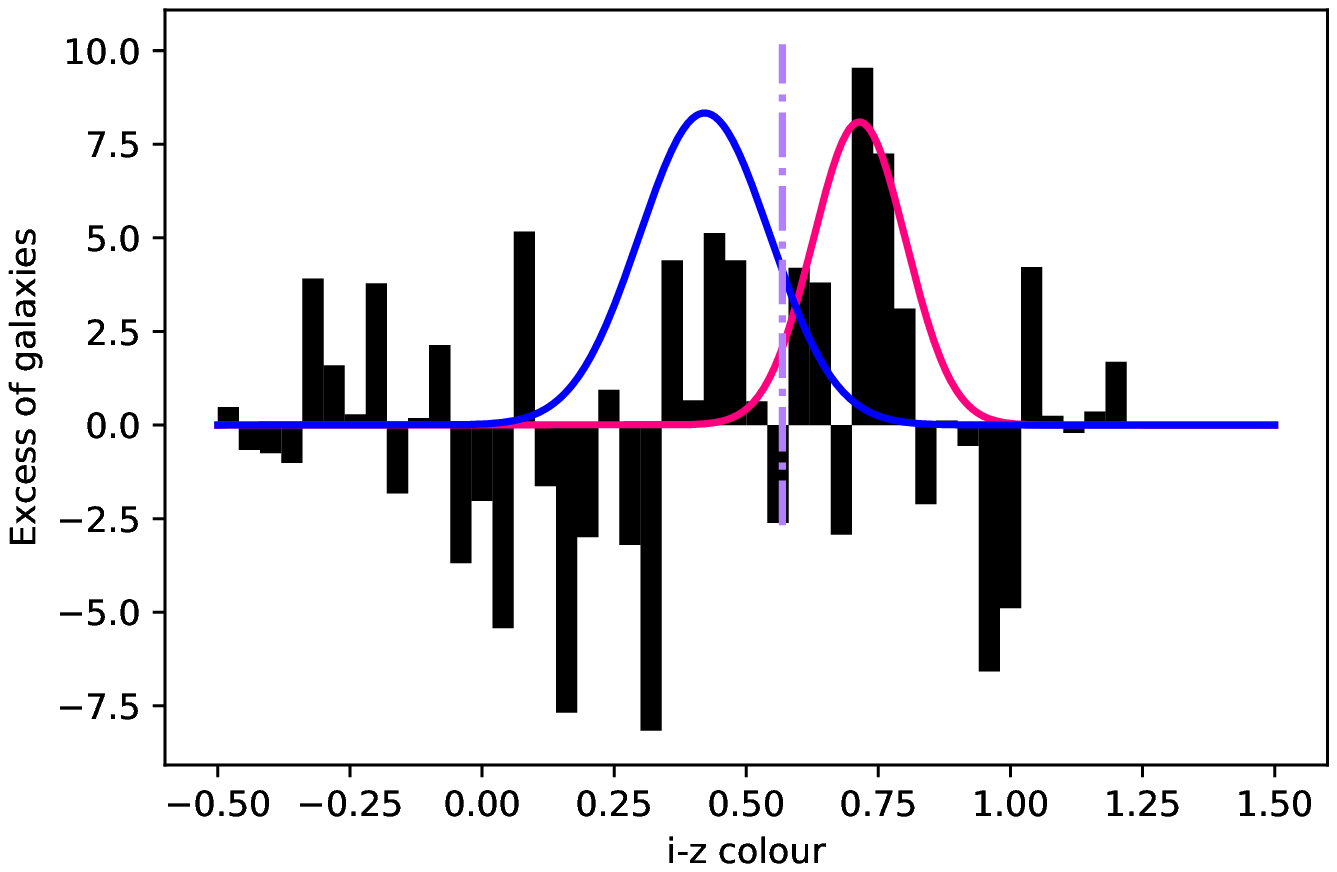}
\end{subfigure}
\hfill
\begin{subfigure}{0.3\textwidth}
\includegraphics[width=6cm]{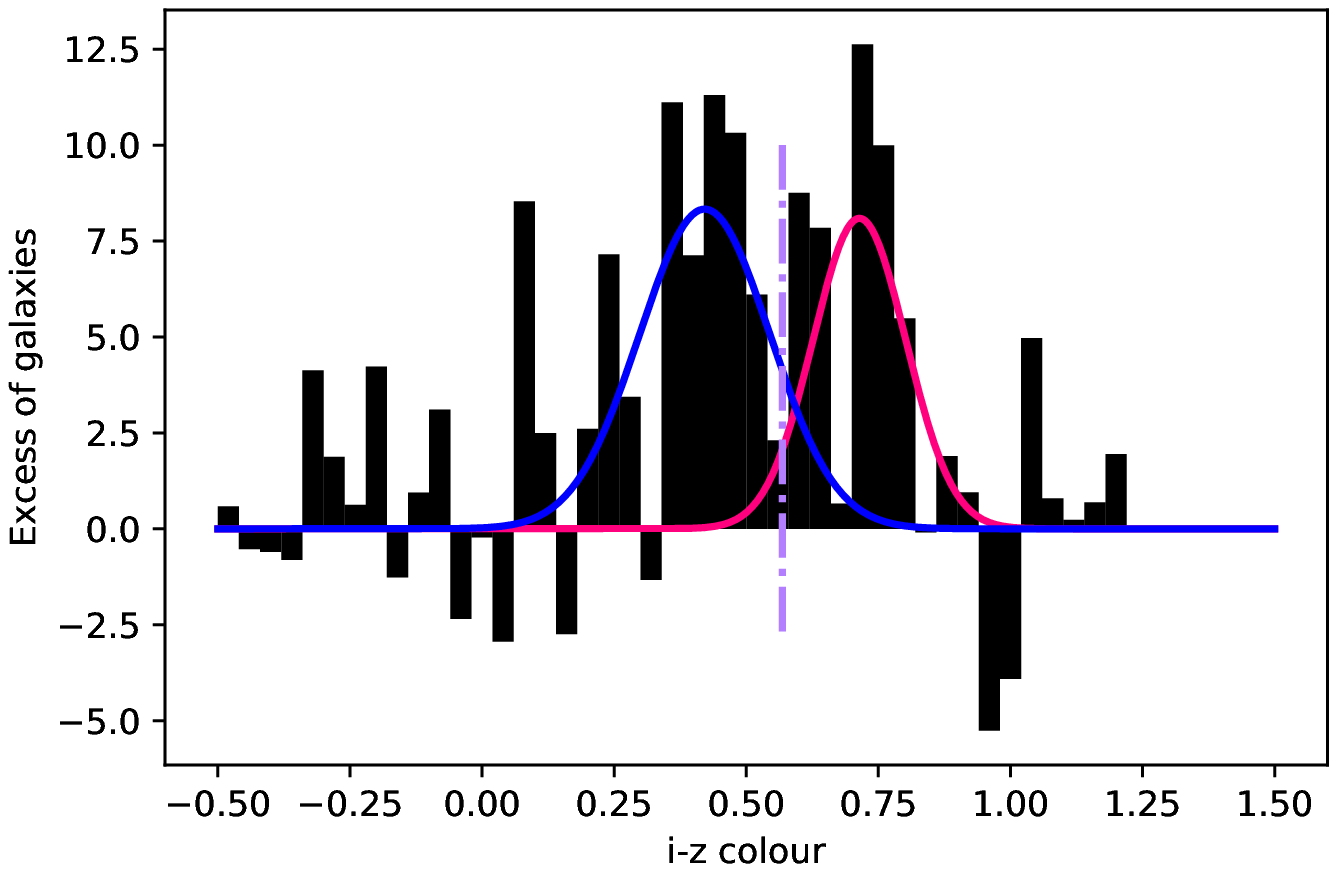}
\end{subfigure}
\begin{subfigure}{0.3\textwidth}
\includegraphics[width=6cm]{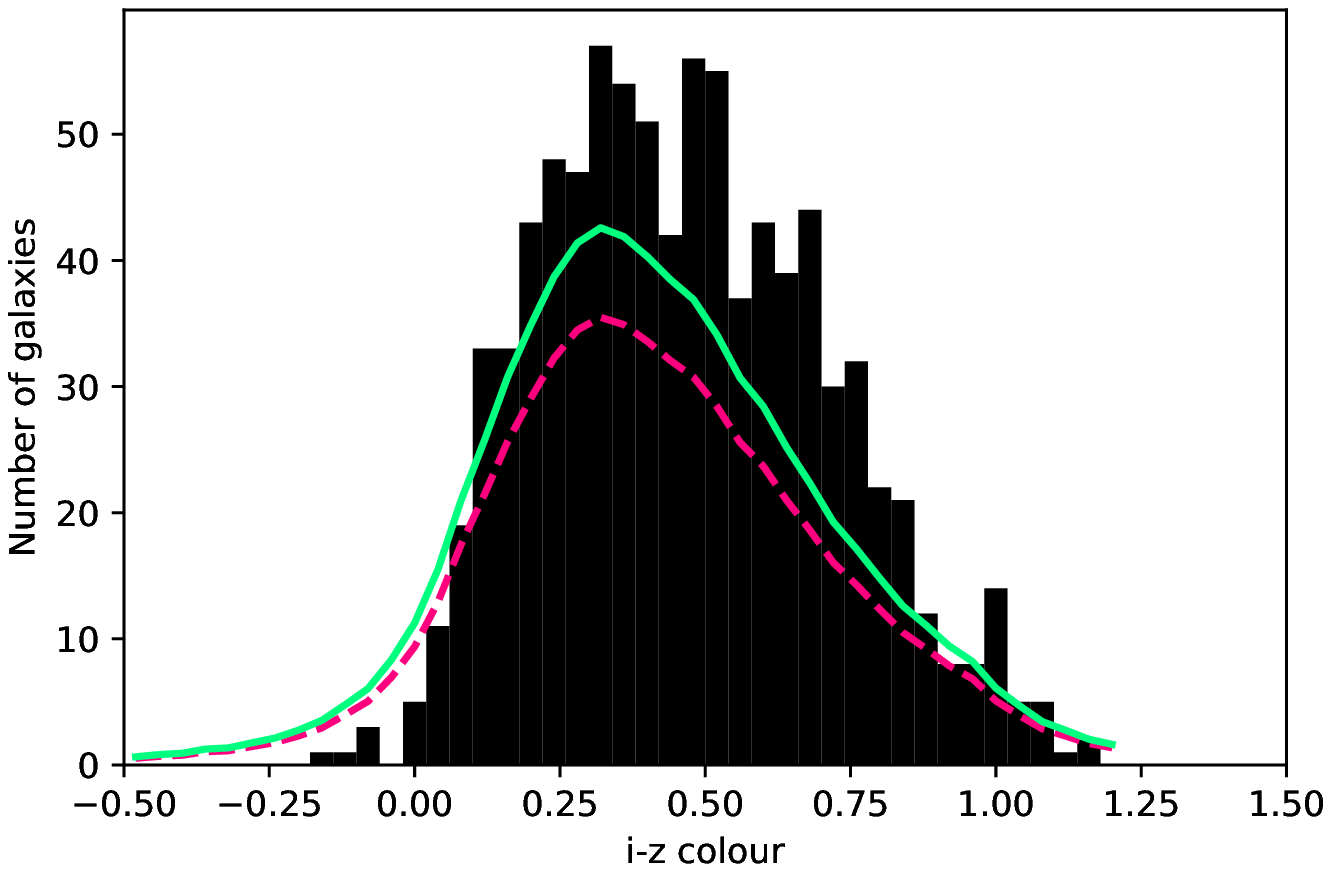}
\end{subfigure}
\hfill
\begin{subfigure}{0.3\textwidth}
\includegraphics[width=6cm]{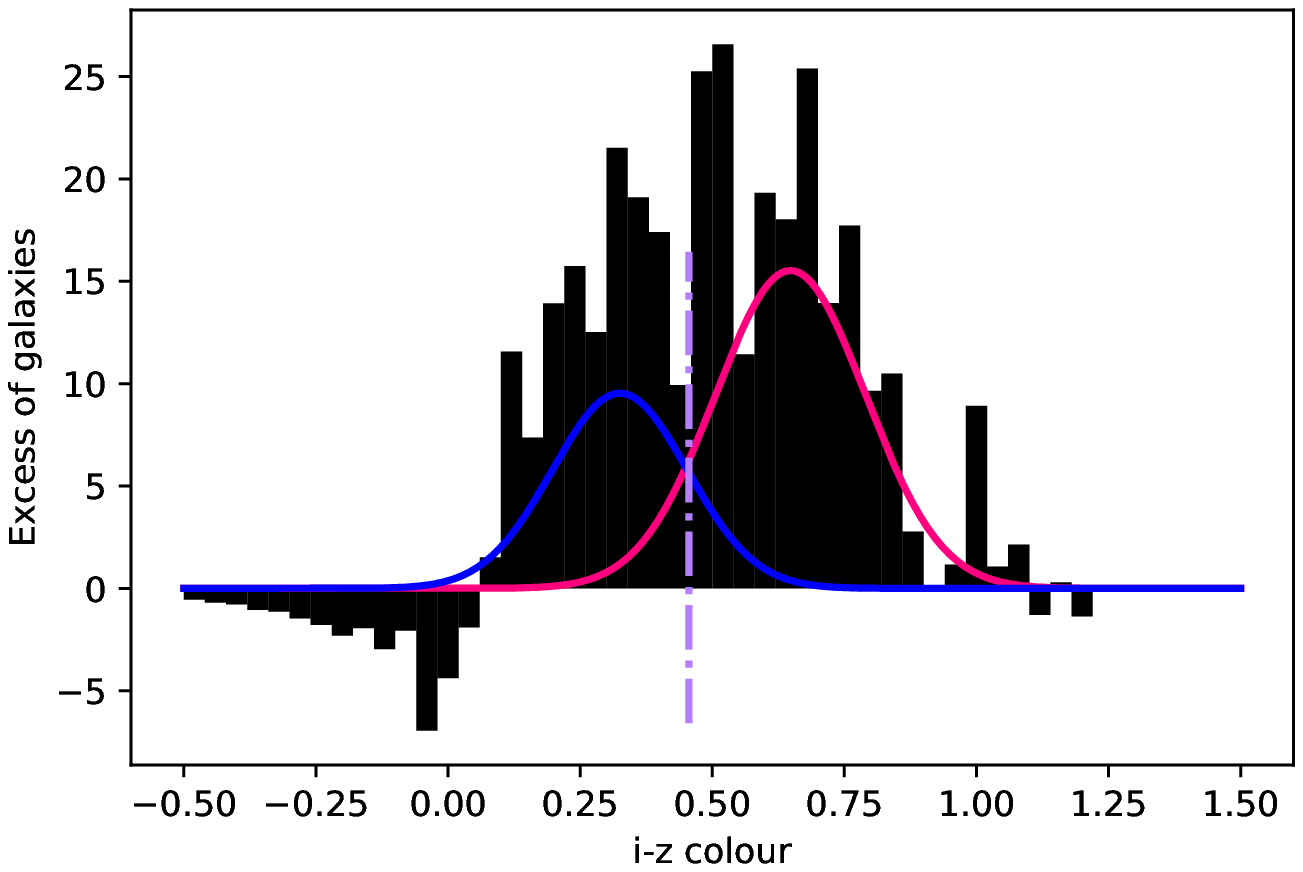}
\end{subfigure}
\hfill
\begin{subfigure}{0.3\textwidth}
\includegraphics[width=6cm]{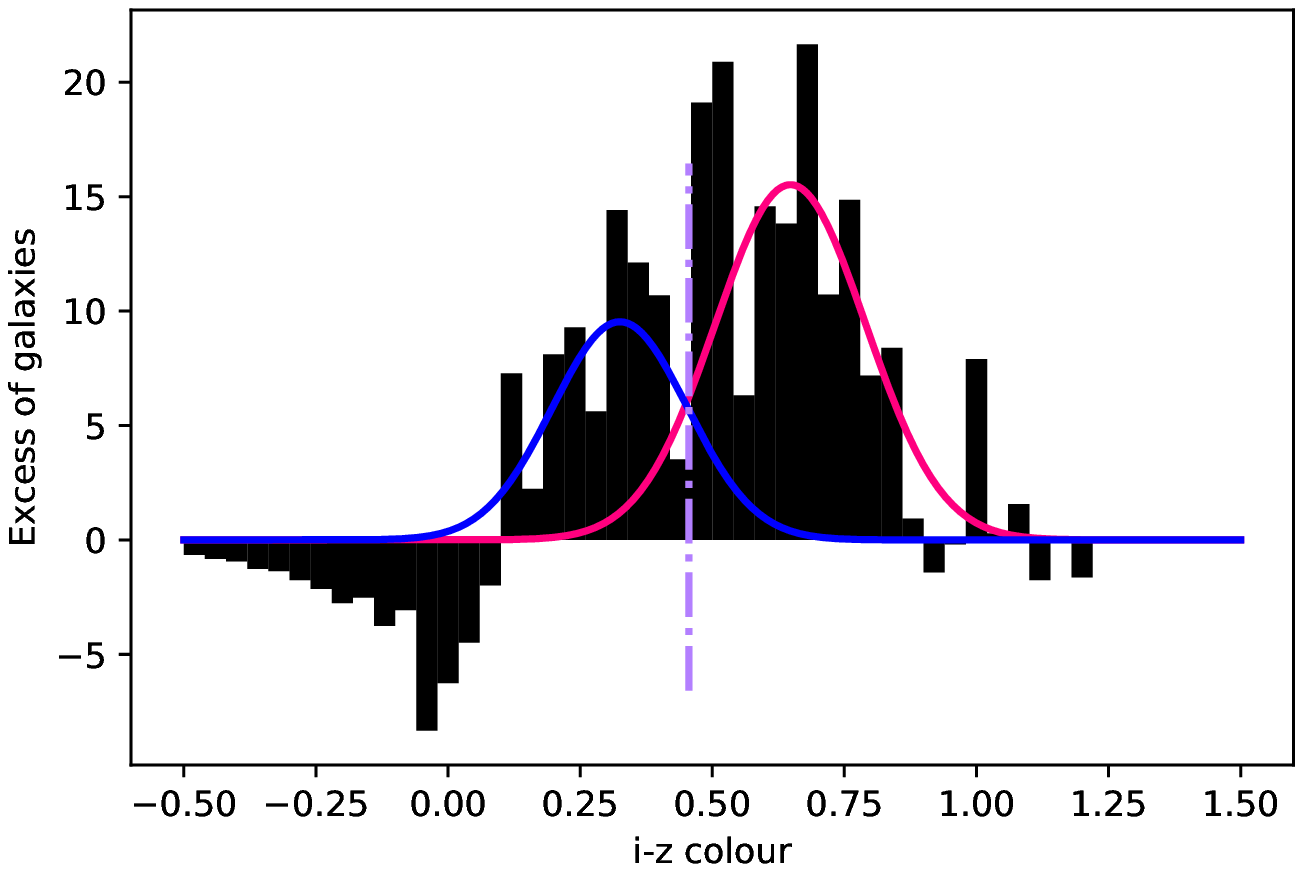}
\end{subfigure}
\caption{Illustration of two steps of the first method. Left panels: i-z colour histograms in two fields-of-view (candidate 13 and candidate 8) where the default background (pink line) is either too high or too low. The adjusted background is overplotted in green. Middle panels: Resulting colour distribution is the background is left unadjusted. For comparative purpose, the Gaussian models of the red sequence and the blue cloud are shown in pink and blue respectively, although they have been computed with an adjusted background. The mauve dash-dotted line is the \lq boundary\rq~
 used in method 2, 3, and 4. Right panels: Colour distribution once the background has been adjusted.}
\label{fig_quenching_bck}
\end{figure*}


\begin{figure*}
\begin{subfigure}{0.3\textwidth}
\includegraphics[width=6cm]{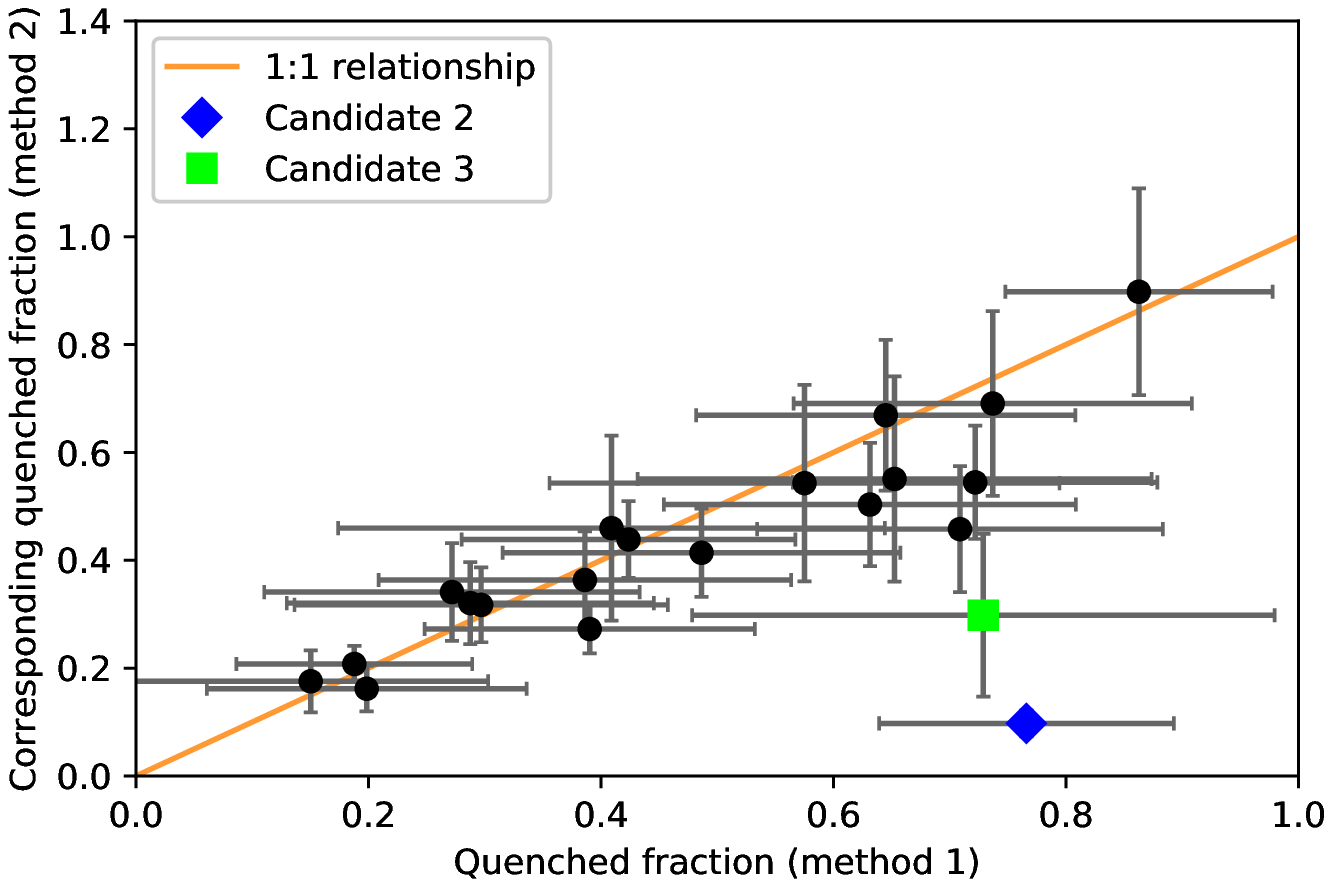}
\end{subfigure}
\hfill
\begin{subfigure}{0.3\textwidth}
\includegraphics[width=6cm]{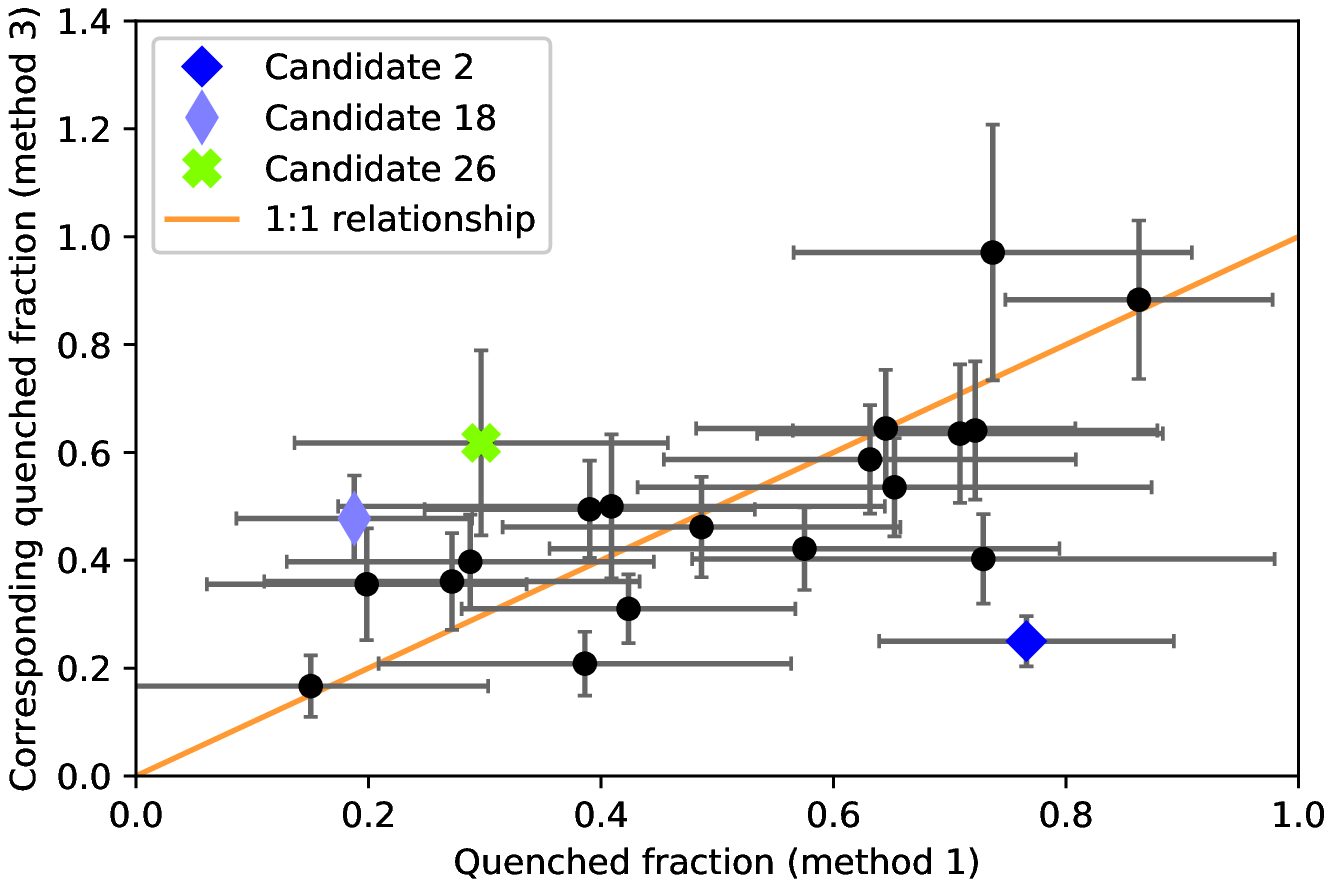}
\end{subfigure}
\hfill
\begin{subfigure}{0.3\textwidth}
\includegraphics[width=6cm]{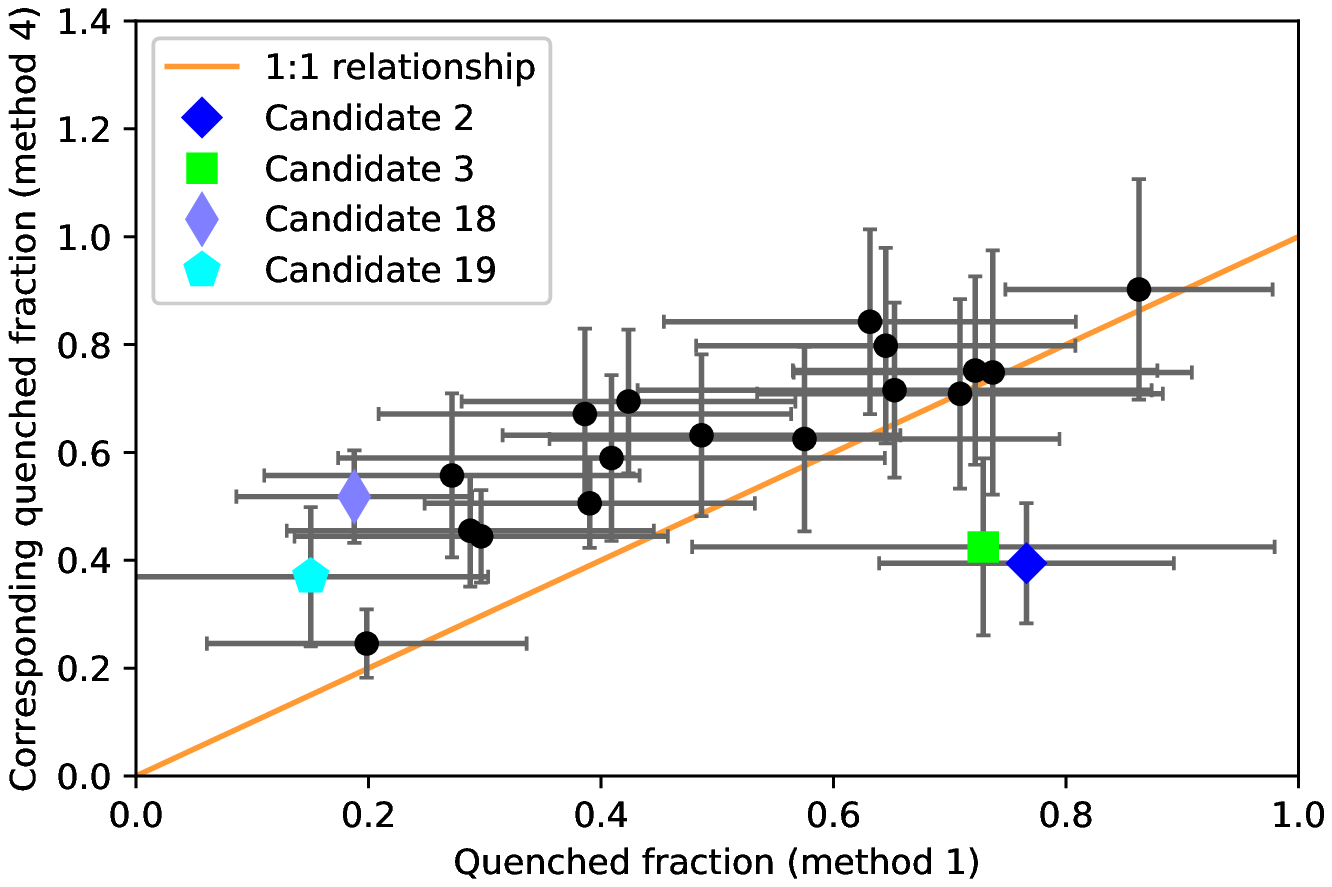}
\end{subfigure}
\caption{Comparison between the quenched fractions obtained by methods 1 and 2 (left panel), methods 1 and 3 (middle panel), or methods 1 and 4 (right panel). The error bars are based on the Poissonian uncertainties on the number of quenched galaxies in the cluster and on the number of galaxies in the considered field-of-view. The background uncertainty included is estimated to 5\% of the unscaled background subtraction, since we adjusted the subtracted background by steps of 10\%. Clusters that have quenching ratios above or below 1.5 times the standard deviation of $\frac{Q_{i}}{Q_{1}}$ are highlighted.}
\label{fig_comparison}
\end{figure*}


The first step consists of defining the appropriate area within which
to select field-of-view galaxies for each cluster. Some studies
\citep[e.g.][]{wetzel_galaxy_2012,pintos-castro_evolution_2019} analyse
the quenched/star-forming fraction up to several virial
radii. However, since several of our fields appear to contain secondary
overdensities, we choose to restrict our analysis closer to the cluster
centre, namely within a radius of 1.35 Mpc around the X-ray coordinates. A
radius of 1.35 Mpc corresponds to approximately 1.5 times the value of $r_{500}$ for a $1\times
10^{14} ~\mathrm{M_\odot}$ galaxy cluster \citep{chen_statistics_2007}
and to an angle of 2.76 arcminutes at $z=1$. We further select
galaxies with a photometric redshift between 0.60 and 2.04, both in
the cluster field and background catalogue, and sample the resulting
distributions into 0.04 wide bins in $i-z$ colour.

We then correct this field-of-view distribution using the same method
as applied in Eq. \ref{bck_subtraction}. Visual inspection
reveals that the resulting colour distribution is bimodal (see Fig.
\ref{fig_quenching_bck}). We model this bimodal colour distribution
using two Gaussian functions \---\ one representing the red sequence
and the other the blue cloud. In some fields we note that the
background correction is either too large or too small to yield to a
clear bimodal distribution, resulting in poor fits or even in a non-convergent fitting
algorithm (candidate 7). In such cases, we adjusted the background correction by up
to 20\% before determining whether the adjusted background results in
an improved fit. Figure \ref{fig_quenching_bck} presents the effect of an unadjusted background on the colour distribution on  two typical field. Although both the too high and too low cases are presented, the former concerns only four clusters. One of these clusters is confirmed (candidate 3) and two others (including the case presented in Fig. \ref{fig_quenching_bck}) are among the sample of other photometric studies \citep[see][]{olsen_galaxy_2007,wen_galaxy_2011,licitra_redgold_2016}. Thus, most of them are unlikely to be false detections. 

Three of the eight clusters with an increased background lie in overlaps between two VIDEO footprints and have thus, according to \citet{jarvis_vista_2013} longer exposure times. Moreover, two additional clusters are in the XMM1 field, which is part of the ultra-deep layer of HSC-SPP survey and thus possess 1 to 1.5 mag deeper photometry in the i and z band \citep[see][]{adams_rest-frame_2020}.


The quenched fraction is then computed as the integral of the
best-fitting red sequence Gaussian profile, divided by the sum of the
integrals of the blue cloud and red sequence Gaussian profiles.

Some clusters were removed from the analysis. We removed 
1, 4, 10, 21, and 24 
since there are indications of
more than one overdensity along the line-of-sight to each of these
X-ray sources. Candidate clusters at $z\geq1.4$ were also removed, resulting in a sample of 21 candidate clusters.


Next, we investigate whether the Gaussian modelling produces consistent average star formation histories, i.e. whether the red sequence colour can be described adequately by a simple stellar population model. We used Flexible Stellar Population Models
\citep[\textsc{FSPS}; see][]{conroy_propagation_2009,conroy_propagation_2010,zenodo_binding},
modelling the red sequence stellar population with an exponentially
decreasing star formation rate characterised by an $e$-folding
timescale $\tau$, displaying solar metallicity and no dust. We
generated a grid of models with two free parameters: we tested 100
formation redshifts between 4 and 16, and 100 $\tau$ values between 0.1
and 2.5 Gyr for a total of 10000 models. The model providing the best fit has a formation redshift of 16 with a characteristic time of 1.19
Gyr and a reduced $\chi^2_\nu=0.76$, corresponding to a rejection probability of 24\%. The red sequence displayed on the CMD plot in Fig. \ref{fig_candidates} is based on this model. The associated uncertainty is the standard deviation of the difference of red sequence colours, as predicted by the model, and the Gaussian modelling results. We also used this fit
to compute the evolutionary and K-correction associated with the
characteristic luminosity used in method 4.

\subsection{Other quenching methods}\label{ssec_other_methods}

The second method to identify quenched galaxies employs the same colour binning and
background subtraction procedure as employed by Gaussian method
described above. We specify the colour where the blue and red
Gaussian models are equal as the \textit{boundary} between the red
sequence and the blue cloud. We further restrict the colour interval
by rejecting any galaxy redder than 2.5 times the standard deviation
of the red Gaussian or bluer than 2.5 times the standard deviation of the
blue Gaussian. Where those boundaries fall within a colour bin we
perform a fractional calculation within the affected bins.

%
The third method does not include the area-corrected background
subtraction used in the first two methods but instead selects cluster
members based on their photometric redshift. Any galaxy within the
field-of-view and with a redshift consistent with the cluster mean redshift 
plus or minus 1.5 times its standard deviation (calculated in Sect. \ref{ssec_overdensities_confirmation}) is considered as a cluster
member. We then selected the red sequence and blue cloud members using
the colour boundaries defined in method two.

All the methods described above employ the 5$\sigma$ limit of the VIDEO catalogue in J and
$\mathrm{K_s}$ bands as a magnitude selection
threshold. The fourth quenching method is essentially the
same as method 2 yet with an evolving magnitude limit based upon the
values of $\mathrm{L^\ast}$ in J and $\mathrm{K_s}$ bands computed using our best-fitting \textsc{FSPS} model for the red
sequence, i.e. including both an evolutionary and $k$-correction (see Fig. \ref{fig_quench_z}). As in Sect. \ref{ssec_finding_clusters}, we assume a characteristic absolute magnitude of -22.26 at $z=0$ \citep[see also][]{cirasuolo_new_2010}.



A comparison between method 1 and the other methods is
presented in Fig. \ref{fig_comparison}. Each method generates similar quenched fractions compared to method 1. Within the limited variation of the approaches taken by each method, this indicates that the quenched fractions are robust from method to method. The few clusters for which the quenched fractions obtained by method 1 are more than 1.5 times the standard deviation above or below the mean are identified and highlighted. For now on, we will focus on method 4 results.

\subsection{Quenching results}\label{ssec_quenching_result}

\begin{figure*}
\centering
\begin{subfigure}{0.45\textwidth}
\includegraphics[width=8cm]{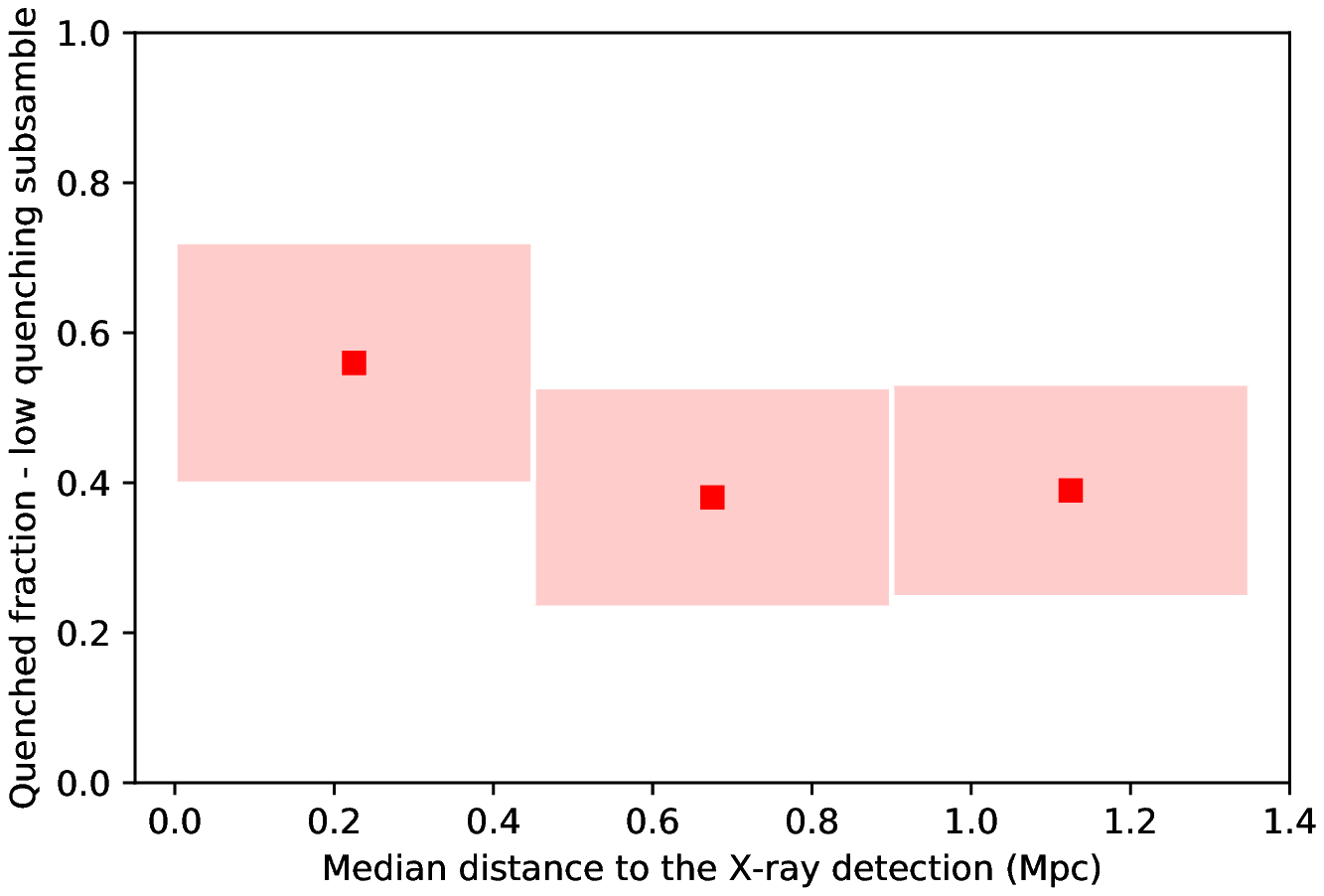}
\end{subfigure}
~\qquad
\begin{subfigure}{0.45\textwidth}
\includegraphics[width=8cm]{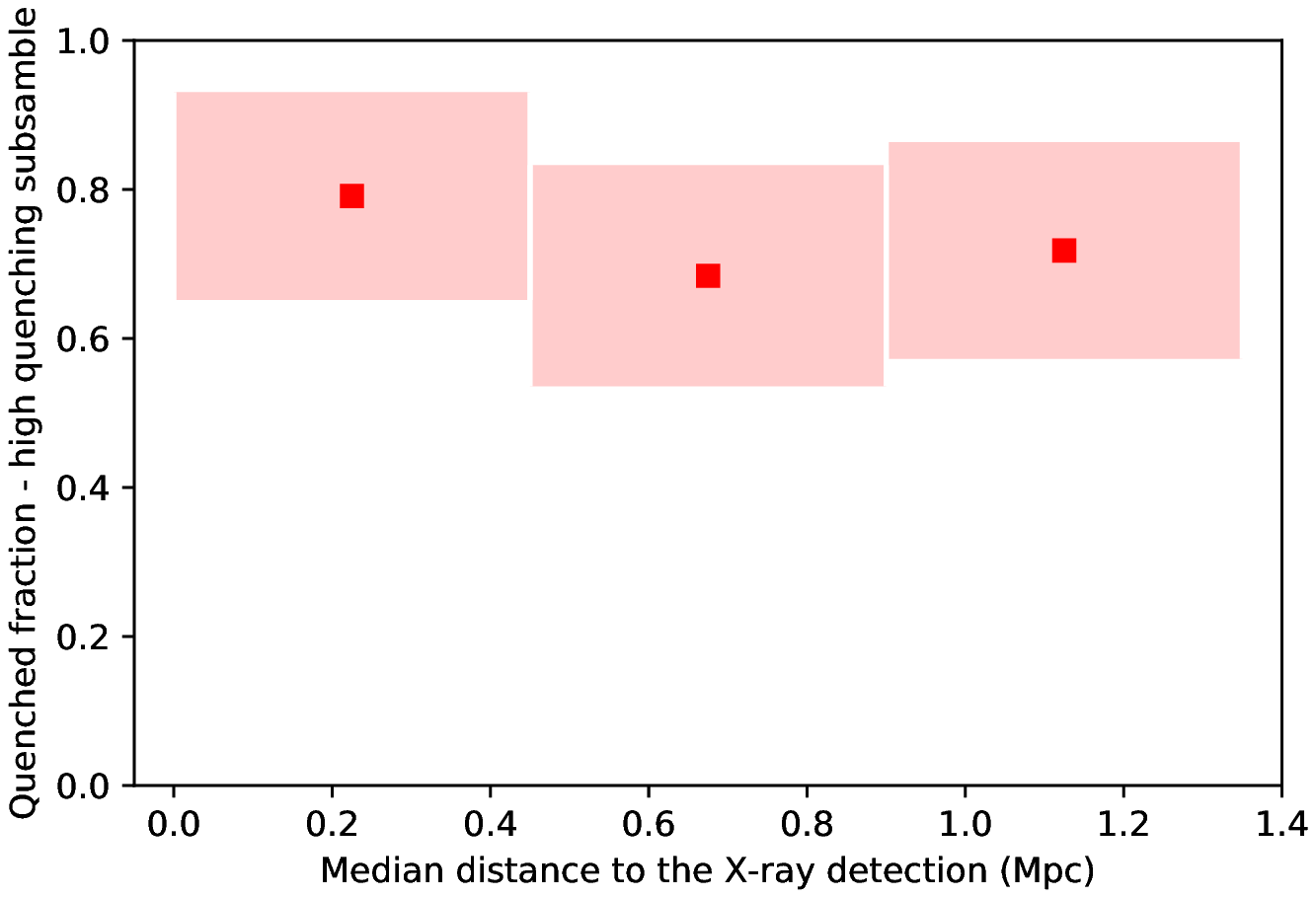}
\end{subfigure}
\\

\centering
\begin{subfigure}{0.45\textwidth}
\includegraphics[width=8cm]{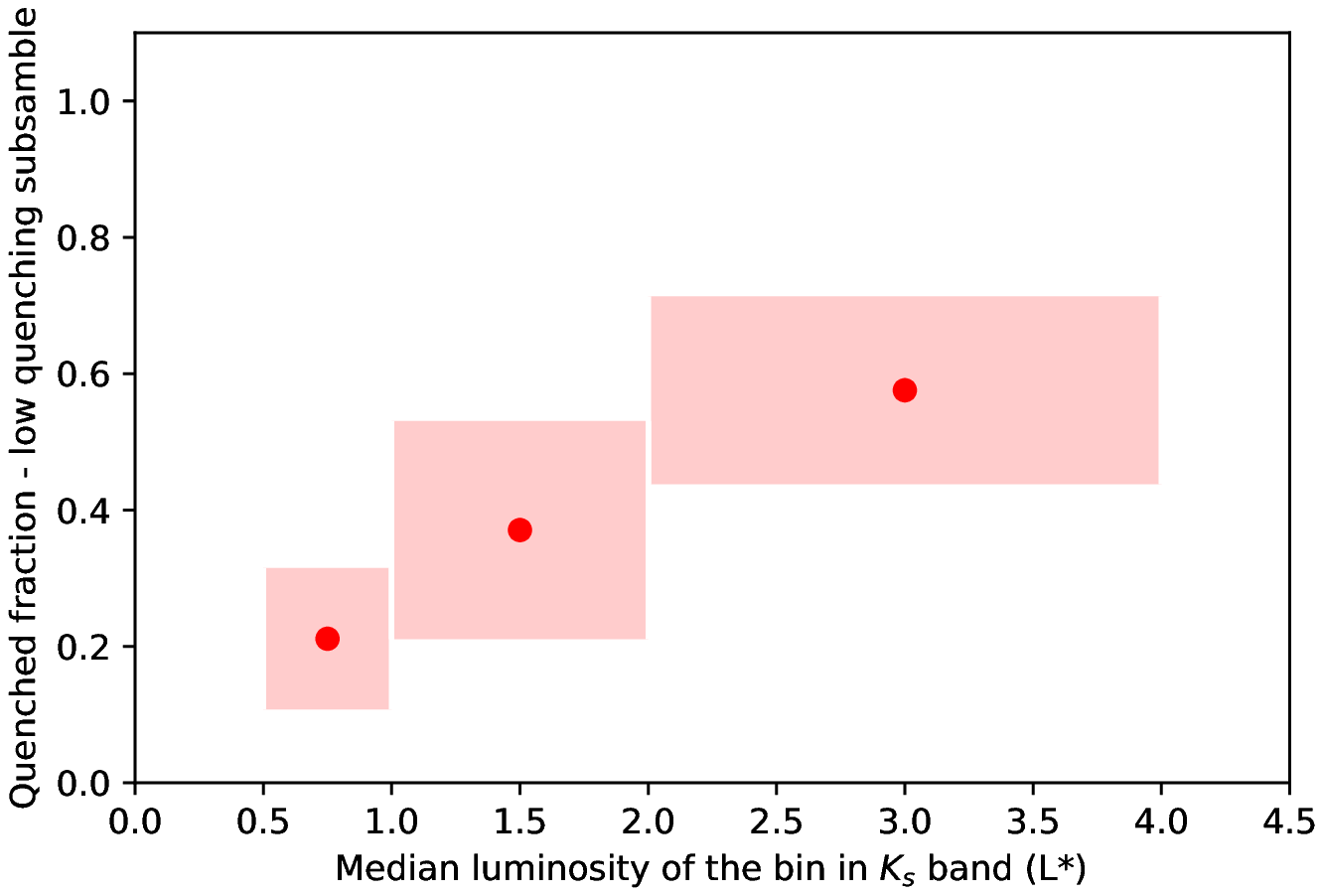}
\end{subfigure}
~\qquad
\begin{subfigure}{0.45\textwidth}
\includegraphics[width=8cm]{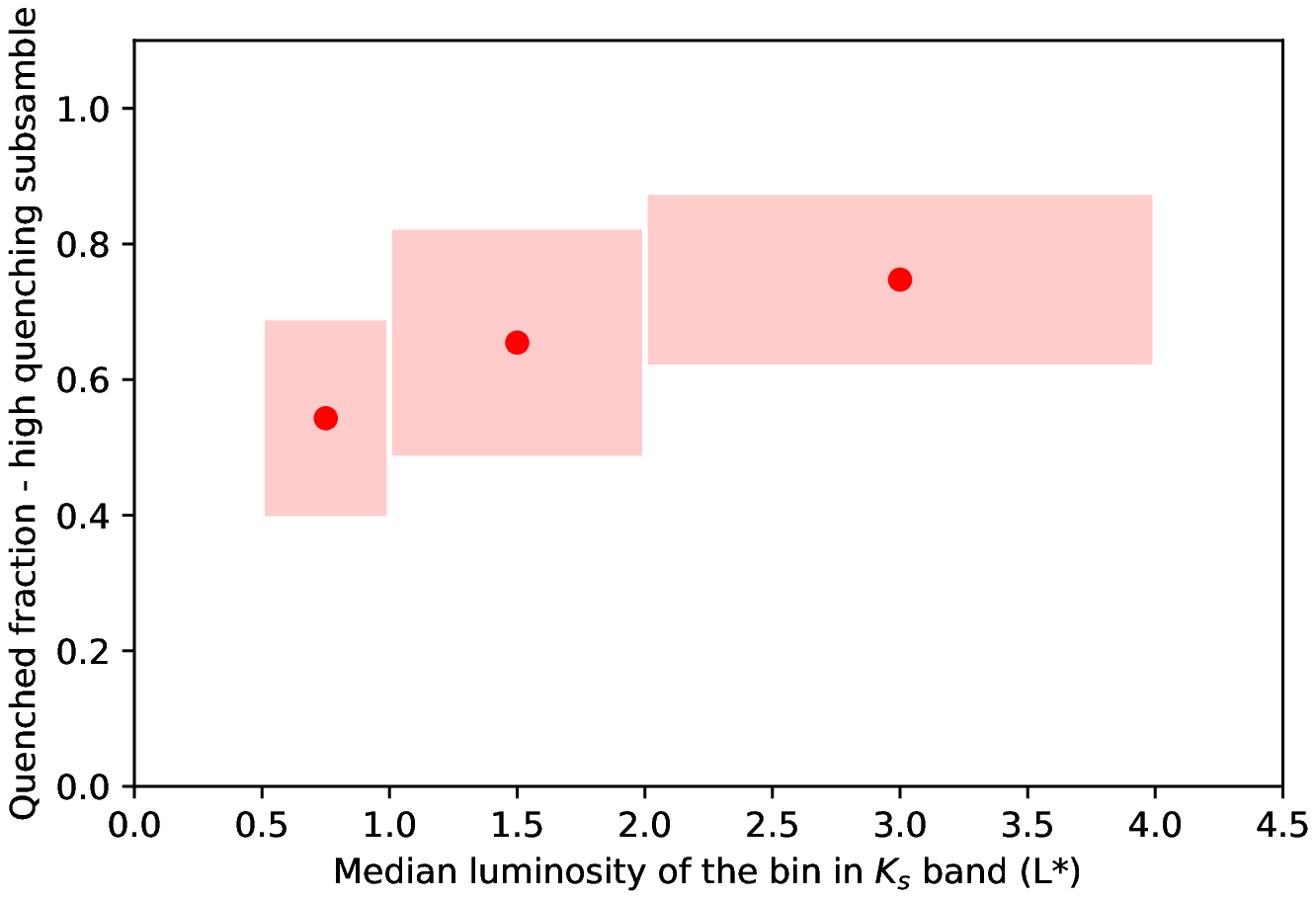}
\end{subfigure}
\caption{Top row: Mean quenched fraction for each distance bin, for the low quenching (left) and high quenching (right) candidate clusters. Symbols mark the mean quenched fraction while the shaded regions represent the bin size (x axis) and the standard deviation (y axis) of the quenched fraction. Bottom row: Mean quenched fraction for each luminosity bin according to method 2/4 (method 2 is equivalent to method 4 in this context). Luminosities are expressed in term of $\mathrm{L^\ast}$. We stress that $\mathrm{L^\ast}$ absolute magnitude changes with redshift to reflect the passive evolution of a quenched galaxy. In the interval $0.8 \leq z < 1.4$, the absolute magnitude $\mathrm{M^\ast}$ varies from -22.87 to -23.14 in the $\mathrm{K_s}$ band and the corresponding stellar masses from $1.45\times 10^{11}~\mathrm{M_\odot}$ to $1.47\times 10^{11}~\mathrm{M_\odot}$. Again, clusters are divided into a low quenching (left) and a high quenching (right) groups. These plots only include clusters below $z=1.4$, in order to mitigate the selection effects of the catalogue 5$\sigma$ limit.}

\label{fig_quenching_profile}
\end{figure*}

Figure \ref{fig_quench_z} shows the quenched fraction results of the
fourth method and indicates that there is a wide
variety of quenching levels within the interval $0.8\leq z\leq 1.2$. The
wide range of computed quenched fractions is nominally consistent with
the expectation expressed in Sect. \ref{sec_intro} that the X-ray selection of
galaxy clusters should be less biased to the properties of their
member galaxies than optical/IR overdensity methods. 

We next test whether, despite the range of quenched fractions, there
is a consistent variation of quenched fraction as a function of
cluster-centric distance within the sample of clusters as a whole. We
computed the quenched fraction 
for each cluster within three equal radial annuli out to 1.35 Mpc. We
then computed the mean cluster quenched fractions as a function of
radius into two groups based upon the total quenched fraction, i.e. we
stacked those clusters quenched at the level greater than the median
quenched fraction into one group and those quenched at less than the
median level into another.

We then investigated if the quenched fraction depends upon galaxy
luminosity. We computed the quenched fraction in three bins of $\mathrm{J/K_s}$
luminosities (0.5 to 1 $\mathrm{L^\ast}$, 1 to 2 $\mathrm{L^\ast}$, and 2
to 4 $\mathrm{L^\ast}$). 
Once again, the results for each cluster are
stacked on the basis of their overall quenching level. This test is
limited to $z<1.4$ because beyond this 0.5 $\mathrm{L^\ast}$ falls below
the 5$\sigma$ magnitude cut. 

%
%

Figure \ref{fig_quenching_profile} displays no significant trend between the mean quenched fraction and the cluster-centric radius, especially for the highly quenched half of the sample. We note however a slight increase of the quenched fraction toward the centre of the lowly quenched cluster, suggesting a weak dependence of the quenched fraction with the distance. The bottom panels of Fig. \ref{fig_quenching_profile} show that more luminous galaxies are more quenched, although the variation is less pronounced in highly quenched clusters.


\section{Brightest cluster galaxies}\label{sec_BCG_ssp}



To identify candidate BCGs within each galaxy cluster we noted
initially that 80\% of BCGs are located within 0.1 virial radii of the
cluster X-ray emission peak \citep{lin_k-band_2004}. They are usually
\---\ but interestingly not always \citep[e.g.][]{lange_brightest_2018} \---\ the most luminous galaxy in
the cluster. We therefore selected galaxies
within 225 kpc of the X-ray detection coordinates where this distance
corresponds to approximately 15\% of the virial radius of a $1\times
10^{14} ~\mathrm{M_\odot}$ cluster (we added an extra 5\% to account for possible offsets between the X-ray best-fit model centres and the X-ray emission peaks). 
We then applied a luminosity cut
to select galaxies brighter than 3 $\mathrm{L^\ast}$ in J and $\mathrm{K_s}$ bands (using the model computed in
Sect. \ref{ssec_first_method} to compute evolutionary effects and a
full $k$-correction). If the luminosity cut resulted in less than three
candidate BCGs, we extended the cut to 2.5 $\mathrm{L^\ast}$. If there still
remained less than three galaxies, we enlarged our search radius to 450
kpc. We implemented this three candidate limit because we noticed that
some fields contained faint stars misclassified as galaxies or quasars
in the VIDEO catalogue.

To identify spurious candidates we next performed a visual check of
the BCG candidates, paying special attention to the brightest and
second brightest candidates. We also flagged (but did not remove)
candidates with unreliable photometry, such as candidates within the
halo of a star or blended sources. We then created
$\mathrm{i}-\mathrm{z}$, $\mathrm{z}-\mathrm{J}$, and $\mathrm{J}-\mathrm{K_s}$ colour magnitude
diagrams of the central 225 kpc of each cluster to determine which candidate BCGs displayed colours consistent with the cluster redshift. 
This step also conferred the benefit that it
identified galaxies with magnitudes and colours comparable to the 2
most luminous candidates that might have been missed by the previous
cuts. We then selected up to three potential BCGs within each
cluster. In most clusters considered, a single BCG candidate is
prominent. In cases with more than one candidate BCG, we used the
$\mathrm{K_s}$ band luminosities to select the most likely BCG,
followed by projected cluster-centric distance if $\mathrm{K_s}$ band
magnitudes were similar. When possible, we used the CESAM spectroscopic database to check the redshift of our selected BCG candidates and make adjustments in the case of obvious inconsistencies with the estimated cluster redshift. 

%

We compared our BCG list to \citet{wen_galaxy_2011}, \citetalias{lavoie_xxl_2016}, and \citet[][hereafter \citetalias{ricci_xxl_2018}]{ricci_xxl_2018}, although we have respectively only 15, 1, and 5 clusters in common. Those three studies have slightly different BCG selection criteria: \citet{wen_galaxy_2011} use the i, $\mathrm{i^\ast}$, or the r band depending on the data available, while \citetalias{lavoie_xxl_2016} and \citetalias{ricci_xxl_2018} respectively use z and i' band photometry as their main selection criterion. Each uses larger search radii and, in the case of \citetalias{ricci_xxl_2018}, a stricter photometric redshift cut.

We found the same BCGs for 11 of the 15 shared candidates with \cite{wen_galaxy_2011}. When our selection disagrees, \cite{wen_galaxy_2011} BCG is either fainter than our selected candidate (candidates 13 and 18), possesses an incompatible spectroscopic redshift (candidate 12), or is an obvious foreground galaxy (candidate 1). For candidate 20, the only common candidate with \citetalias[][]{lavoie_xxl_2016}, our BCG selection agrees. In the case of clusters in common with \citetalias{ricci_xxl_2018}, 4 of the 5 common clusters have matching BCGs (candidates 3, 8, 20, and 21). For candidate 2, they selected our second-best BCG candidate, which, unlike our chosen candidate, possesses a spectroscopic redshift. However, \citetalias{ricci_xxl_2018} preferred candidate is fainter than our chosen candidate in i, z, J, and $\mathrm{K_s}$ band.

As a final step, we employed \textsc{FSPS} to generate a suite of stellar
population models which we compared to the $\mathrm{J-K_s}$ colours of
our total sample of BCGs as a function of redshift. We tested 100
different formation redshifts between 3 and 16, in combination with
100 different metallicities between 0.1 and 5 $\mathrm{Z_\odot}$. We
assumed a instantaneous starburst, no dust extinction, and a Salpeter
initial mass function \citep{salpeter_luminosity_1955}. We then tested
the effect of including a third free parameter, using ten different
formation redshifts and 50 metallicities within the same intervals as
above. We first tested the effect of dust, with 25 dust obscuration
percentages covering 0 to 99.3\%, calculated as a power law with a
-0.7 index. We also investigated the effect of permitting a dust-free
exponentially decreasing star formation rate (SFR) employing 25 $\tau$
values between 0.004 and 1 Gyr.


\begin{figure*}
\centering
\begin{subfigure}{0.3\textwidth}
\includegraphics[width=6cm]{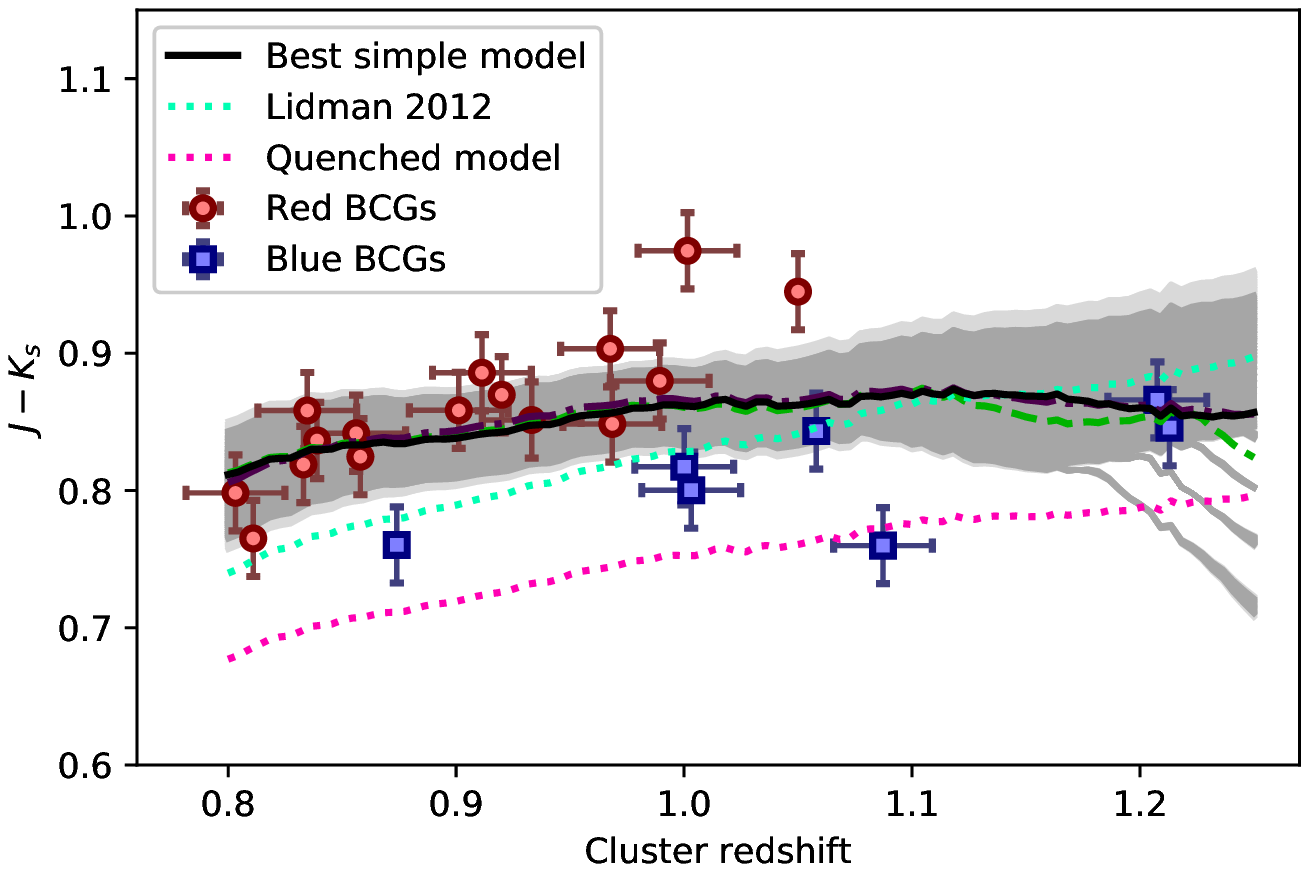}
\end{subfigure}
\hfill
\begin{subfigure}{0.3\textwidth}
\includegraphics[width=6cm]{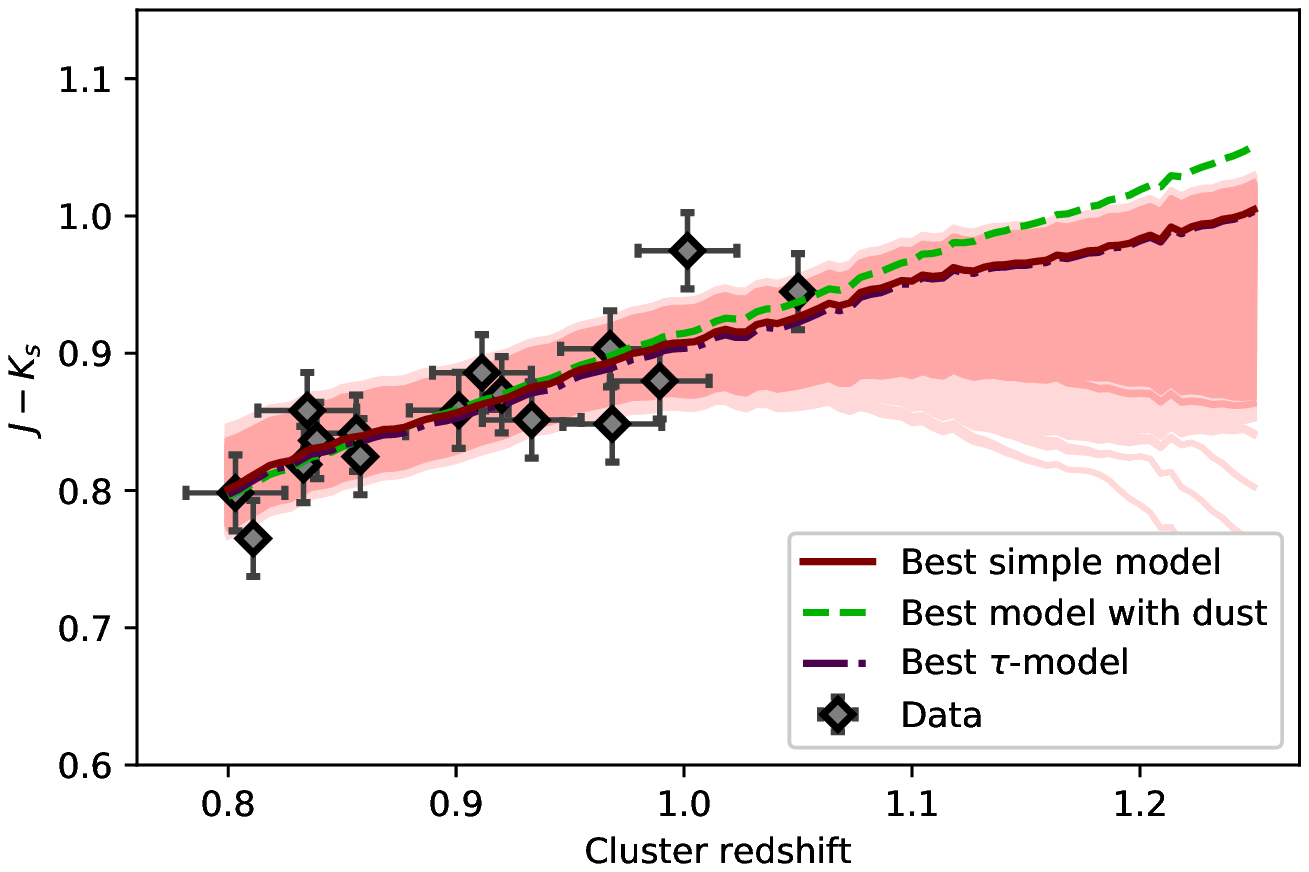}
\end{subfigure}
\hfill
\begin{subfigure}{0.3\textwidth}
\includegraphics[width=6cm]{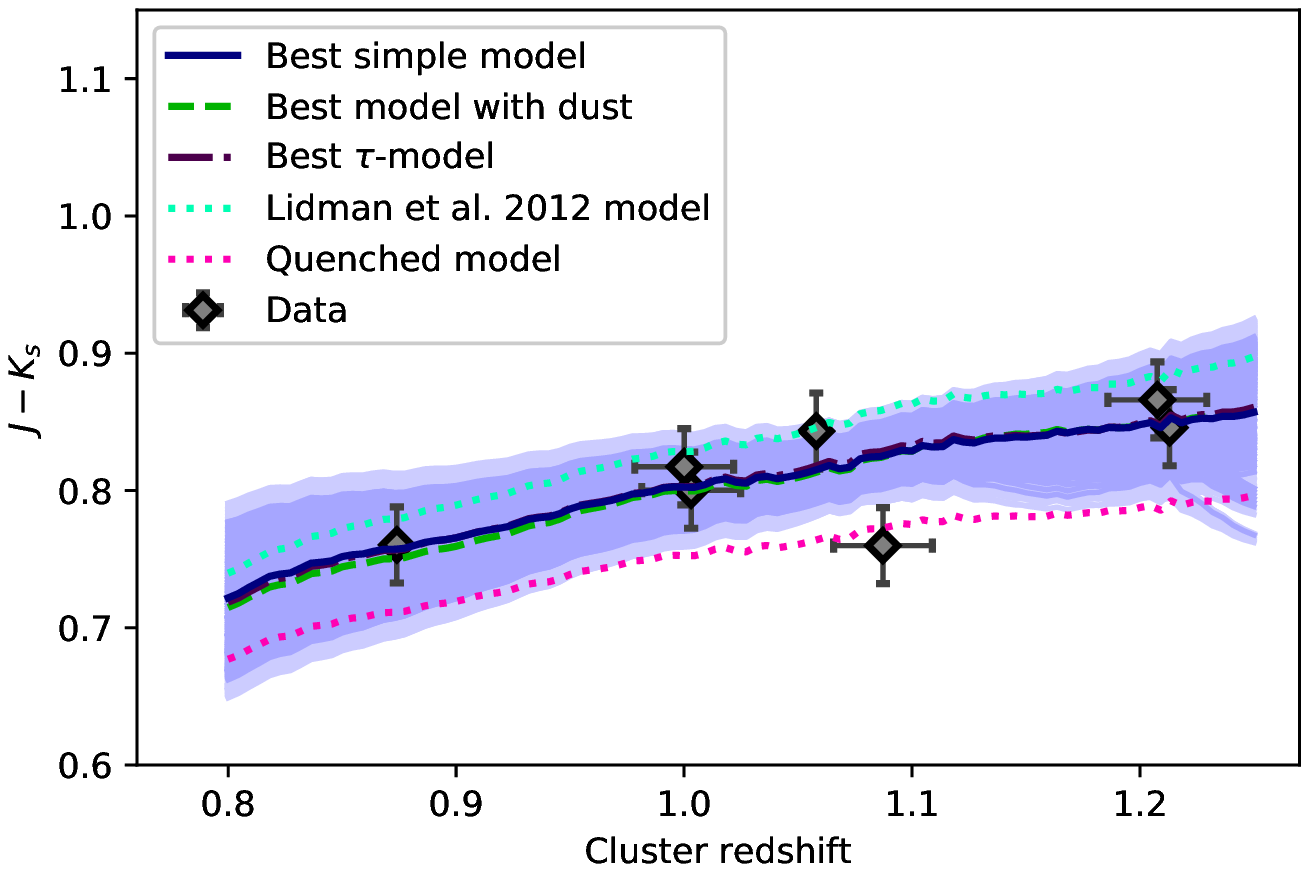}
\end{subfigure}
\caption{
Best fits for different subsamples. In each case, three fits have been
tested: a simple model, in which the metallicity and the formation
redshift (assuming an instantaneous starburst) are allowed to vary,
and two other models where an additional parameter is allowed to vary:
dust content in one case (green dashed line) or the characteristic
timescale of the star formation (assuming a exponentially decreasing
star formation rate, instead of an instantaneous starburst; purple
dash-dotted line). For comparative purposes, we display also
\citet{lidman_evidence_2012} best fit model and our red sequence fit (respectively the dotted cyan and dotted magenta lines; see Sect. \ref{ssec_first_method}) on two panels. The shaded region correspond to the 95\% confidence region for
the simplest model tested and the darker zone to the 68\% confidence
region. Left panel: Best fits for all BCGs at $0.8\leq z\leq 1.2$, except candidate 10 and 3
BCGs with known photometric problems. Red galaxies are part of the red
sample, and blue galaxies are part of the blue sample. Middle panel:
Best fits for the red BCGs subsample. Right panel: Best fits for the
blue subsample.}
\label{fig_best_fits_BCG}
\end{figure*}

We limited the stellar population modelling to BCGs drawn from clusters
in the interval $0.8 \leq z< 1.4$ and also removed BCGs with unreliable
photometry. We also excluded candidate 10 because no suitable BCG candidate
was identified. We thus performed our fits with a sample of 23 BCGs. 
The models described above generates fits characterised by reduced $\chi^2$ (hereafter noted $\chi^2_\nu$) approximately equal to 3.5. The left panel of Fig. \ref{fig_best_fits_BCG} presents these best
fitting models and their associated confidence intervals. An examination
of that Fig. indicates that the single model fits might well be
averaging two groups of BCGs, namely a group with red $\mathrm{J-K_s}$
colours and a comparatively bluer group. We will refer to these
groupings as red and blue BCGs. None of the BCGs in the blue sample is bluer than the red sequence model, displayed as a magenta dotted line in Fig. \ref{fig_best_fits_BCG}. This discrepancy is discussed further in Sect. \ref{ssec_BCG_bimodal}.

\begin{figure*}
\begin{subfigure}{0.3\textwidth}
\includegraphics[width=6cm]{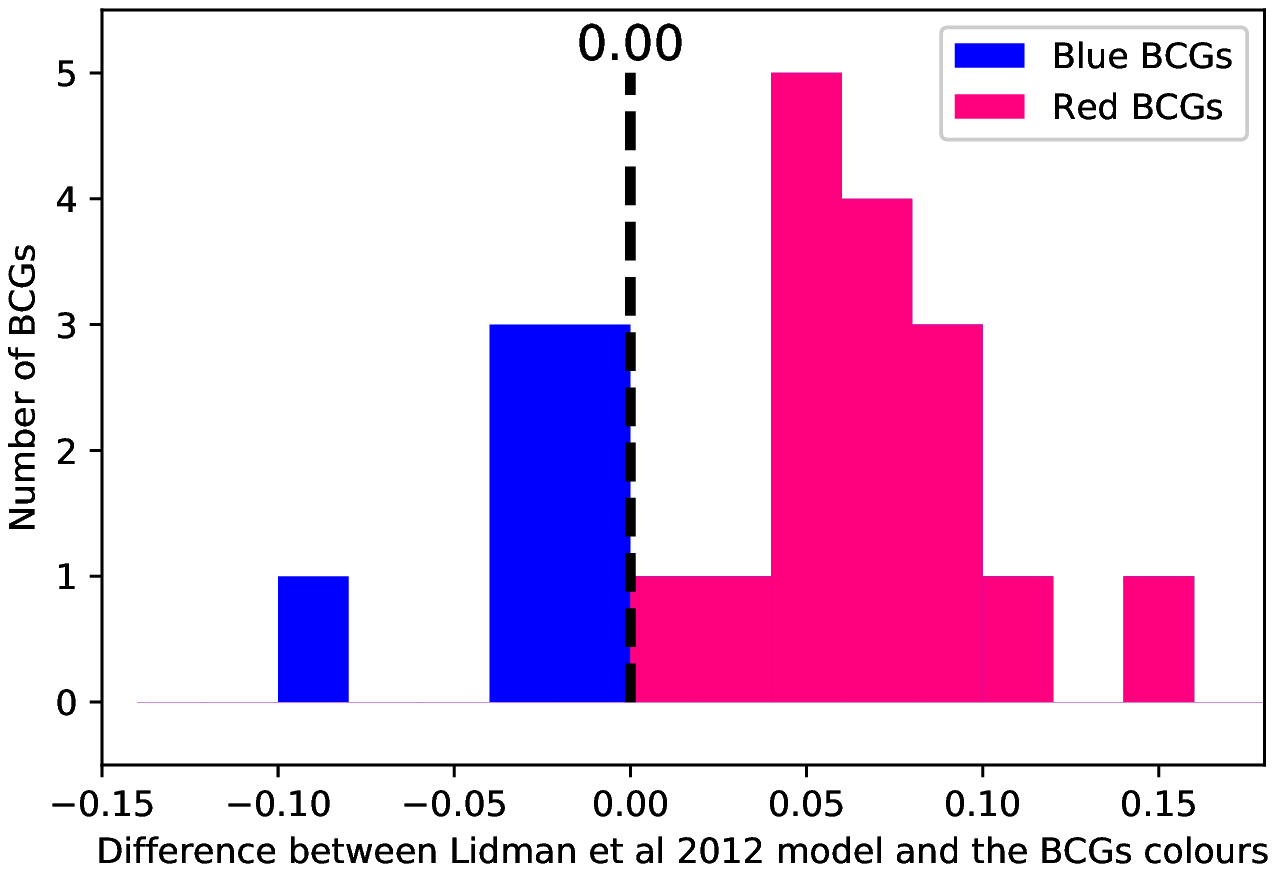}
\end{subfigure}
\hfill
\begin{subfigure}{0.3\textwidth}
\includegraphics[width=6cm]{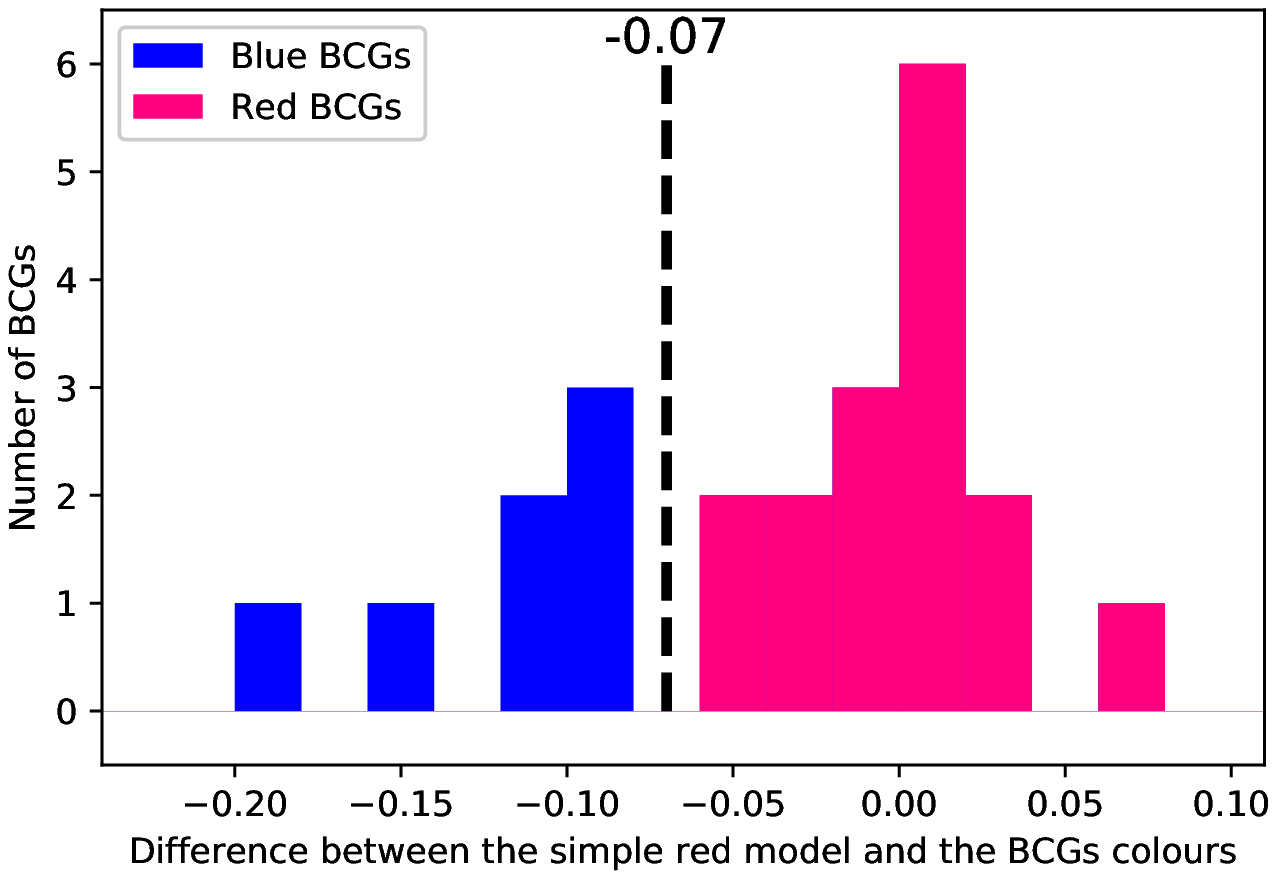}
\end{subfigure}
\hfill
\begin{subfigure}{0.3\textwidth}
\includegraphics[width=6cm]{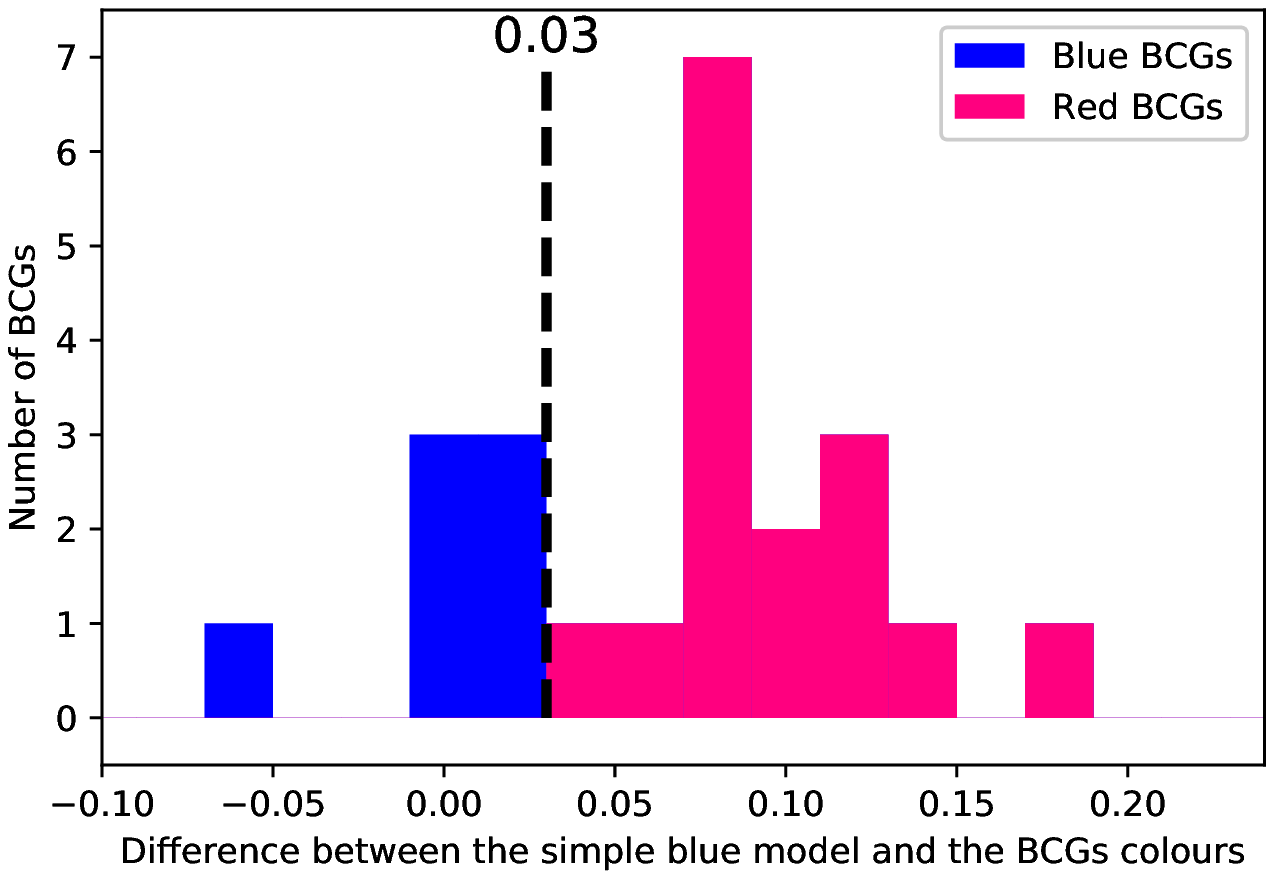}
\end{subfigure}
\caption{Division of the blue and red BCGs based on their colour
 difference with a reference model, the dashed line representing the
 limit between the two groups. Left: Division based on
 \citet{lidman_evidence_2012} model. Middle: Assessment of our sample
 bimodality based on the red BCGs best fit. Right: Same, based on the
 blue BCGs best fit}
\label{fig_division_BCGs}
\end{figure*}

We find that none of the fitted stellar population models appears to
capture the overall slope of colour versus redshift for the combined
BCG sample. However, we noted with interest that the stellar
population model of \citet{lidman_evidence_2012} appears to
provide an effective method to segregate the red and blue BCGs.
We therefore segregated the red and blue BCGs employing the colour
difference with respect to the \citet{lidman_evidence_2012} model
colours as a function of redshift (see Fig. \ref{fig_division_BCGs},
left panel). Having split the BCGs into red and blue on the basis of
this criterion we then applied the \textsc{FSPS} fitting procedure described
above to both populations and performed a final check of the BCG
colour evolution versus redshift with respect to the best-fitting red
and blue models (Fig. \ref{fig_division_BCGs}, centre and right
panels). All three approaches generate the same division between red
and blue BCGs.


\begin{table*}
\caption{Summary of the stellar population fits obtained for the red BCGs} 
\label{table_fits_red}
\centering
\begin{tabular}{l c c c c c}
\hline
Fit & $\chi^2_\nu$ & $z_0$ & Dust absorption$^a$ & Metallicity & $\tau^b$ \\ 
 & & & (\%) & ($\mathrm{Z_\odot}$) & (Gyr) \\
\hline
No dust, varying $z_0$ and Z & 1.140 & 14.214 & 0 & 1.44 & 0 \\
Varying $z_0$, Z, and dust & 1.155 & 11.030 & 61.28 & 0.40 & 0 \\
No dust, varying $z_0$, Z, and $\tau$ & 1.250 & 11.030 & 0 & 1.40 & 0.010 \\

\hline
\multicolumn{6}{l}{$^a$ Fraction of starlight absorbed by dust.}\\
\multicolumn{6}{l}{$^b$ Characteristic timescale for an exponentially decreasing star formation rate.}\\
\end{tabular}
\end{table*}

\begin{table*}
\caption{Summary of the stellar population fits obtained for the blue BCGs} 
\label{table_fits_blue}
\centering
\begin{tabular}{l c c c c c}
\hline
Fit & $\chi^2_\nu$ & $z_0$ & Dust absorption$^a$ & Metallicity & $\tau^b$ \\ 
 & & & (\%) & ($\mathrm{Z_\odot}$) & (Gyr) \\
\hline
No dust, varying $z_0$ and Z & 1.497 & 4.426 & 0 & 1.24 & 0 \\
Varying $z_0$, Z, and dust & 1.920 & 3.613 & 6.67 & 1.10 & 0 \\
No dust, varying $z_0$, Z, and $\tau$ & 1.954 & 5.241 & 0 & 1.20 & 0.317\\

\hline
\multicolumn{6}{l}{$^a$ Fraction of starlight absorbed by dust.}\\
\multicolumn{6}{l}{$^b$ Characteristic timescale for an exponentially decreasing star formation rate.}\\

\end{tabular}
\end{table*}

Treating the red and blue BCGs as separate populations provides a
significant improvement to the quality of the stellar population fits
(see Fig. \ref{fig_best_fits_BCG},
middle and right panels).

Tables \ref{table_fits_red} and \ref{table_fits_blue} give the $\chi^2_\nu$ and best parameter fits for every model tested. The uncertainties are based on the 1$\sigma$ photometry errors.
Despite relatively good $\chi^2_\nu$ (1.140 for the simplest model), none of the red BCG fits seem able to capture the slope of the J-$\mathrm{K_s}$ colours of the high-redshift half of the the sample. 
Both the simple and dusty models perform similarly, but the plausibility of the latter seems questionable, since it features a starlight dust absorption percentage of 61\%. 

For the blue BCGs, there is no visible slope discrepancy between the data and the model. However, the $\chi^2_\nu$ are larger than in the red BCGs case, probably because none of the applied models include a term for intrinsic scatter. 
In fact, the 95\% confidence interval of the simplest model (which has also the minimum $\chi^2_\nu$) is barely spawning the variety of colours observed in the blue BCGs. This may indicates that the blue BCG population is too diverse to be represented by a single stellar population model.

\begin{table*}
\caption{Positions and stellar masses of the red BCGs} 
\label{table_mass_red}
\centering
\begin{tabular}{c l c c c}
\hline
\# & Candidate name & BCG RA \& Dec & z & BCG stellar masses \\ 
& & (degrees) & & ($10^{11}~\mathrm{M_\odot}$)\\
\hline
1 & 3XLSS J022222.9-044043 & 35.5854 -4.6857 & 0.80$^a$ & $2.6\pm 0.4$\\
2 & XLSSC 184 & 35.3169 -4.2079 & 0.81 & $3.5\pm 0.6$\\
3 & XLSSC 071 & 35.6420 -4.9655 & 0.83 & $7\pm 1$\\
4 & 3XLSS J022432.9-044742 & 36.1398 -4.7940 & 0.83$^a$ & $4.6\pm 0.8$\\
5 & 3XLSS J022135.2-051811 & 35.3985 -5.3052 & 0.84 & $5.6\pm 0.9$\\
6 & XLSSU J021947.4-050841 & 34.9458 -5.1399 & 0.86$^a$ & $6\pm 1$\\
7 & XLSSC 15 & 35.9273 -5.0336 & 0.86 & $8\pm 1$\\
9 & 3XLSS J022557.1-042845 & 36.4802 -4.4804 & 0.90$^a$ & $3.8\pm 0.7$\\
11 & 3XLSS J021945.5-044831 & 34.9362 -4.8124 & 0.91$^a$ & $5\pm 1$\\
12 & XLSSU J022530.3-042544 & 36.3738 -4.4295 & 0.92 & $4.1\pm 0.8$\\ 
13 & 3XLSS J022804.6-045351 & 37.0203 -4.9056 & 0.93$^a$ & $7\pm 1$\\ 
14 & XLSSU J022051.0-050958 & 35.2108 -5.1620 & 0.97$^a$ & $8\pm 2$\\
15 & 3XLSS J022103.0-045524 & 35.2634 -4.9222 & 0.97$^a$ & $3.4\pm 0.7$\\
16 & 3XLSS J022739.0-045830 & 36.9073 -4.9698 & 0.99$^a$ & $3.9\pm 0.8$\\
18 & XLSSC 044 & 36.1345 -4.2287 & 1.00$^a$ & $6\pm 1$\\ 
20 & XLSSC 029 & 36.0175 -4.2240 & 1.05 & $6\pm 1$\\
\hline
\multicolumn{5}{l}{$^a$ Photometric redshift. The uncertainties are $\pm 0.02$.}
\end{tabular}
\end{table*}
%

\begin{table*}
\caption{Positions and stellar masses of the blue BCGs} 
\label{table_mass_blue}
\centering
\begin{tabular}{c l c c c }
\hline
\# & Candidate name & BCG RA \& Dec & z & BCG stellar masses \\ 
& & (degrees) & & ($10^{11}~\mathrm{M_\odot}$)\\
\hline

8  & XLSSC 064 & 34.6335 -5.0165 & 0.87 & $2.6\pm 0.5$\\
17 & 3XLSS J022044.7-041713 & 35.1953 -4.2900 & 1.00$^a$ & $2.5\pm 0.5$\\
19 & XLSSC 124 & 34.4272 -4.8658 & 1.00$^a$ & $3.7\pm 0.8$\\
21 & XLSSC 005 & 36.7872 -4.2988 & 1.06 & $2.0\pm 0.4$\\
23 & 3XLSS J022027.0-043538 & 35.1173 -4.6041 & 1.09$^a$ & $3.7\pm 0.8$\\
25 & XLSSC 141 & 34.3478 -4.6696 & 1.21$^a$ & $1.6\pm 0.4$\\
26 & XLSSC 046 & 35.7636 -4.6043 & 1.21 & $3.7\pm 0.9$\\
\hline
\multicolumn{5}{l}{$^a$ Photometric redshift. The uncertainties are $\pm 0.02$.}
\end{tabular}
\end{table*}

%

Nevertheless, we computed the BCG stellar masses for the red and the blue BCGs, using the parameters from the simple model (i.e the instantaneous starburst model with no dust). 
We adopt a similar approach than \citet{lidman_evidence_2012}, computing the mass as the quotient between the observed and the modelled flux density in $\mathrm{K_s}$ band, correcting with the model stellar mass. To estimate the uncertainties, we computed the $\mathrm{K_s}$ band fluxes of every model enclosed in the 68\% $\chi^2$ confidence region. We then take their standard deviation as the uncertainty on the flux and propagated this error to the mass.

Results are presented in Tables \ref{table_mass_red} and \ref{table_mass_blue}. The mean masses are $5\pm 1\times 10^{11}~\mathrm{M_\odot}$ and $2.8\pm 0.6\times 10^{11}~\mathrm{M_\odot}$ for the red and blue BCGs respectively.


\begin{figure}[t]
\includegraphics[width=8cm]{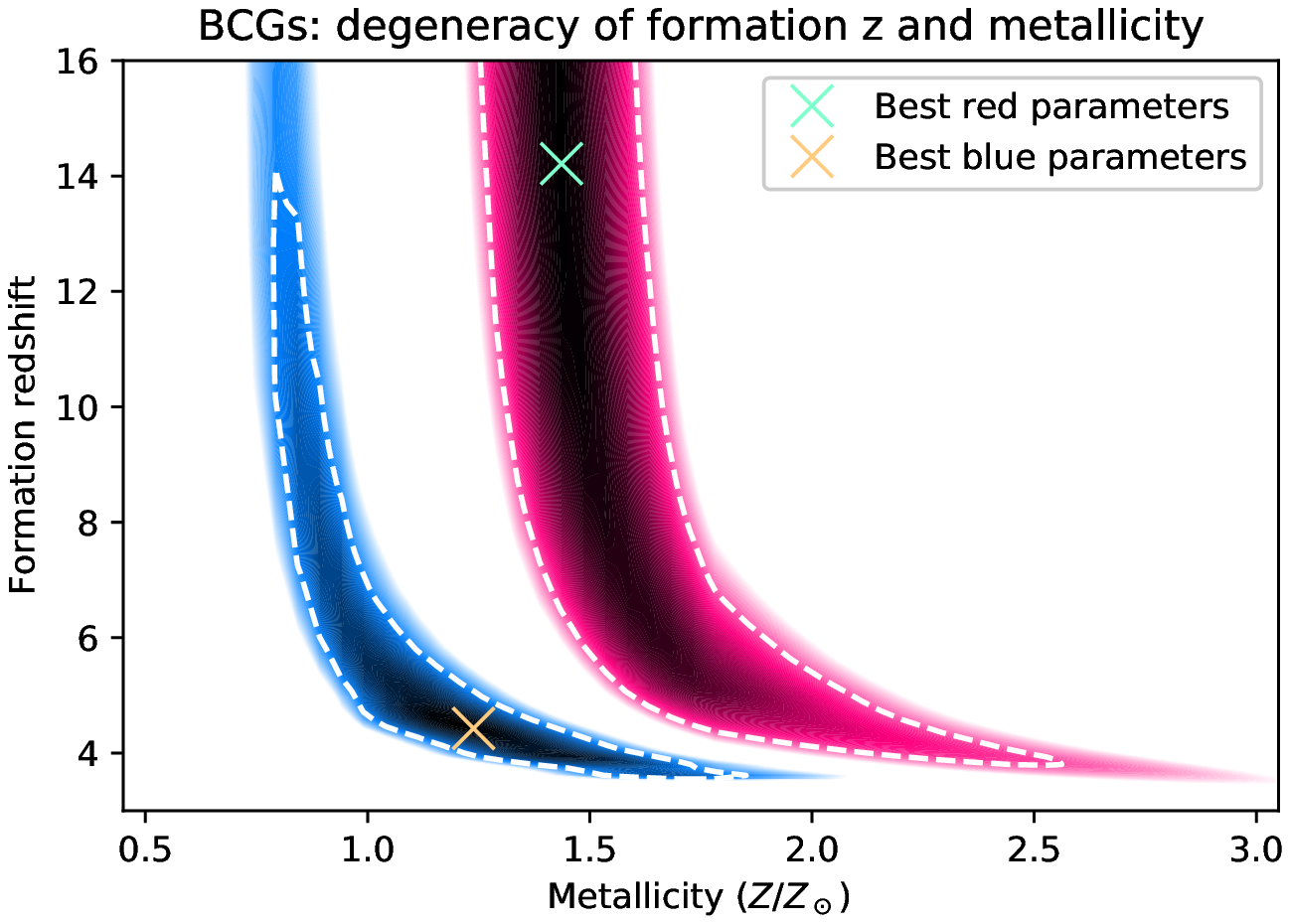}
\caption{Contour plots for the red and blue BCG subsamples showing the
$\chi^2_\nu$ value as a function of the two varying parameters
 (formation redshift and metallicity). Dashed white curves correspond to the 68\%
 confidence limit. The coloured regions correspond to the 98\%
 confidence interval. The tested metallicity range is $Z=0.1~Z_\odot$ to $Z=5~Z_\odot$ but is restricted here to
 $Z=0.5~Z_\odot$ to $Z=3.5~Z_\odot$. The tested formation redshift 
 range is $z_{form}=3$ to $z_{form}=16$. The cross displays 
 the location of the best fit parameters.}
 
\label{fig_degeneracy_BCG}
\end{figure}

\section{Discussion}\label{sec_discuss}

\subsection{Cluster quenched fractions}\label{ssec_quenching_level}

Figure \ref{fig_comparison} presents a wide variety of quenched
fractions, with no clear trend, despite that, at higher
redshifts, only the most luminous (massive) galaxies are visible. The
fourth method of computing the quenched fraction incorporates a
luminosity cut specifically intended to mitigate this bias. However,
although this method generates slightly higher quenched fractions (except for the two lowest redshift clusters in our sample; see Fig. \ref{fig_comparison}), it does not otherwise affect the apparent diversity of quenched fractions. There is thus no evidence of evolution with redshift in Fig. \ref{fig_quench_z}.


The results presented in Sect. \ref{ssec_quenching_result} demonstrate a link between quenched fraction and the $\mathrm{K_s}$ band
luminosity, although the link seems weaker for the highly quenched half of our sample. $\mathrm{K_s}$ band luminosities are a proxy for galaxy stellar masses (assuming dust extinction is negligible). Such a link has been observed before \citep[e.g.][]{peng_mass_2010,peng_mass_2012,muzzin_gemini_2012,balogh_evidence_2016,kawinwanichakij_effect_2017,jian_first_2018,pintos-castro_evolution_2019}, although whether this is due to mass-dependent galaxy evolution and/or to environment remains an open question.


We observe no significant evidence that the quenching fraction varies with the cluster-centric distance, although the quenched fraction is slightly higher toward the centre of the lowly quenched clusters in our sample \---\ an observation in agreement with previous studies \citep[e.g.][]{muzzin_gemini_2012,pintos-castro_evolution_2019}. The more quenched half of our sample possesses levels of quenching above 70\% in all bins, which might be because only galaxies above $\mathrm{L^\ast}$ are included in this profile and that bright galaxies tend to be slightly more quenched at all radii.

\subsection{A bimodal Brightest Cluster Galaxy population?}\label{ssec_BCG_bimodal}

The $J-K_s$ colours of the candidate BCGs in the distant cluster
sample are not consistent with the properties of a single,
passively-evolving stellar population. There is tentative evidence
for a bimodal distribution of BCG colours. Separating the sample
employing the fiducial stellar population model of \citet{lidman_evidence_2012}, we
find that the colours of the redder BCGs in our sample are consistent
with an older, instantaneous burst of metal-rich stars whereas the
bluer BCG colours are more diverse and possibly related to a younger
component in the stellar population.

Despite this variety of colours, only one of the BCGs is bluer than the red sequence model computed in Sect. \ref{ssec_first_method} (see also Fig. \ref{fig_best_fits_BCG}). This might indicate that BCG stellar populations have different properties (such as metallicity) than the average quenched galaxies in cluster, as suggested by some low-redshift studies \citep[e.g.][]{von_der_linden_how_2007,loubser_stellar_2009}.

Figure \ref{fig_degeneracy_BCG} shows the degeneracy between
metallicity and formation redshift as applied to the stellar
population models describing the two BCG samples. We observe that the 98\% 
confidence regions of the two BCG samples do not overlap, supporting
the idea that the stellar populations of the blue and red BCGs are
indeed different. The confidence intervals for the blue sample are
centred on lower metallicities than the red sample regions while
formation redshift also tends to be lower \---\ although
the confidence regions of both models extend over a wide range of
redshifts.

Based on the stellar population analysis, one might infer that blue
BCGs have formed more recently than the red ones. However,
\citet{li_how_2007} report that when more than one stellar population
is present, the age and metallicity obtained by colours alone might be
biased toward the younger and more metal-rich stars. 
Therefore, the blue BCGs might be bluer because they have experienced either an extended star formation episode or more than one bursts in the past. In such scenarios, differences in the duration of the star formation or in the relative importances and epochs of the secondary bursts would account for the spread in colour observed in the bluer BCGs.

One factor that may trigger a star formation episode is a cooling
flow. Several studies have reported the existence of blue, moderately
star-forming BCGs in cool core clusters at low to moderate redshifts
\citep[e.g.][]{egami_spitzer_2006,bildfell_resurrecting_2008,stott_near-infrared_2008,loubser_stellar_2009,loubser_regulation_2016,pipino_evidence_2009,rawle_relation_2012,green_multiwavelength_2016}. Since
X-ray selected clusters are biased toward relaxed clusters
\citep{rossetti_cool-core_2017}, this might partially explain why the
blue BCGs represent a significant part of our sample.

Alternatively, statistical studies based on infrared 
\citep[e.g.][]{webb_star_2015,bonaventura_red_2017} or SZ-selected clusters
\citep[e.g.][]{mcdonald_star-forming_2016} have shown that in-situ
star formation is an important mechanism at $z\gtrsim 1$, possibly
triggered by galaxy interactions
\citep{mcdonald_star-forming_2016}. This suggests that a range of BCG
colours might be part of every high-redshift sample, regardless of the
sample selection.

One limitation of the current analysis is that the fitted stellar
population models are unable to accommodate the slope of the high-redshift end of the red
sample J-$\mathrm{K_s}$ colour variation with redshift (see Fig.
\ref{fig_best_fits_BCG}) and that no BCGs are red beyond $z=1.05$. This suggest that the stars making up the
red BCGs formed later or evolve faster than predicted by our
model. The former is supported by the degeneracy between formation
redshift and metallicity. The best-fitting formation redshift,
$z_{form}\sim 12$, might be considered high compared with the
predictions of the most recent simulations
\citep[e.g.][]{ragone-figueroa_bcg_2018,rennehan_rapid_2020},
but \---\ alternatively \---\ consistent with recent observations
\citep{hashimoto_onset_2018,willis_spectroscopic_2020}.

Dust extinction might be evoked to explain why these objects are so
red. Indeed, a fit including dust provides an equivalent
statistical description of the red BCGs (see Table
\ref{table_fits_red}), although we note that the colour of a model
stellar population is degenerate between star formation history,
metallicity, and dust obscuration. For this reason, we prefer a simple,
dust free model for the analysis of the red BCG population.


Ultimately, the current data available for the BCGs sample are unable
to provide a definitive explanation of the blue/red dichotomy. The
acquisition of rest-frame optical spectroscopy of several BCGs of both
groups is needed in order to get a more accurate and thorough picture
of the star formation history of these objects \citep[e.g.][]{lonoce_old_2015,lonoce_stellar_2020,belli_kmos3d_2017,saracco_age_2019}. 

\section{Summary}\label{sec_conclu}

We have identified a sample of 35 X-ray-selected distant galaxy
clusters in the XXL-N/VIDEO field and performed a preliminary analysis
of their galaxy populations. Clusters are selected as extended X-ray
sources (C1, C2, or AC) coincident with overdensities of bright
galaxies in photometric redshift space. Of the 35 candidate clusters
at $z_{phot}\geq 0.8$, ten are spectroscopically confirmed clusters, and
a further 15 are presented here for the first time. The ten remaining
clusters have been detected previously but never confirmed.

The sample of clusters displays a wide variety of quenched fractions,
a result nominally consistent with the assertion made in Sect. 1
that selecting clusters on the ICM properties is insensitive to the
star formation history of their member galaxies. The relationship between galaxy luminosity and quenched fraction appears to be
in place at $z\sim 1$ although we do not observe a significant variation of the quenched fraction with the cluster-centric radius.

The sample of BCGs is inconsistent with a single stellar population
model. The observed distribution is bimodal in colour and is
consistent with an old, passive population and a possibly younger,relatively bluer, and more diverse 
population. Although we are unable to provide a definite explanation
for this split, we suggest that the blue BCGs may have experienced either an extended or
more than one star-formation episode.

\section*{Acknowledgements}\label{sec_thanks}
AT is supported by the Natural Sciences and Engineering Research Council of Canada (NSERC) Postgraduate Scholarship-Doctoral Program. JPW acknowledges support from the NSERC Discovery Grant program. 
The Saclay group acknowledges long-term support from the Centre National d'Etudes Spatiales (CNES). This work was supported by the Programme National Cosmology et Galaxies (PNCG) of CNRS/INSU with INP and IN2P3, co-funded by CEA and CNES. NA acknowledges funding from the Science and Technology Facilities Council (STFC) Grant Code ST/R505006/1. 
RAAB is supported by the Glasstone Foundation, MJJ and RAAB acknowledge support from the Oxford Hintze Centre for Astrophysical Surveys which is funded through generous support from the Hintze Family Charitable Foundation and the award of the STFC consolidated grant (ST/N000919/1). 
XXL is an international project based around an XMM Very Large Programme surveying two 25 $\mathrm{deg^2}$ extragalactic fields at a depth of $\sim~6 \times 10{-15} ~\mathrm{erg ~cm^{-2} ~s^{-1}}$ in the [0.5-2] keV band for point-like sources. The XXL website is \url{http://irfu.cea.fr/xxl}. Multi-band information and spectroscopic follow-up of the X-ray sources are obtained through a number of survey programmes, summarised at \url{http://xxlmultiwave.pbworks.com/}. This work is based on data products from observations made with ESO Telescopes at the La Silla Paranal Observatory under ESO programme ID 179.A- 2006 and on data products produced by the Cambridge Astronomy Survey Unit on behalf of the VIDEO consortium. This research has made use of the NASA/IPAC Extragalactic Database (NED), which is funded by the National Aeronautics and Space Administration and operated by the California Institute of Technology. 



\bibpunct{(}{)}{;}{a}{}{,} 
\bibliographystyle{aa} 
\bibliography{ref_doc}

\begin{appendix}

\section{Notes on individual objects}\label{sec_weirdo}
\subsection*{Candidates 1 and 24}\label{ssec_18_581}
Candidate 1/24 was observed by Gemini South in 2010 and classified as a distant cluster (John Stott, personal communication), but, to our knowledge, the results were never published. We observed two spikes in the photometric redshift space of this source, one at $z_{phot}=0.80\pm 0.02$ and one at $z_{phot}=1.19\pm 0.02$. Furthermore, the X-ray emission associated with this detection is complex with several brighter spots. This suggest that there might be two distant clusters in the same line-of-sight.

\subsection*{Candidate 5}\label{ssec_18_1175}
The BCG of this $z=0.84$ cluster has a possible companion 0.454 magnitudes less luminous in $\mathrm{ K_s}$ band. The optical centre of both galaxies lie at a projected separation of 58.2 kpc (7.48 arcsec). The projected separation between the BCG and the X-ray coordinates is 27.4 kpc and 43.6 kpc for the companion. Neither the BCG nor the companion exhibit signs of interactions. We were able to confirm this cluster with archival spectroscopic observations stored in the CESAM database \citepalias{adami_xxl_2018}.

\subsection*{Candidates 6 and 27}\label{ssec_faint_cand}
Both candidates 6 and 27 are faint detection in the vicinity of bright point sources. Thus, in these two fields, we increased the number of  X-ray contours from 10 to 25. Candidate 6 is at $z=0.86\pm 0.02$ and candidate 27 is at $z=1.44 \pm 0.14$.

\subsection*{Candidate 10}\label{ssec_18_155}
Despite appearing as an isolated spike in the redshift space, candidate 10 seems to be constituted of two or more overdensities: the background subtracted colour diagram of candidate 10 exhibit four \lq bumps\rq~rather than two. Similarly, none of our BCG candidates seems reliable: the best one exhibit a very red J-$\mathrm{K_s}$ colour, more consistent with $z\sim 1$ than $z\sim 0.9$ and indeed has photometric redshift of $z~0.98$. The two other BCGs candidates both sit at more than 750 kpc of projected separation with the X-ray coordinates. 
Since candidate 10 meets our detection criteria, we included it in our list of candidate clusters, but not in our analysis.

\subsection*{Candidate 30}\label{ssec_18_1344}
Candidate 30 has a photometric redshift of $1.48\pm 0.14$ which make it one of our newly discovered very high-z cluster. Unfortunately, the photometry of the BCG and of another central galaxy are unreliable because they are in the halo of a foreground star.


\subsection*{Candidate 33}\label{ssec_17_1463}
Among the candidate clusters presented here for the first time, candidate 33 is probably the second most distant one, with a photometric redshift of $1.79\pm 0.14$. This is again a robust (significance between 4.5 and 5.5 galaxies) detection in the photometric space. X-ray emission is regular, and the distance between its coordinates and the BCG is 60.8 kpc. The BCG is extremely red: its Z-J and J-$\mathrm{K_s}$ colours are 2.45 and 1.59 respectively while the mean colours of the others $z\gtrsim 1.5$ BCGs with reliable photometry (i.e. excluding candidate 31) are 1.45 and 0.90. Interestingly, the i-z colour discrepancy between candidate 33 and the other high-z BCGs is smaller in i-z: 0.88 for candidate 33 compared to 0.63. Such colours might originate from the presence of dust in this object, but more data are needed to confirm.


\subsection*{Candidate 35}\label{ssec_08_866}
This candidate cluster is probably the farthest object in our sample, with a photometric redshift of $1.93\pm 0.14$. This X-ray detection is associated with multiple spikes in the photometric redshift space. Since the photometric redshift does not perform very well at $z\sim 2$, some of these spikes might be associated with the main overdensity. However the presence of a significant spike at $z\sim 1$ suggests the presence of at least another overdensity along the line-of-sight.


\subsection*{Other candidate clusters at $z\gtrsim 1.4$}\label{ssec_others_high-z}
There are four other new candidate clusters at $z\gtrsim 1.4$ in our sample: candidates 28, 29, 31, and  32 sitting at redshifts of $1.45\pm 0.14$, $1.48\pm 0.14$, $1.54\pm 0.14$, and $1.57\pm 0.14$. Candidate 32 is a weaker X-ray detection than the 3 others but also features the most significant photometric redshift spike.


\section{Candidate clusters}\label{sec_images}
\begin{figure*}
\centering
\begin{subfigure}{0.3\textwidth}
\includegraphics[width=5.25cm]{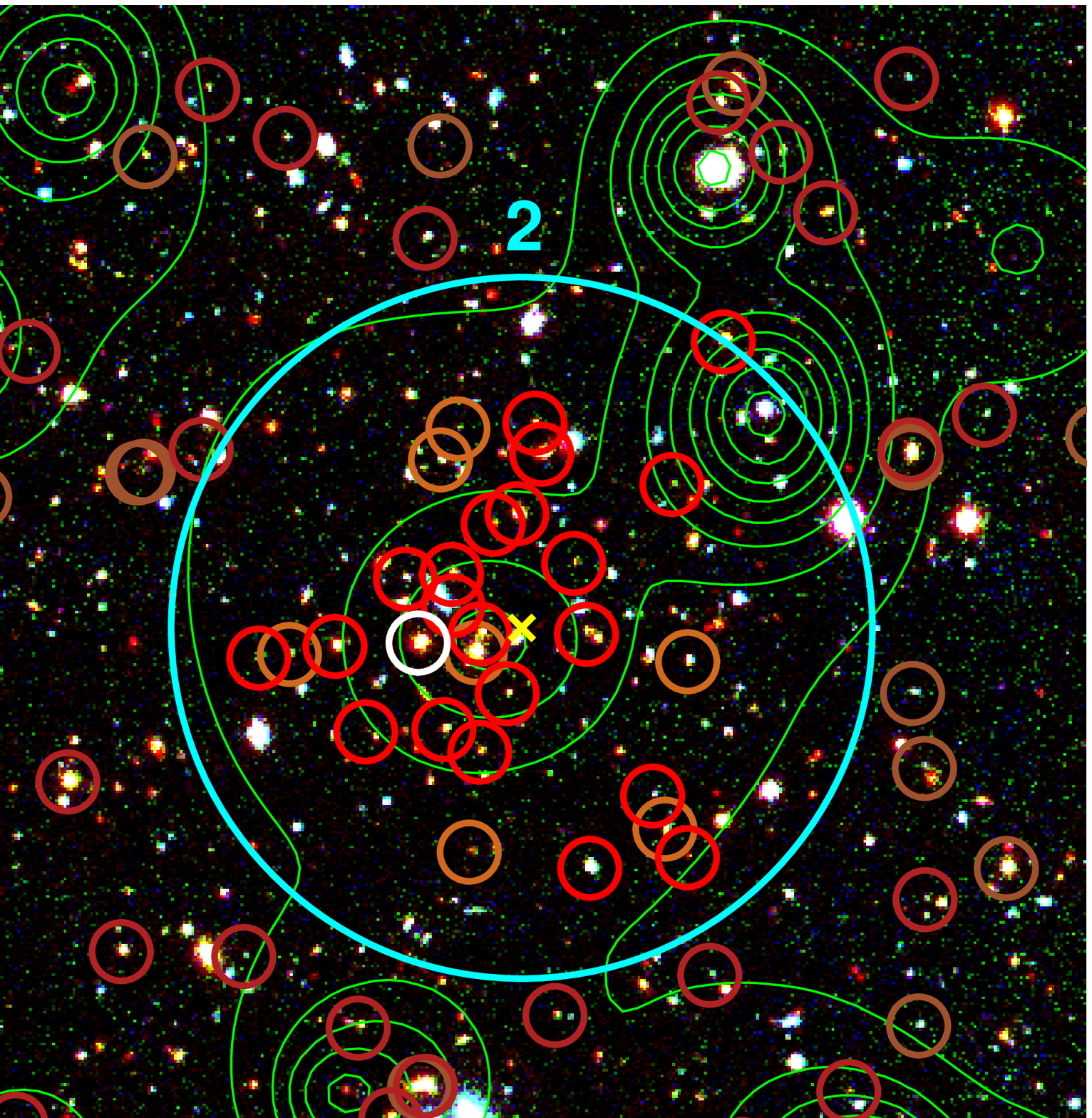}
\end{subfigure}
\hfill
\begin{subfigure}{0.3\textwidth}
\includegraphics[width=6cm]{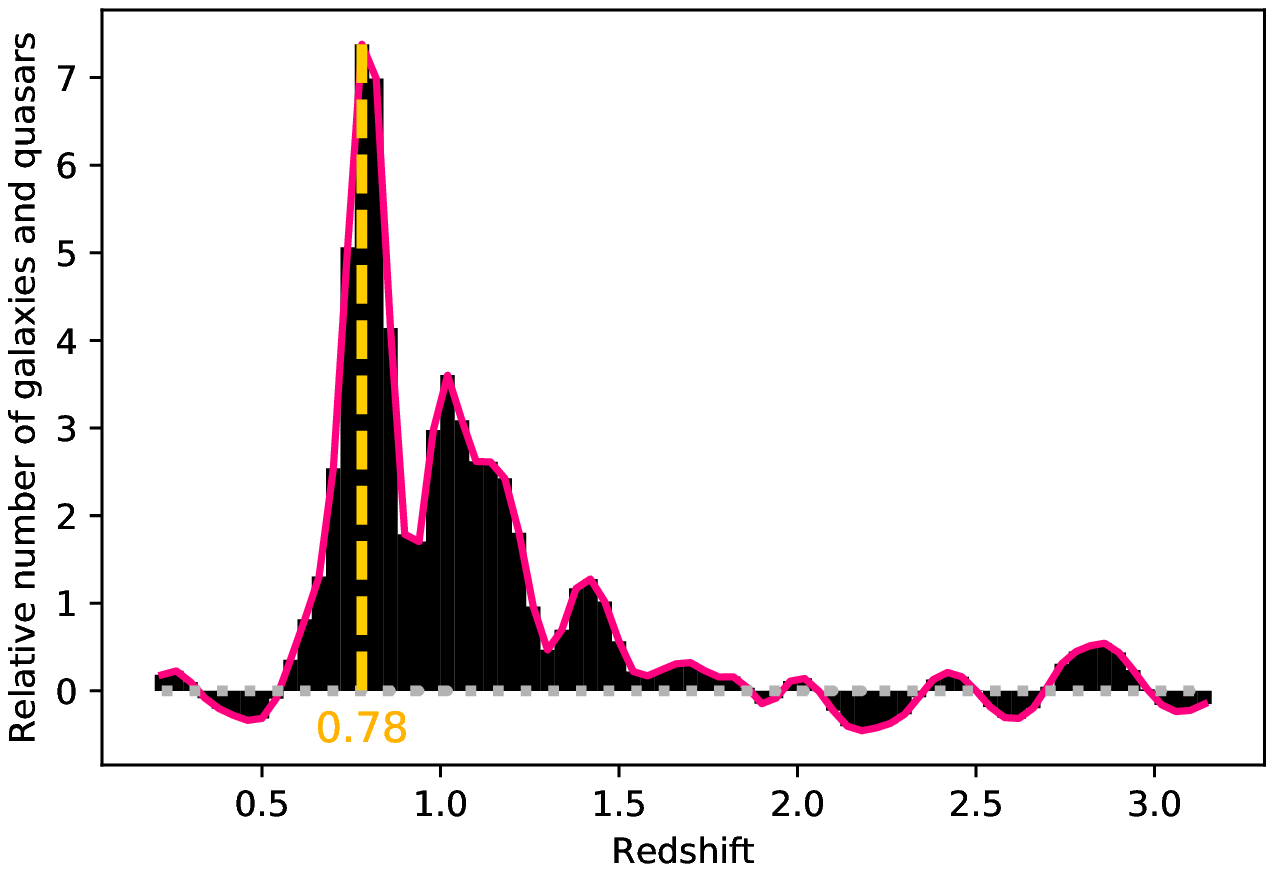}
\end{subfigure}
\hfill
\begin{subfigure}{0.3\textwidth}
\includegraphics[width=6cm]{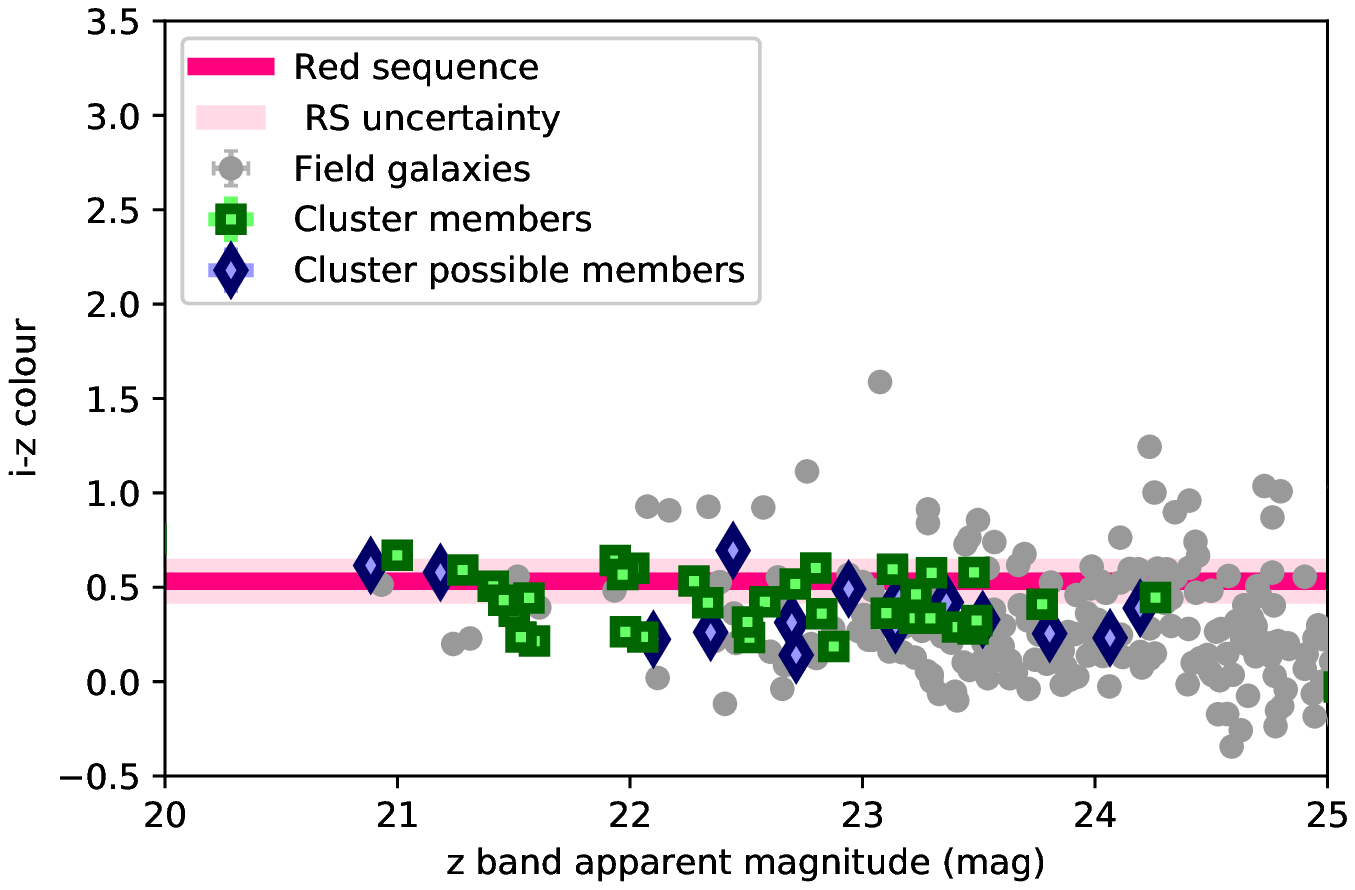}
\end{subfigure}

\begin{subfigure}{0.3\textwidth}
\includegraphics[width=5.25cm]{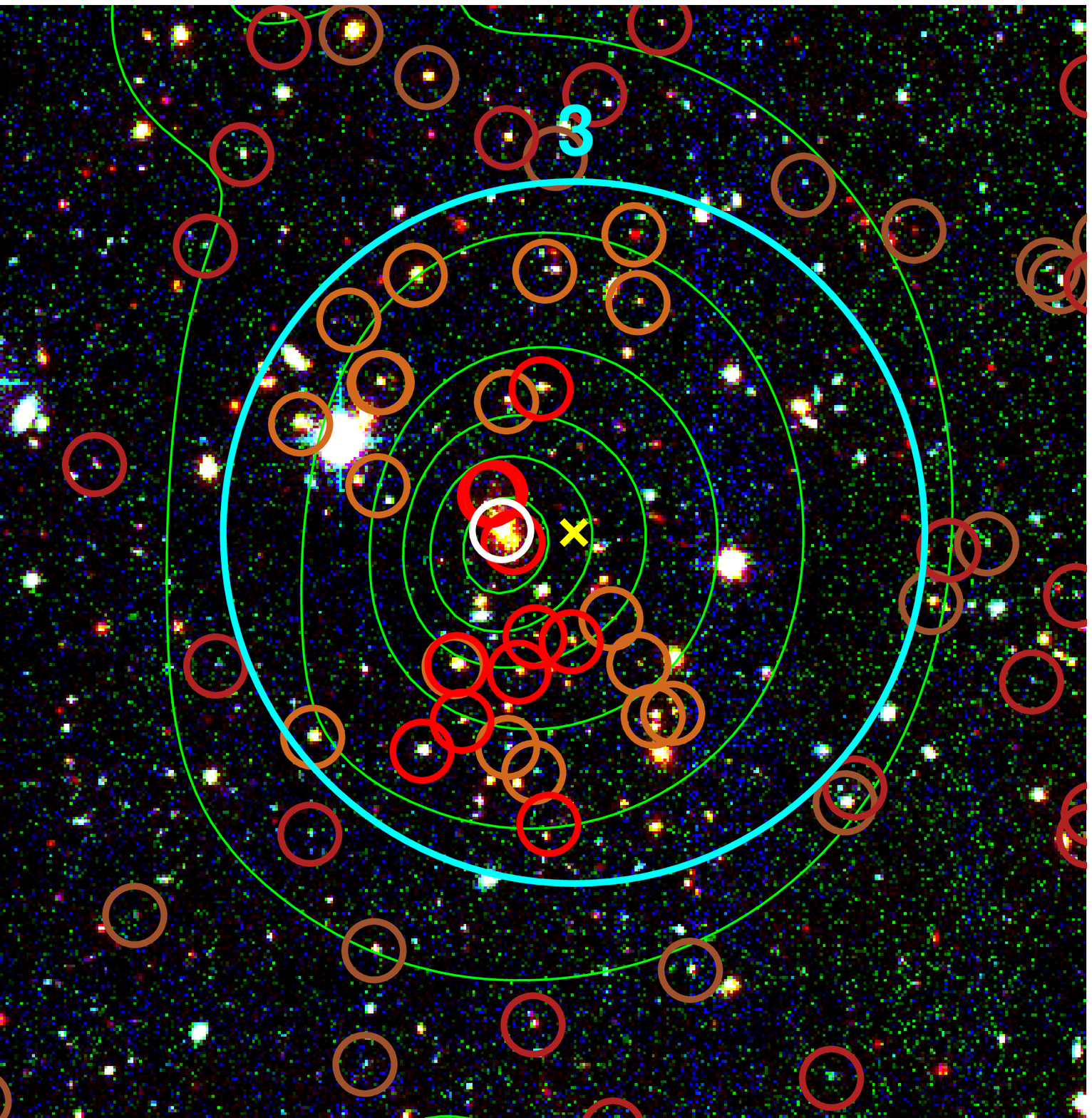}
\end{subfigure}
\hfill
\begin{subfigure}{0.3\textwidth}
\includegraphics[width=6cm]{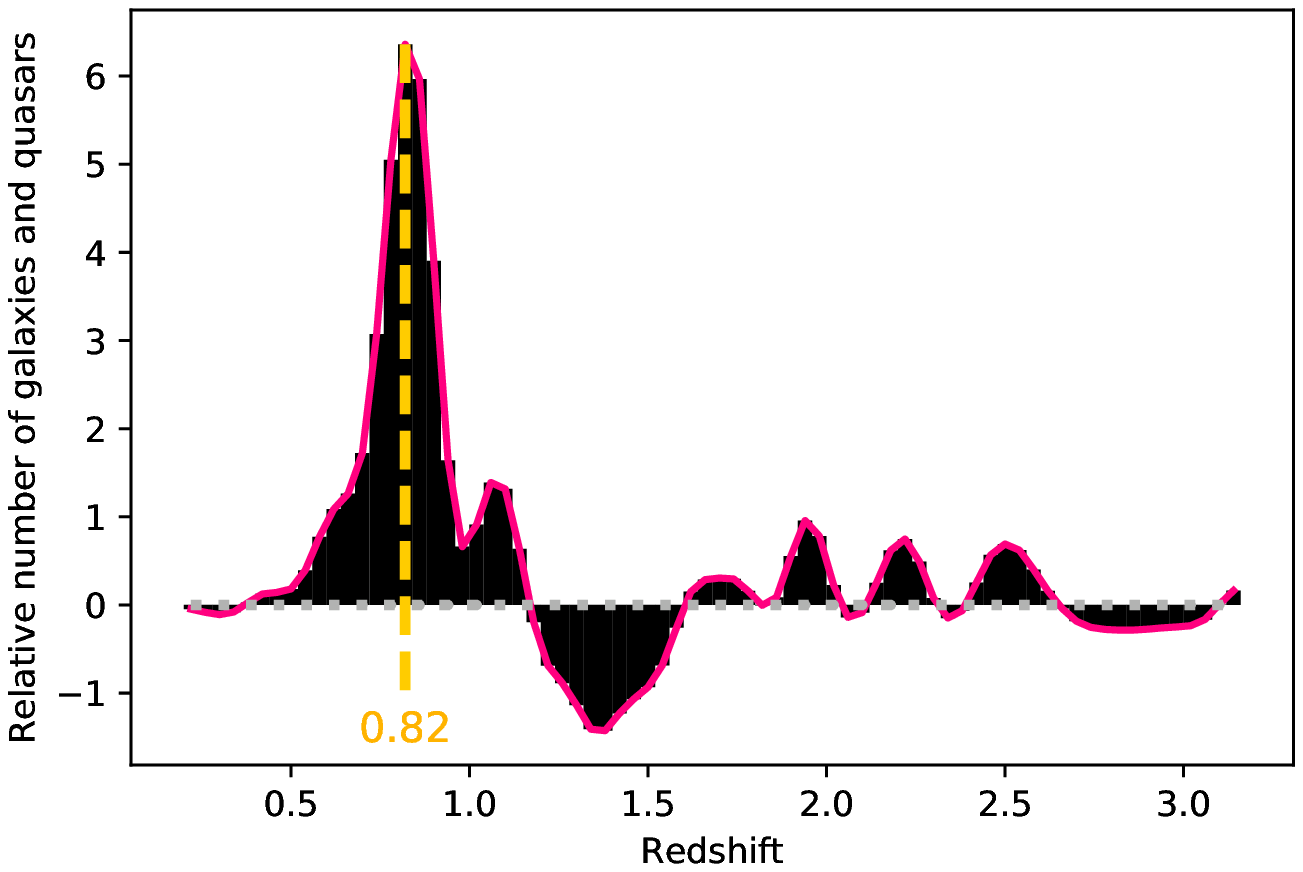}
\end{subfigure}
\hfill
\begin{subfigure}{0.3\textwidth}
\includegraphics[width=6cm]{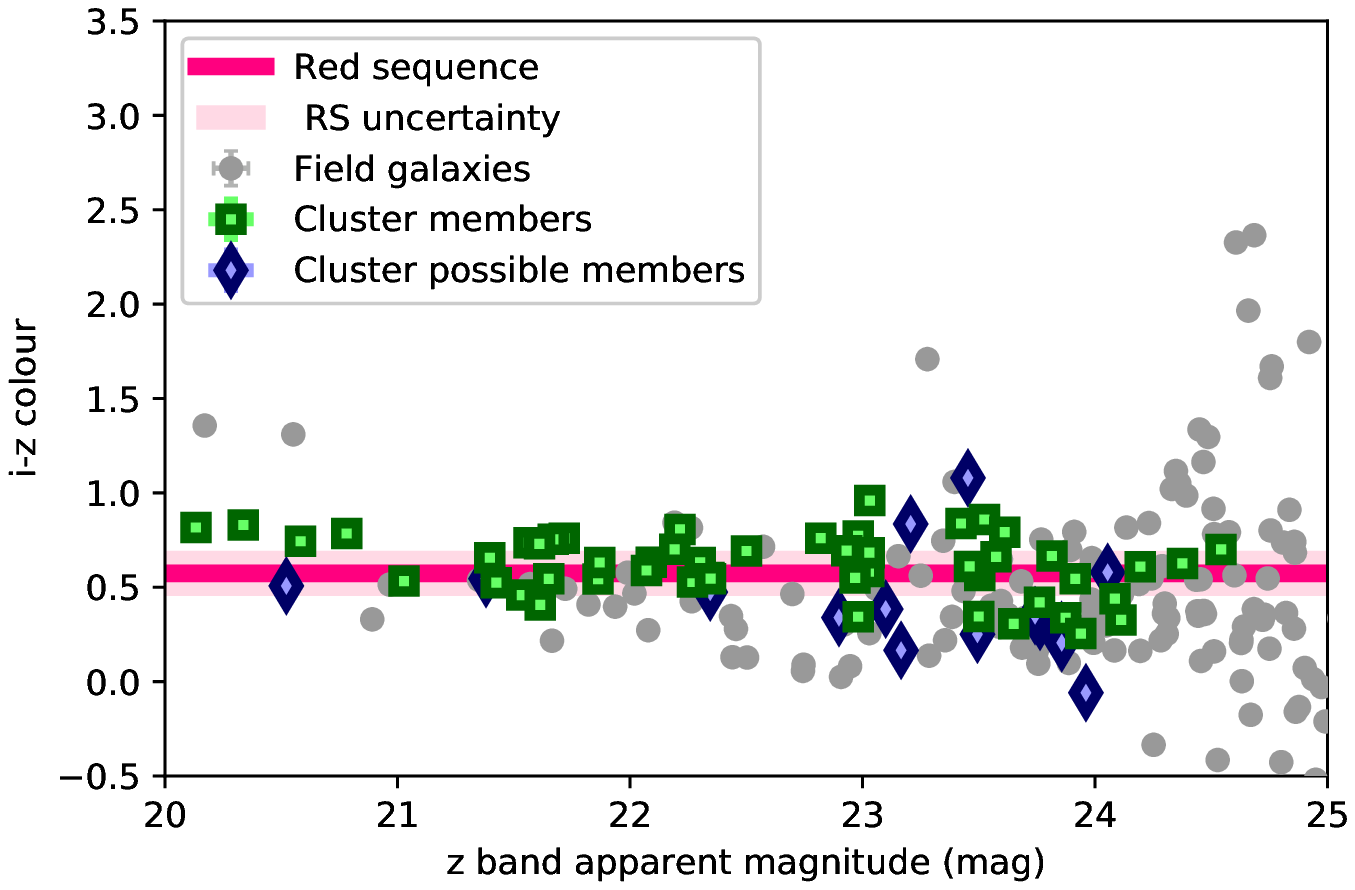}
\end{subfigure}

\begin{subfigure}{0.3\textwidth}
\includegraphics[width=5.25cm]{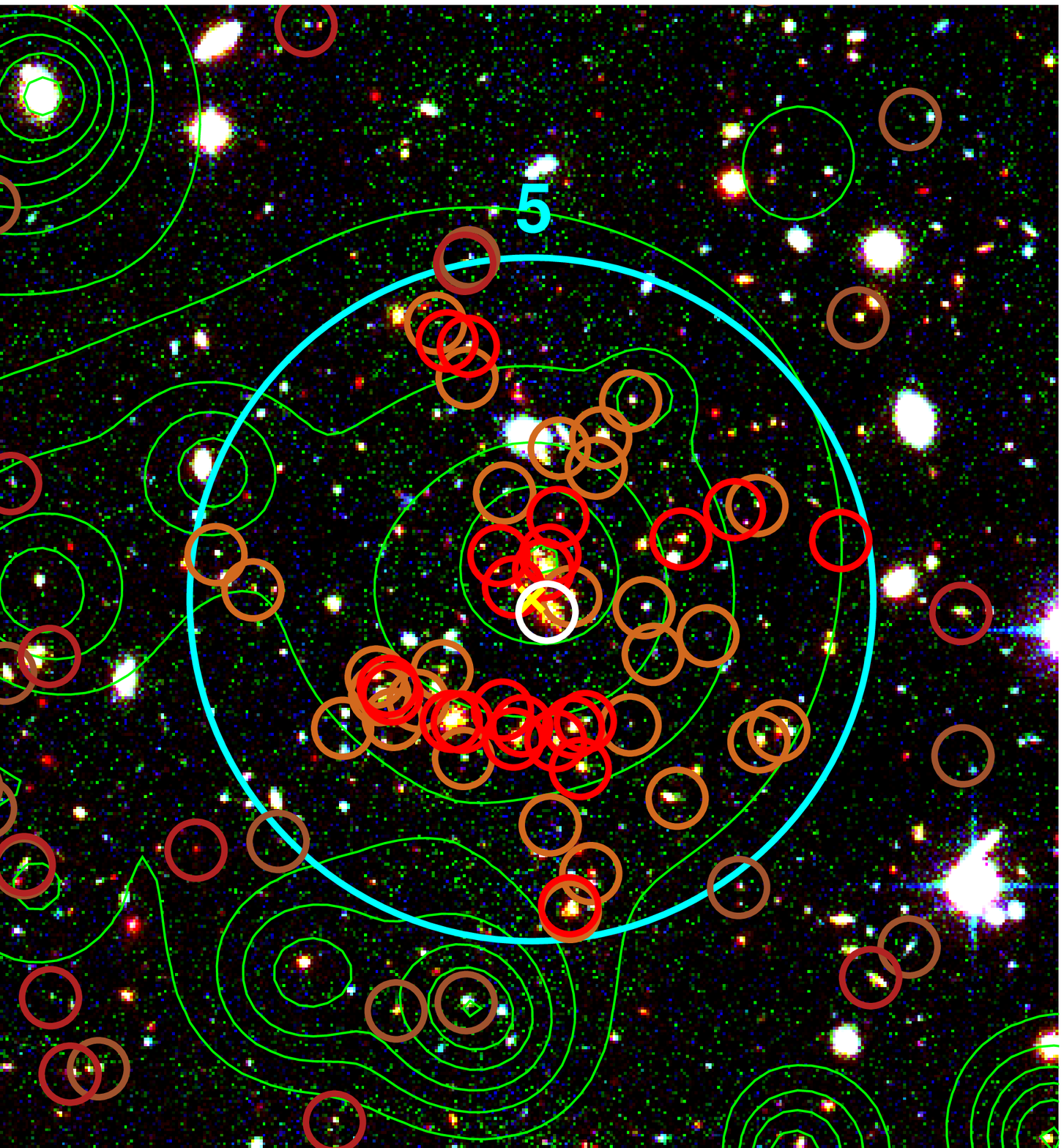}
\end{subfigure}
\hfill
\begin{subfigure}{0.3\textwidth}
\includegraphics[width=6cm]{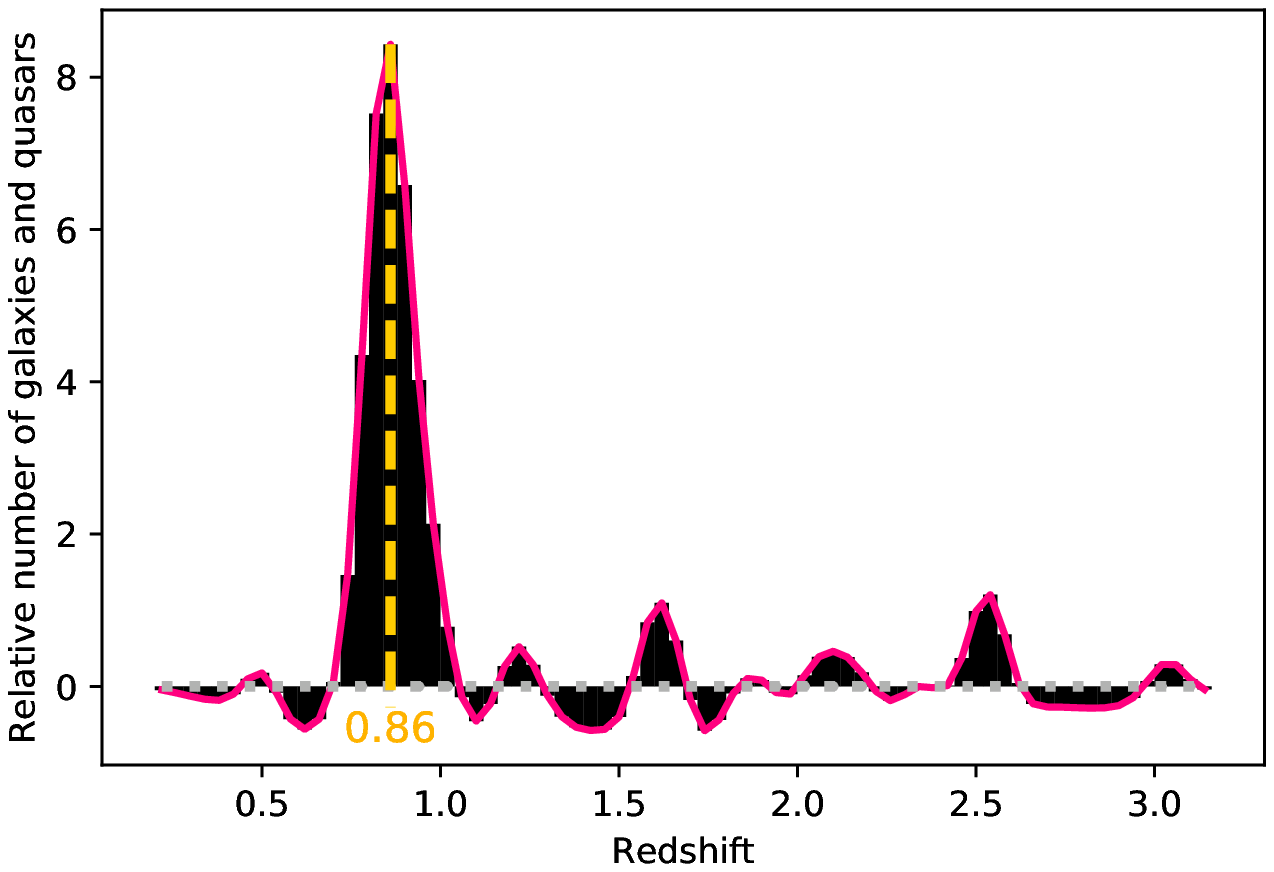}
\end{subfigure}
\hfill
\begin{subfigure}{0.3\textwidth}
\includegraphics[width=6cm]{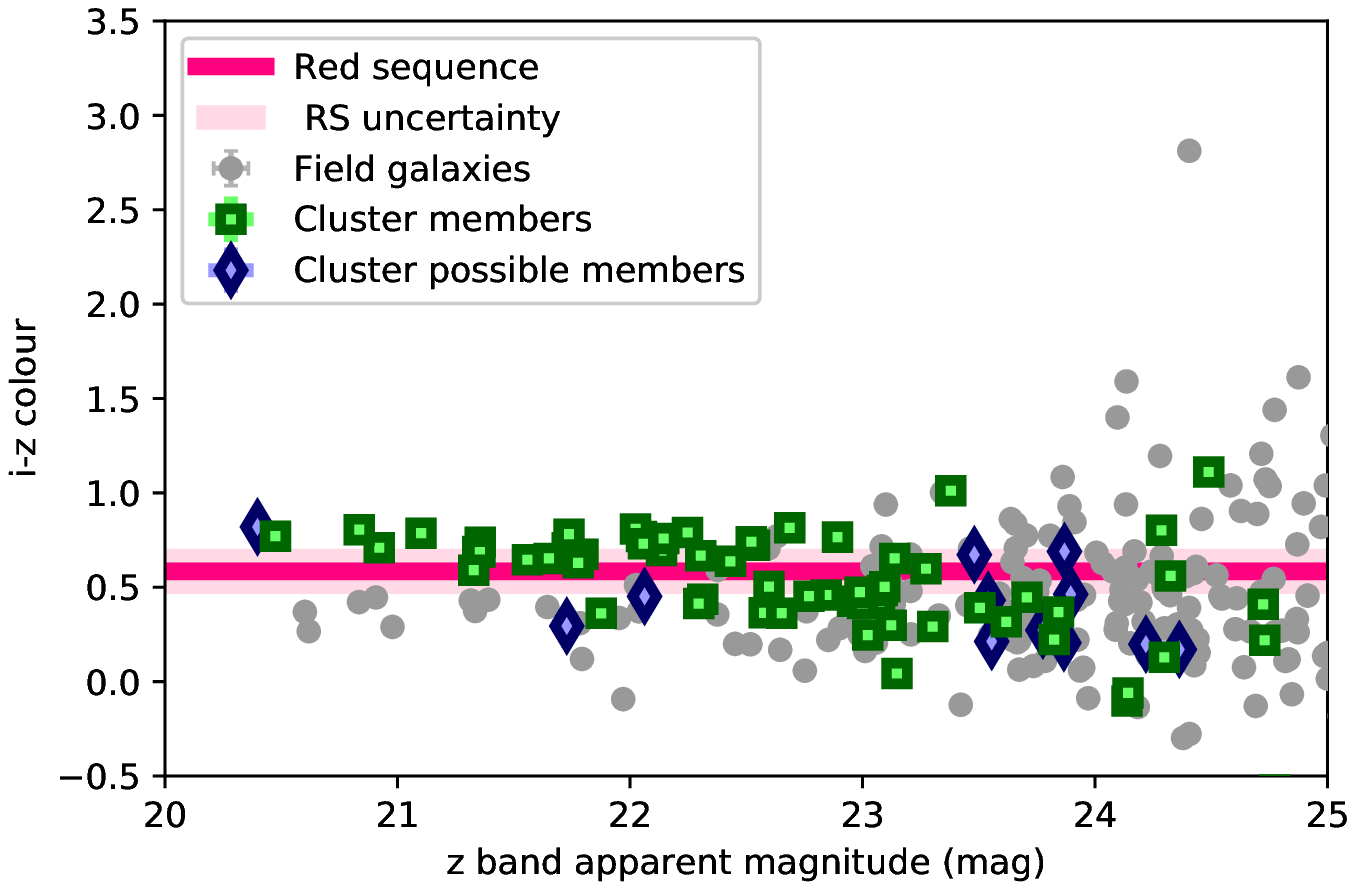}
\end{subfigure}
\caption{Left cols.: Megacam R and I filter and VIDEO H filter images for each confirmed cluster at $z\geq 0.8$ meeting our selection criteria, classified by increasing redshifts. The cyan circles delimit regions within one arcmin of the X-ray best fit model centres which are marked by yellow crosses. The red and brown circles highlight the bright galaxies with a redshift corresponding respectively to the cluster peak redshift $\pm 0.02$ and to the cluster redshift $\pm 0.06$. Darker circles indicate galaxies outside the central region. The BCGs are circled in white. The X-ray contours in green are logarithmically distributed in 10 levels between the maximum and minimum emission observed in a $7\times 7 ~\mathrm{arcmin^2}$ box around the X-ray source. Middle Cols.: Background subtracted and Gaussian filtered redshift distribution of the bright galaxies within the central arcmin, for the corresponding candidates. The dashed line highlights the median redshift of the highest bin in the redshift spike. Its colour assesses the importance of the overdensity, and therefore our confidence in the detection: gold for the most reliable candidates (the highest bin height is above 5.5), grey for the reliable one (above 4.5), and beige for the other. Right Cols.: i-z ($0.8\leq z< 1.2$) or z-J ($z\geq 1.2$) CMD plot of the galaxies above VIDEO 5$\sigma$ limit within 1 arcmin of the centre. The green squares indicate the galaxies with photometric redshifts consistent with the mean redshift plus or minus 1.5 times the standard deviation of the most accurate Gaussian modelling of the redshift spike. The blue lozenges indicate galaxies with redshifts consistent with the sidewings of the most accurate Gaussian model, up to three times the standard deviation. The deep pink lines indicate the colours predicted by the stellar population model computed in Sect. \ref{ssec_first_method}. The light pink region is the standard deviation of the difference of this model with method 1 red sequences.}
\end{figure*}


\newpage

\begin{figure*}
\centering
\ContinuedFloat
\begin{subfigure}{0.3\textwidth}
\includegraphics[width=5.25cm]{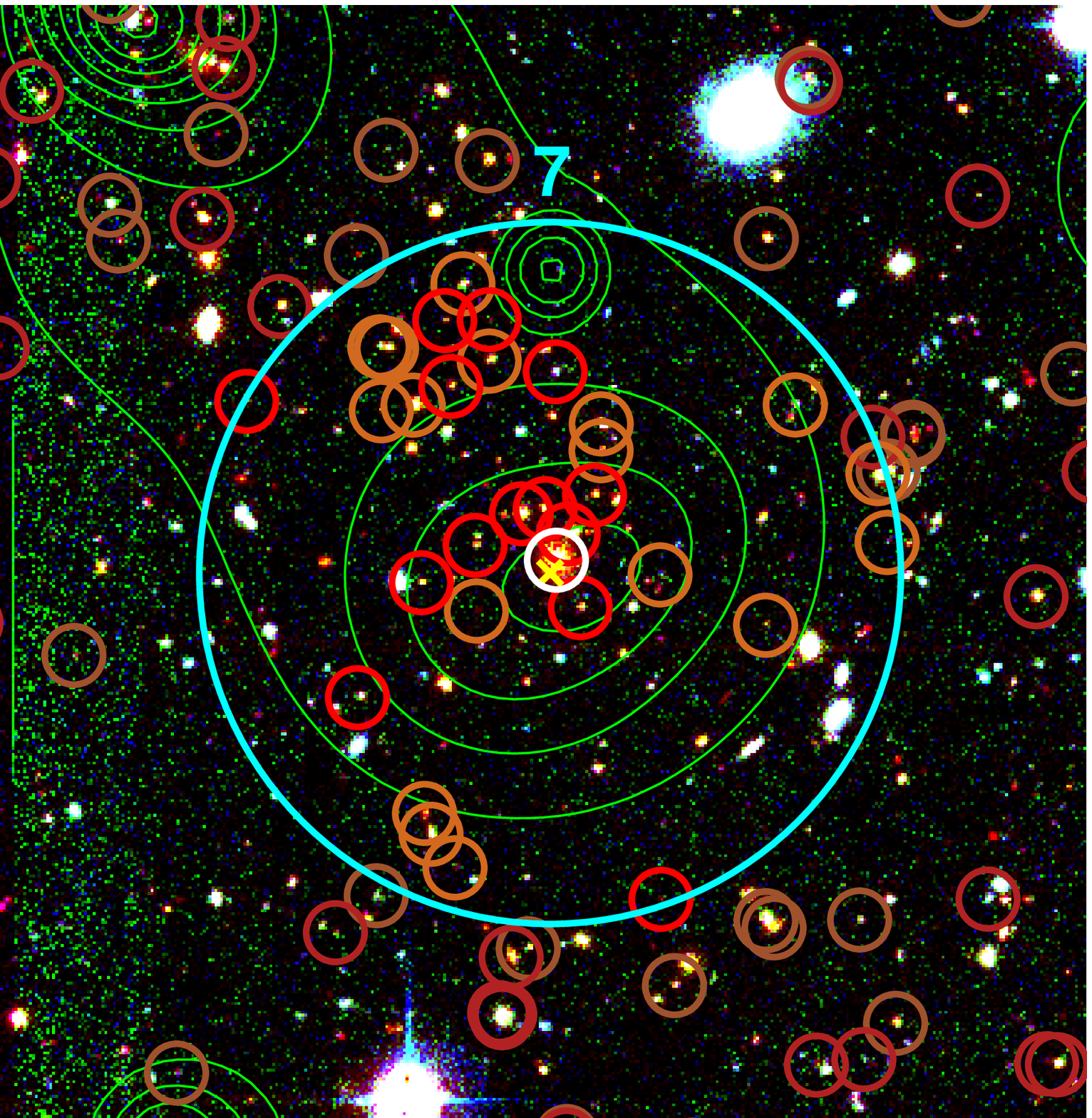}
\end{subfigure}
\hfill
\begin{subfigure}{0.3\textwidth}
\includegraphics[width=6cm]{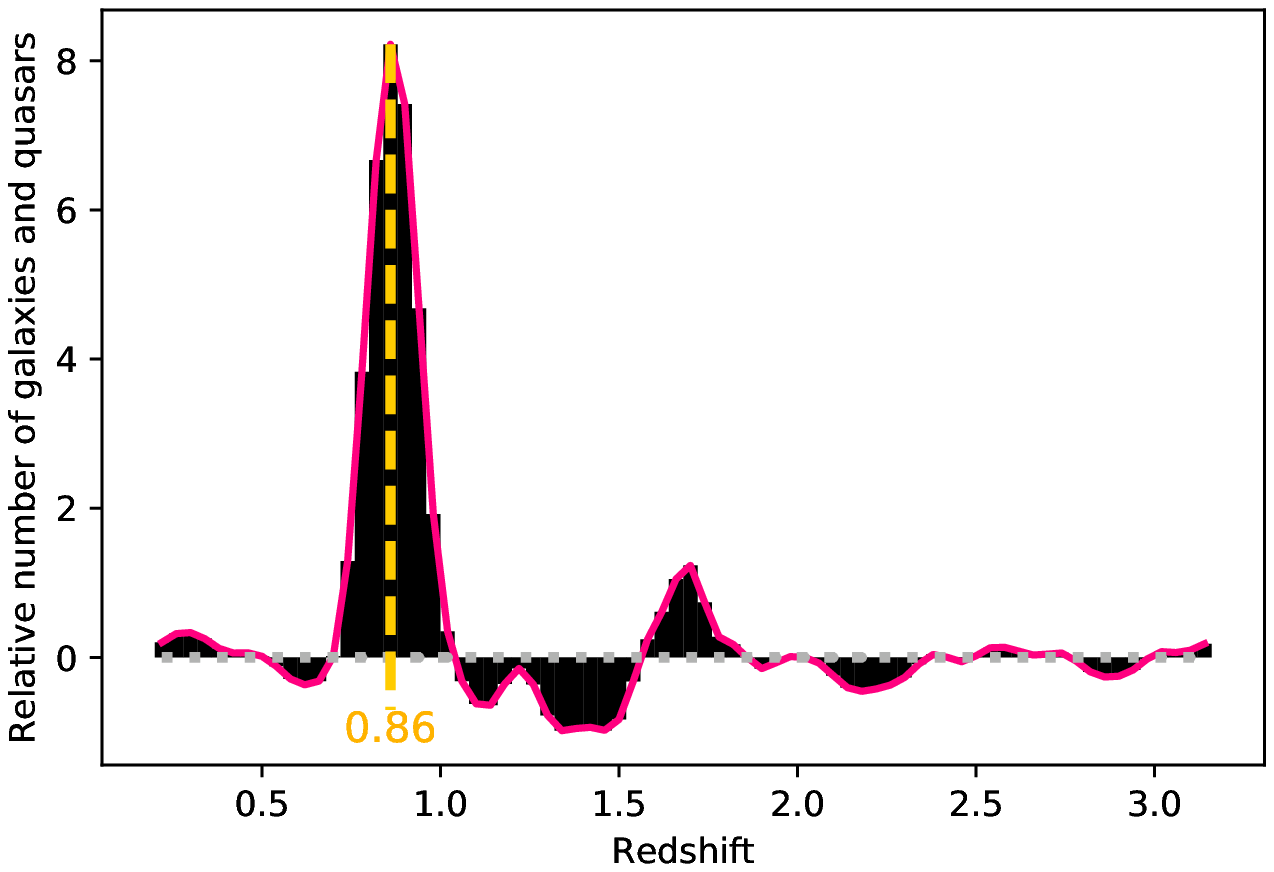}
\end{subfigure}
\hfill
\begin{subfigure}{0.3\textwidth}
\includegraphics[width=6cm]{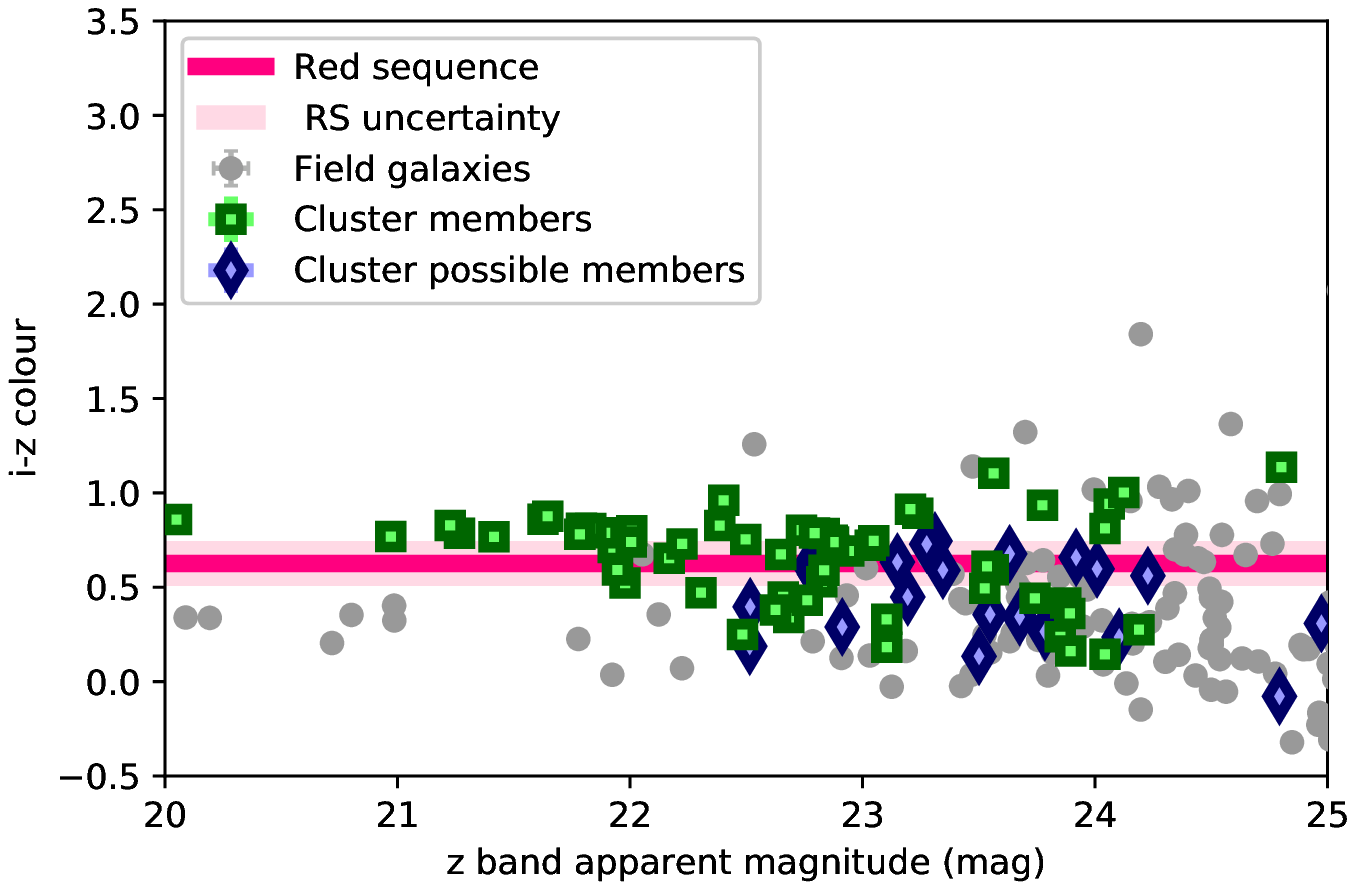}
\end{subfigure}

\centering
\begin{subfigure}{0.3\textwidth}
\includegraphics[width=5.25cm]{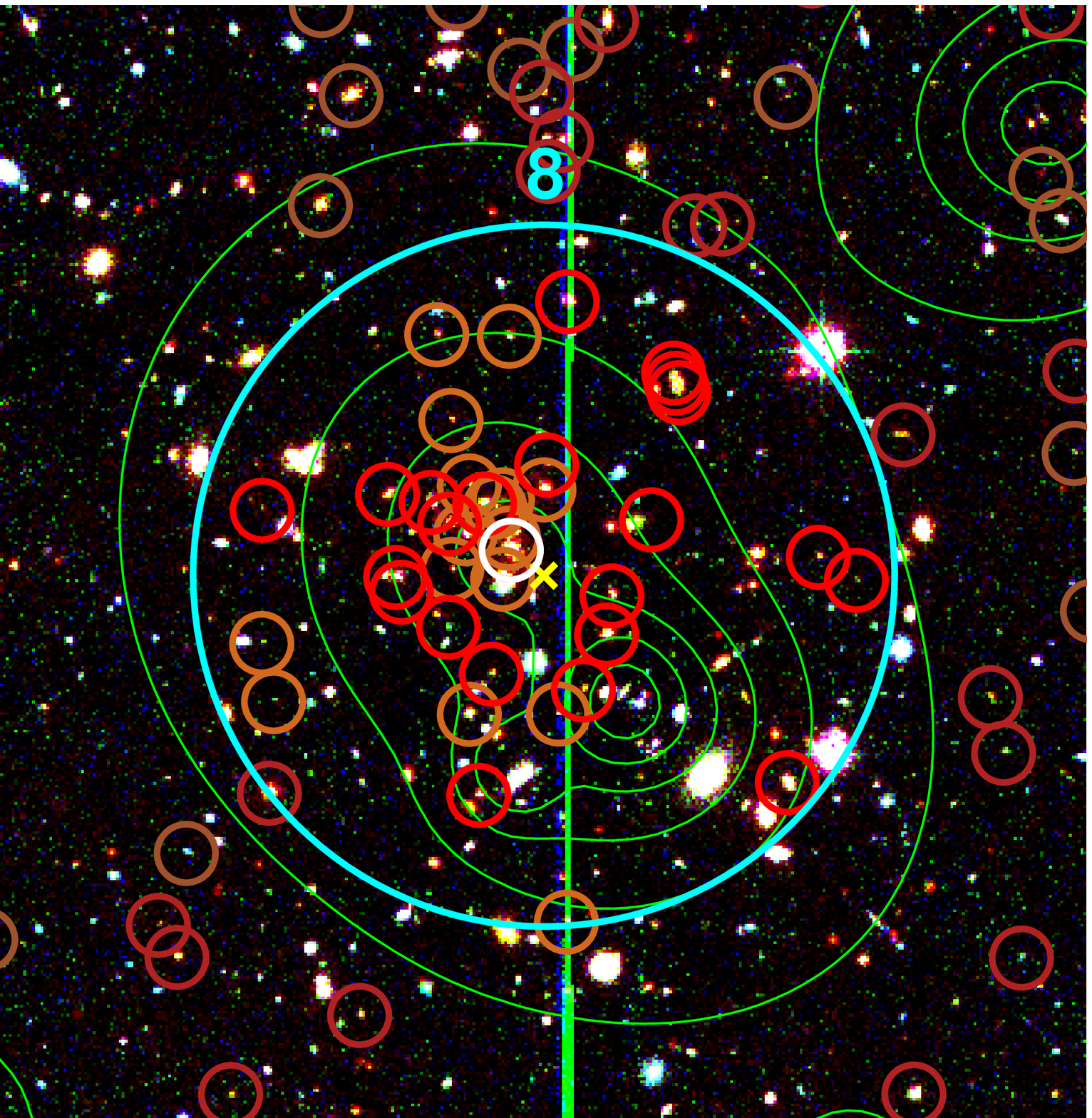}
\end{subfigure}
\hfill
\begin{subfigure}{0.3\textwidth}
\includegraphics[width=6cm]{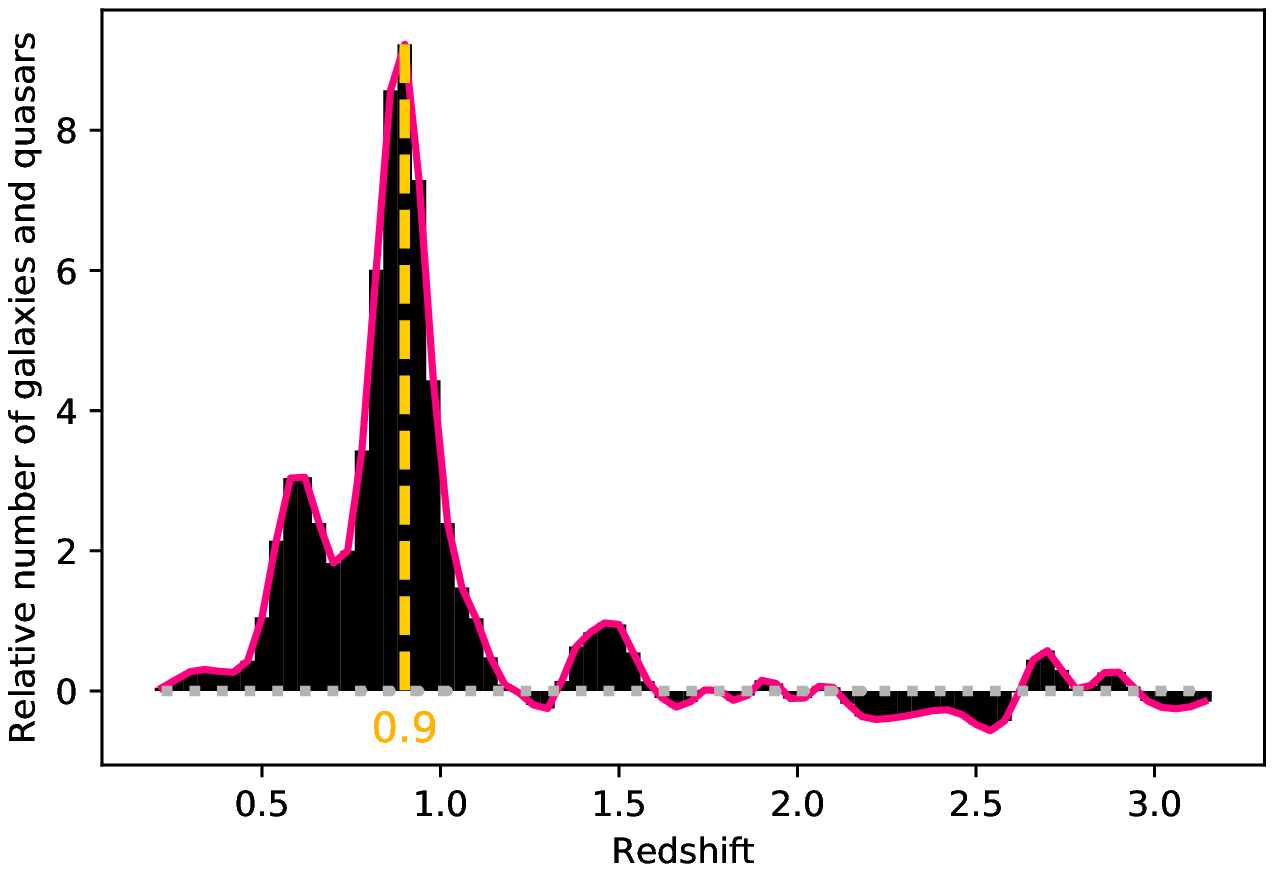}
\end{subfigure}
\hfill
\begin{subfigure}{0.3\textwidth}
\includegraphics[width=6cm]{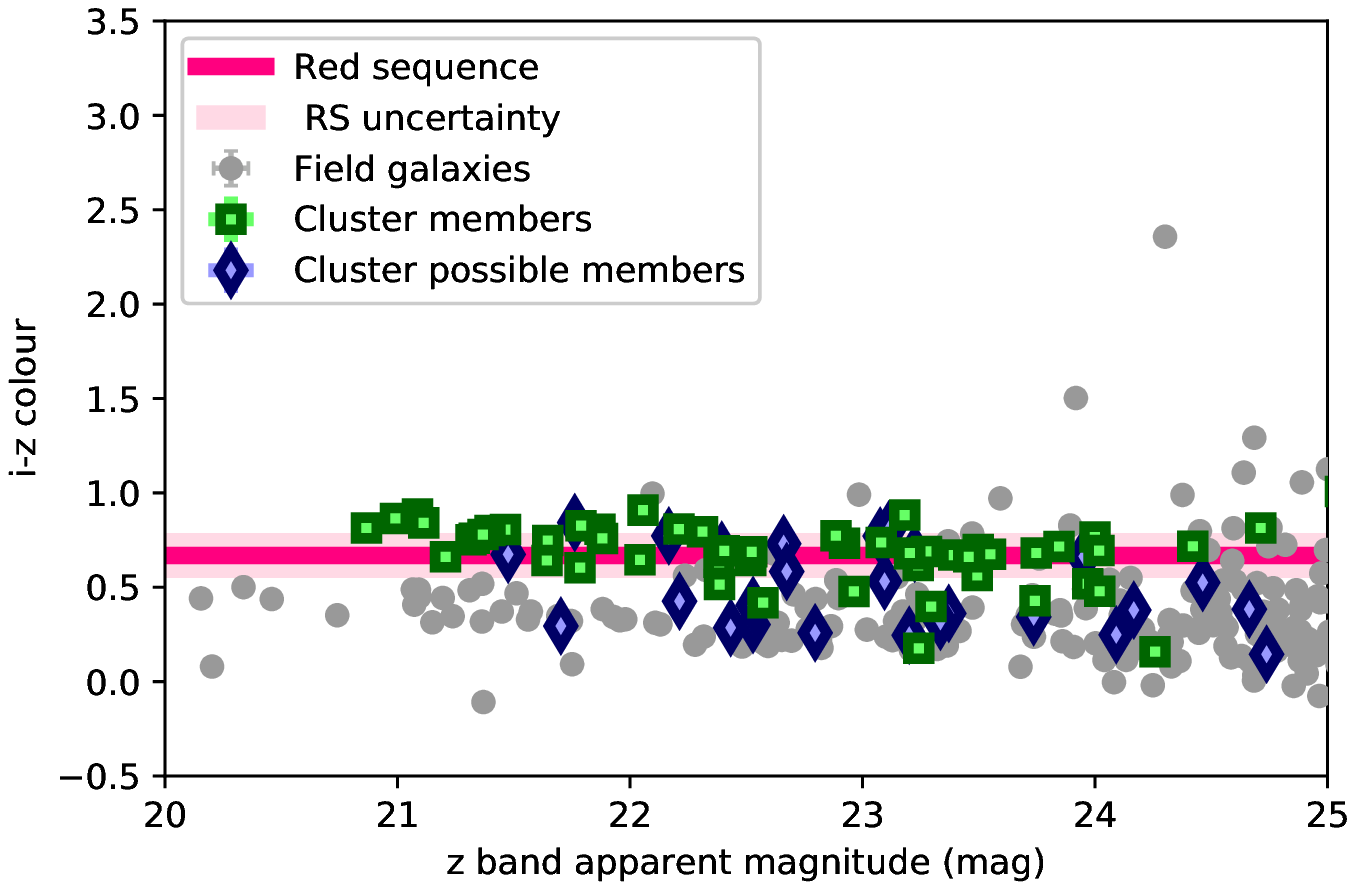}
\end{subfigure}

\centering
\begin{subfigure}{0.3\textwidth}
\includegraphics[width=5.25cm]{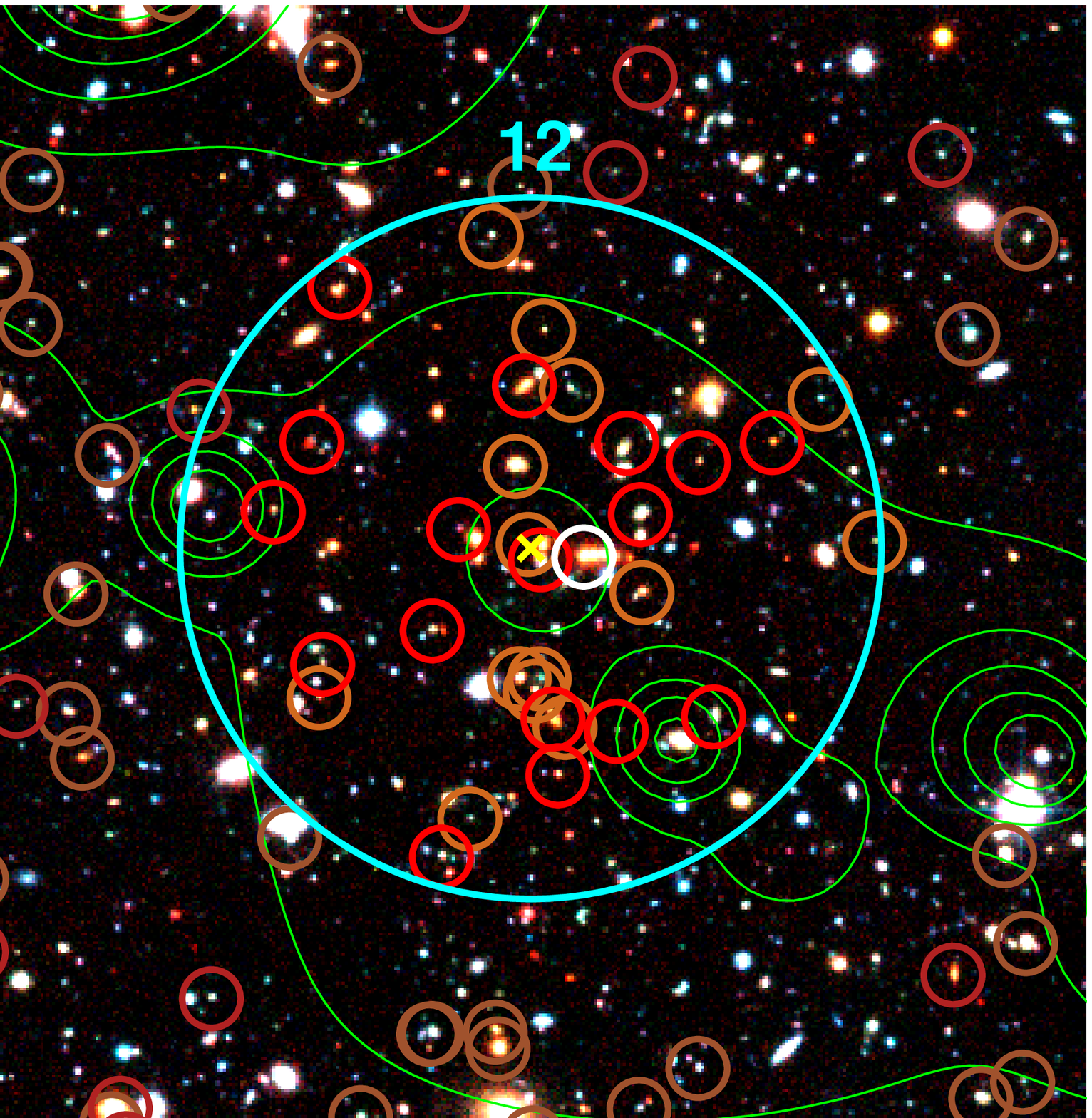}
\end{subfigure}
\hfill
\begin{subfigure}{0.3\textwidth}
\includegraphics[width=6cm]{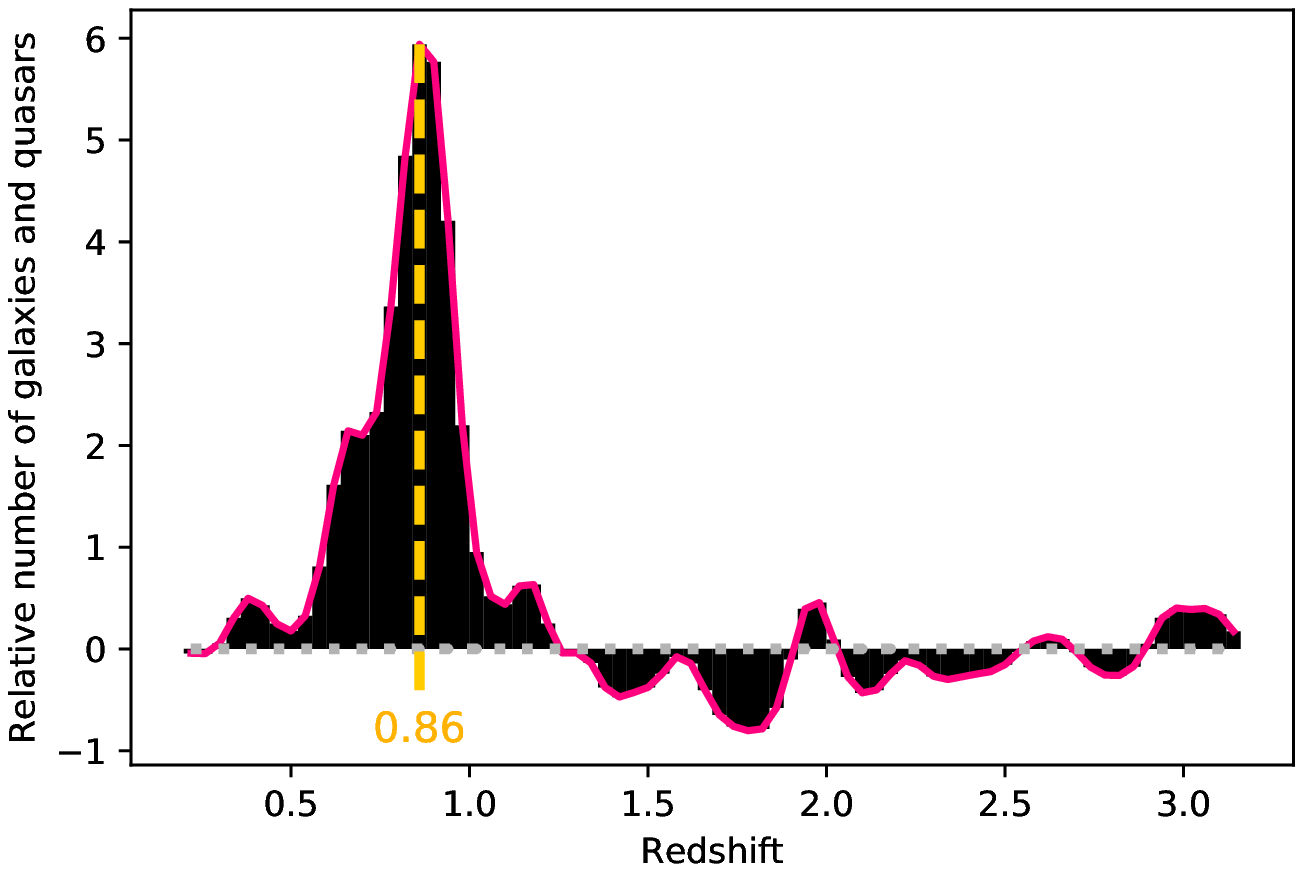}
\end{subfigure}
\hfill
\begin{subfigure}{0.3\textwidth}
\includegraphics[width=6cm]{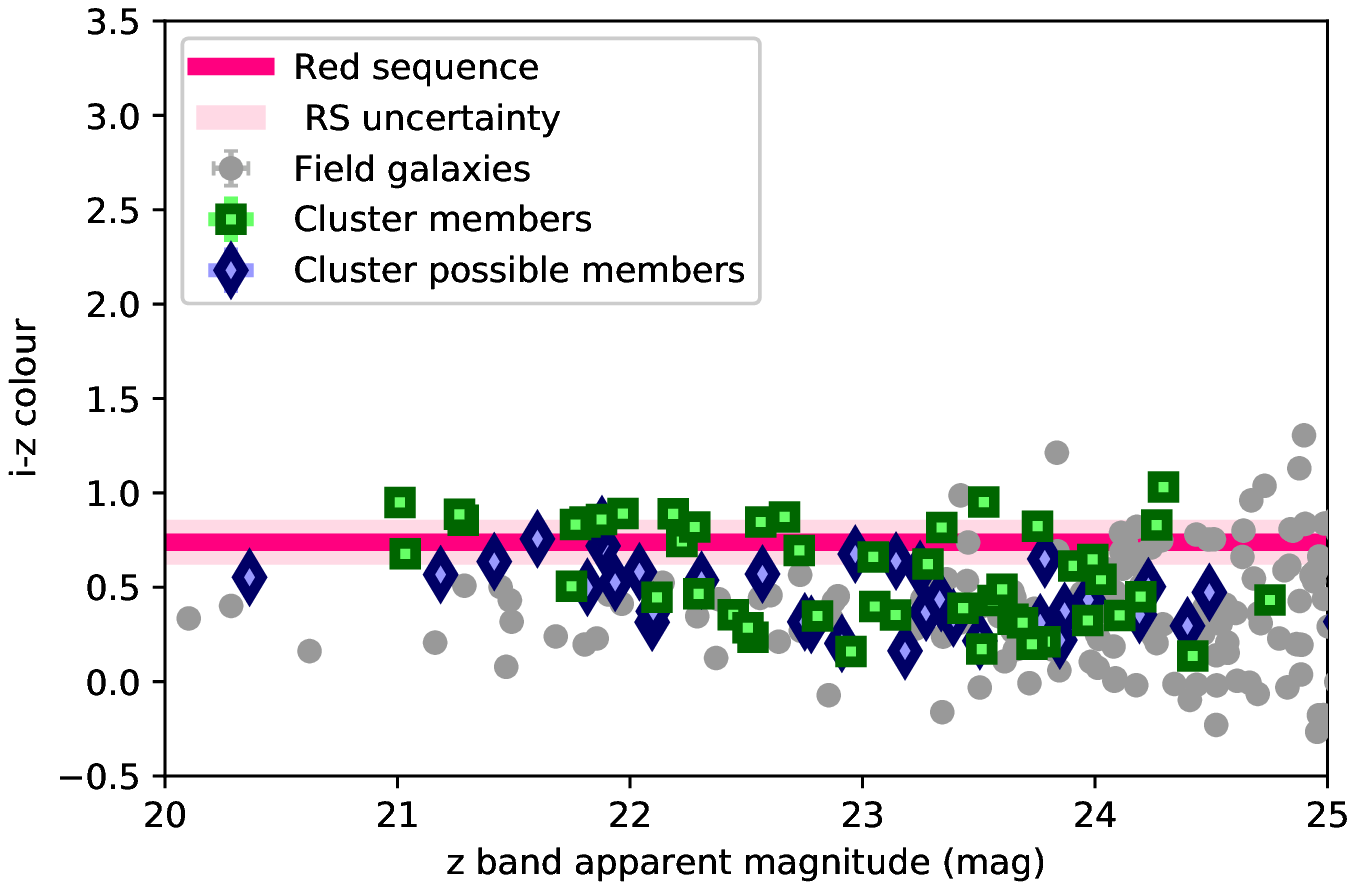}
\end{subfigure}

\centering
\begin{subfigure}{0.3\textwidth}
\includegraphics[width=5.25cm]{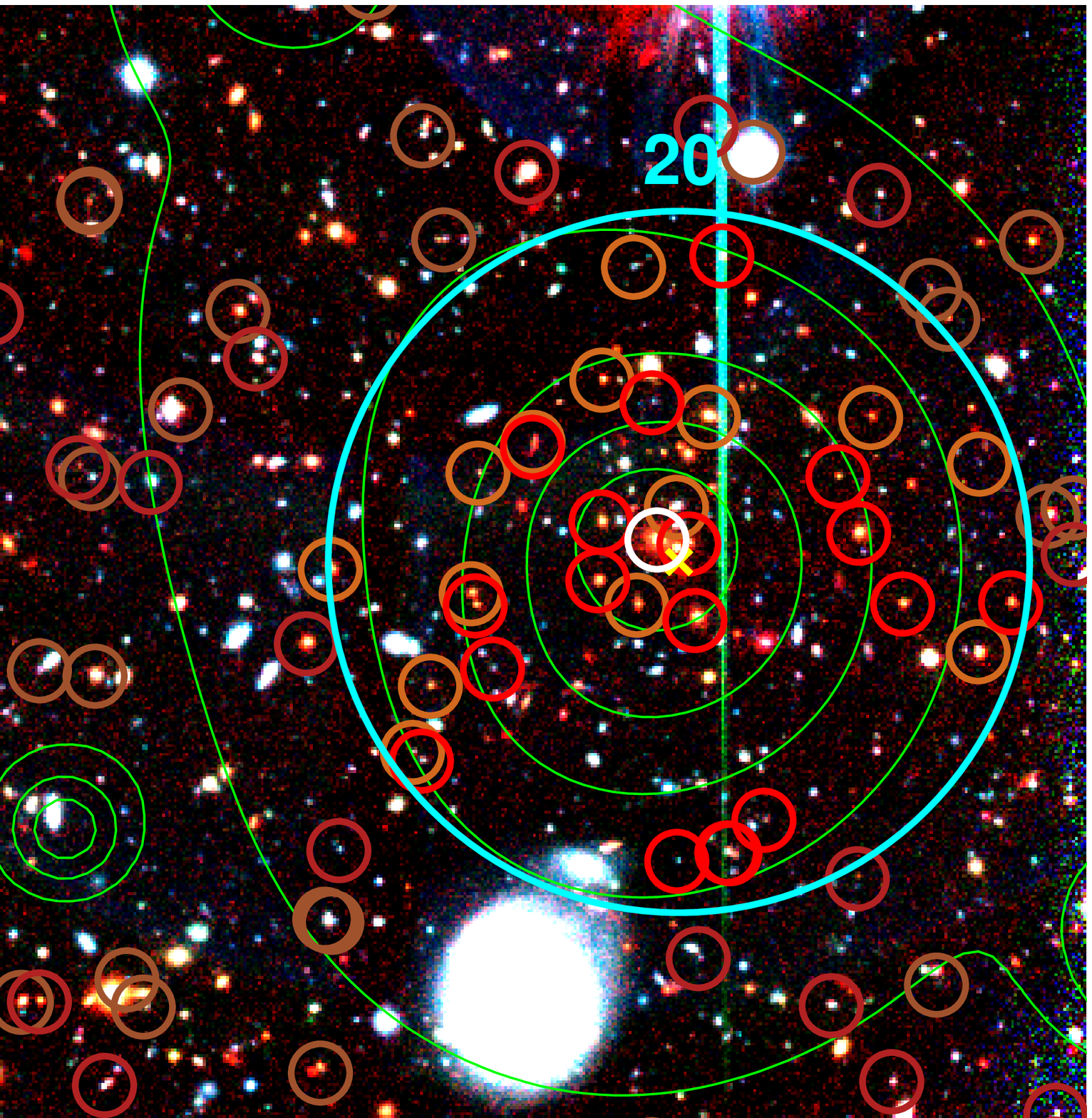}
\end{subfigure}
\hfill
\begin{subfigure}{0.3\textwidth}
\includegraphics[width=6cm]{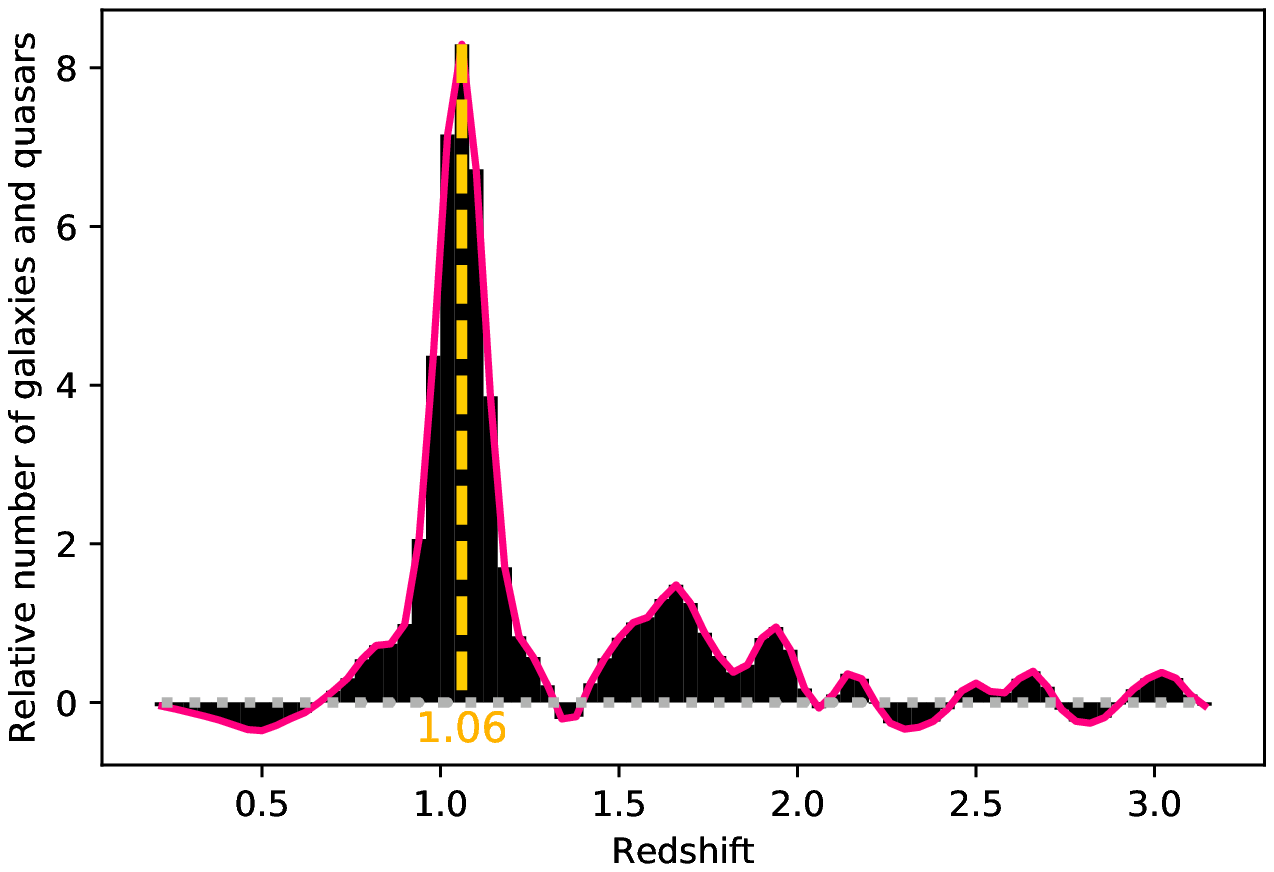}
\end{subfigure}
\hfill
\begin{subfigure}{0.3\textwidth}
\includegraphics[width=6cm]{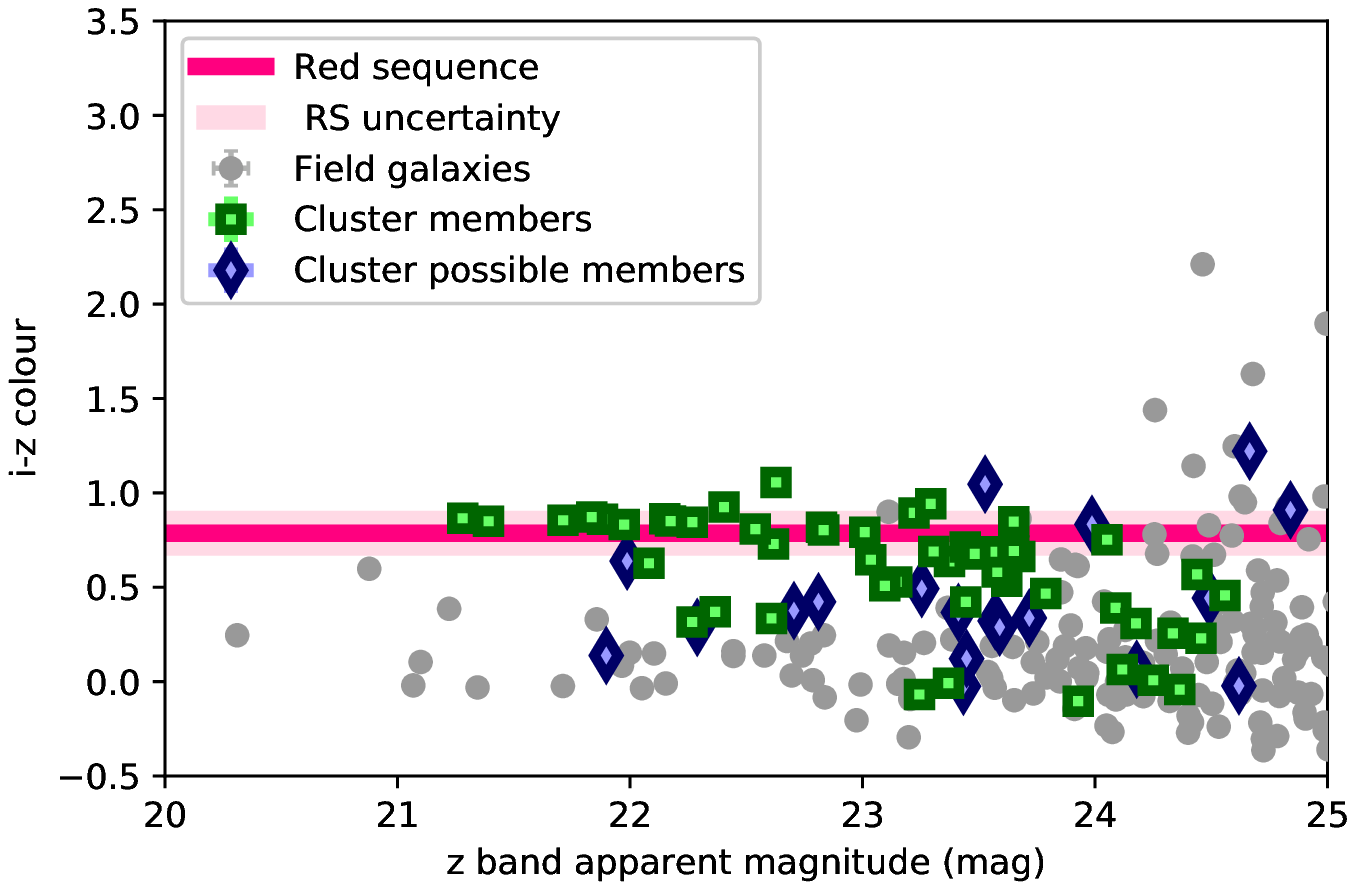}
\end{subfigure}
\caption{\textit{continued}}
\end{figure*}

\newpage

\begin{figure*}
\ContinuedFloat
\centering
\begin{subfigure}{0.3\textwidth}
\includegraphics[width=5.25cm]{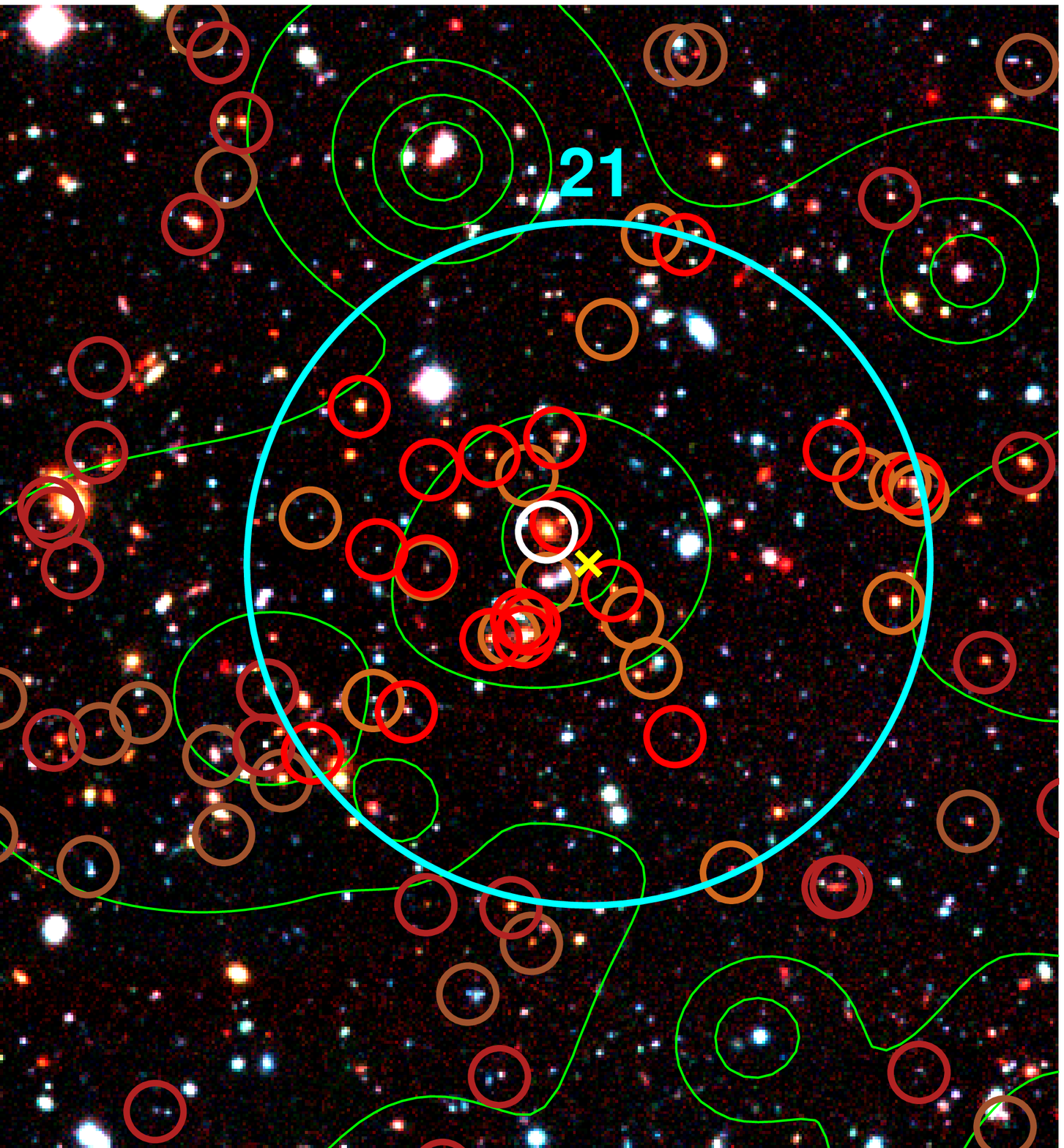}
\end{subfigure}
\hfill
\begin{subfigure}{0.3\textwidth}
\includegraphics[width=6cm]{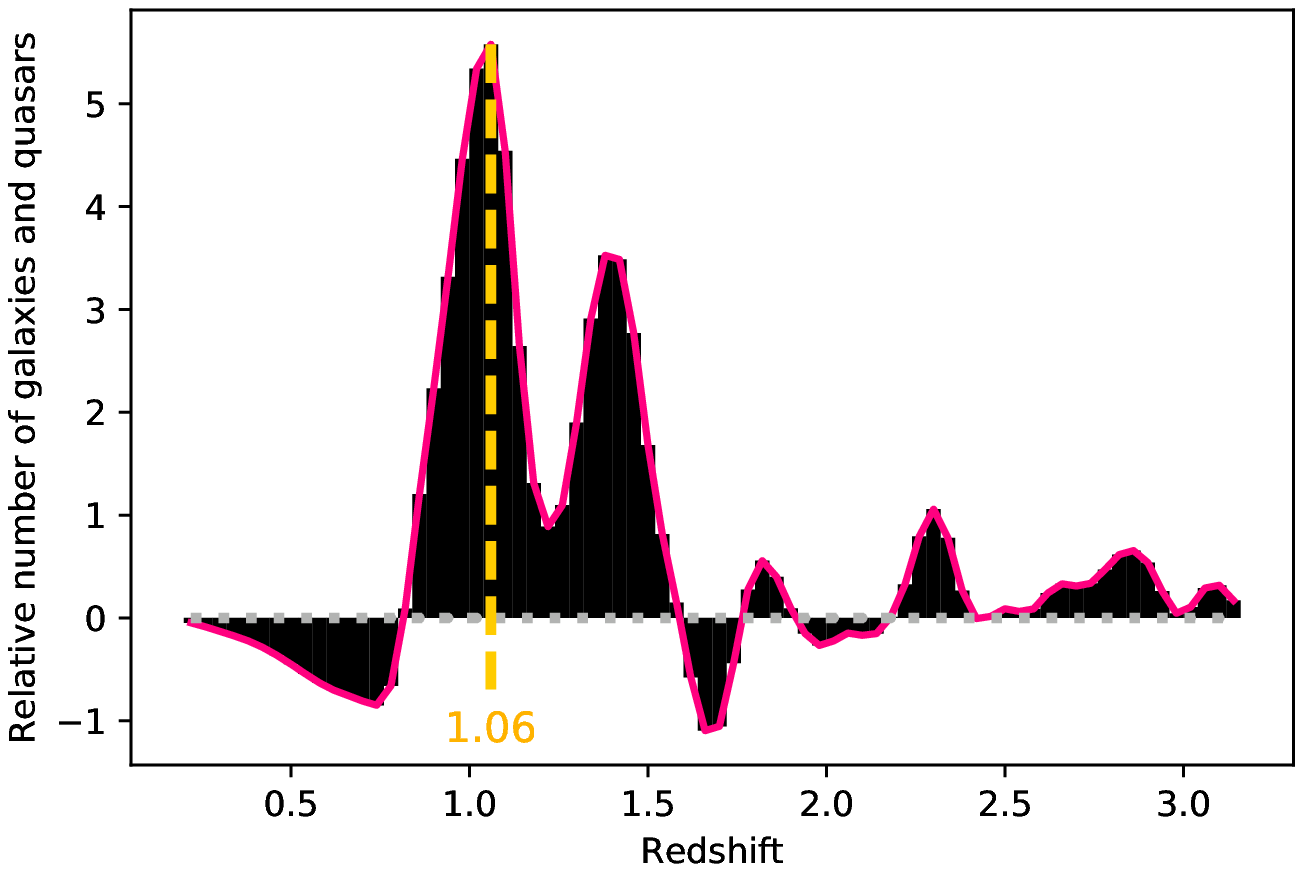}
\end{subfigure}
\hfill
\begin{subfigure}{0.3\textwidth}
\includegraphics[width=6cm]{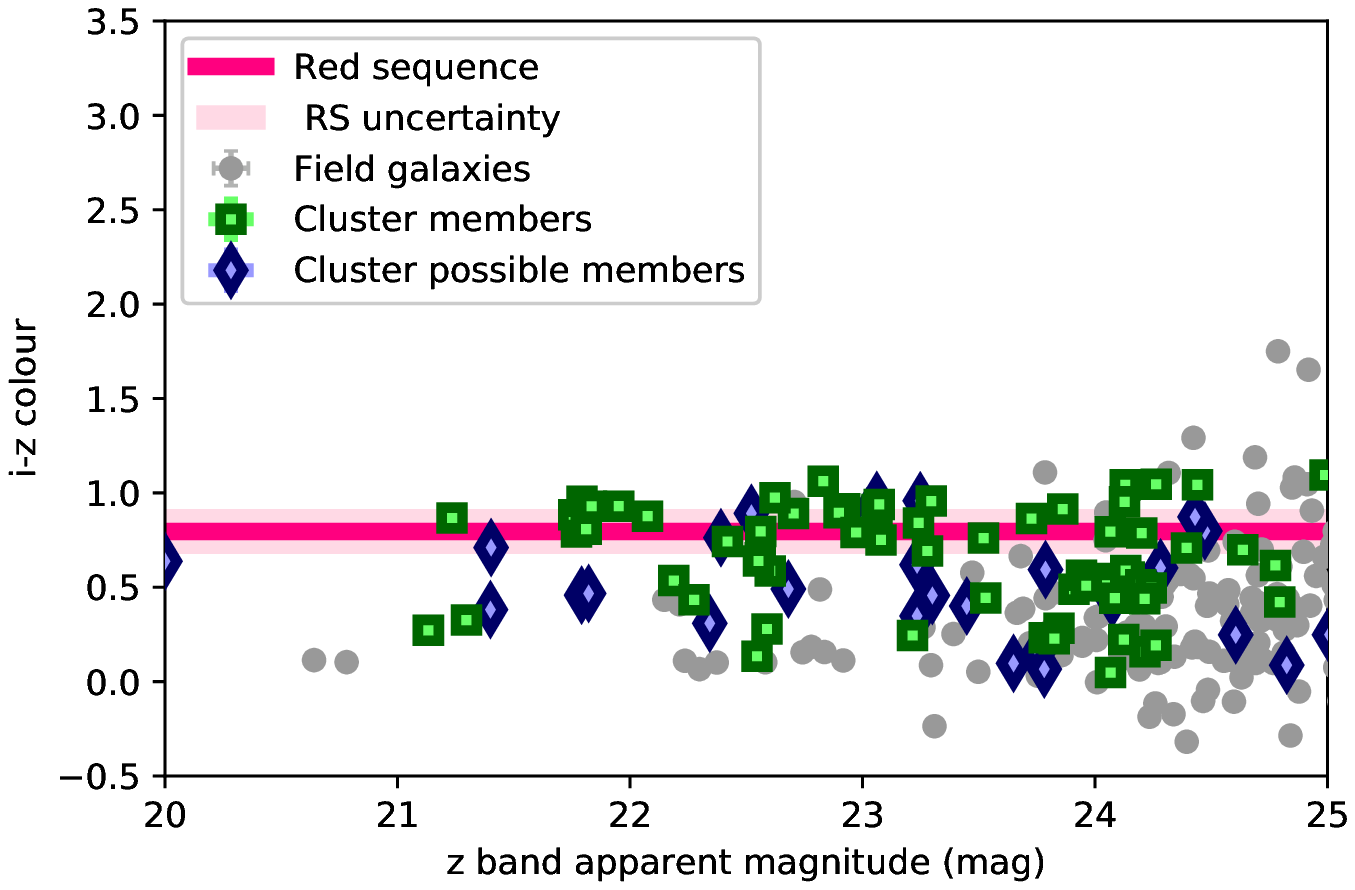}
\end{subfigure}

\begin{subfigure}{0.3\textwidth}
\includegraphics[width=5.25cm]{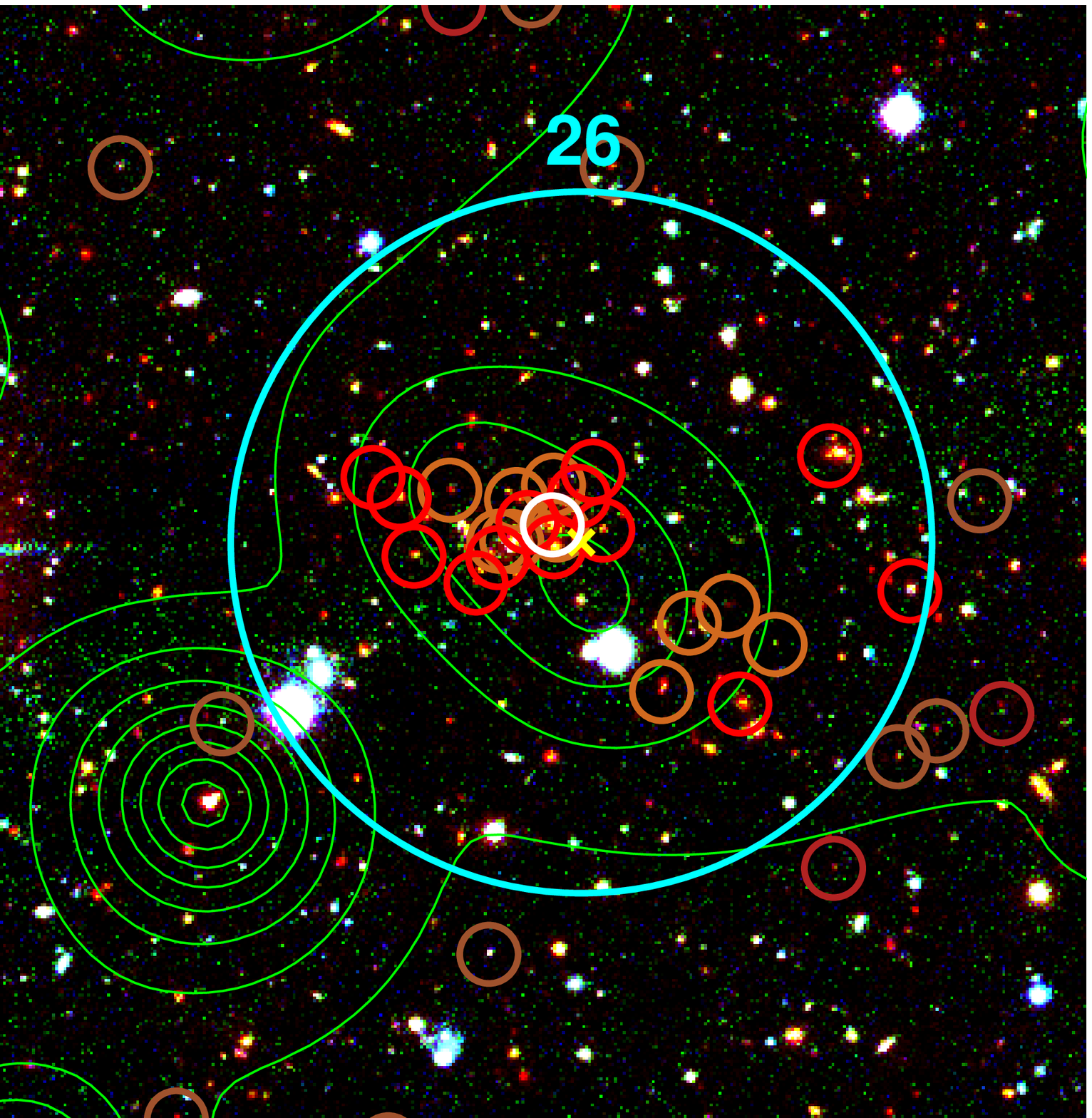}
\end{subfigure}
\hfill
\begin{subfigure}{0.3\textwidth}
\includegraphics[width=6cm]{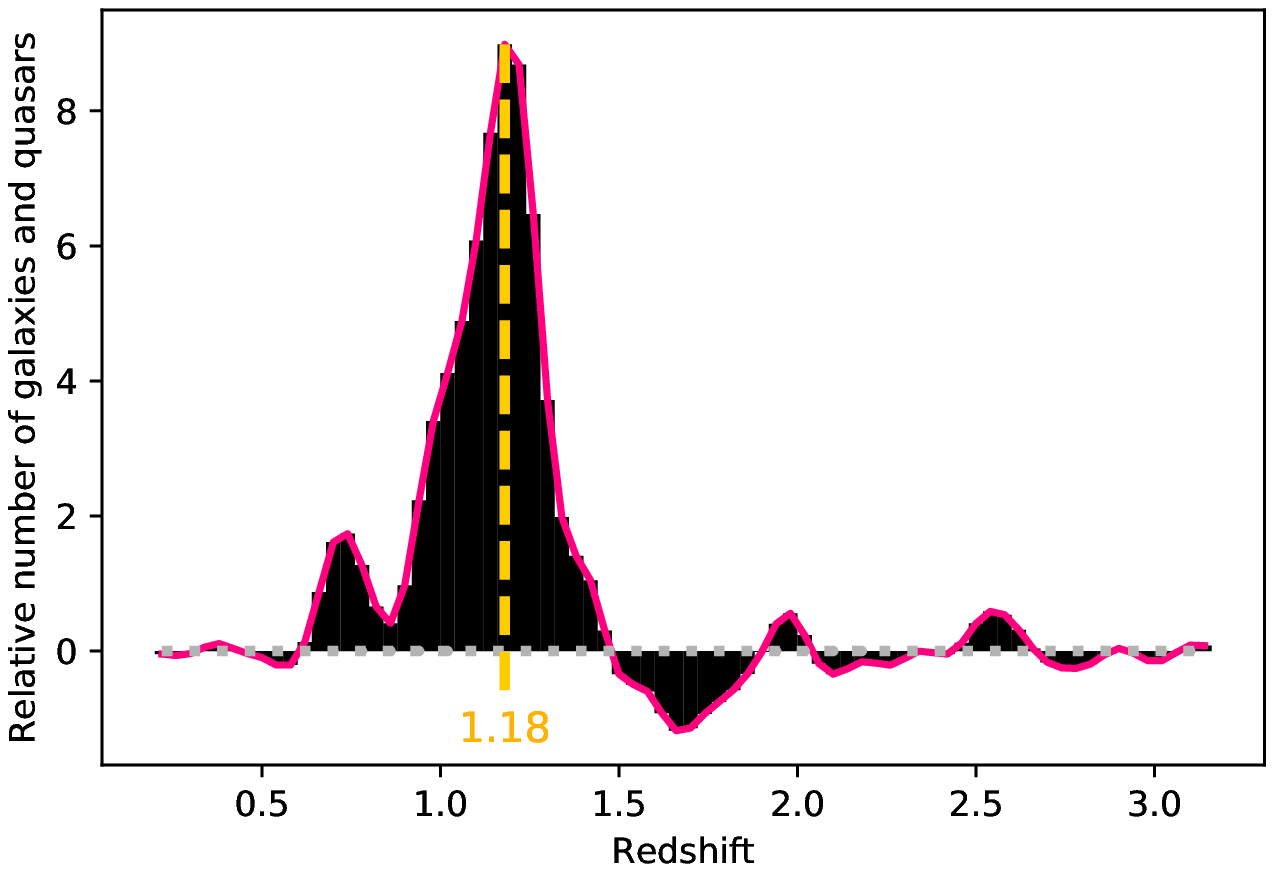}
\end{subfigure}
\hfill
\begin{subfigure}{0.3\textwidth}
\includegraphics[width=6cm]{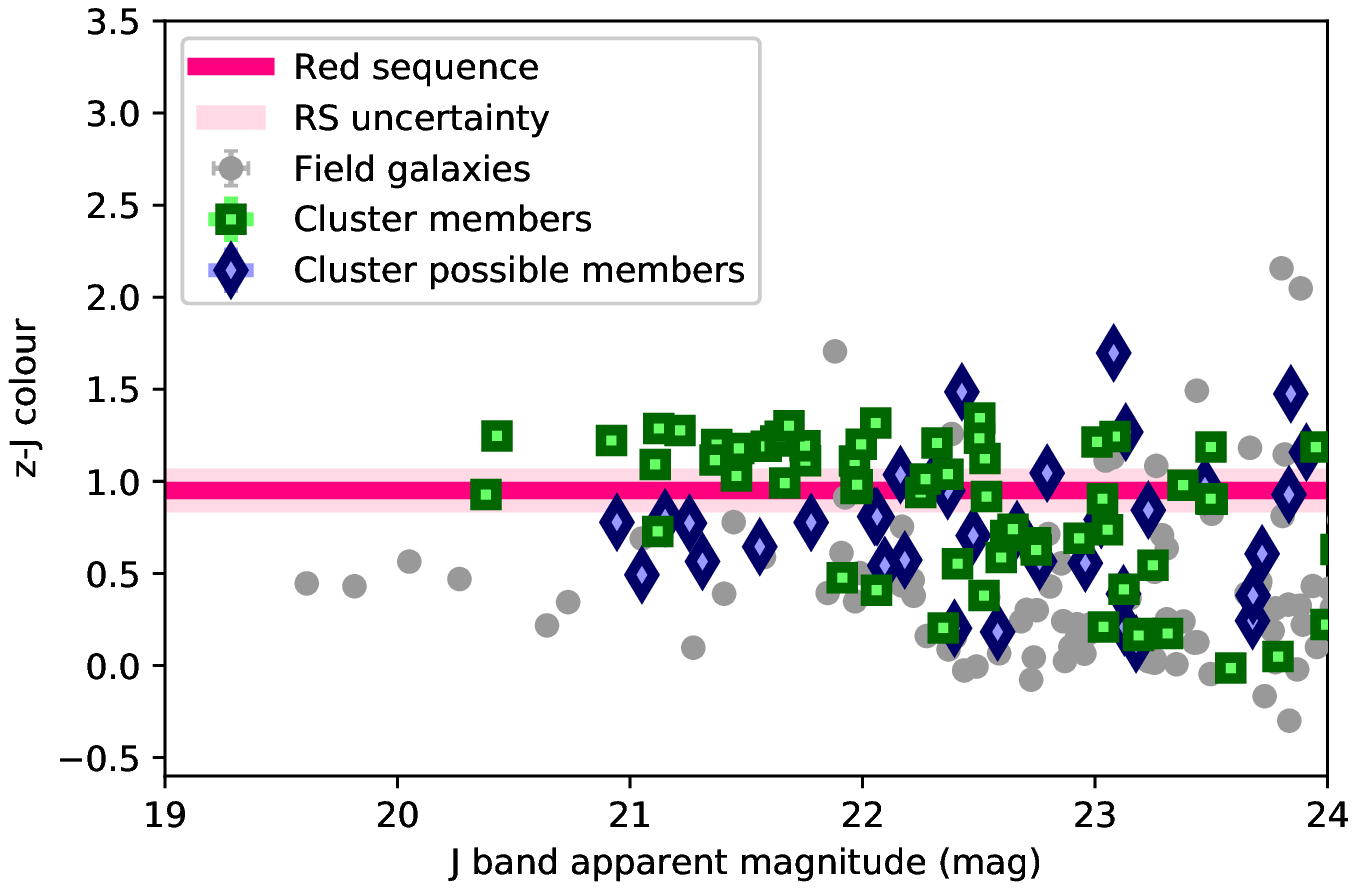}
\end{subfigure}

\centering
\begin{subfigure}{0.3\textwidth}
\includegraphics[width=5.25cm]{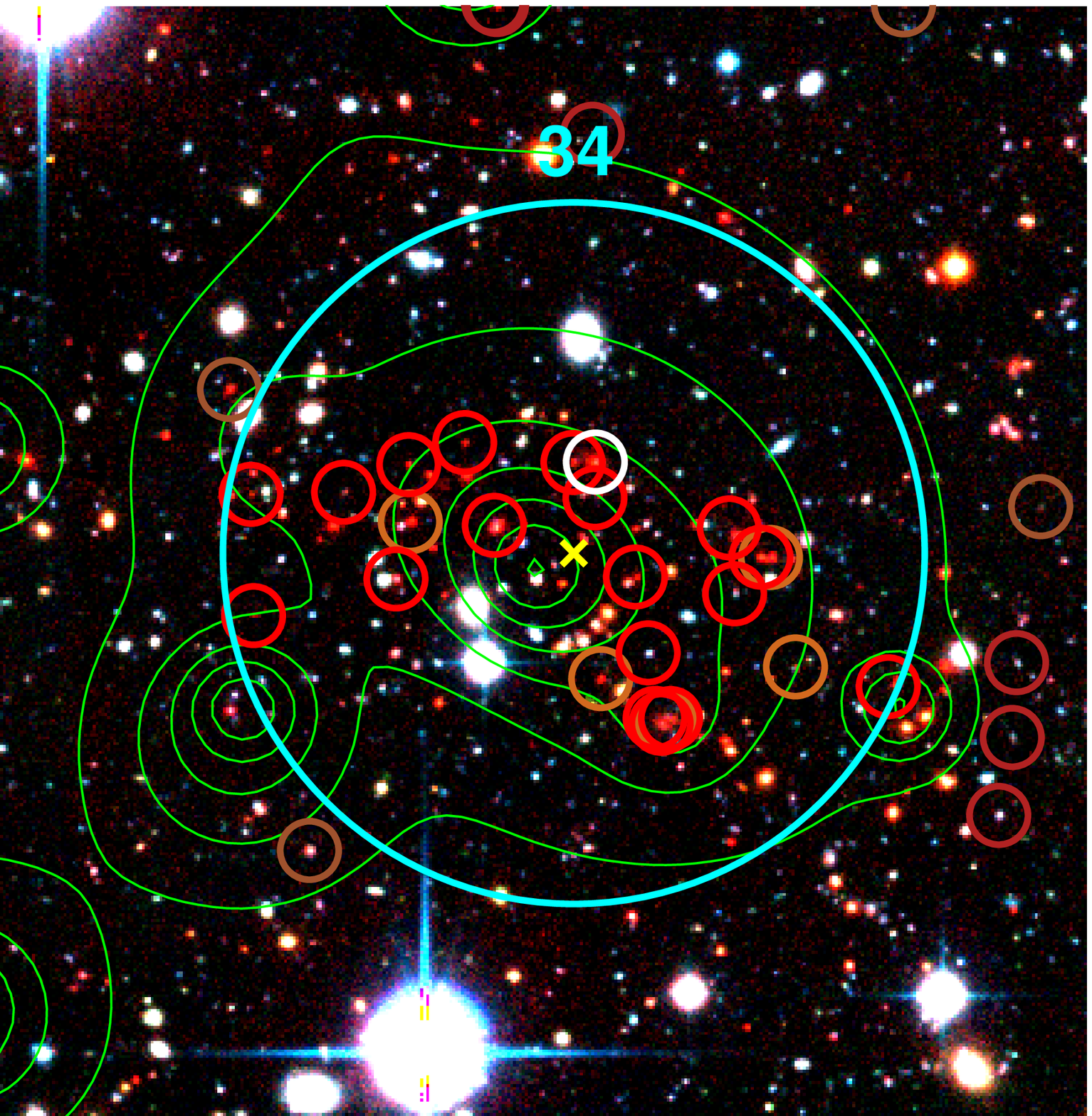}
\end{subfigure}
\hfill
\begin{subfigure}{0.3\textwidth}
\includegraphics[width=6cm]{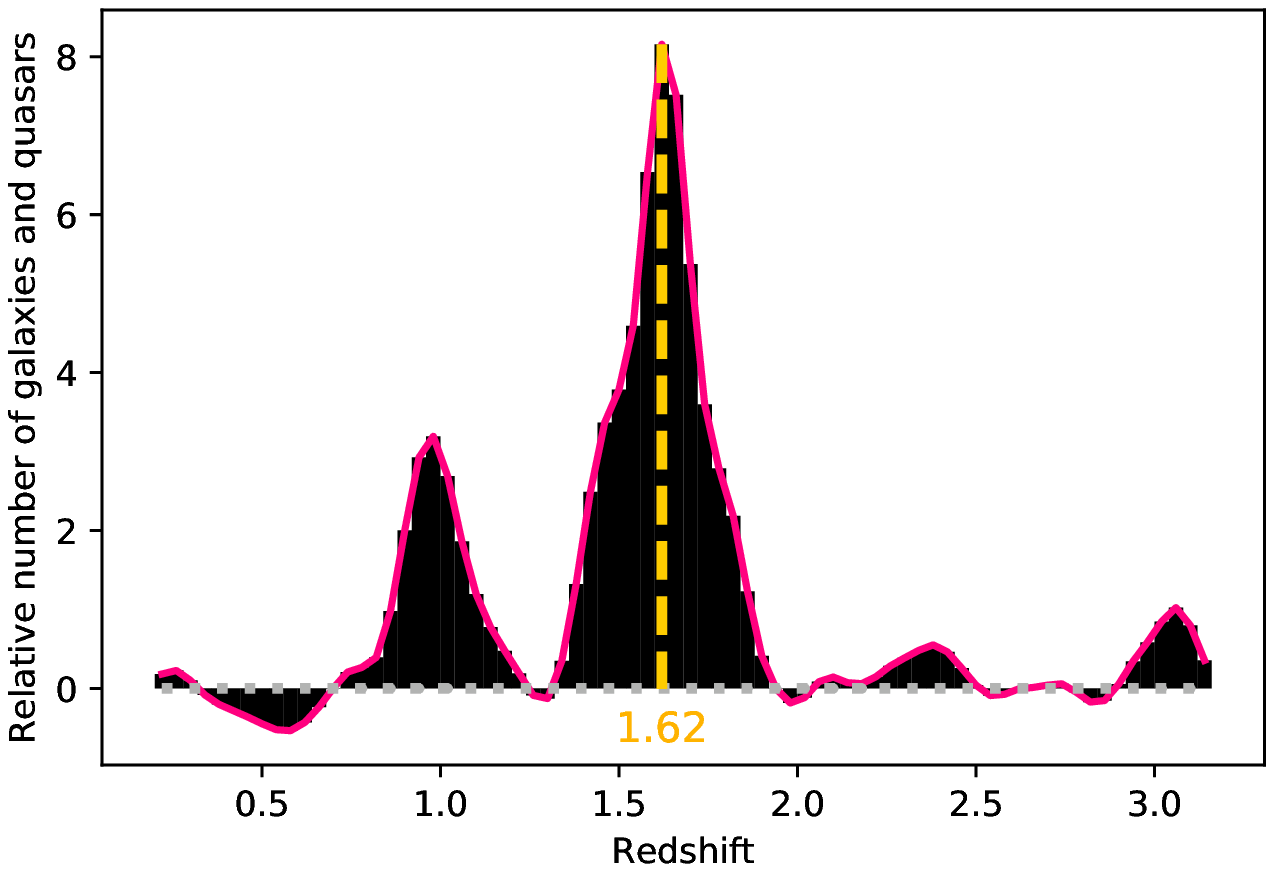}
\end{subfigure}
\hfill
\begin{subfigure}{0.3\textwidth}
\includegraphics[width=6cm]{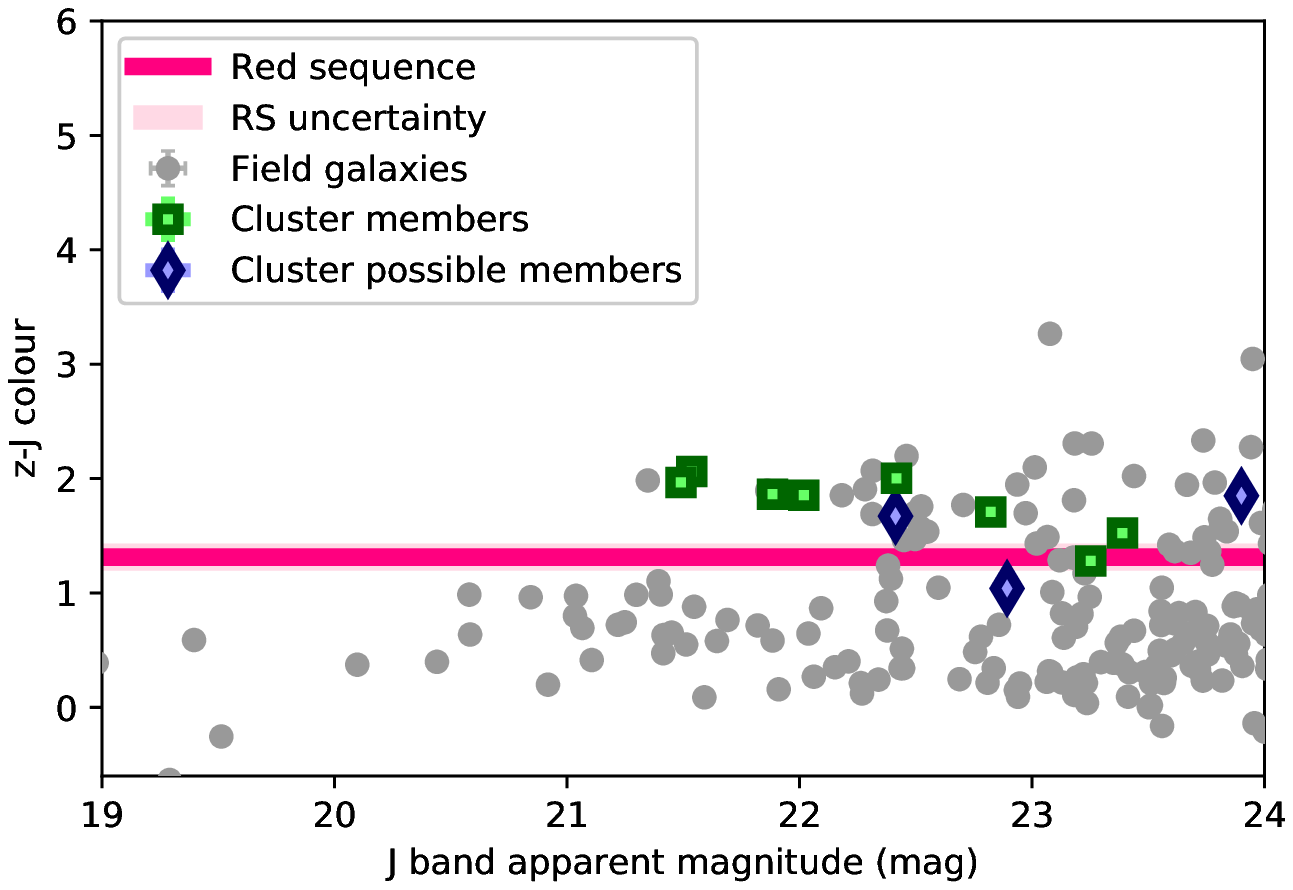}
\end{subfigure}
\caption{\textit{continued}}
\label{fig_known_clusters}
\end{figure*}

\begin{figure*}
\centering
\begin{subfigure}{0.3\textwidth}
\includegraphics[width=5.25cm]{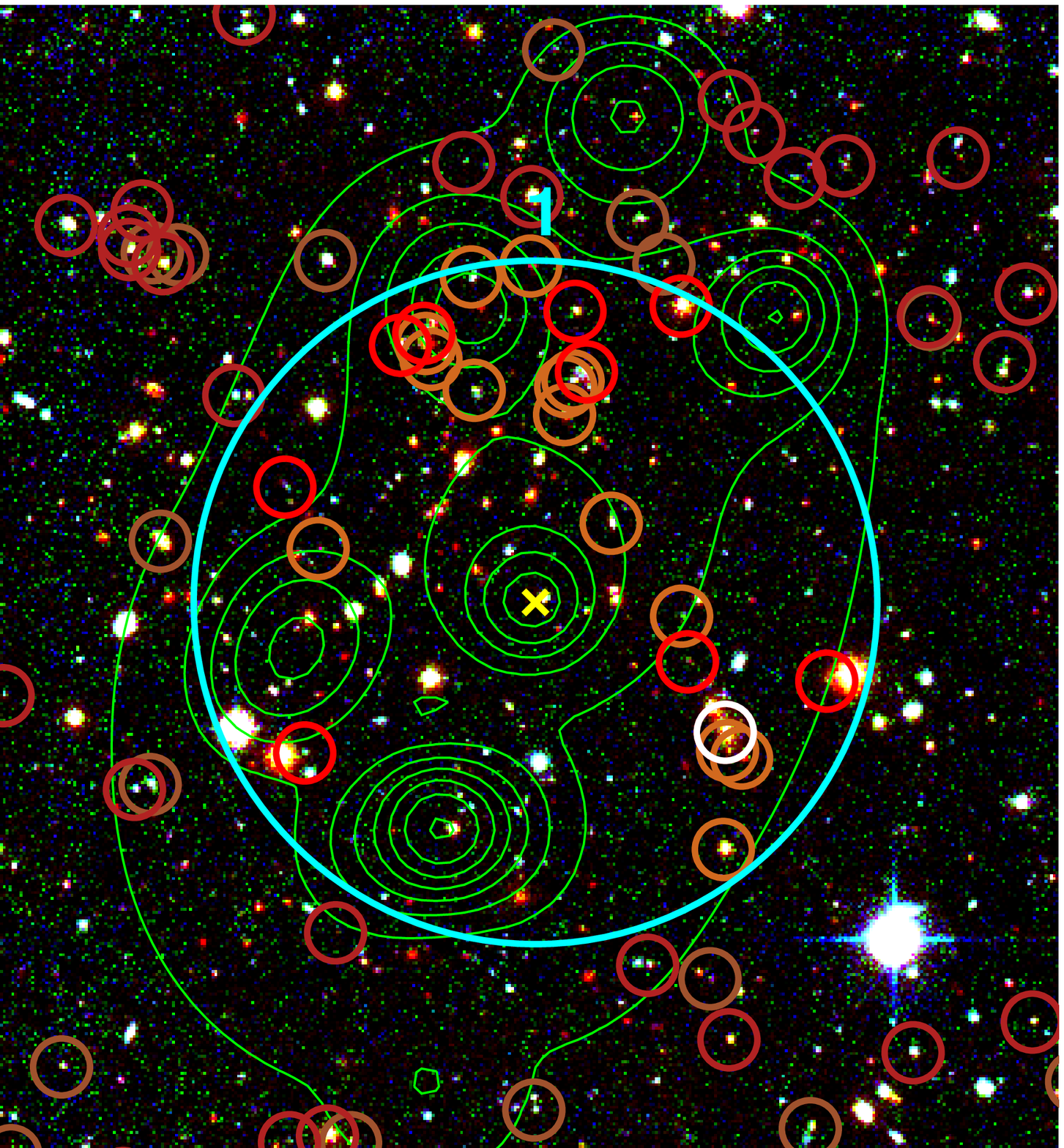}
\end{subfigure}
\hfill
\begin{subfigure}{0.3\textwidth}
\includegraphics[width=6cm]{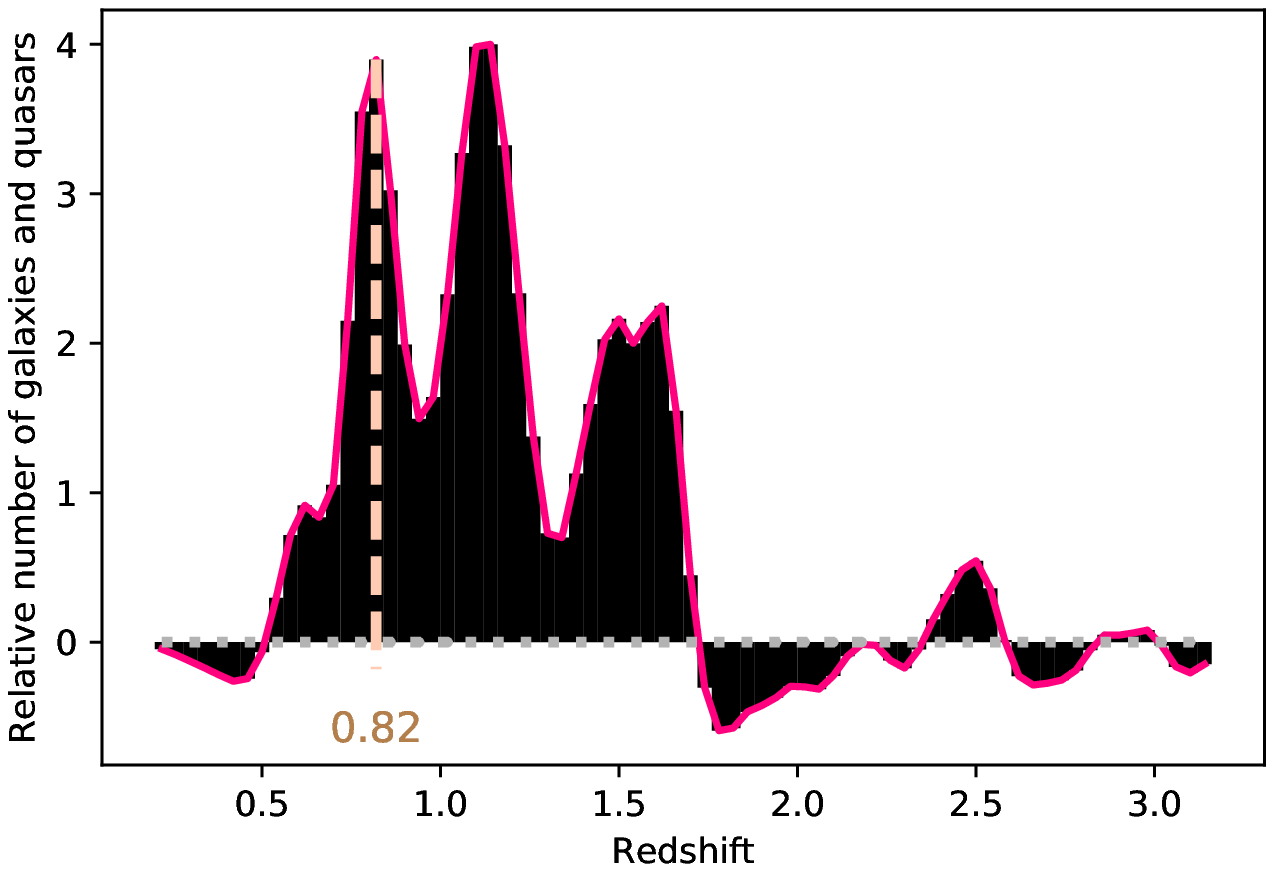}
\end{subfigure}
\hfill
\begin{subfigure}{0.3\textwidth}
\includegraphics[width=6cm]{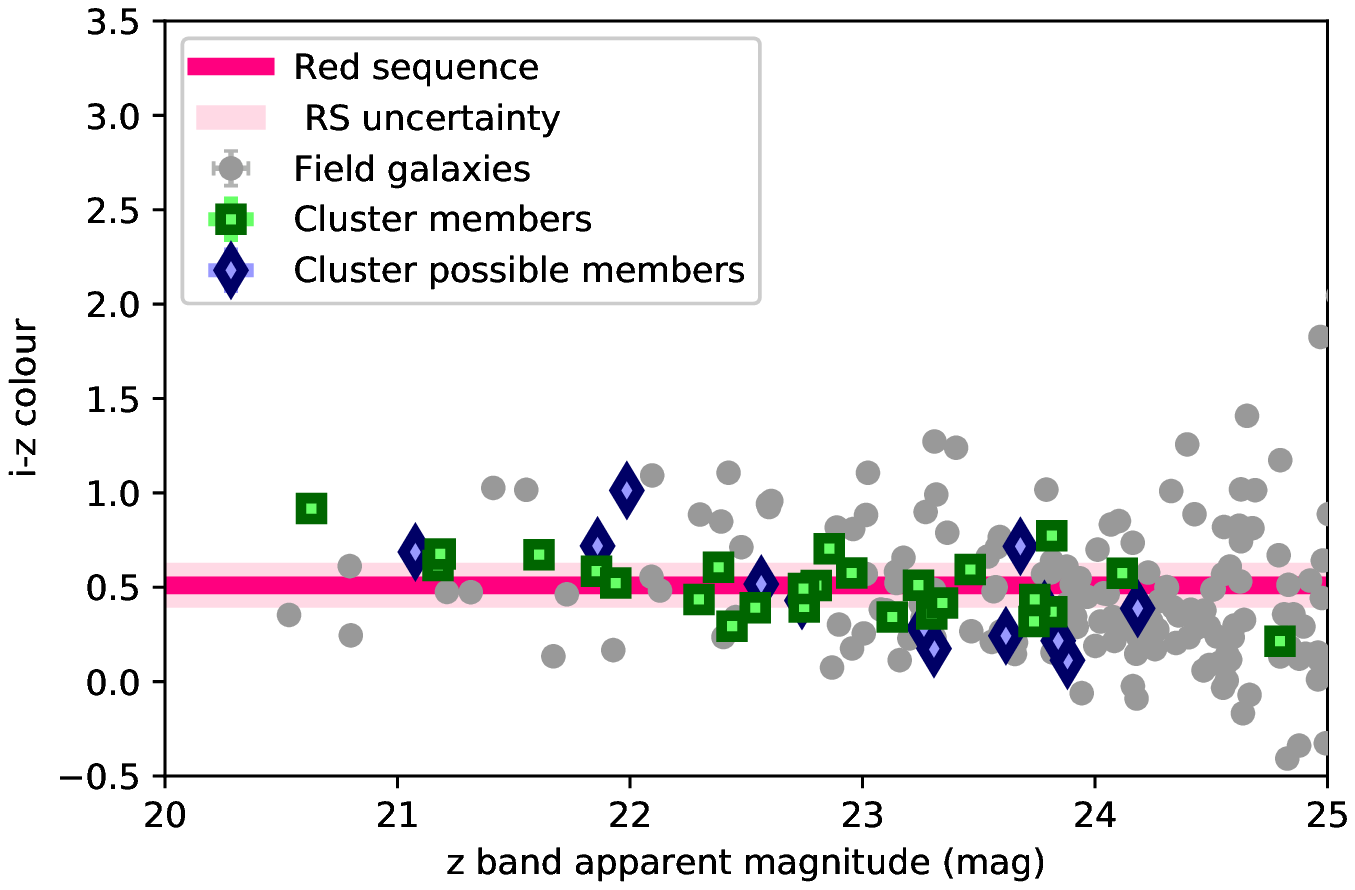}
\end{subfigure}

\centering
\begin{subfigure}{0.3\textwidth}
\includegraphics[width=5.25cm]{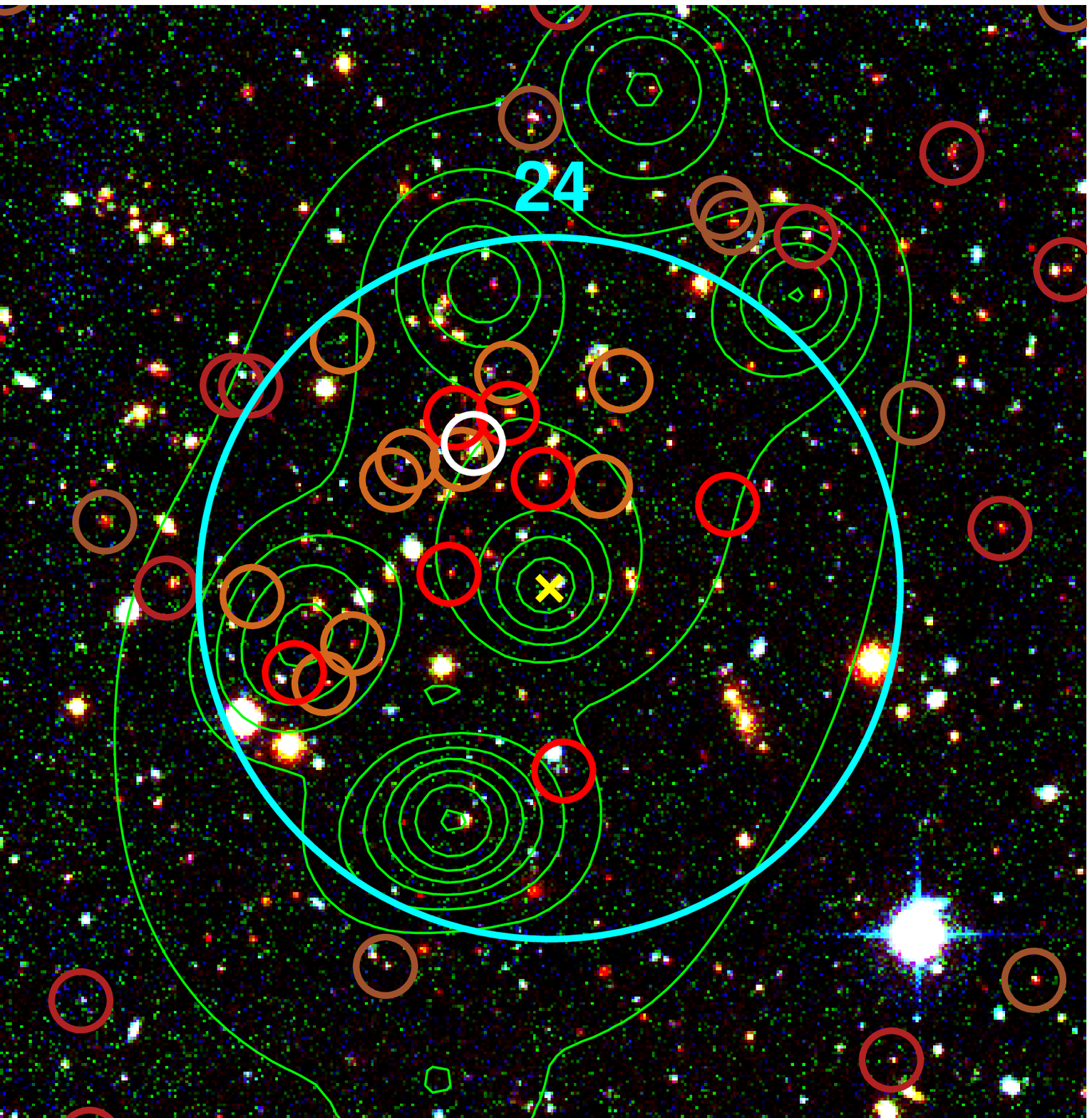}
\end{subfigure}
\hfill
\begin{subfigure}{0.3\textwidth}
\includegraphics[width=6cm]{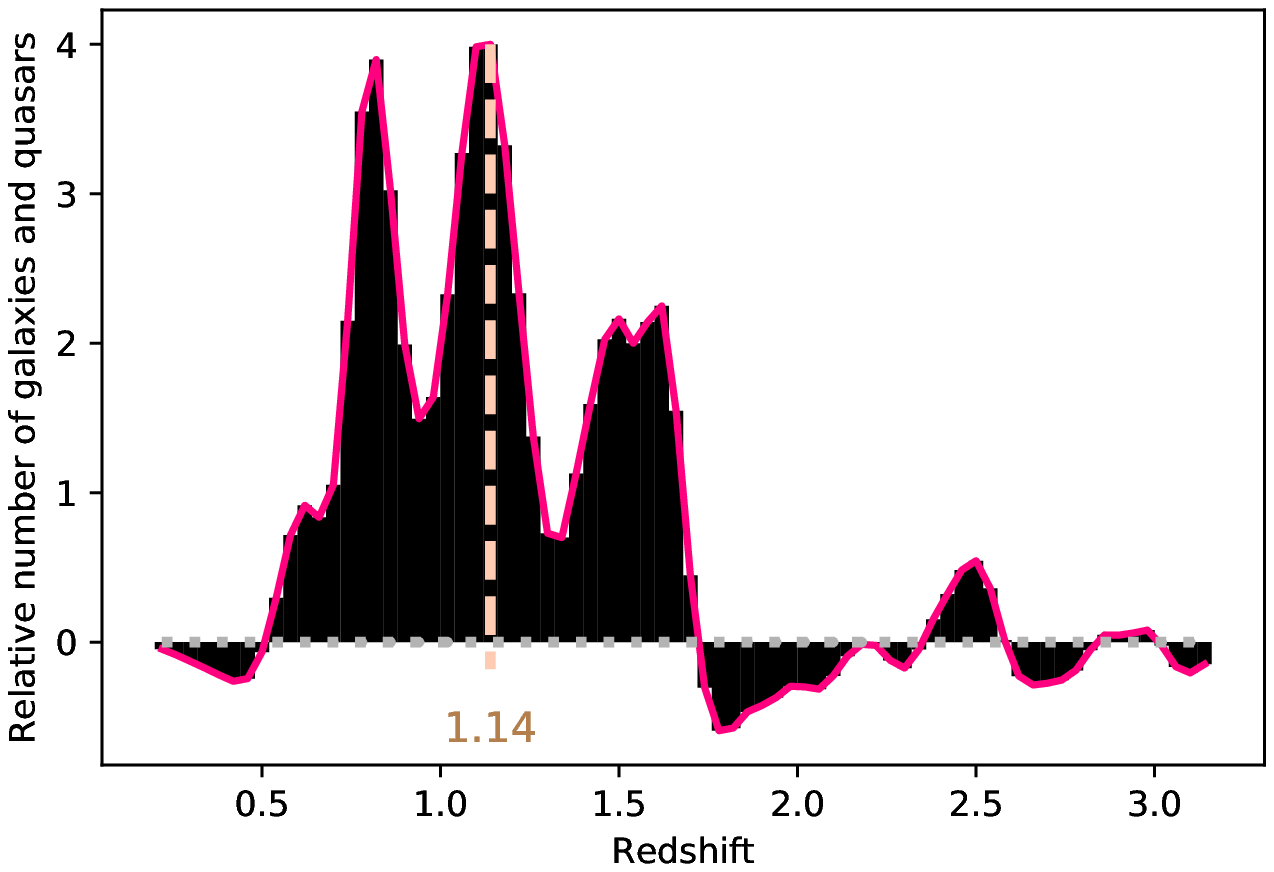}
\end{subfigure}
\hfill
\begin{subfigure}{0.3\textwidth}
\includegraphics[width=6cm]{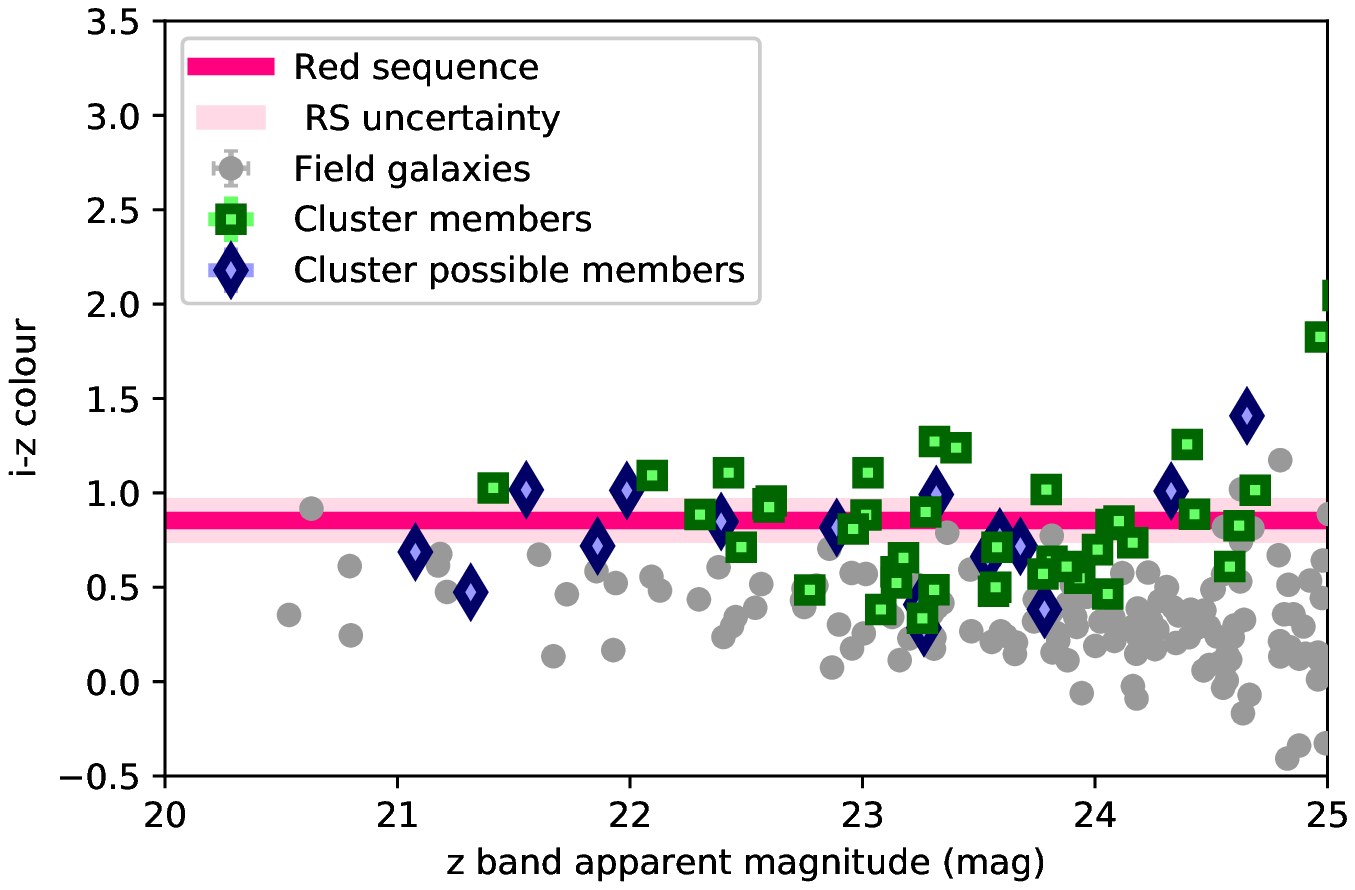}
\end{subfigure}
\caption{Left Cols.: Megacam R and I filter and VIDEO H filter images for the two candidates clusters along the same line-of-sight. Symbols and contours definitions are given in Fig. \ref{fig_known_clusters}. Middle Cols.: Background subtracted and Gaussian filtered redshift distribution of the bright galaxies within the central arcmin, for the corresponding candidates. Bottom Cols.: i-z ($0.8\leq z< 1.2$) or z-J ($z\geq 1.2$) CMD plot of the galaxies above VIDEO 5$\sigma$ limit within 1 arcmin of the centre.}
\label{fig_twin_clusters}
\end{figure*}

\begin{figure*}
\centering
\begin{subfigure}{0.3\textwidth}
\includegraphics[width=5.25cm]{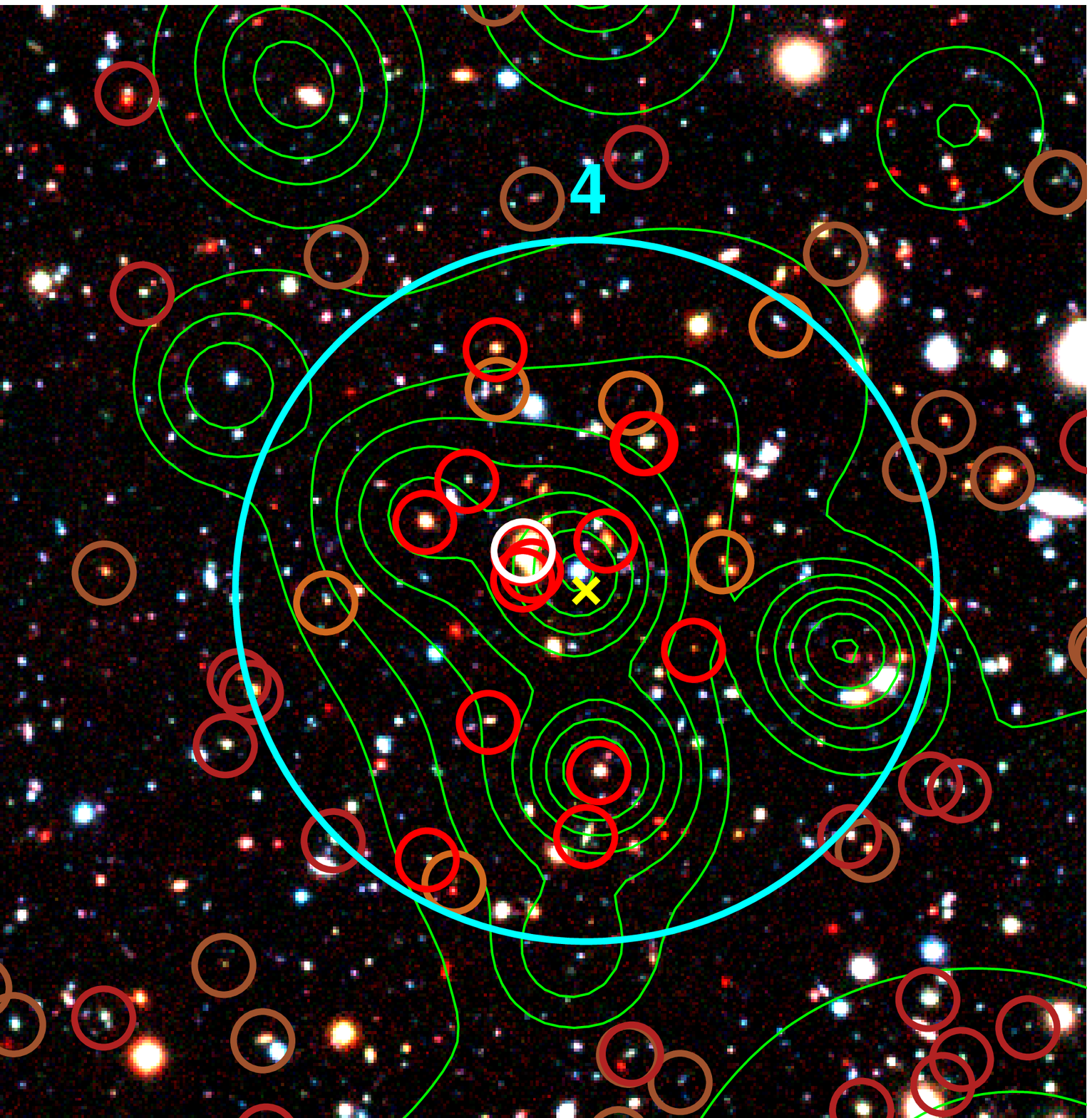}
\end{subfigure}
\hfill
\begin{subfigure}{0.3\textwidth}
\includegraphics[width=6cm]{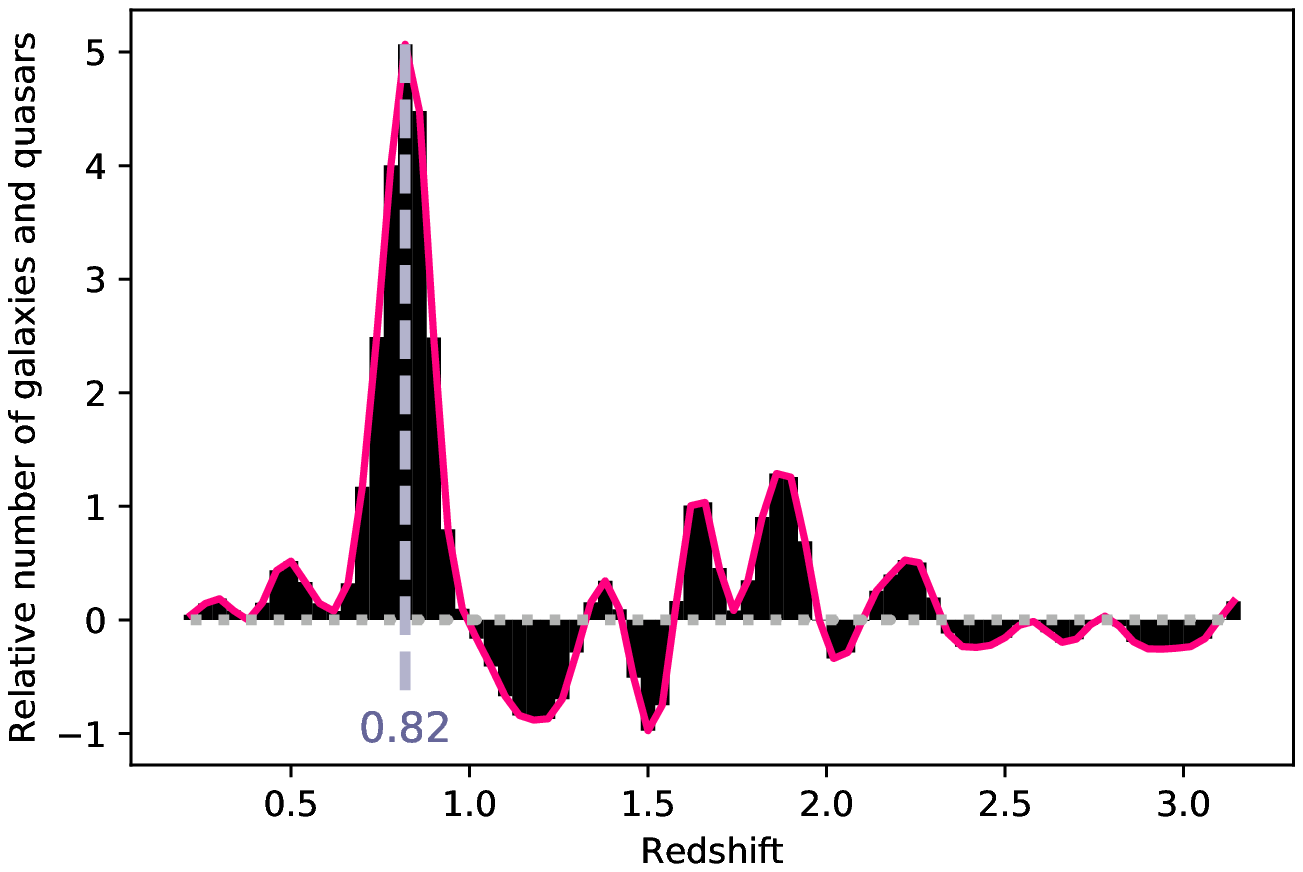}
\end{subfigure}
\hfill
\begin{subfigure}{0.3\textwidth}
\includegraphics[width=6cm]{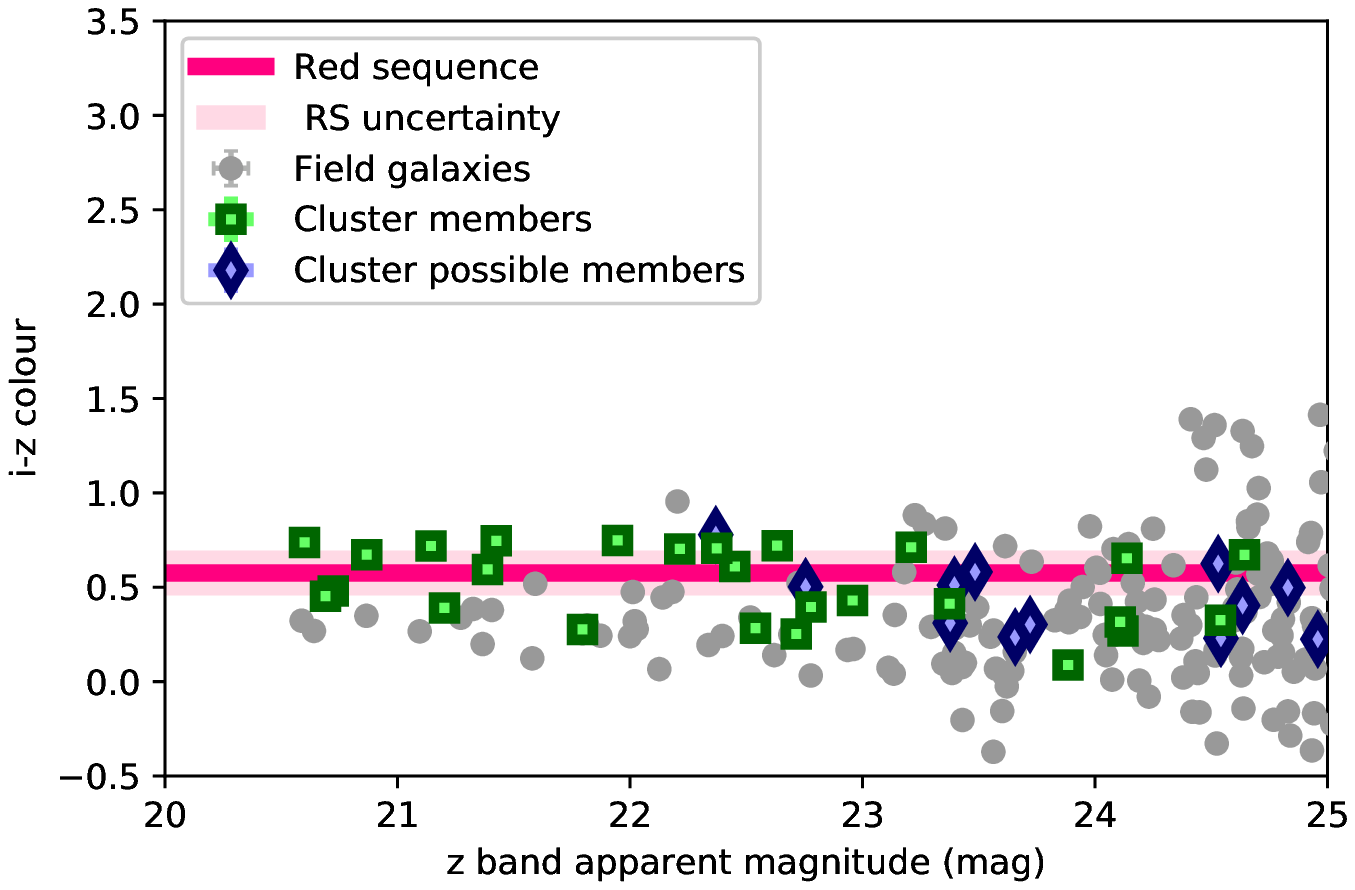}
\end{subfigure}

\centering
\begin{subfigure}{0.3\textwidth}
\includegraphics[width=5.25cm]{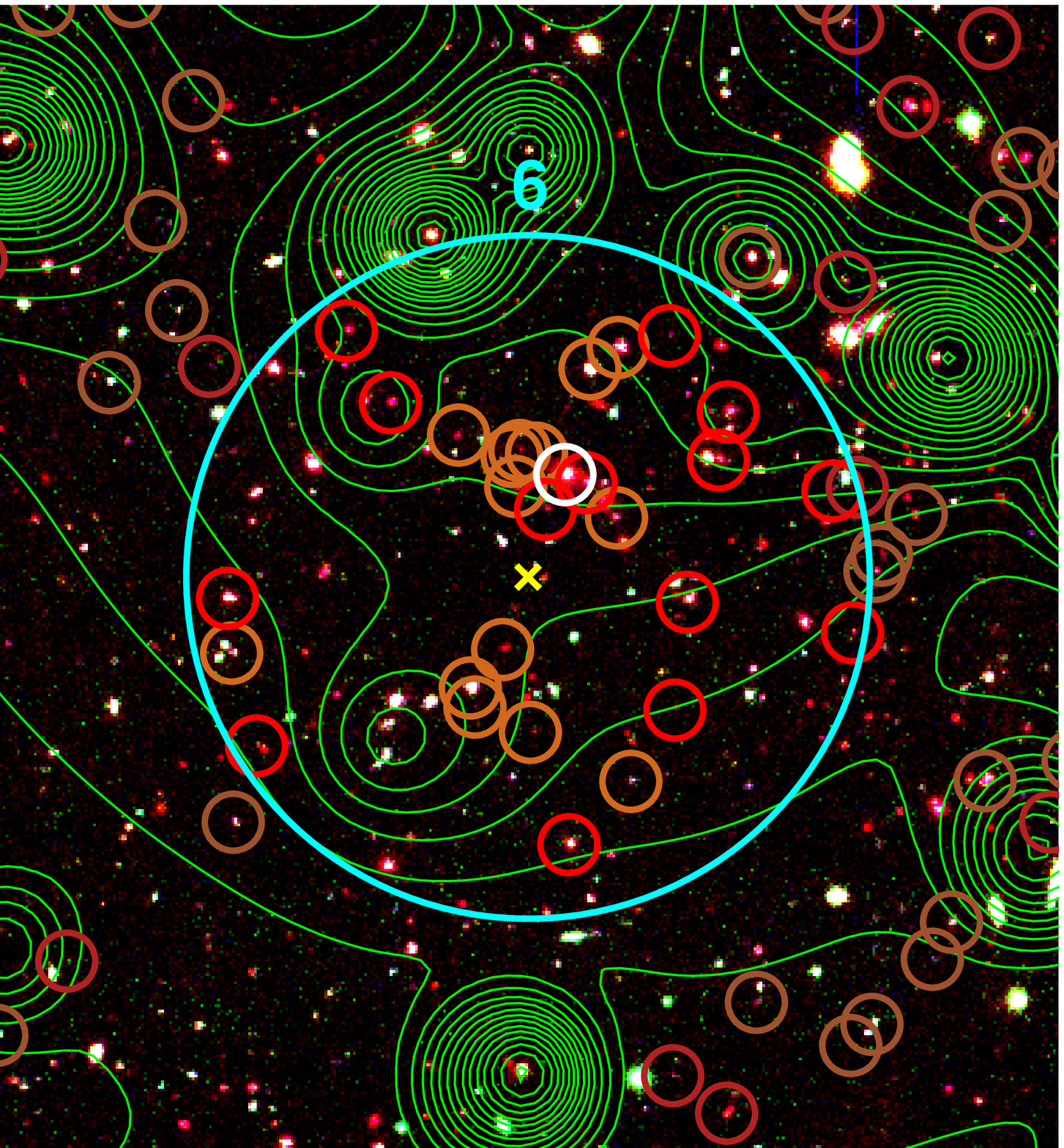}
\end{subfigure}
\hfill
\begin{subfigure}{0.3\textwidth}
\includegraphics[width=6cm]{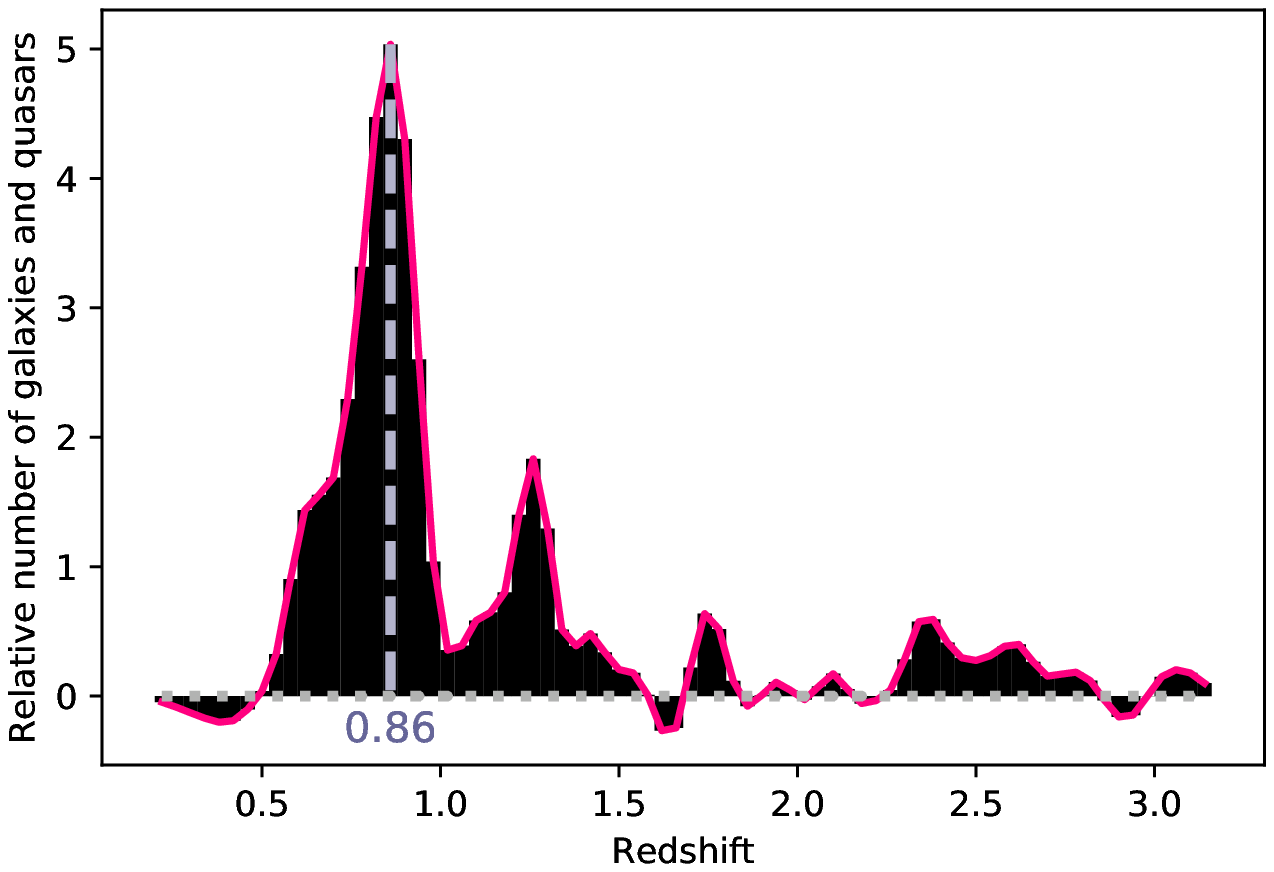}
\end{subfigure}
\hfill
\begin{subfigure}{0.3\textwidth}
\includegraphics[width=6cm]{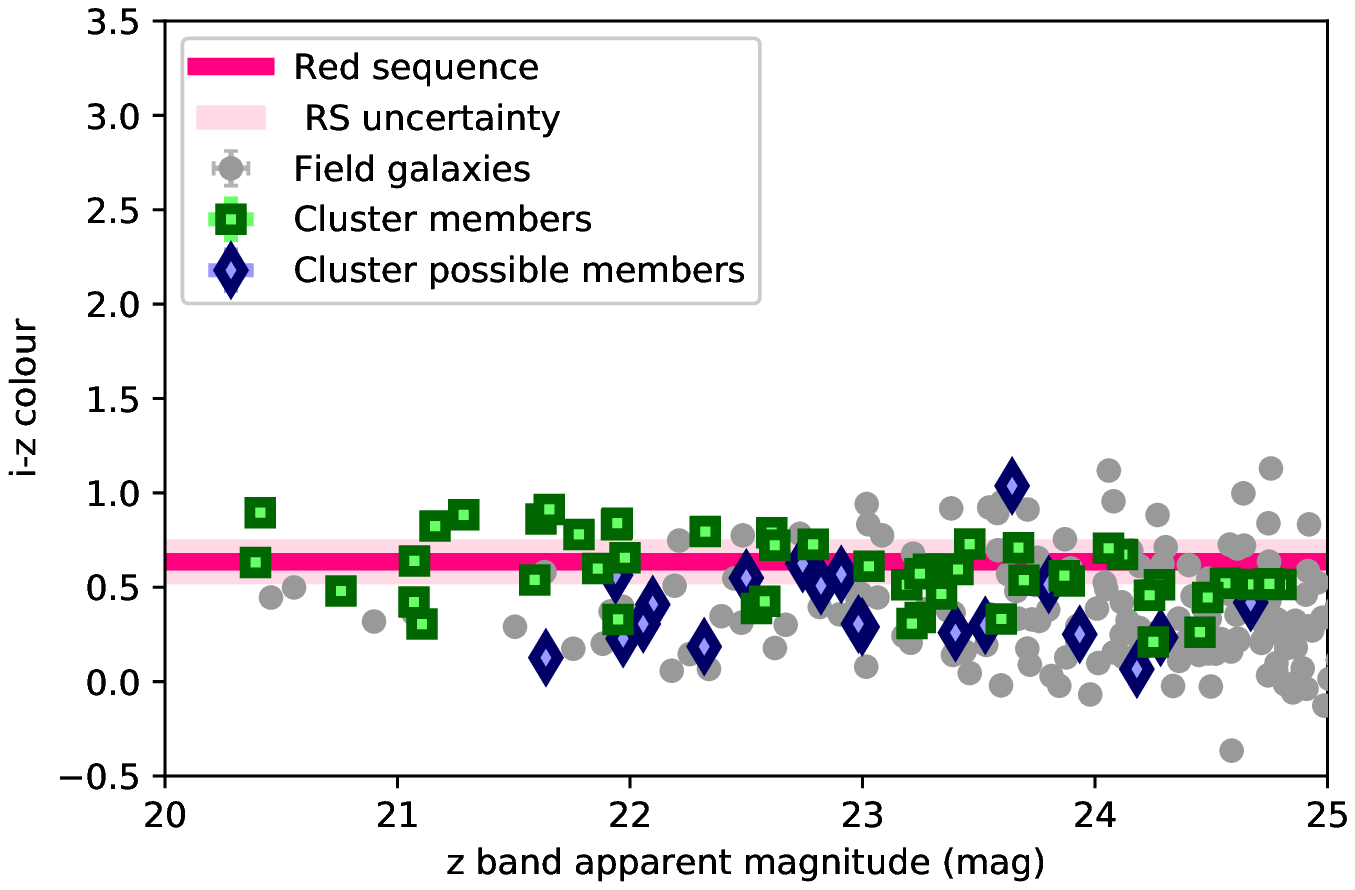}
\end{subfigure}

\centering
\begin{subfigure}{0.3\textwidth}
\includegraphics[width=5.25cm]{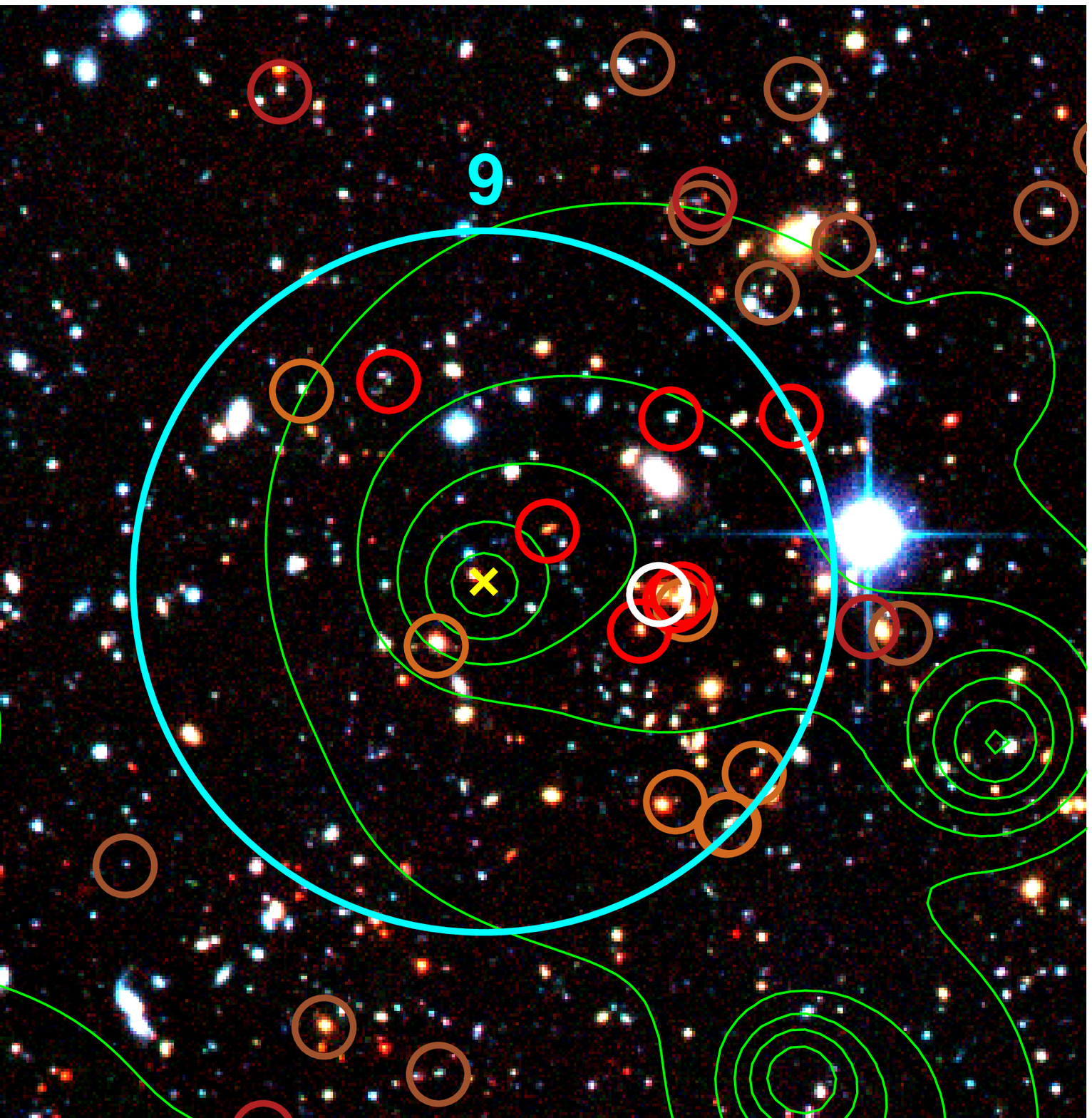}
\end{subfigure}
\hfill
\begin{subfigure}{0.3\textwidth}
\includegraphics[width=6cm]{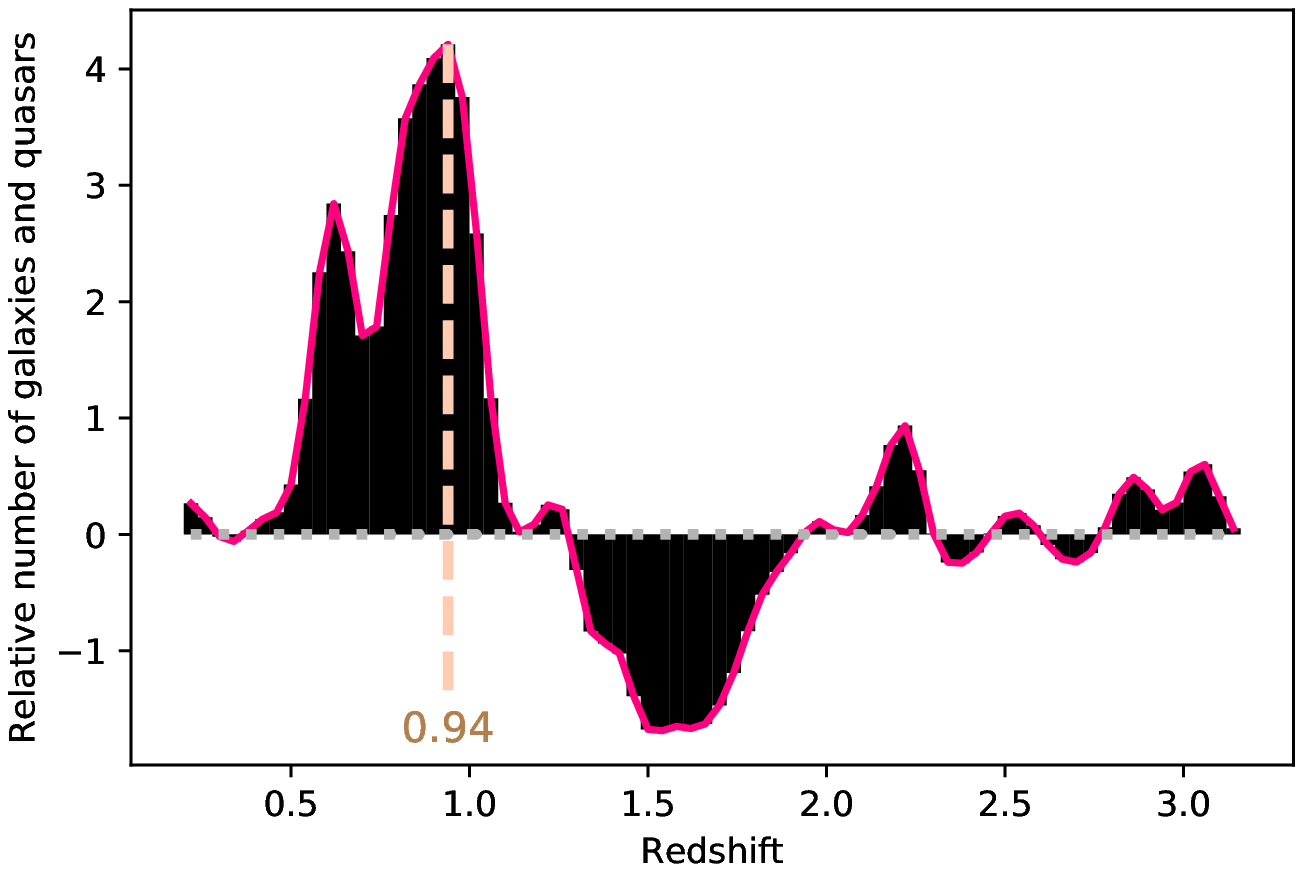}
\end{subfigure}
\hfill
\begin{subfigure}{0.3\textwidth}
\includegraphics[width=6cm]{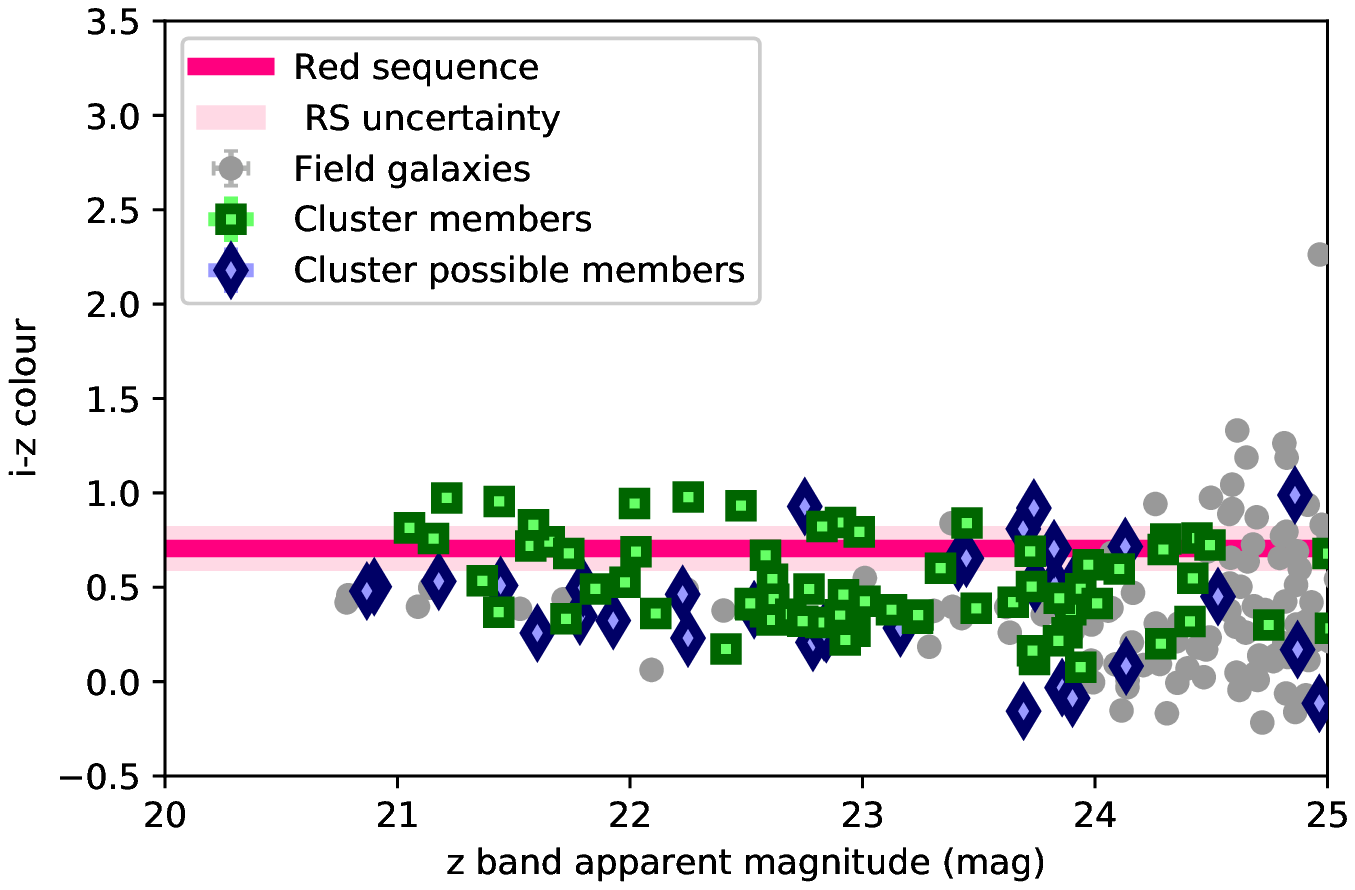}
\end{subfigure}
\centering
\begin{subfigure}{0.3\textwidth}
\includegraphics[width=5.25cm]{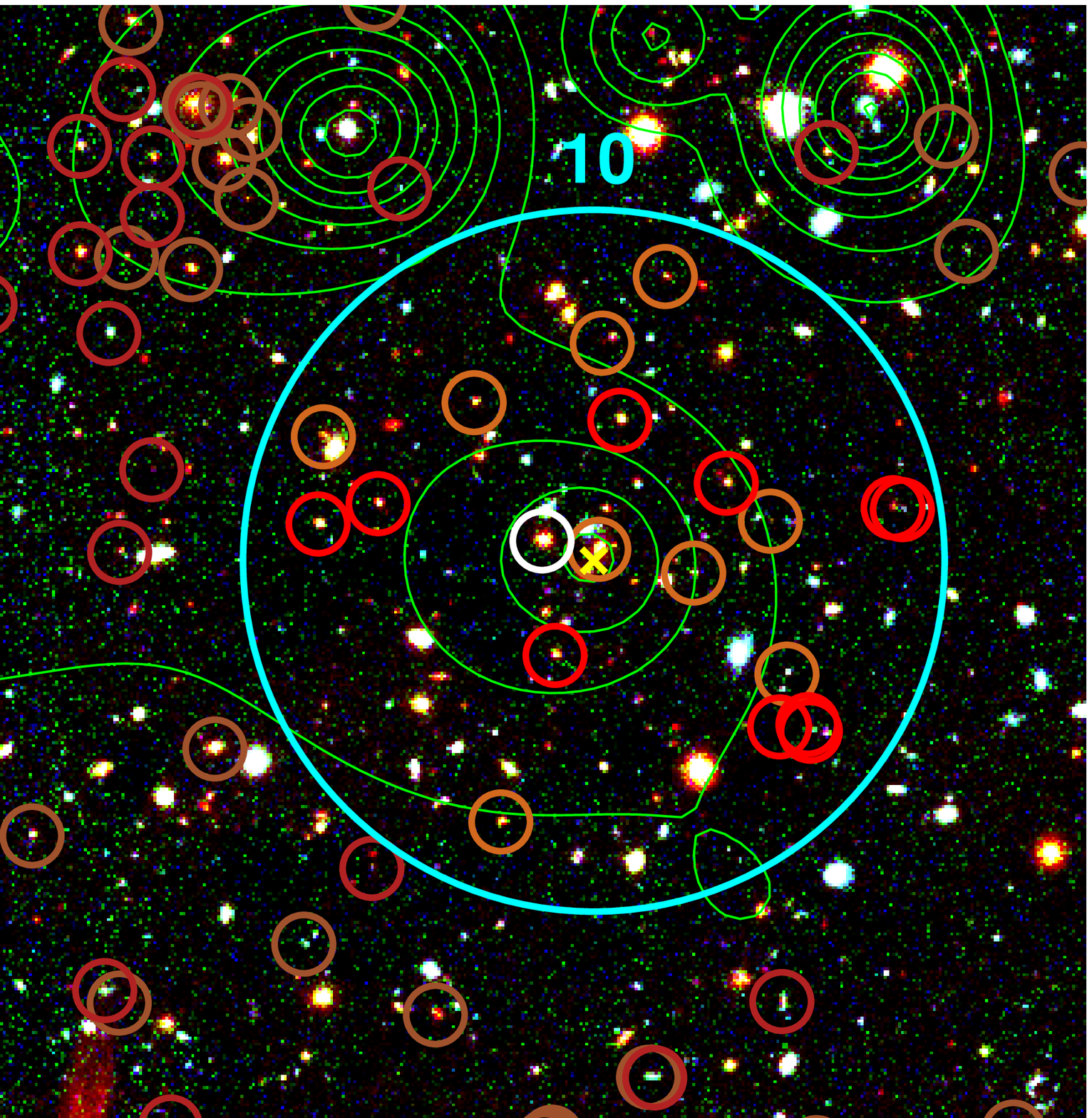}
\end{subfigure}
\hfill
\begin{subfigure}{0.3\textwidth}
\includegraphics[width=6cm]{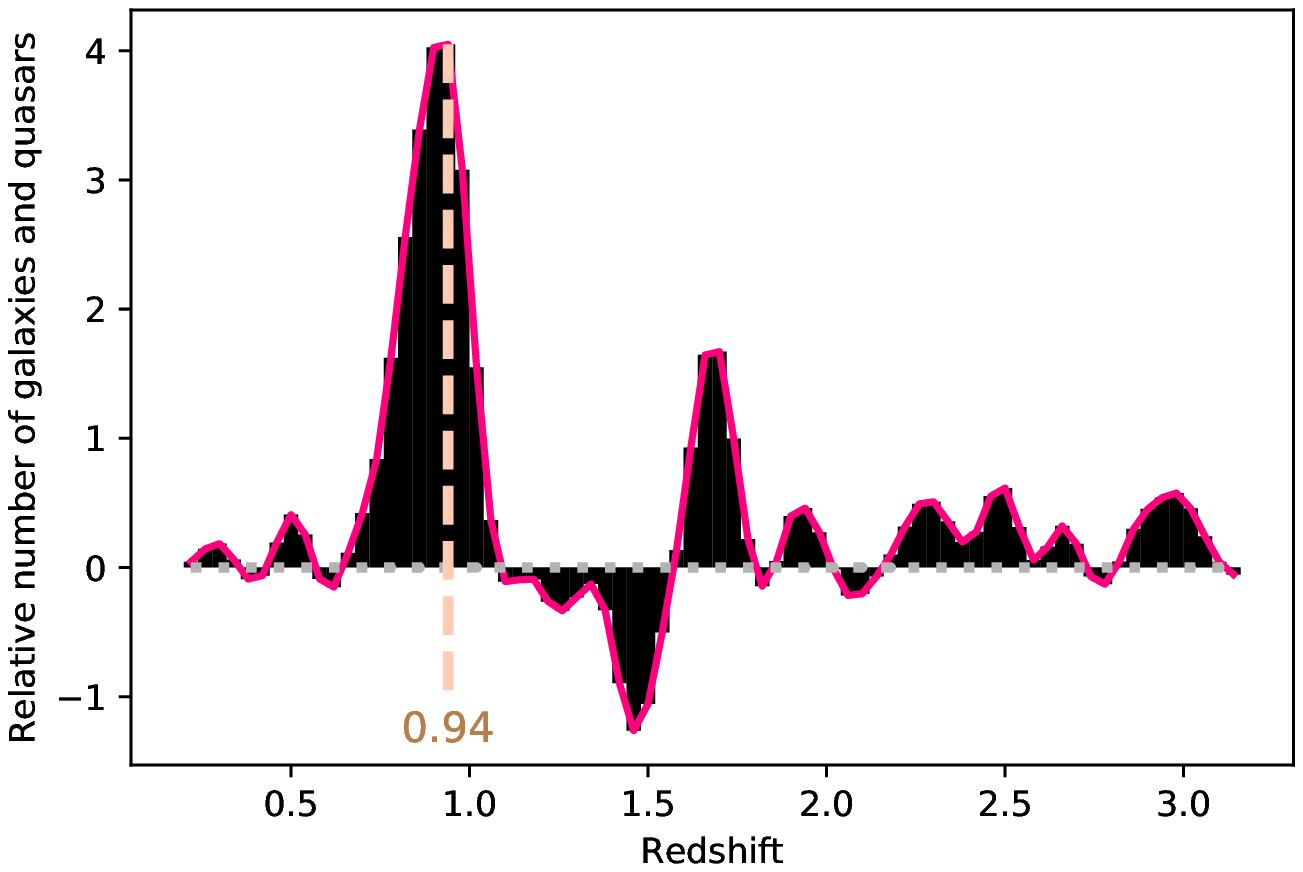}
\end{subfigure}
\hfill
\begin{subfigure}{0.3\textwidth}
\includegraphics[width=6cm]{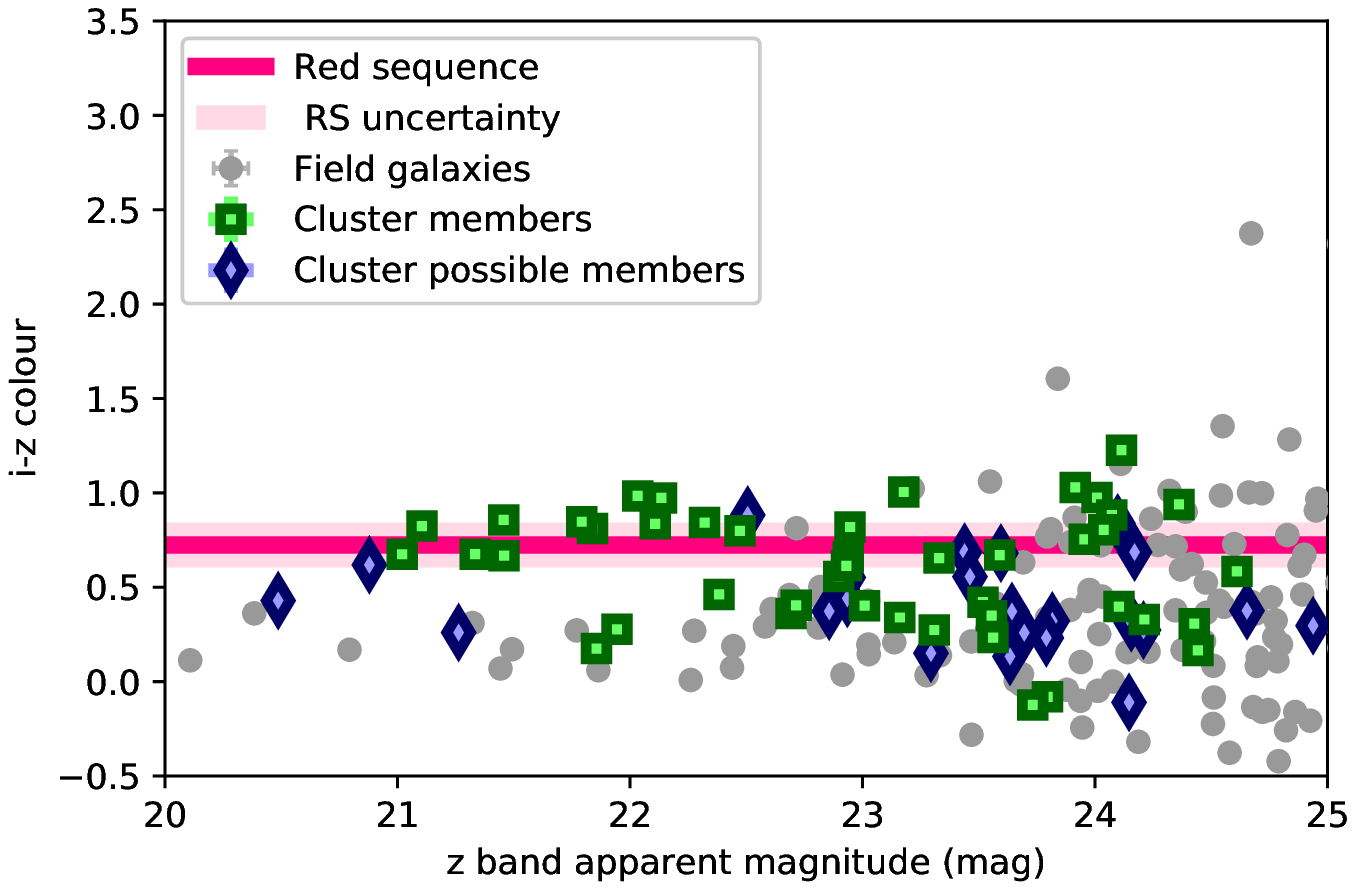}
\end{subfigure}
\caption{Left Cols.: Megacam R and I filter and VIDEO H filter images for the previously detected but unconfirmed candidate clusters at $z\geq 0.8$, classified by increasing redshifts. The X-ray contours in green are logarithmically distributed in 10 levels between the maximum and minimum emission observed in a $7\times 7 ~\mathrm{arcmin^2}$ box around the X-ray source, except for candidate 6 which displays 25 levels based on $4\times 4 ~\mathrm{arcmin^2}$ box. Symbols definitions are given in Fig. \ref{fig_known_clusters}. Middle Cols.: Background subtracted and Gaussian filtered redshift distribution of the bright galaxies within the central arcmin, for the corresponding candidates. Bottom Cols.: i-z ($0.8\leq z< 1.2$) or z-J ($z\geq 1.2$) CMD plot of the galaxies above VIDEO 5$\sigma$ limit within 1 arcmin of the centre.}
\end{figure*}

\newpage

\begin{figure*}
\ContinuedFloat
\centering
\begin{subfigure}{0.3\textwidth}
\includegraphics[width=5.25cm]{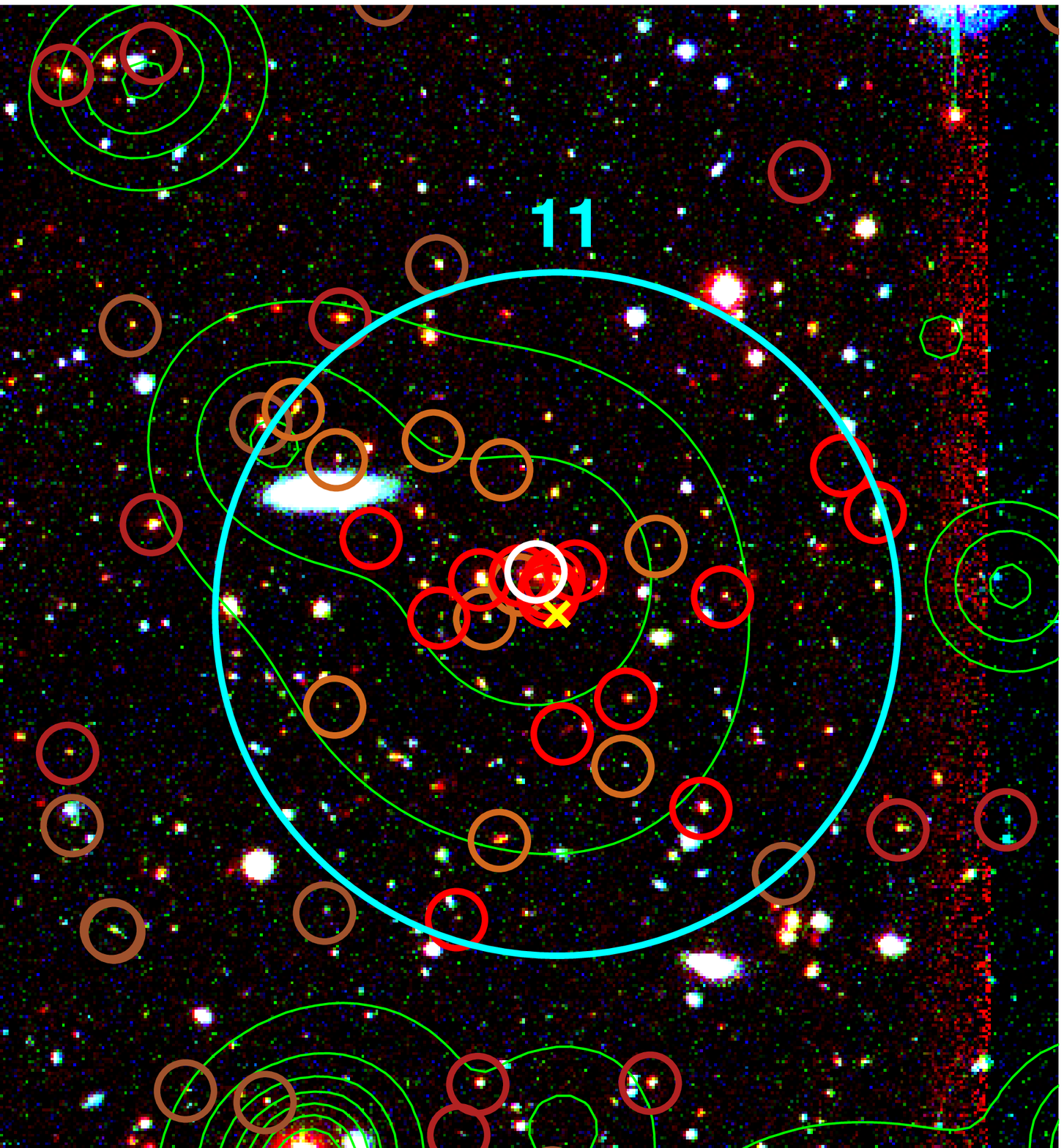}
\end{subfigure}
\hfill
\begin{subfigure}{0.3\textwidth}
\includegraphics[width=6cm]{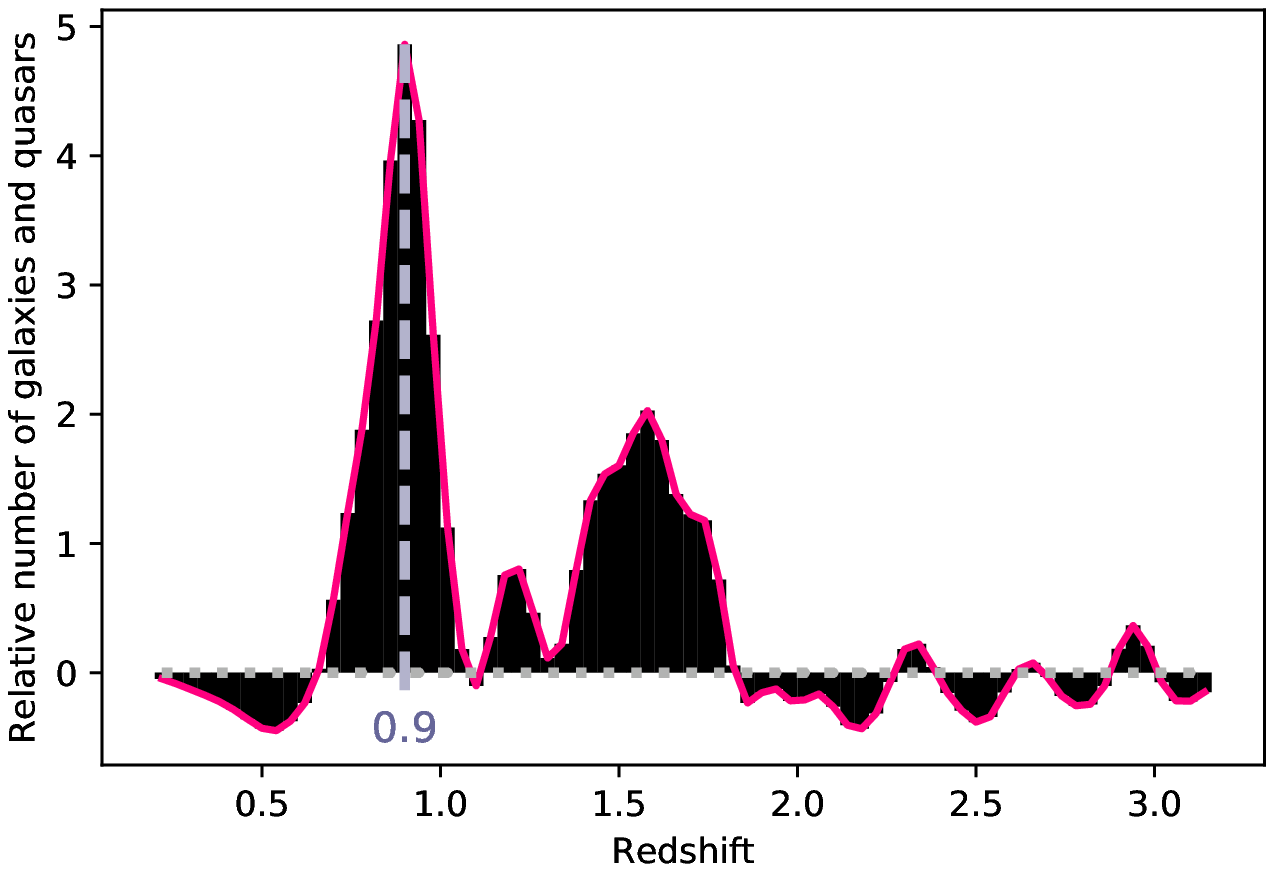}
\end{subfigure}
\hfill
\begin{subfigure}{0.3\textwidth}
\includegraphics[width=6cm]{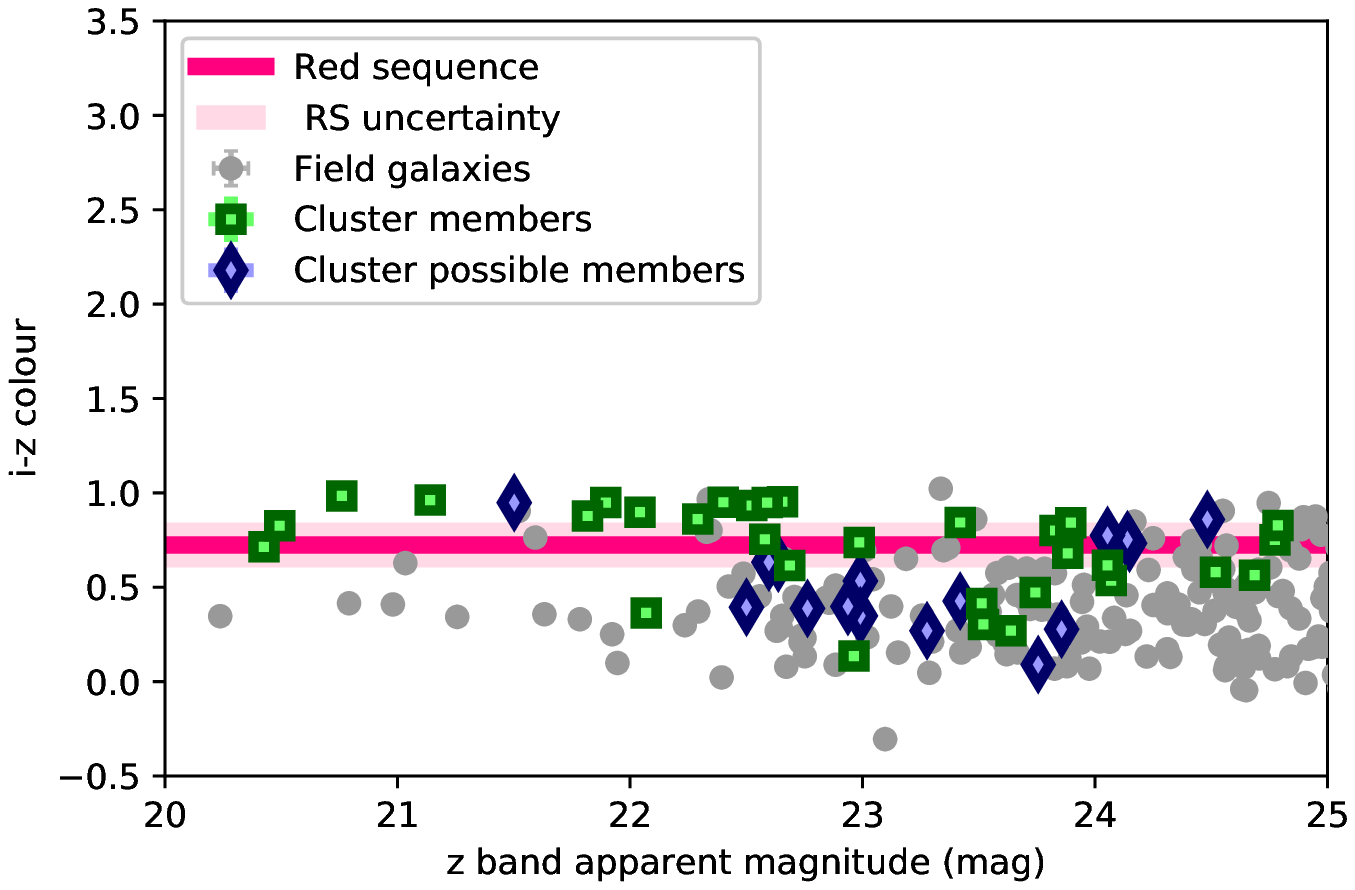}
\end{subfigure}

\centering
\begin{subfigure}{0.3\textwidth}
\includegraphics[width=5.25cm]{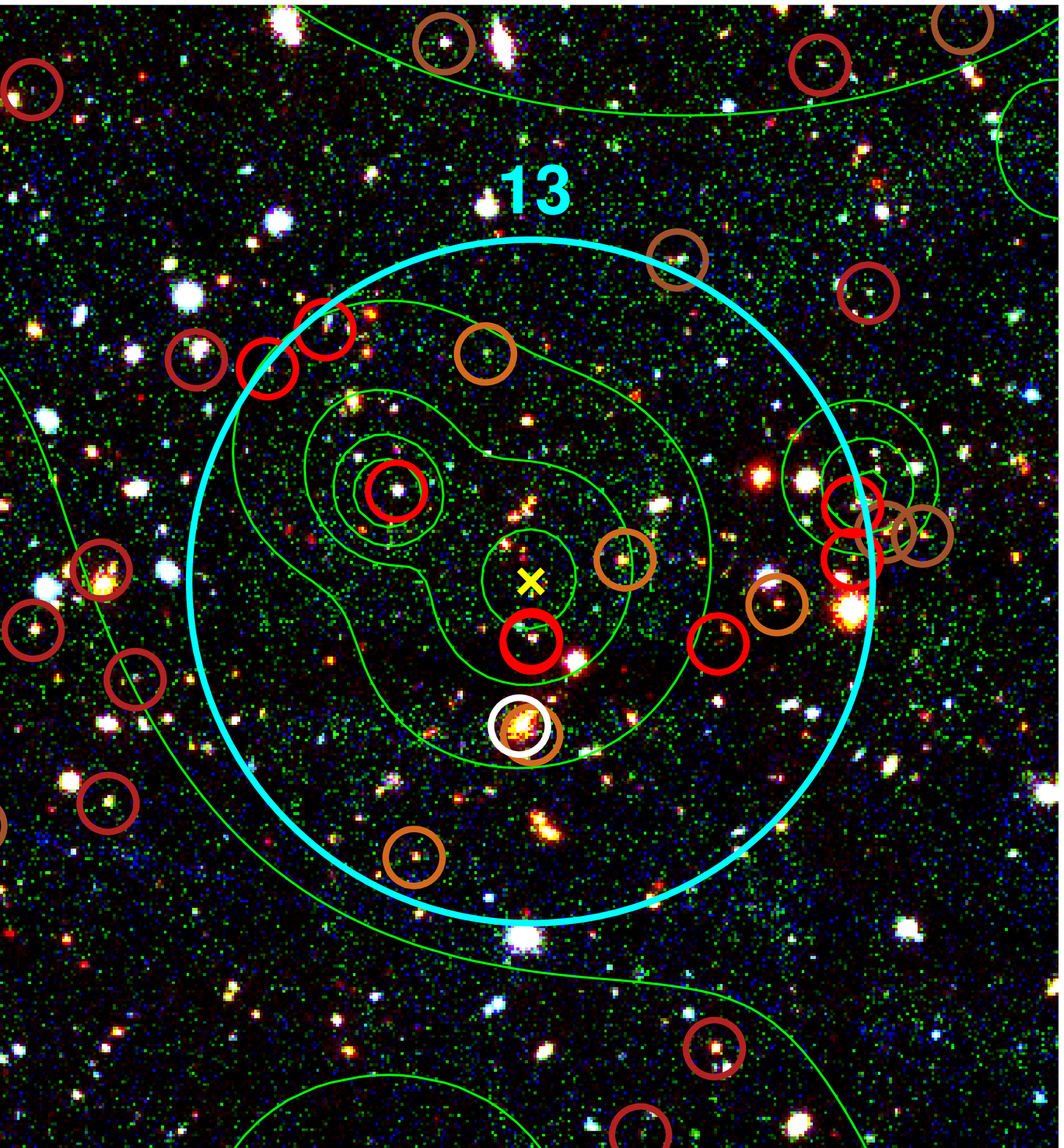}
\end{subfigure}
\hfill
\begin{subfigure}{0.3\textwidth}
\includegraphics[width=6cm]{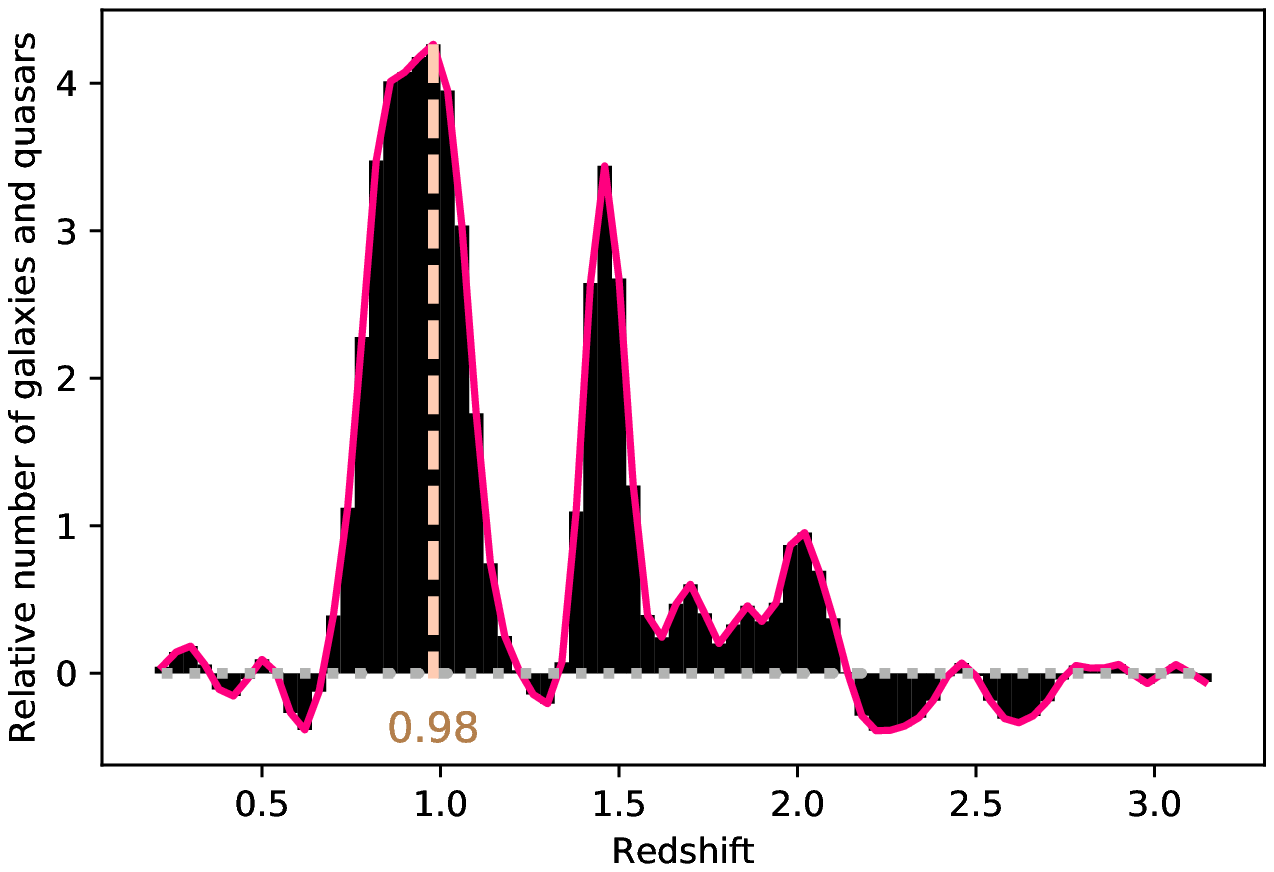}
\end{subfigure}
\hfill
\begin{subfigure}{0.3\textwidth}
\includegraphics[width=6cm]{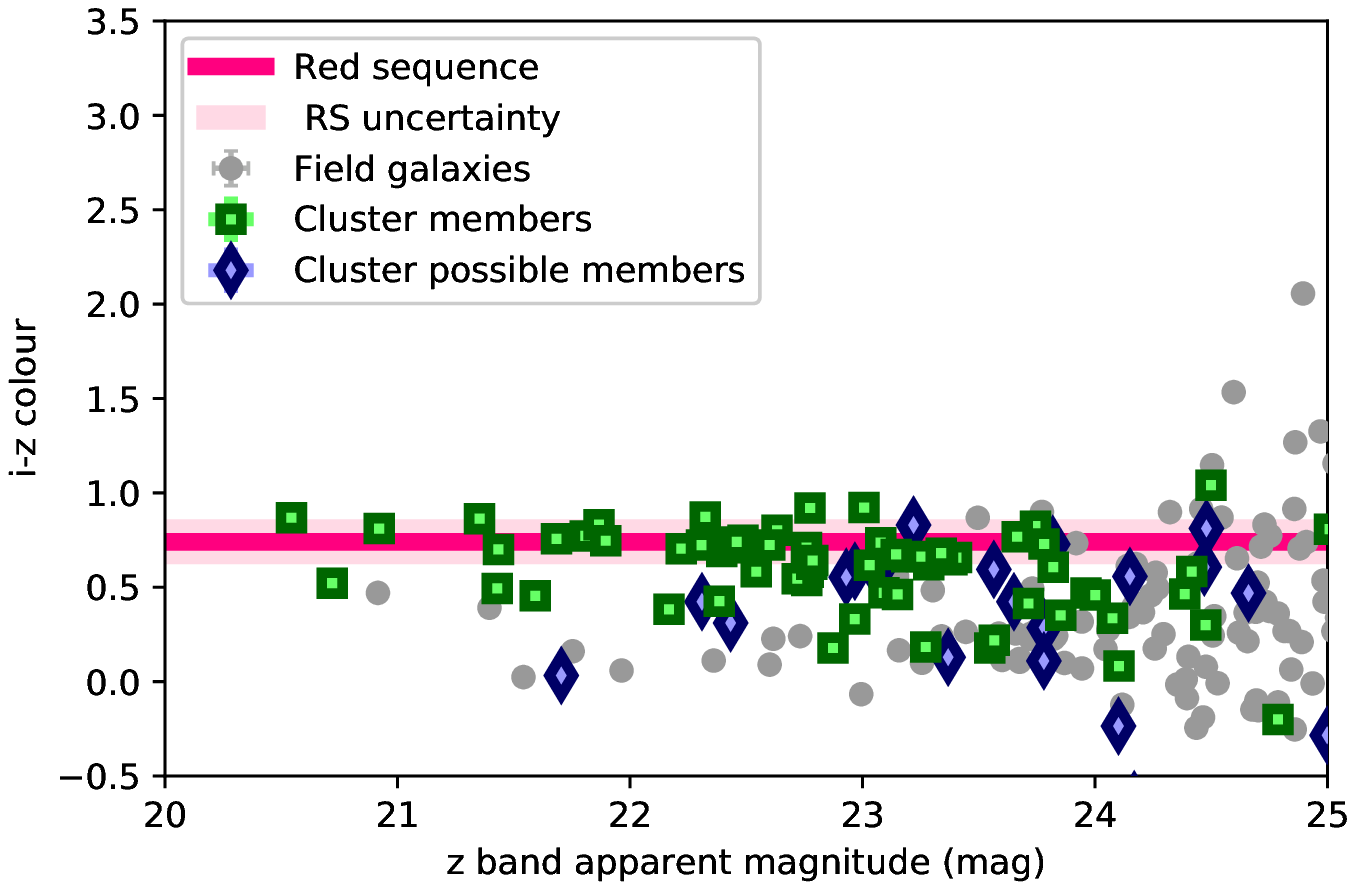}
\end{subfigure}

\centering
\begin{subfigure}{0.3\textwidth}
\includegraphics[width=5.25cm]{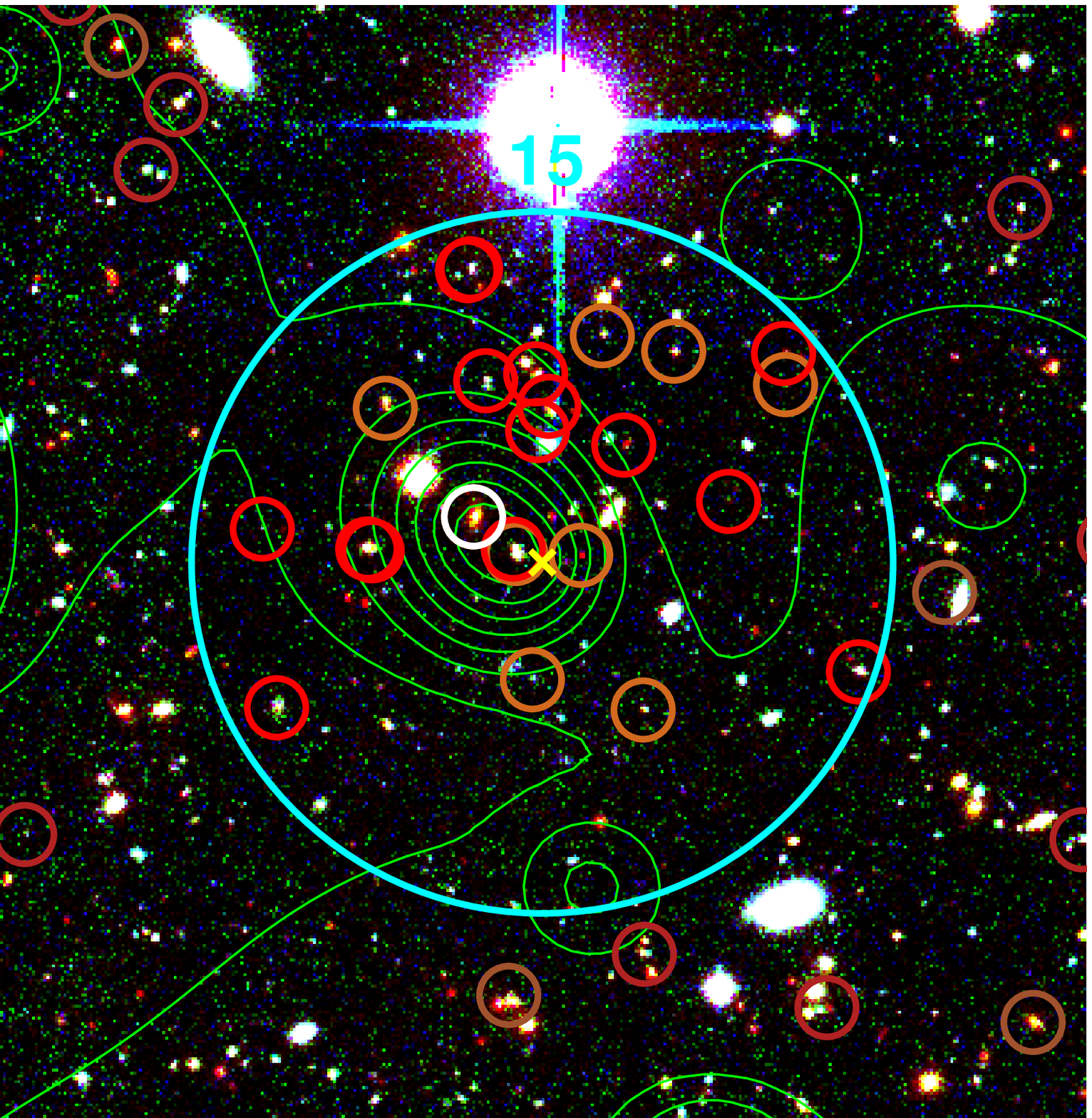}
\end{subfigure}
\hfill
\begin{subfigure}{0.3\textwidth}
\includegraphics[width=6cm]{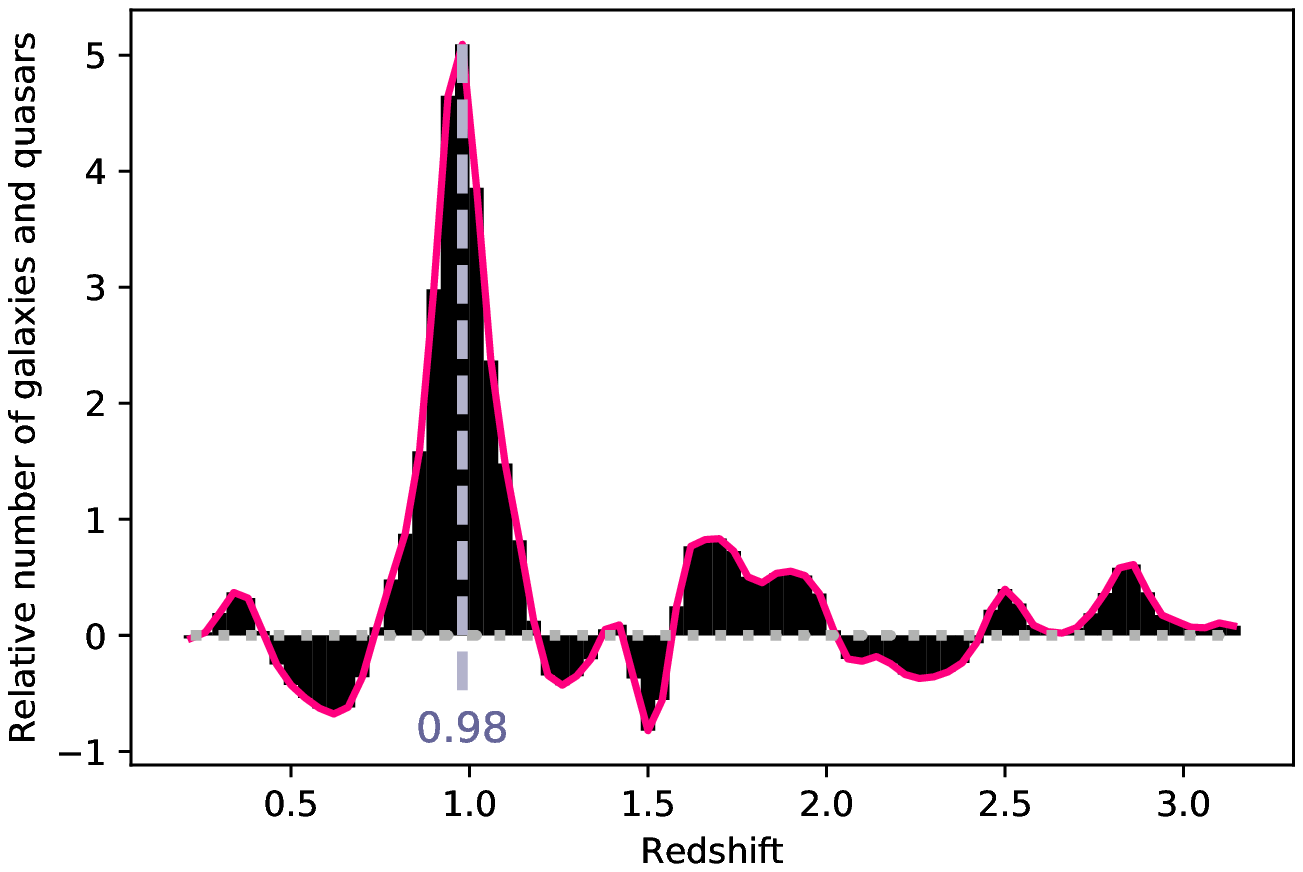}
\end{subfigure}
\hfill
\begin{subfigure}{0.3\textwidth}
\includegraphics[width=6cm]{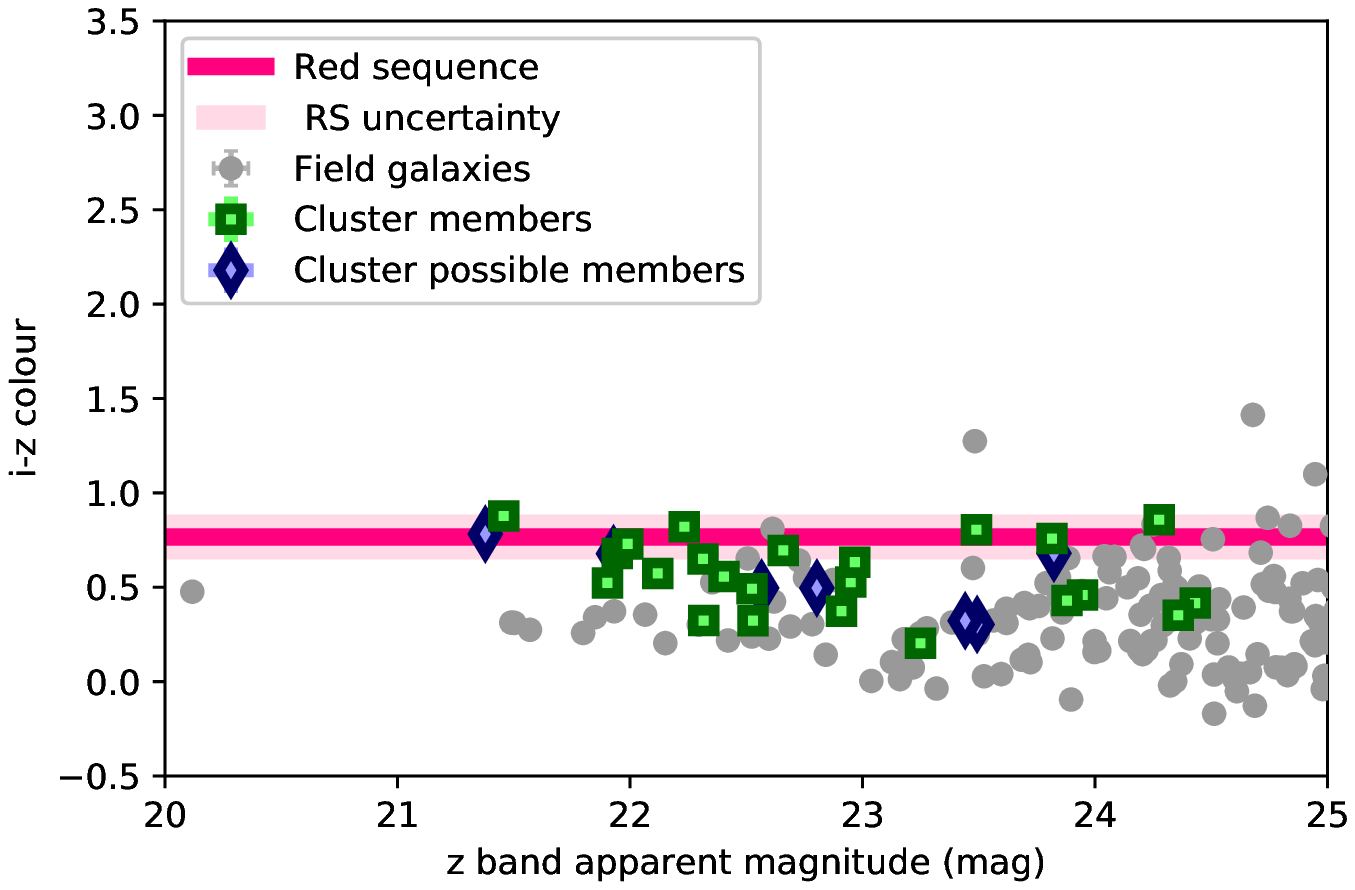}
\end{subfigure}
%
%
%
\centering
\begin{subfigure}{0.3\textwidth}
\includegraphics[width=5.25cm]{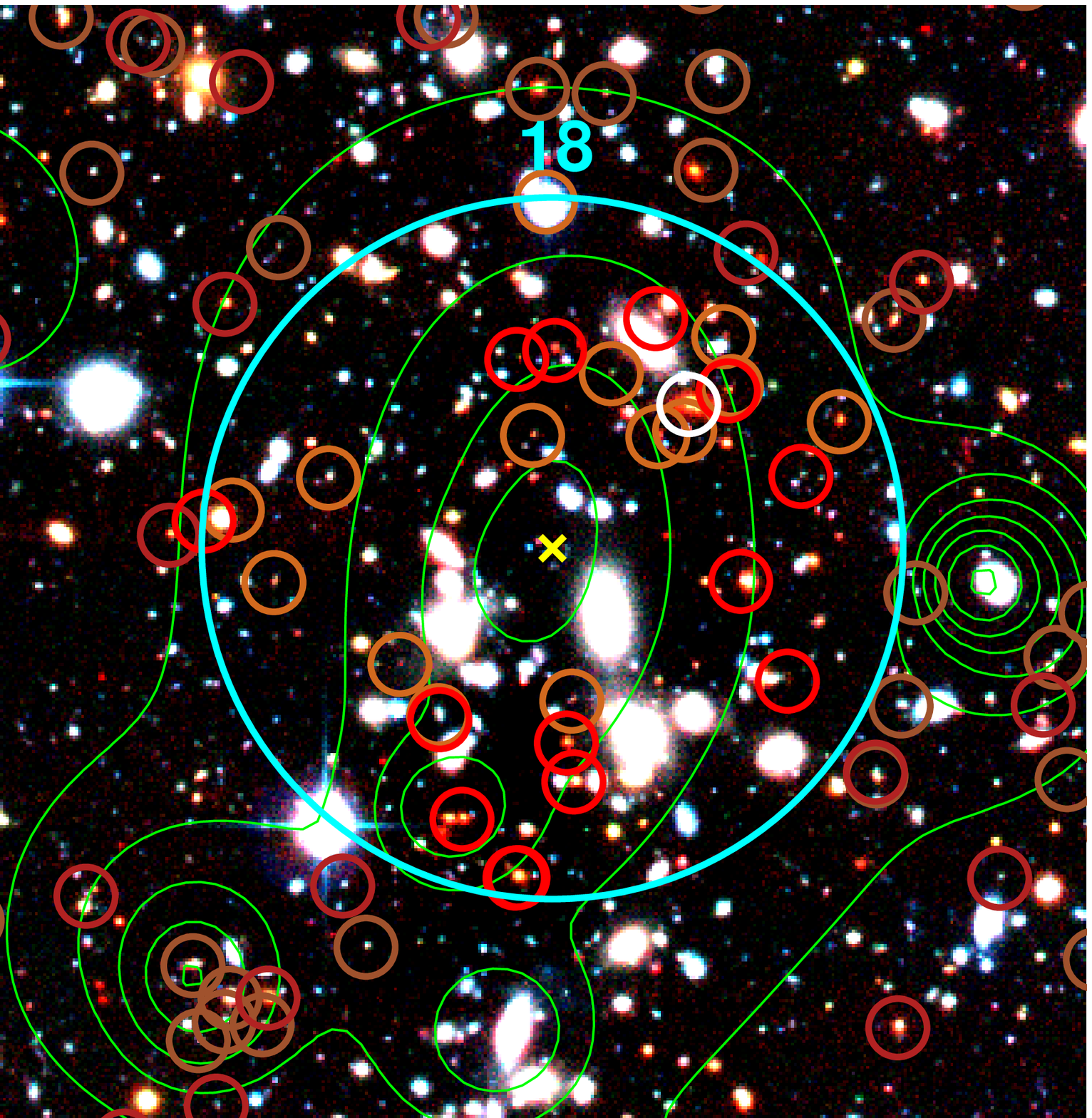}
\end{subfigure}
\hfill
\begin{subfigure}{0.3\textwidth}
\includegraphics[width=6cm]{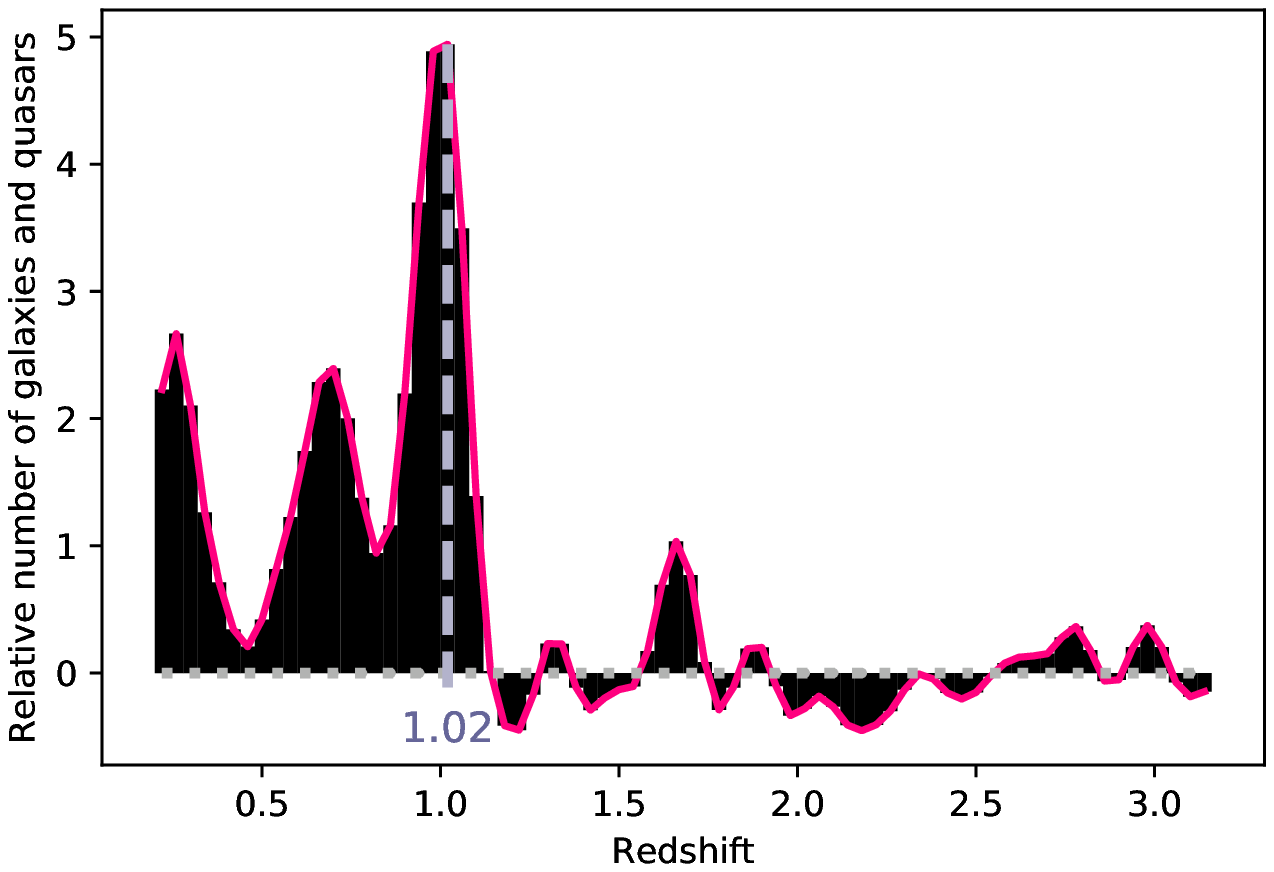}
\end{subfigure}
\hfill
\begin{subfigure}{0.3\textwidth}
\includegraphics[width=6cm]{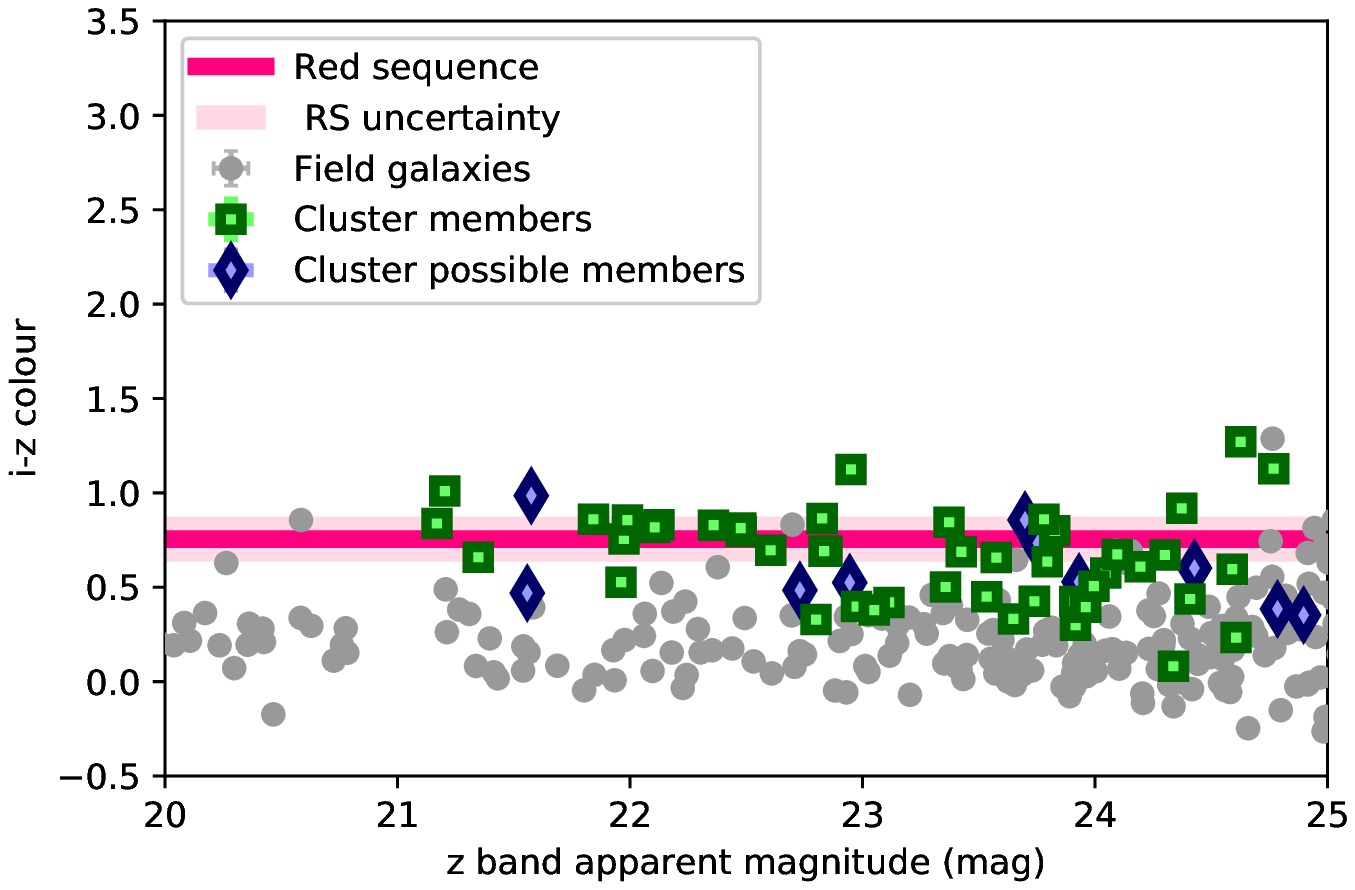}
\end{subfigure}
\caption{\textit{continued}}
\label{fig_candidates_tentatives}
\end{figure*}

%

\begin{figure*}
\centering
\begin{subfigure}{0.3\textwidth}
\includegraphics[width=5.25cm]{Tile-18_994_vtest.eps}
\end{subfigure}
\hfill
\begin{subfigure}{0.3\textwidth}
\includegraphics[width=6cm]{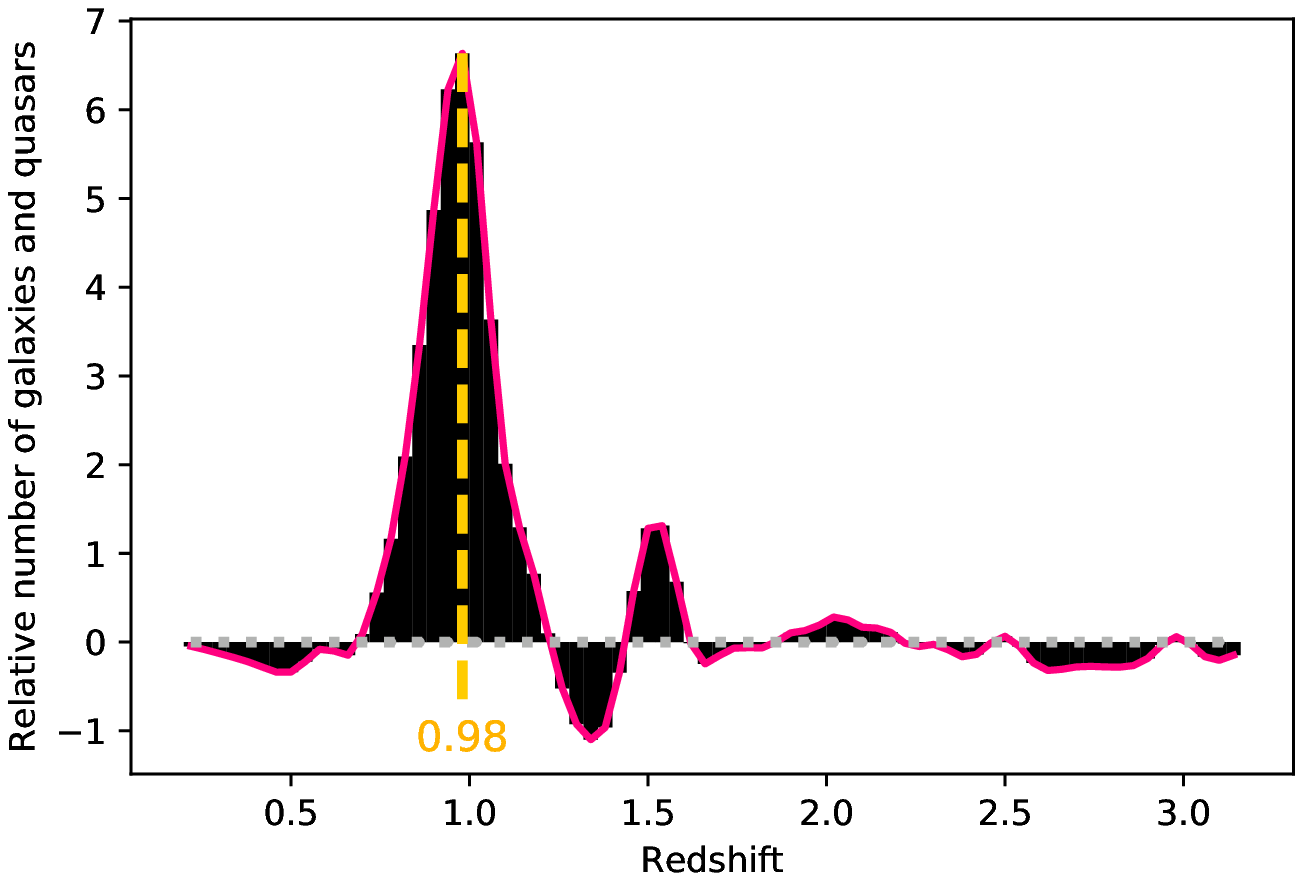}
\end{subfigure}
\hfill
\begin{subfigure}{0.3\textwidth}
\includegraphics[width=6cm]{Tile-18_1002_I-Z_CMD.eps}
\end{subfigure}

\centering
\begin{subfigure}{0.3\textwidth}
\includegraphics[width=5.25cm]{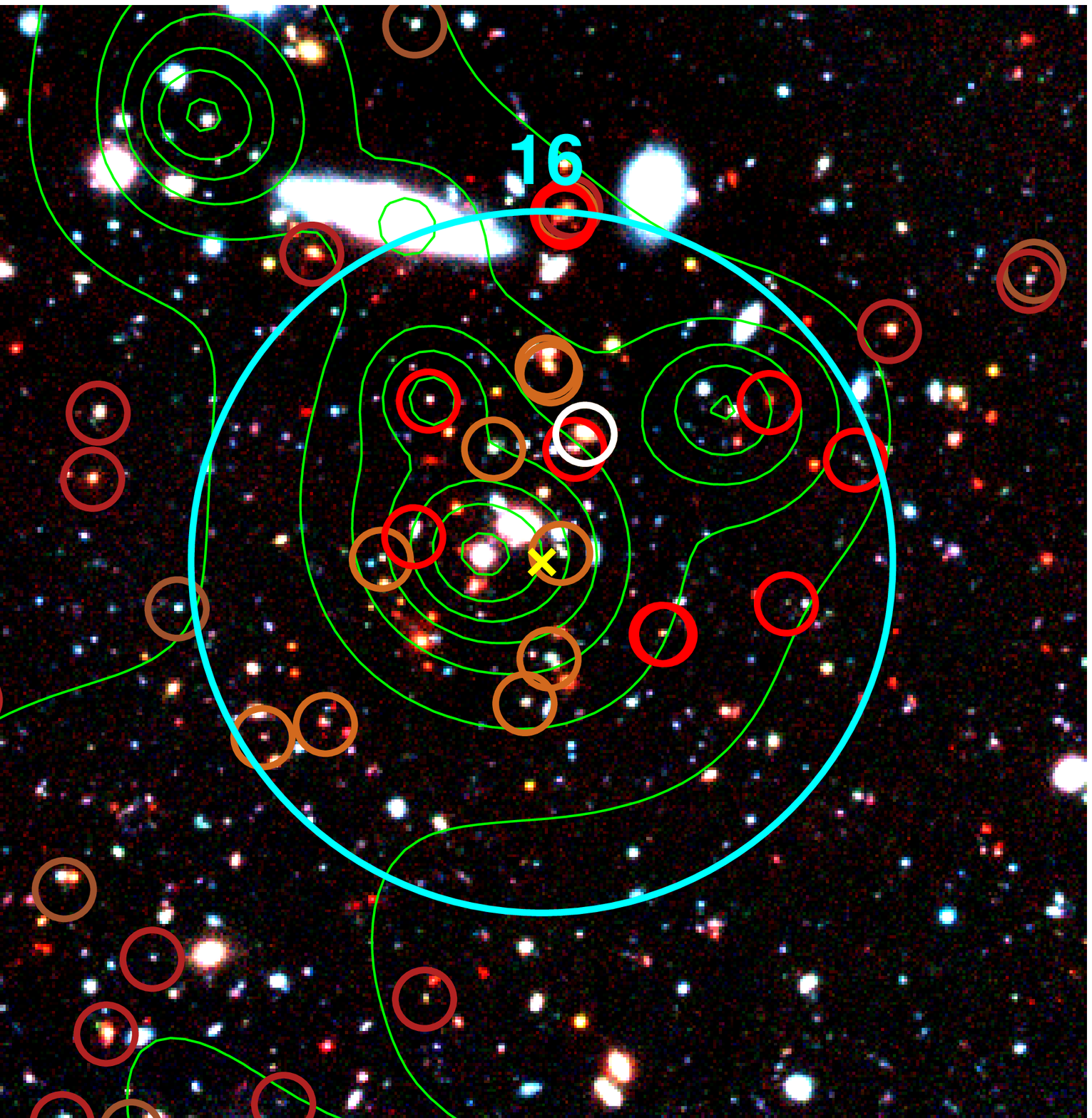}
\end{subfigure}
\hfill
\begin{subfigure}{0.3\textwidth}
\includegraphics[width=6cm]{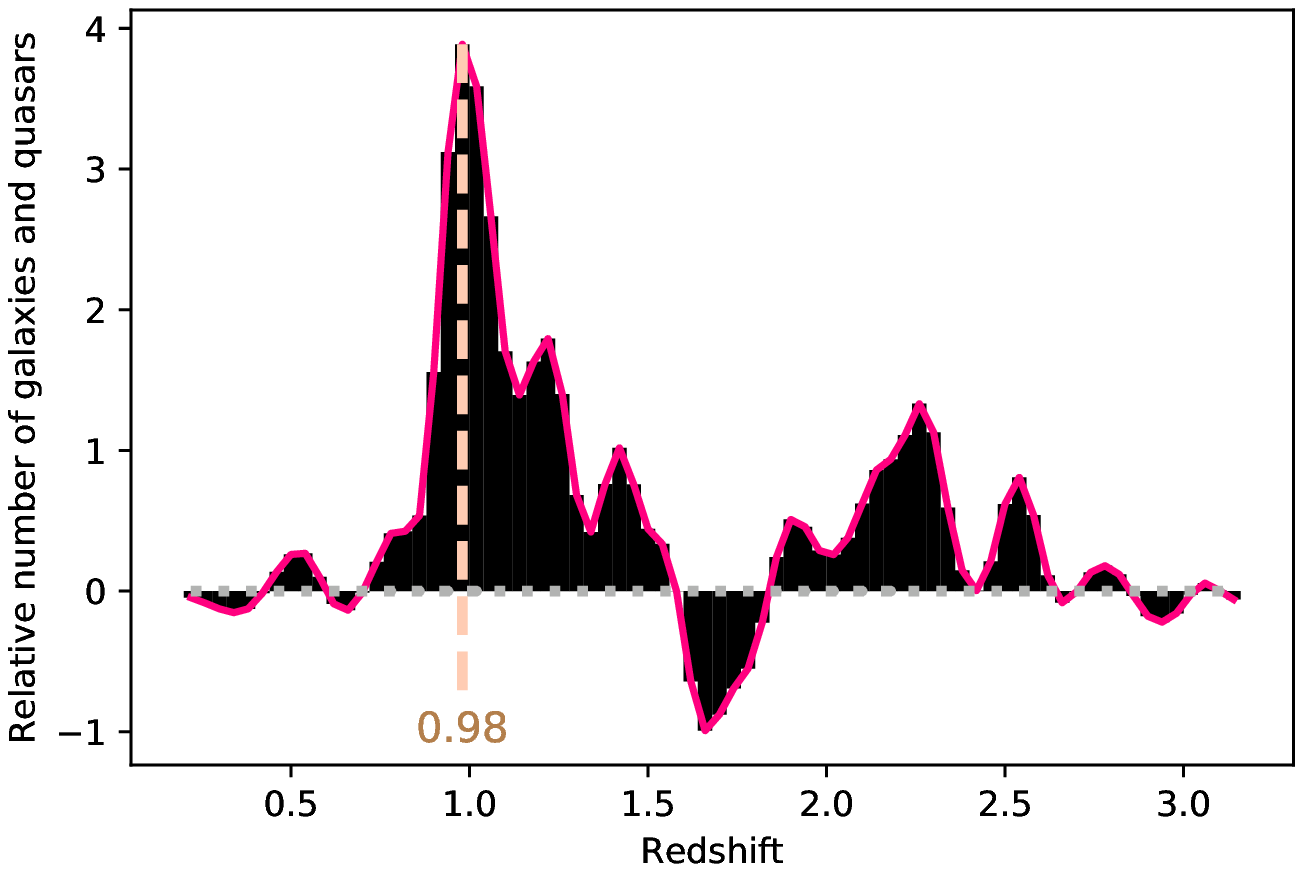}
\end{subfigure}
\hfill
\begin{subfigure}{0.3\textwidth}
\includegraphics[width=6cm]{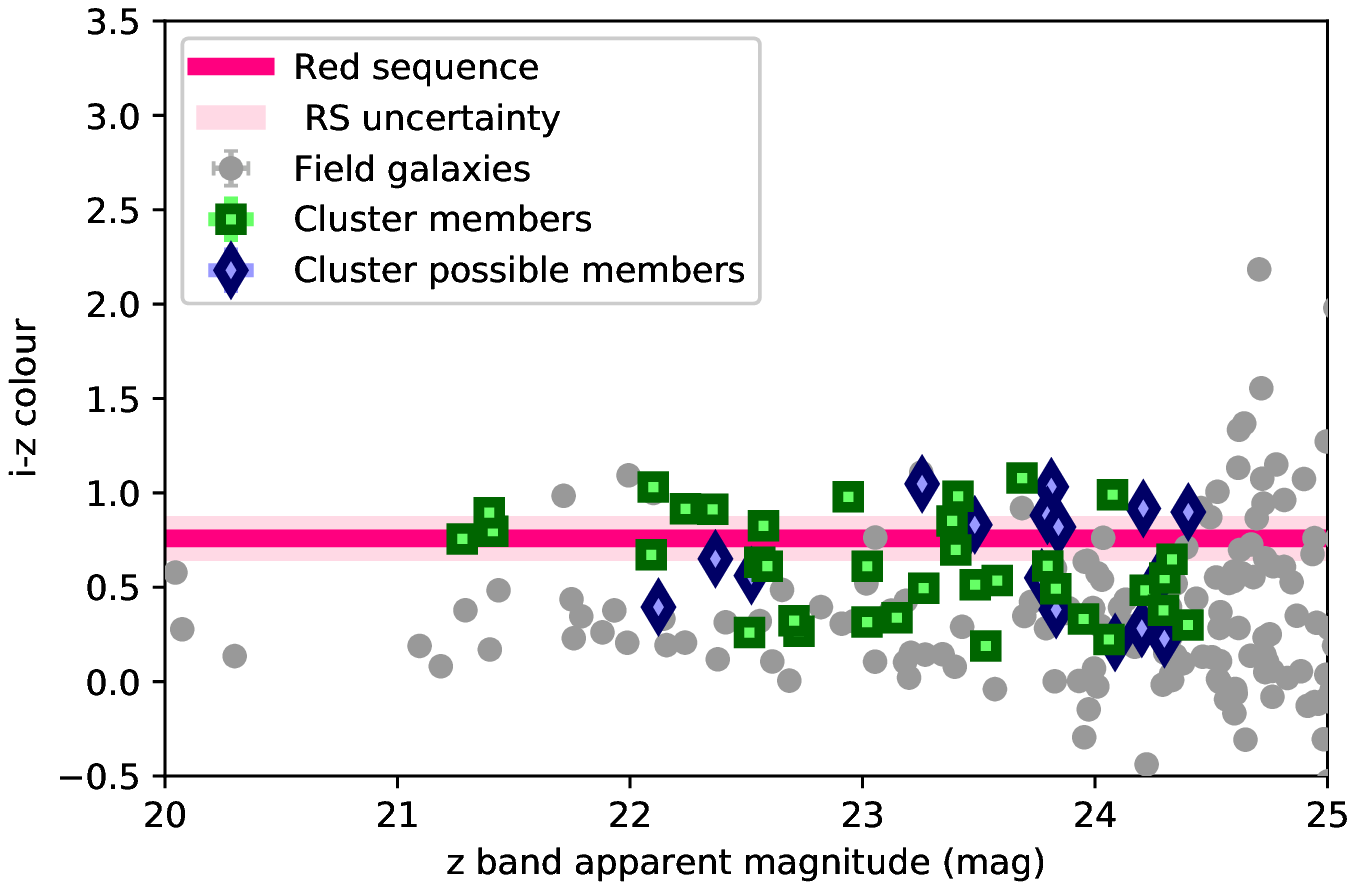}
\end{subfigure}

\centering
\begin{subfigure}{0.3\textwidth}
\includegraphics[width=5.25cm]{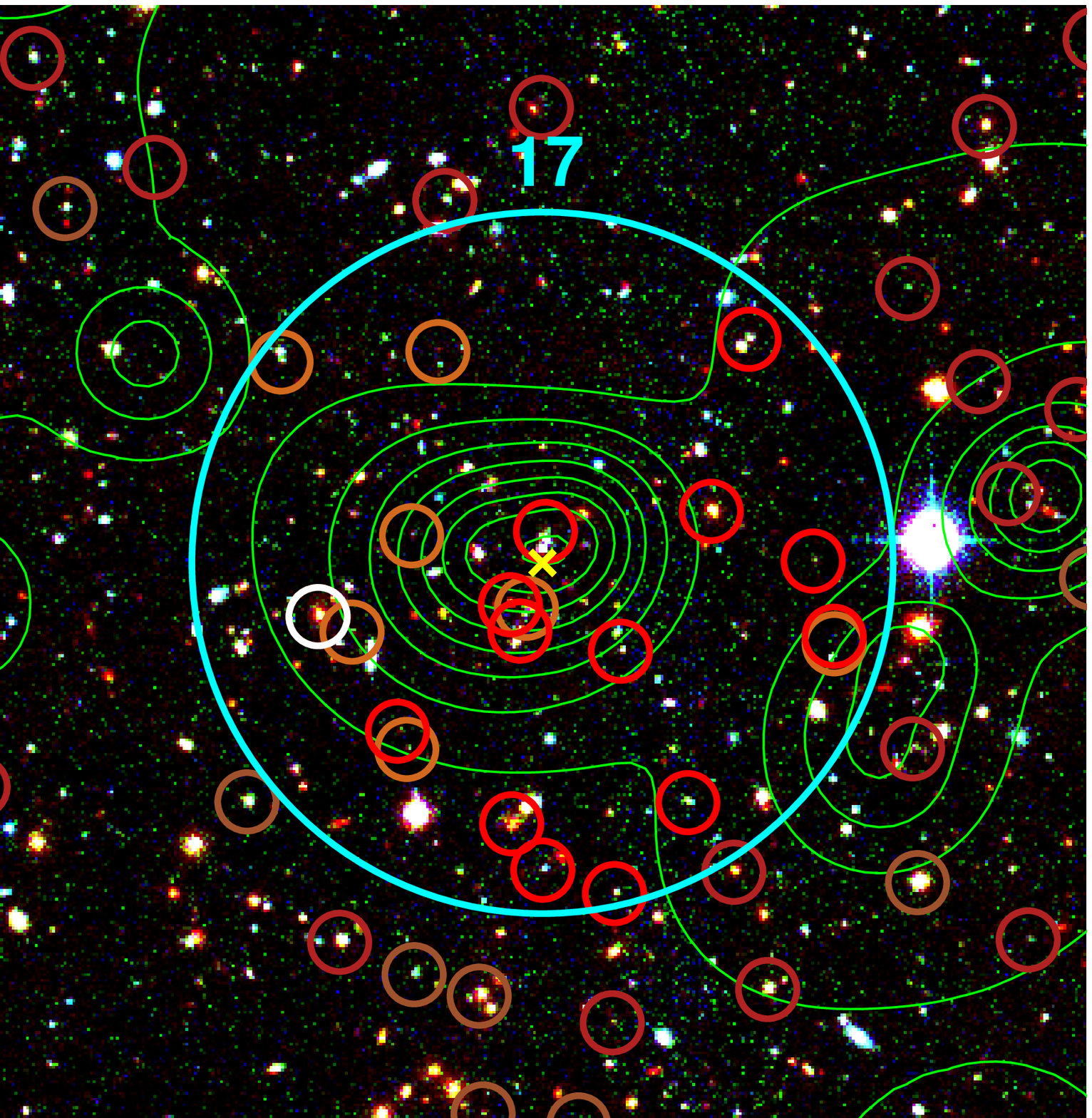}
\end{subfigure}
\hfill
\begin{subfigure}{0.3\textwidth}
\includegraphics[width=6cm]{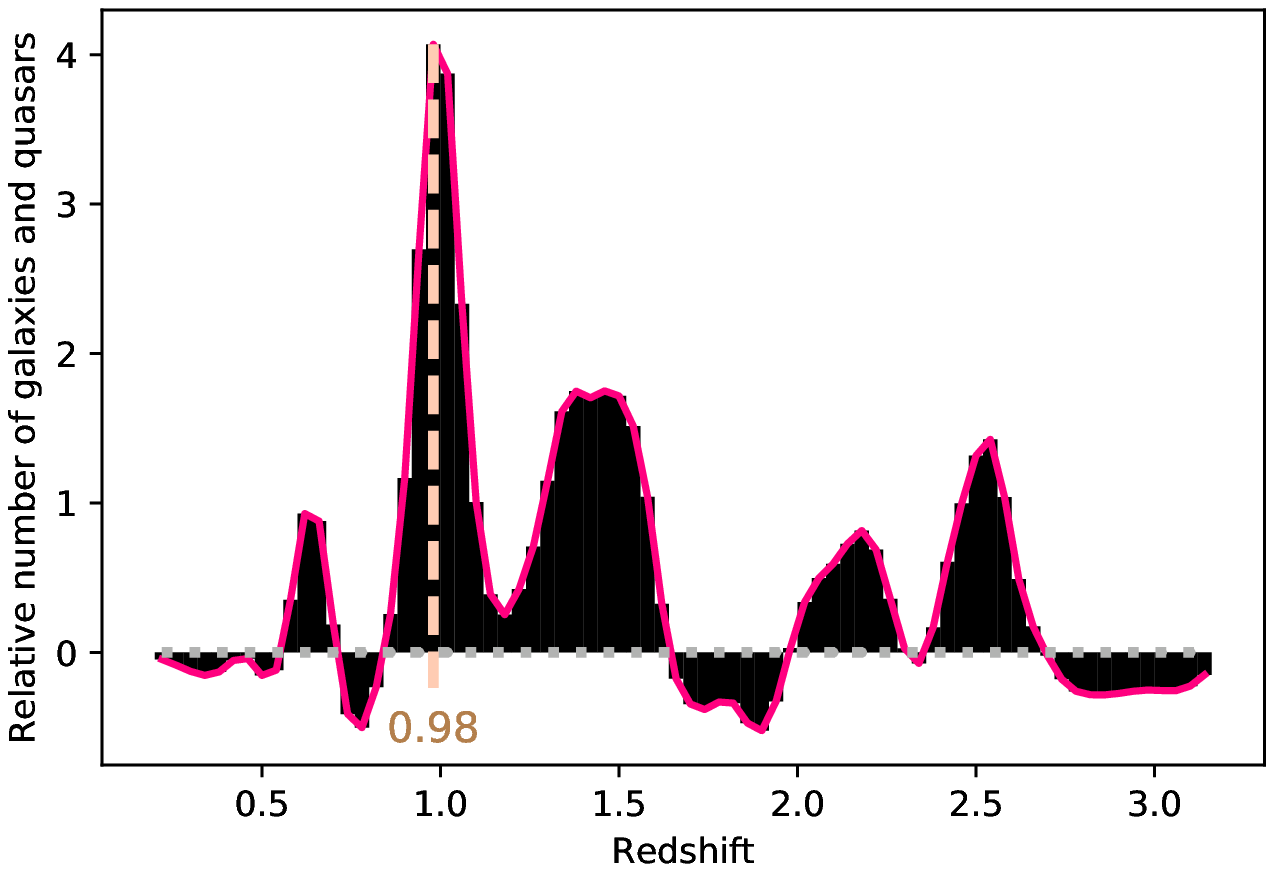}
\end{subfigure}
\hfill
\begin{subfigure}{0.3\textwidth}
\includegraphics[width=6cm]{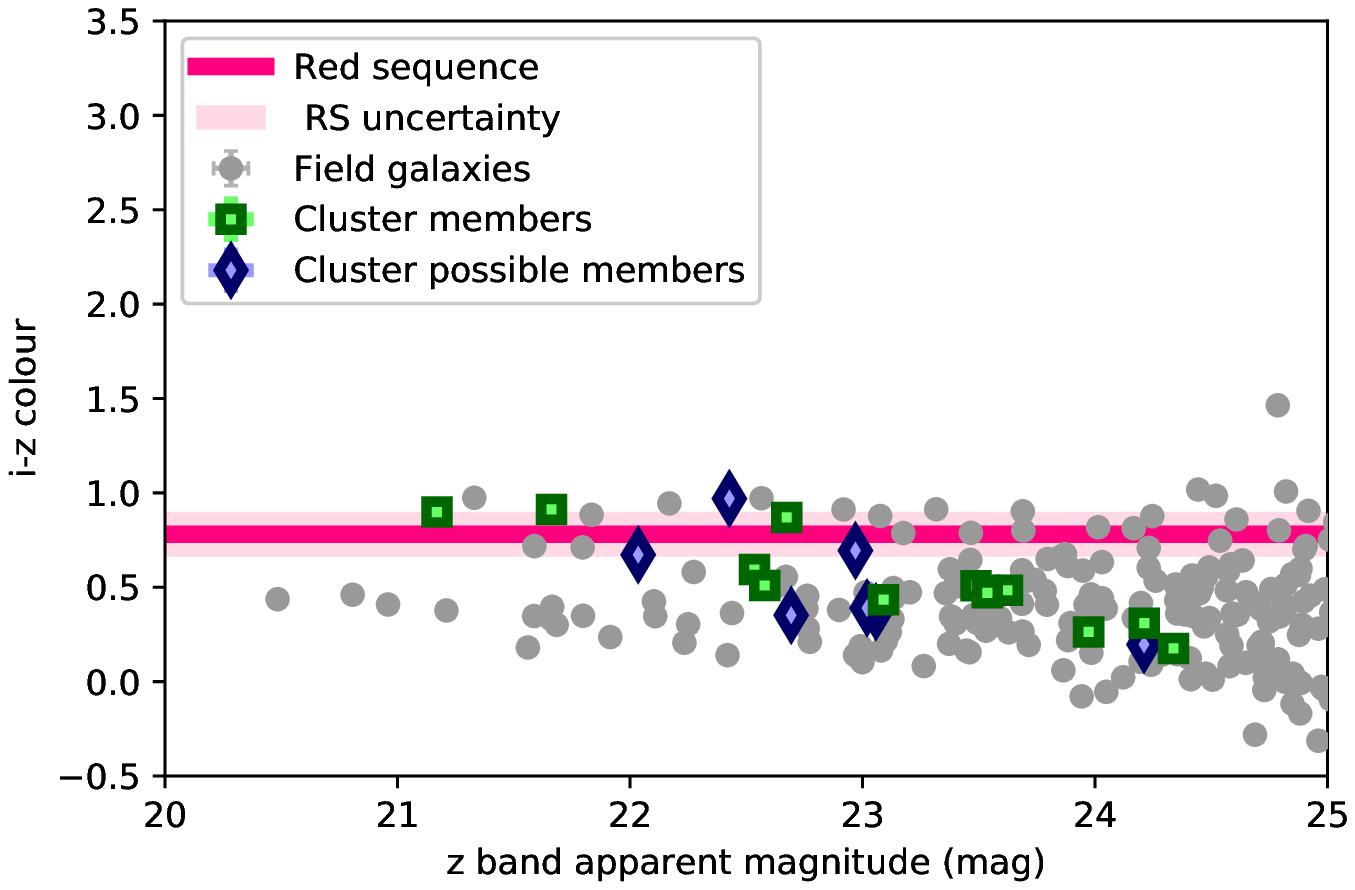}
\end{subfigure}
\centering
\begin{subfigure}{0.3\textwidth}
\includegraphics[width=5.25cm]{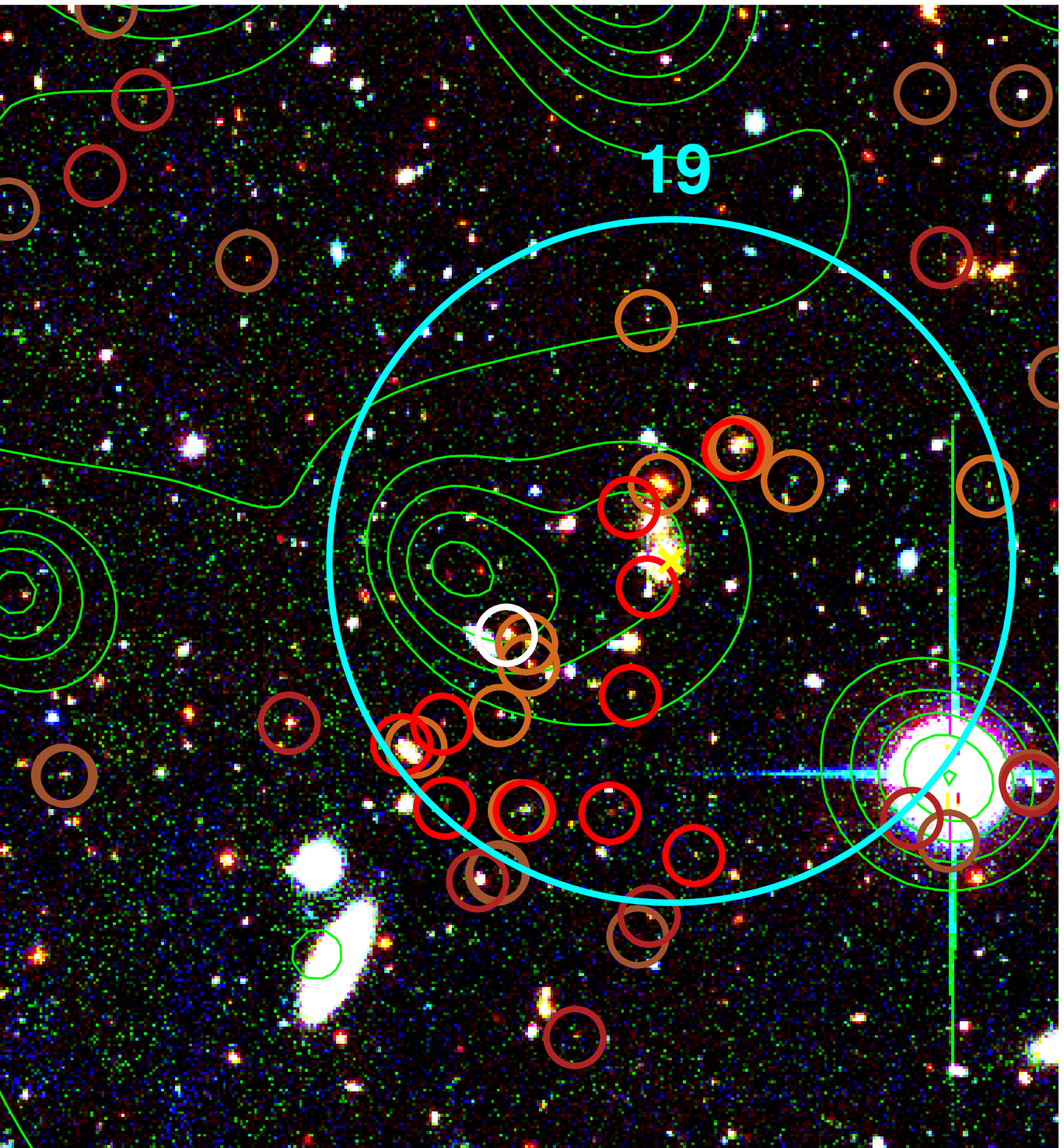}
\end{subfigure}
\hfill
\begin{subfigure}{0.3\textwidth}
\includegraphics[width=6cm]{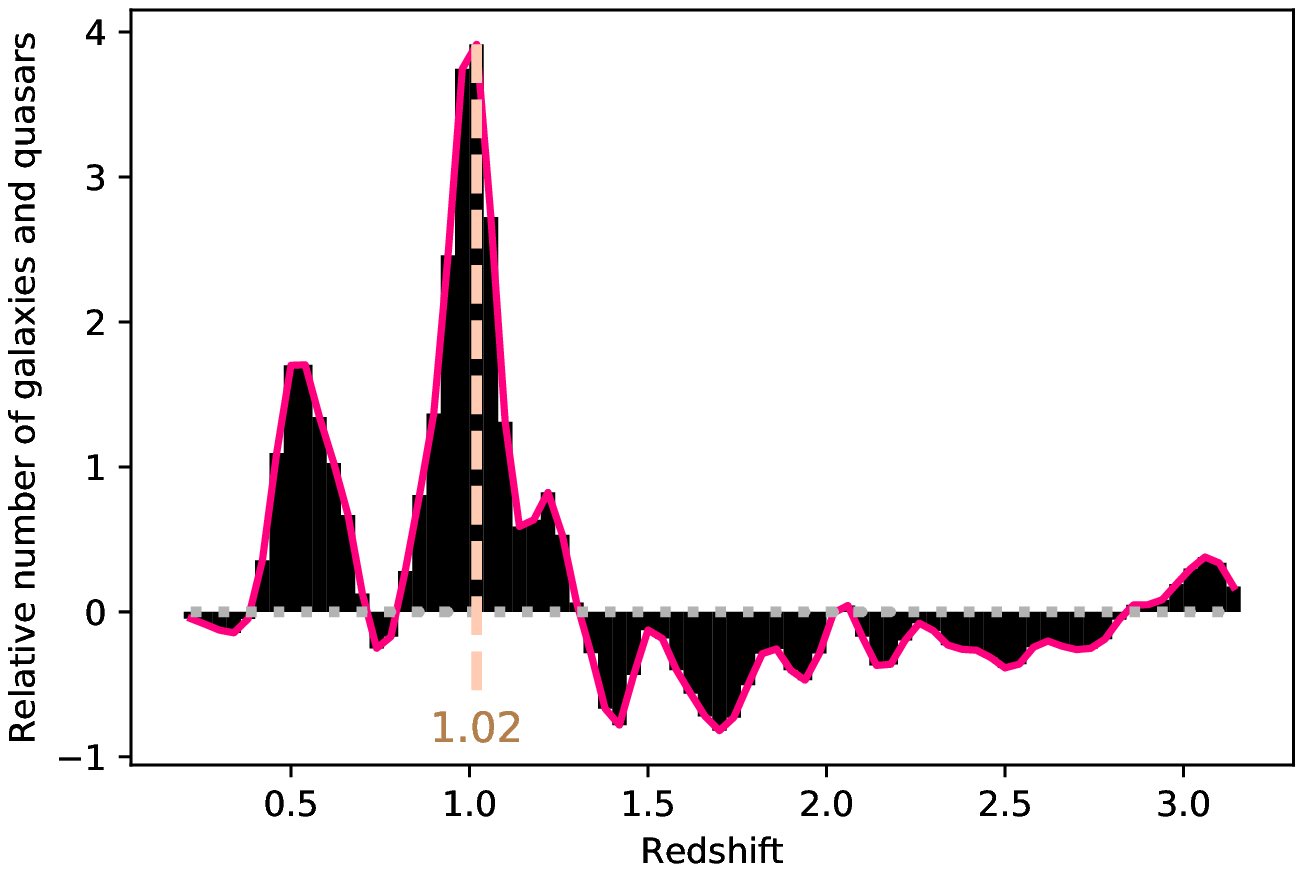}
\end{subfigure}
\hfill
\begin{subfigure}{0.3\textwidth}
\includegraphics[width=6cm]{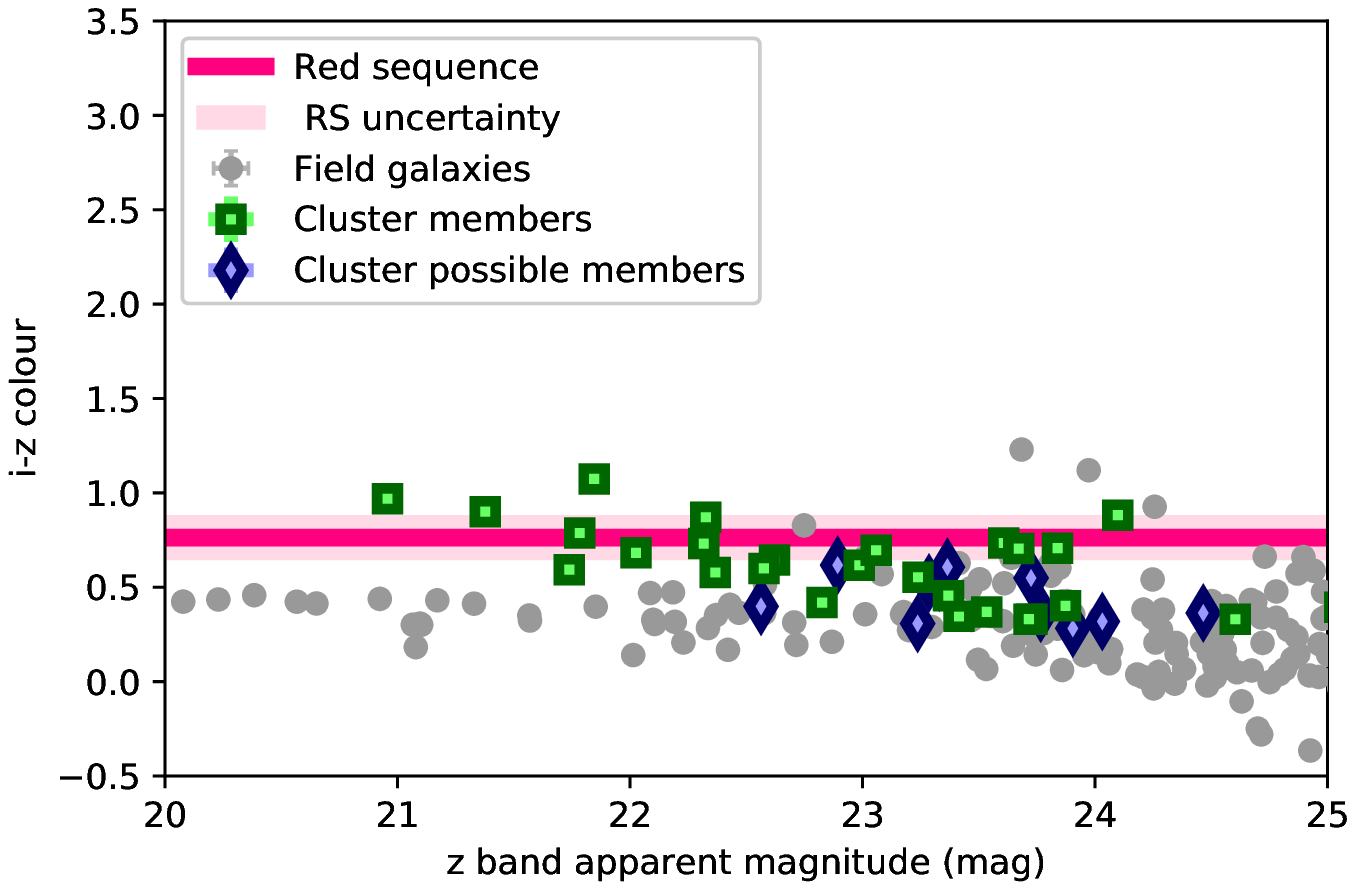}
\end{subfigure}
\caption{Left Cols.: Megacam R and I filter and VIDEO H filter images for the new candidate clusters at $z\geq 0.8$, classified by increasing redshifts. The X-ray contours in green are logarithmically distributed in 10 levels between the maximum and minimum emission observed in a $7\times 7 ~\mathrm{arcmin^2}$ box around the X-ray source, except for candidate 27 which displays 25 levels based on $4\times 4 ~\mathrm{arcmin^2}$ box. Symbols definitions are given in Fig. \ref{fig_known_clusters}. Middle Cols.: Background subtracted and Gaussian filtered redshift distribution of the bright galaxies within the central arcmin, for the corresponding candidates. Bottom Cols.: i-z ($0.8\leq z< 1.2$) or z-J ($z\geq 1.2$) CMD plot of the galaxies above VIDEO 5$\sigma$ limit within 1 arcmin of the centre.}
\end{figure*}

\newpage

\begin{figure*}
\ContinuedFloat
\centering
\begin{subfigure}{0.3\textwidth}
\includegraphics[width=5.25cm]{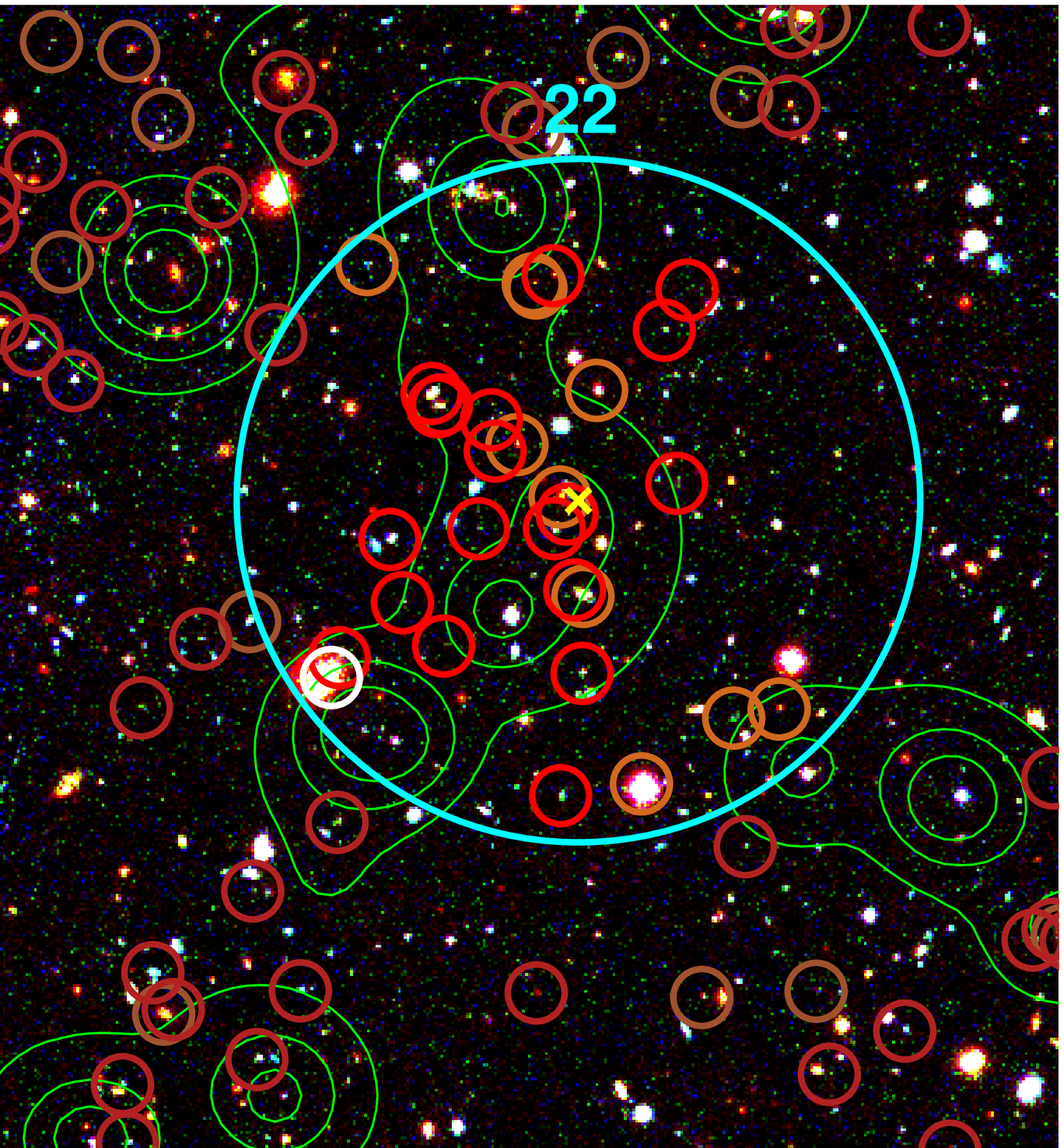}
\end{subfigure}
\hfill
\begin{subfigure}{0.3\textwidth}
\includegraphics[width=6cm]{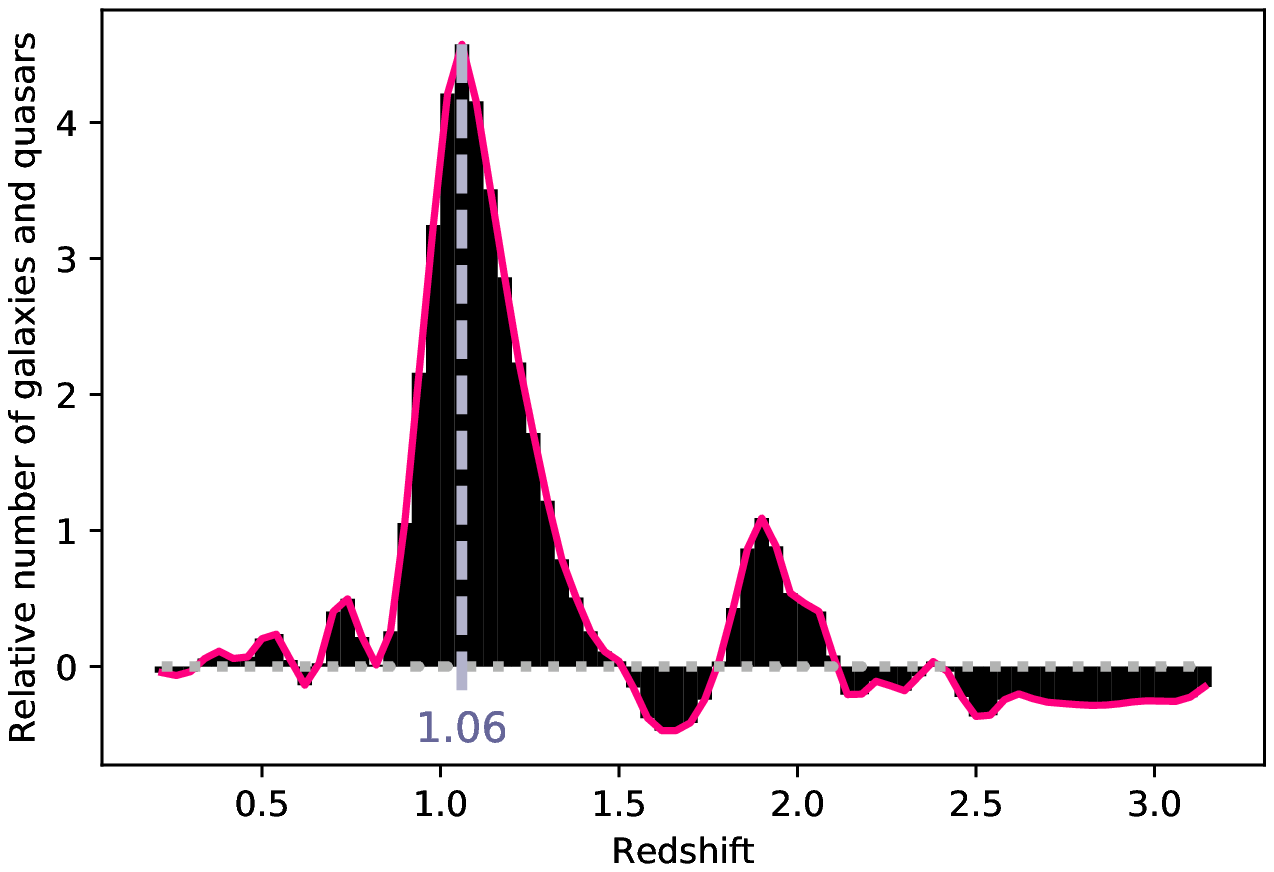}
\end{subfigure}
\hfill
\begin{subfigure}{0.3\textwidth}
\includegraphics[width=6cm]{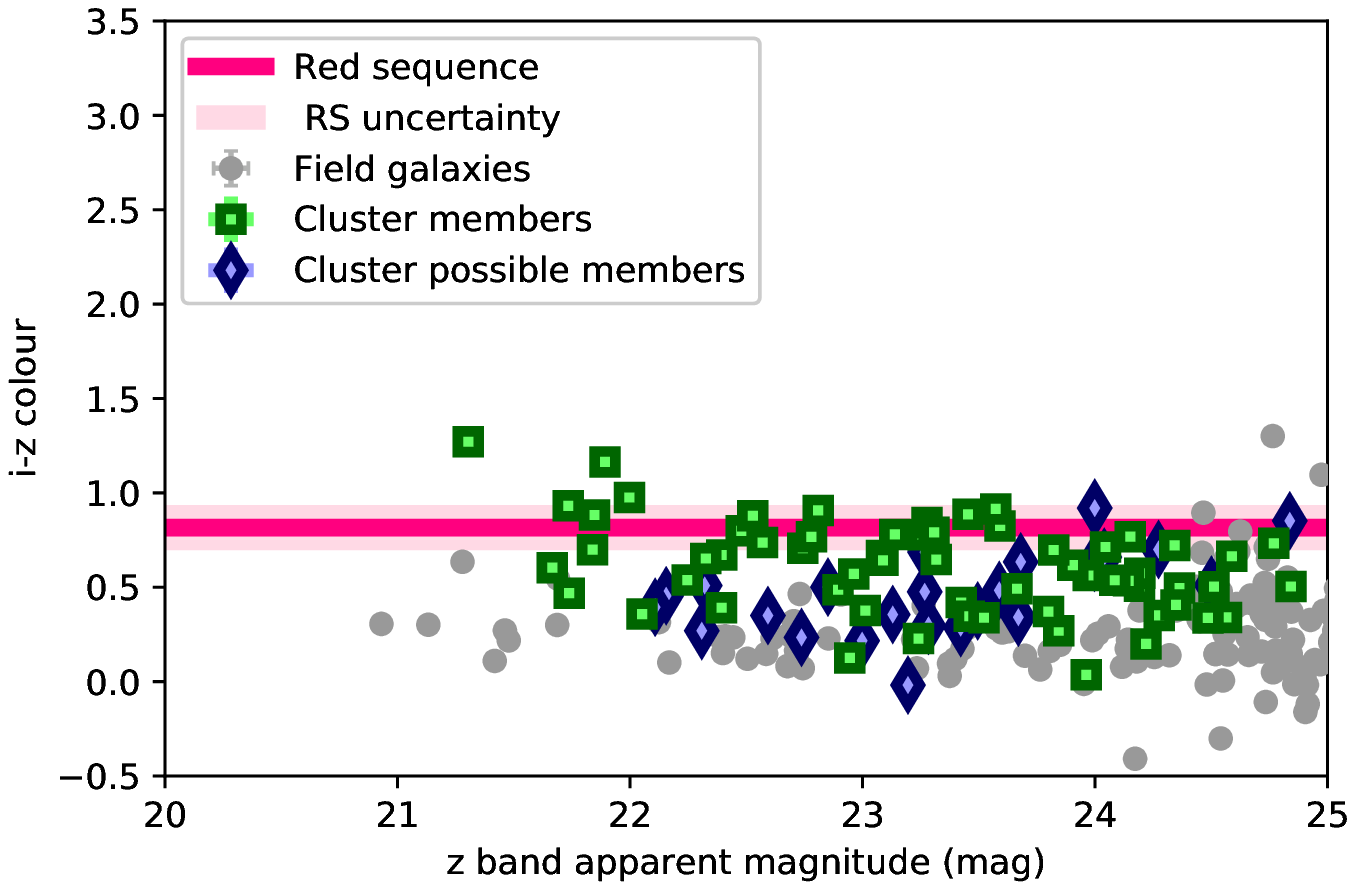}
\end{subfigure}

\centering
\begin{subfigure}{0.3\textwidth}
\includegraphics[width=5.25cm]{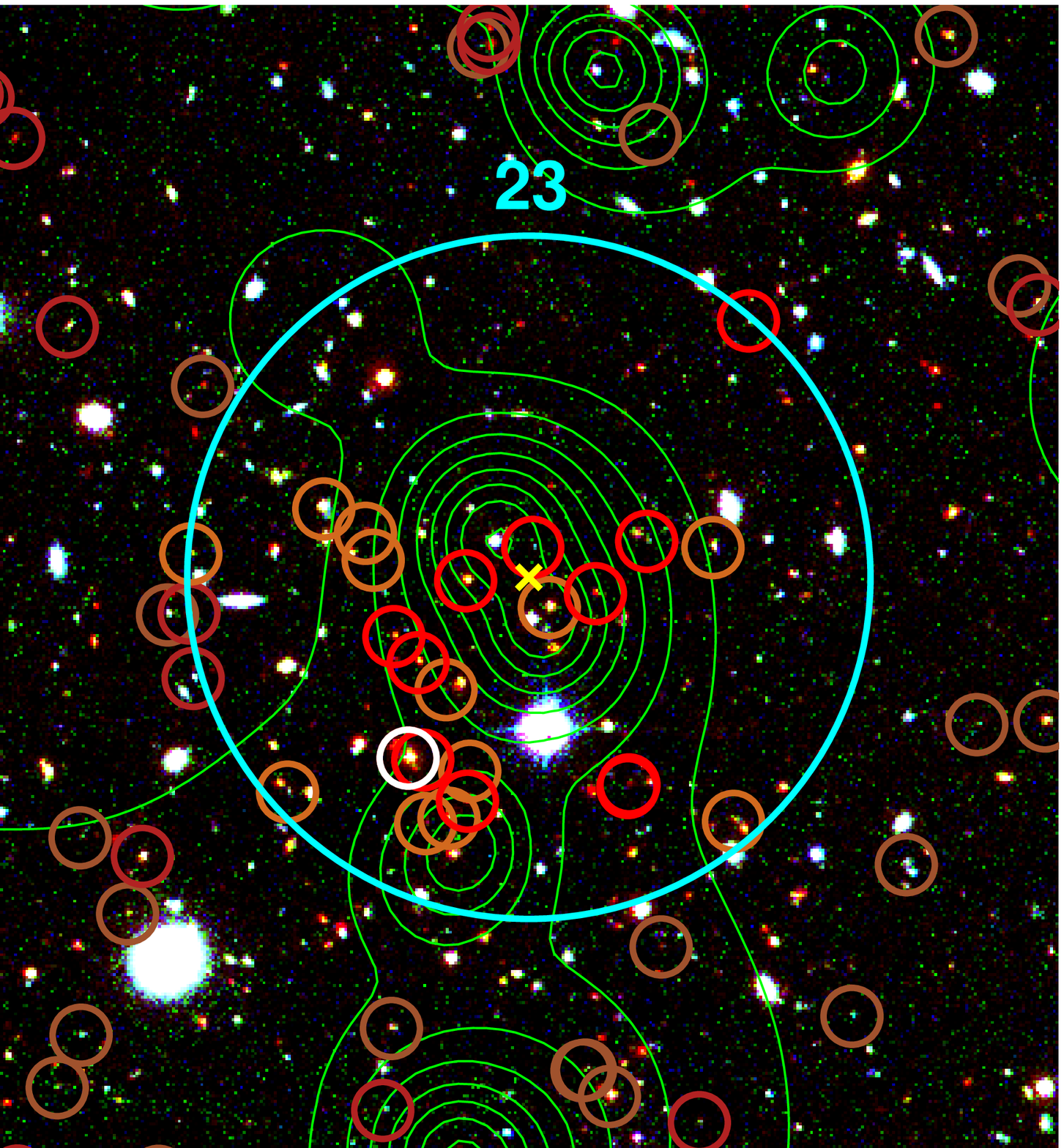}
\end{subfigure}
\hfill
\begin{subfigure}{0.3\textwidth}
\includegraphics[width=6cm]{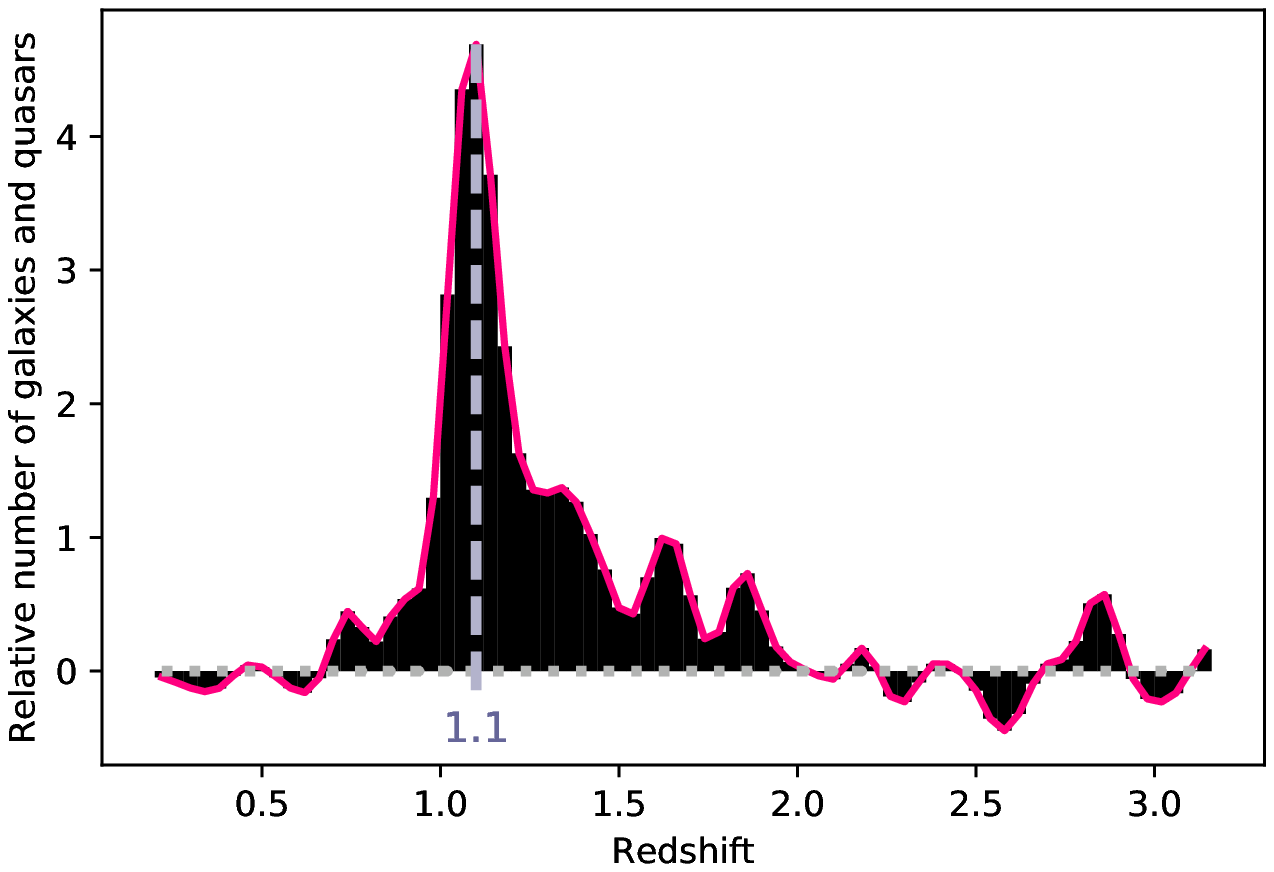}
\end{subfigure}
\hfill
\begin{subfigure}{0.3\textwidth}
\includegraphics[width=6cm]{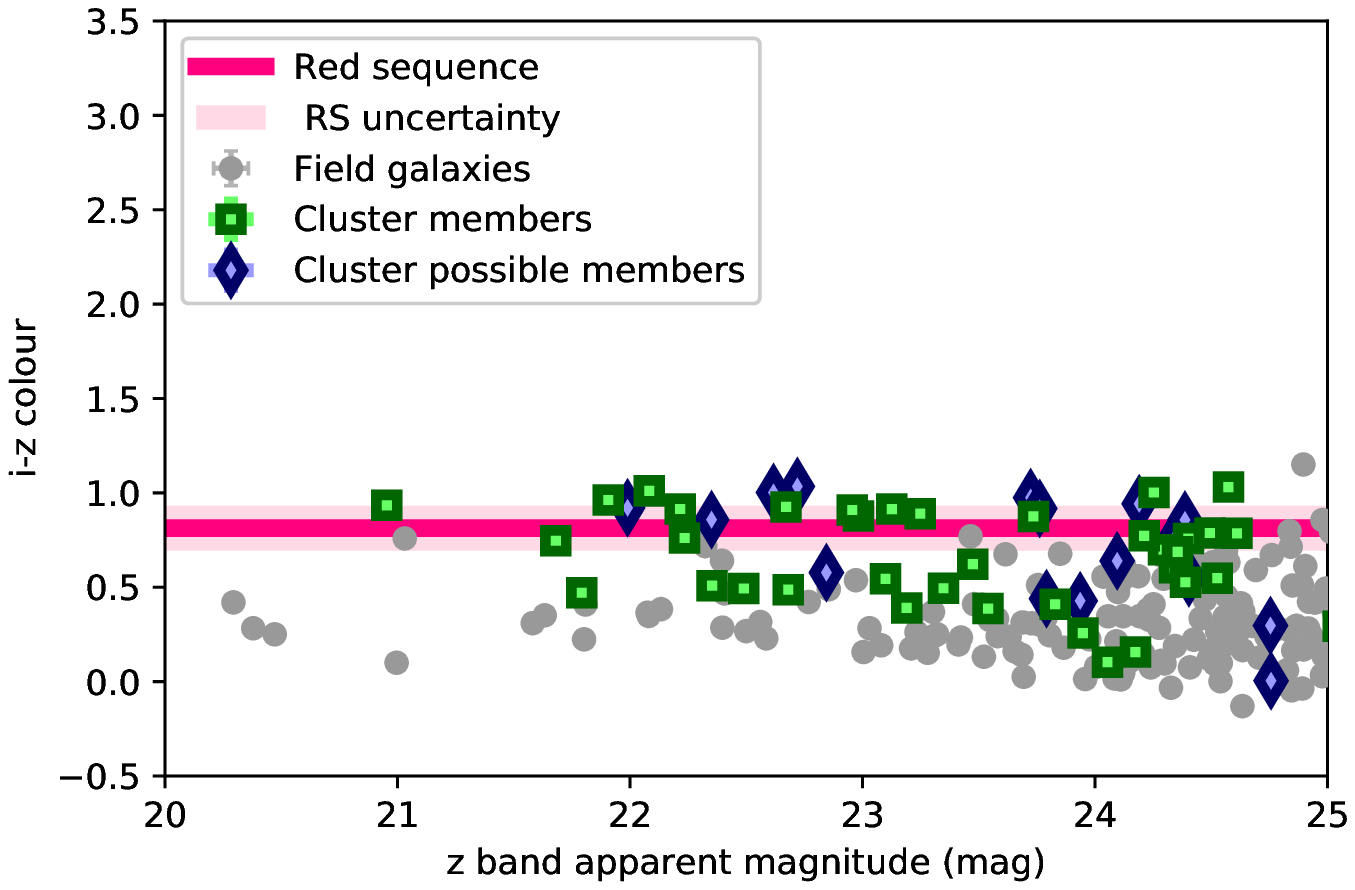}
\end{subfigure}

\centering
\begin{subfigure}{0.3\textwidth}
\includegraphics[width=5.25cm]{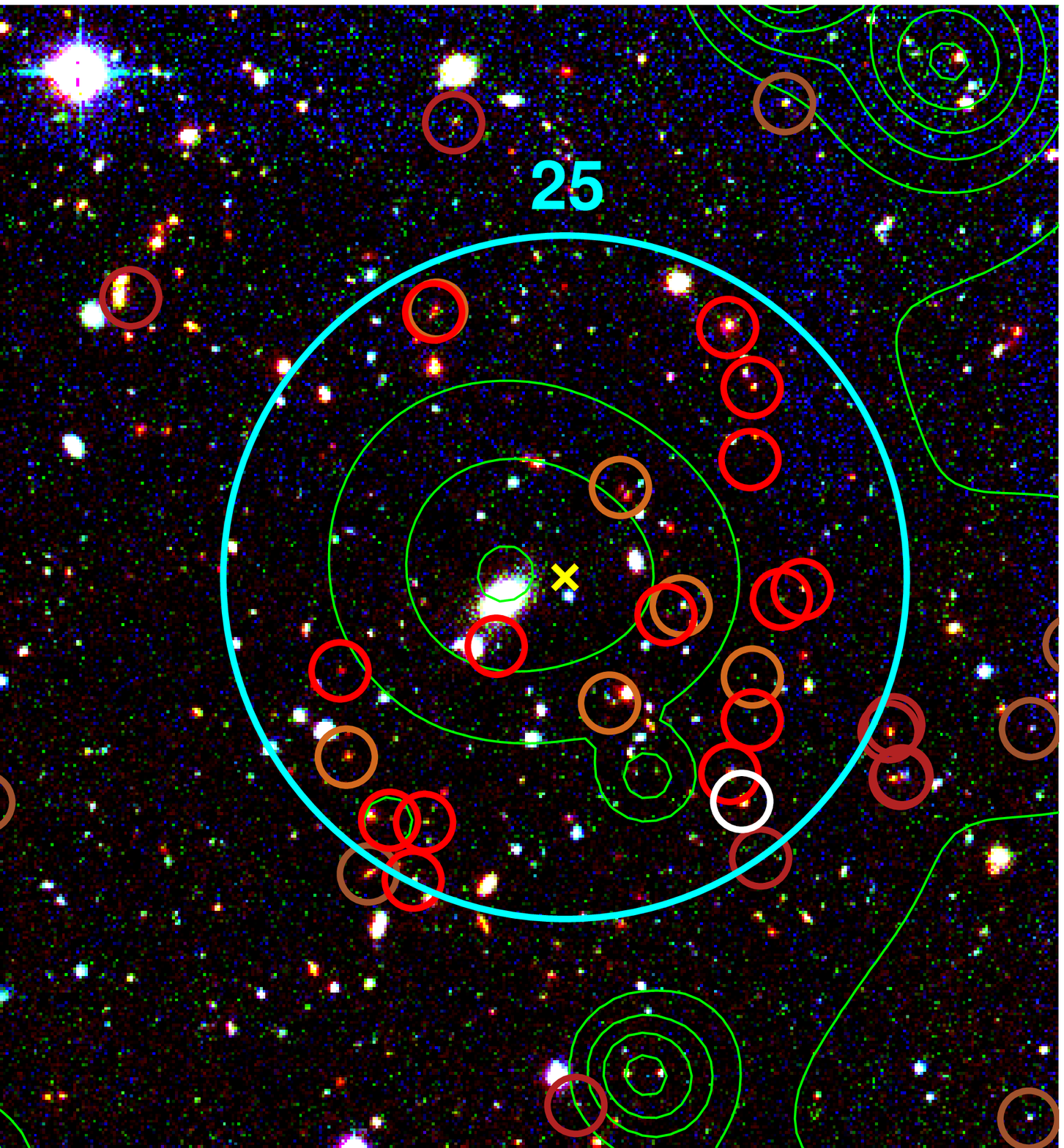}
\end{subfigure}
\hfill
\begin{subfigure}{0.3\textwidth}
\includegraphics[width=6cm]{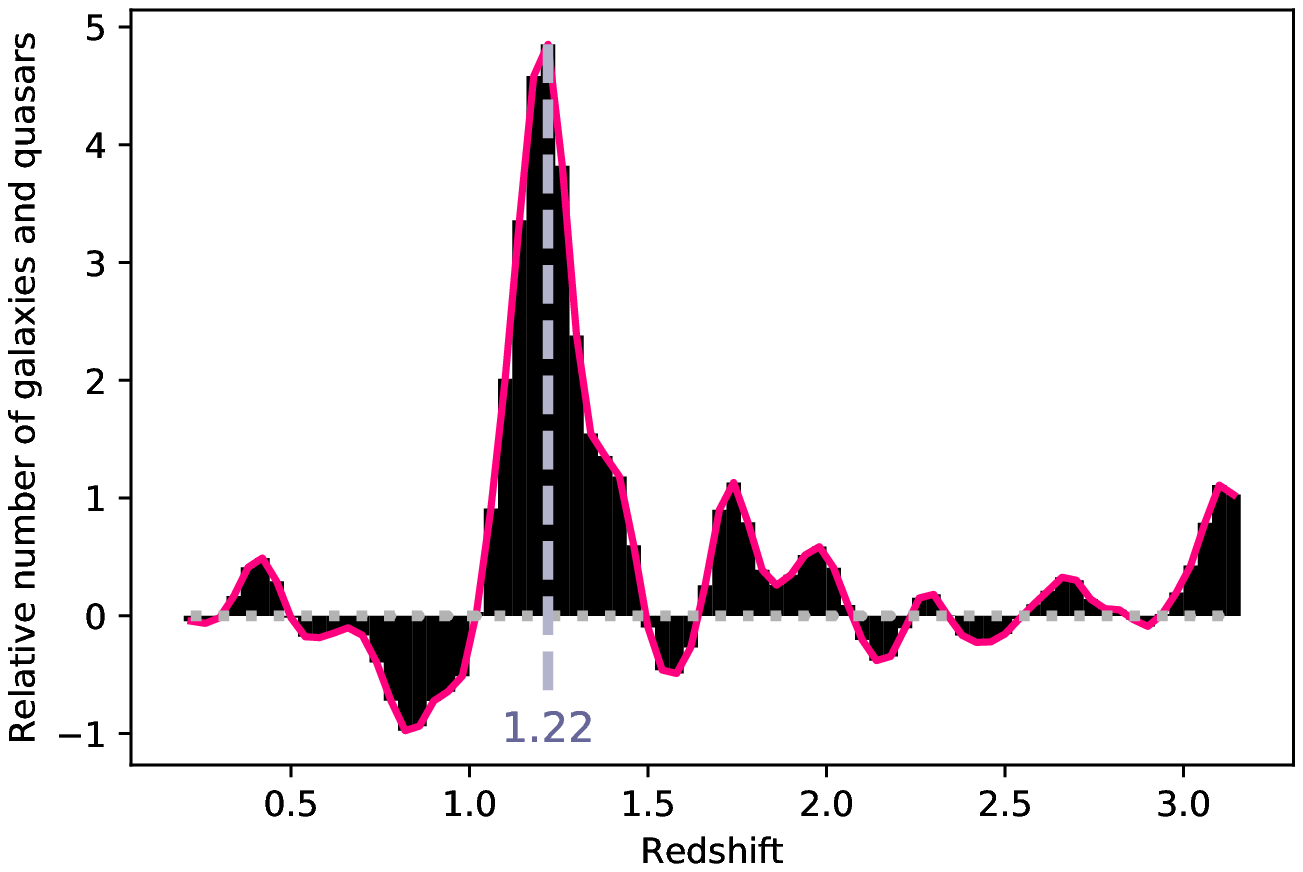}
\end{subfigure}
\hfill
\begin{subfigure}{0.3\textwidth}
\includegraphics[width=6cm]{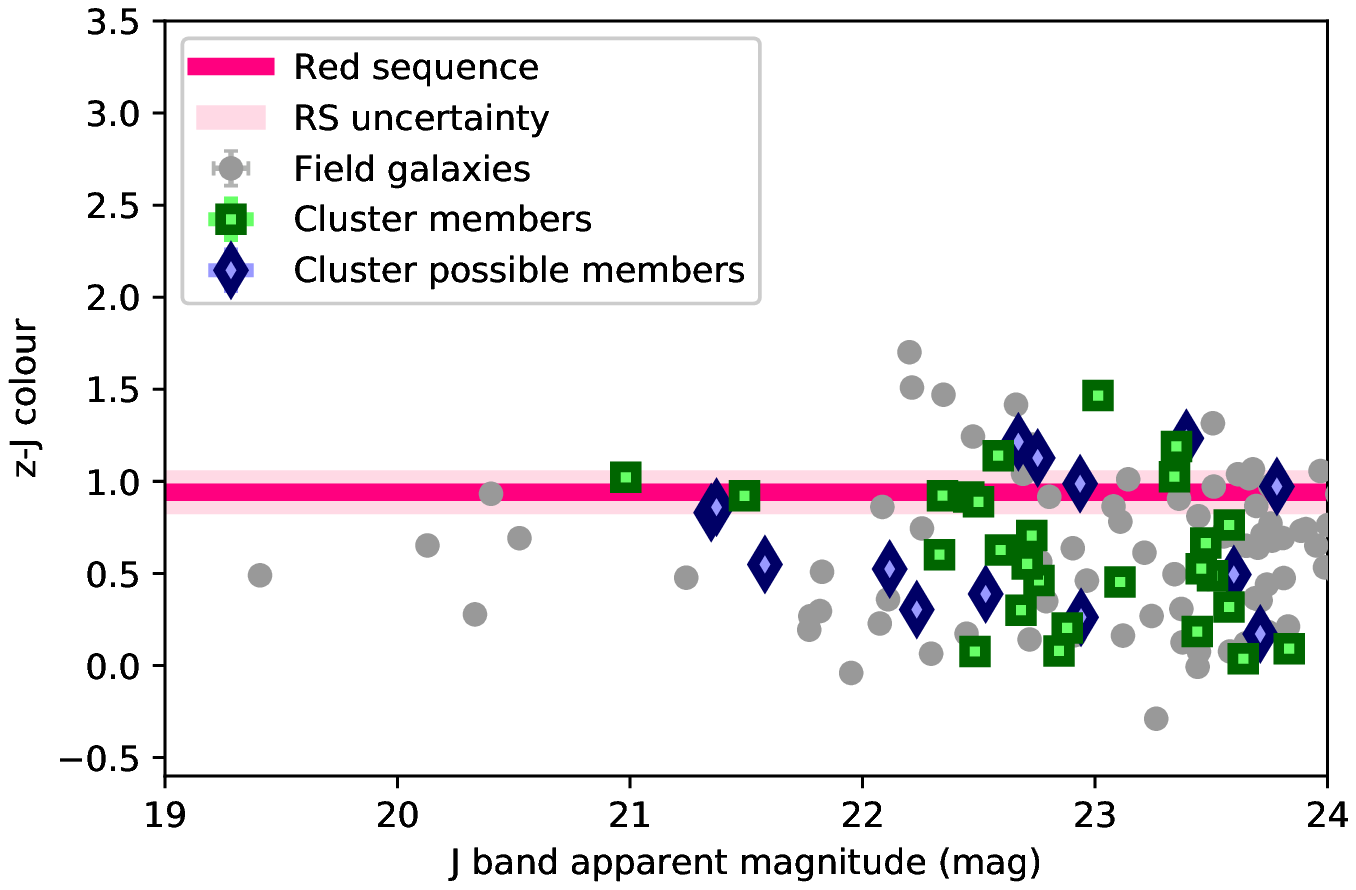}
\end{subfigure}
%
%
%
\centering
\begin{subfigure}{0.3\textwidth}
\includegraphics[width=5.25cm]{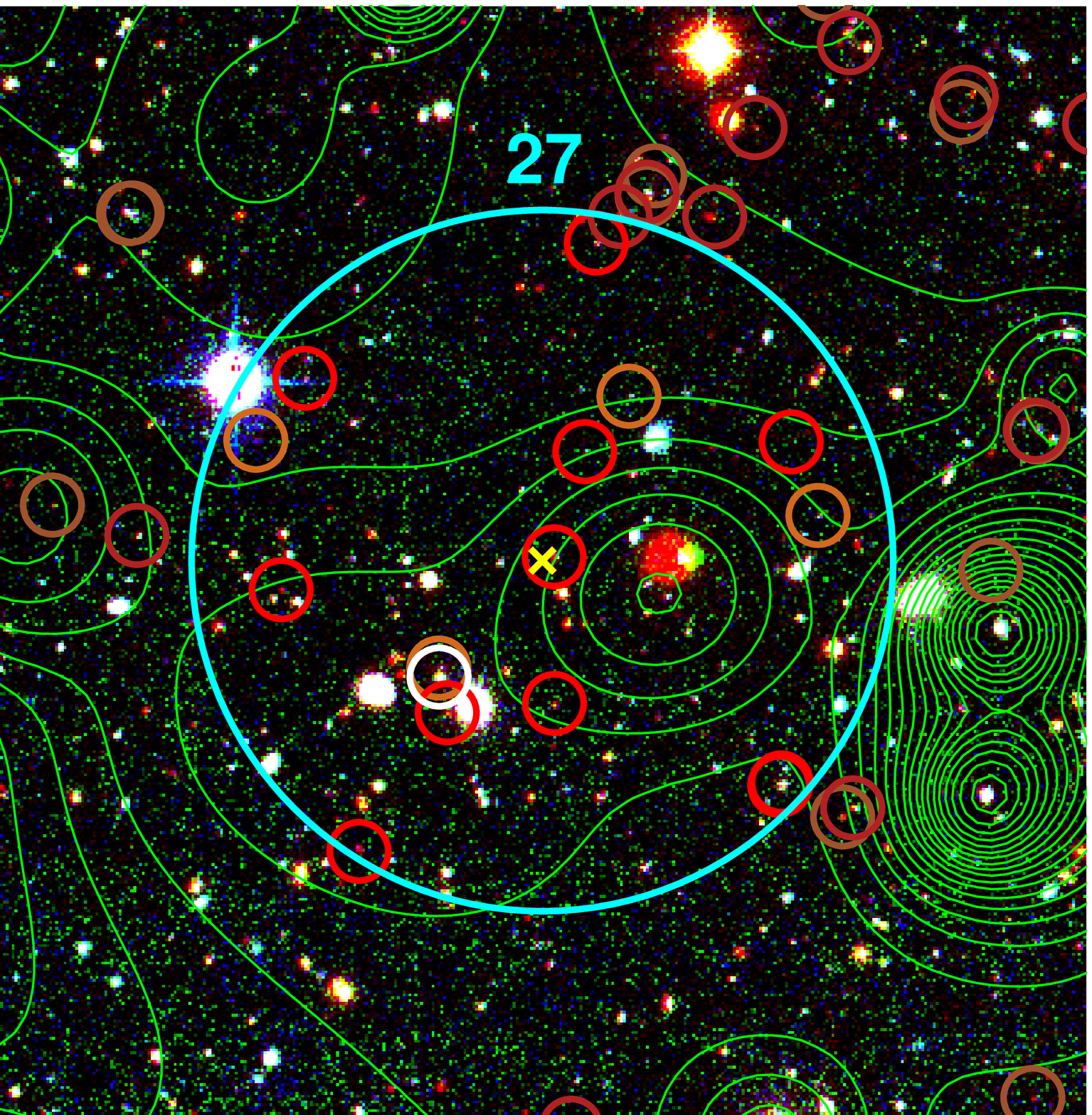}
\end{subfigure}
\hfill
\begin{subfigure}{0.3\textwidth}
\includegraphics[width=6cm]{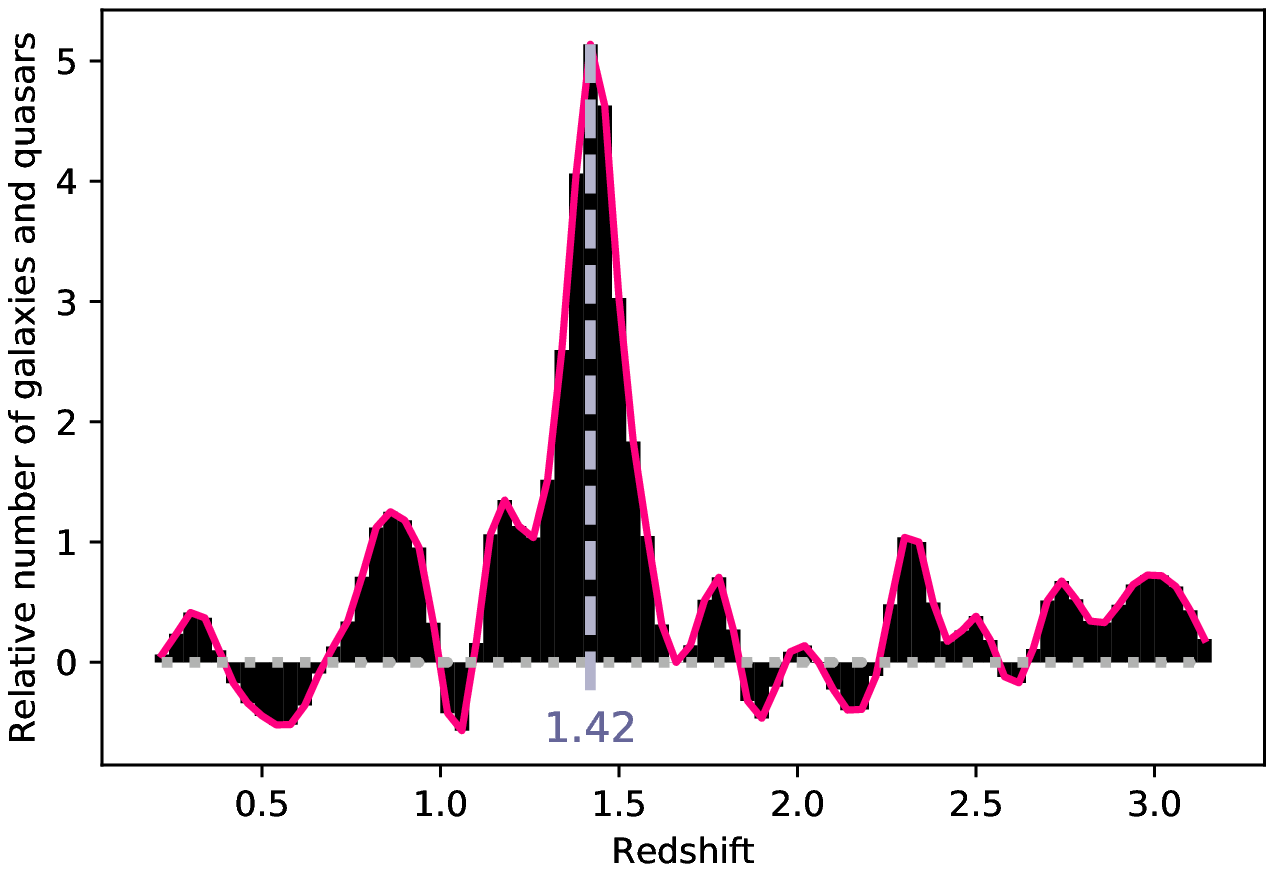}
\end{subfigure}
\hfill
\begin{subfigure}{0.3\textwidth}
\includegraphics[width=6cm]{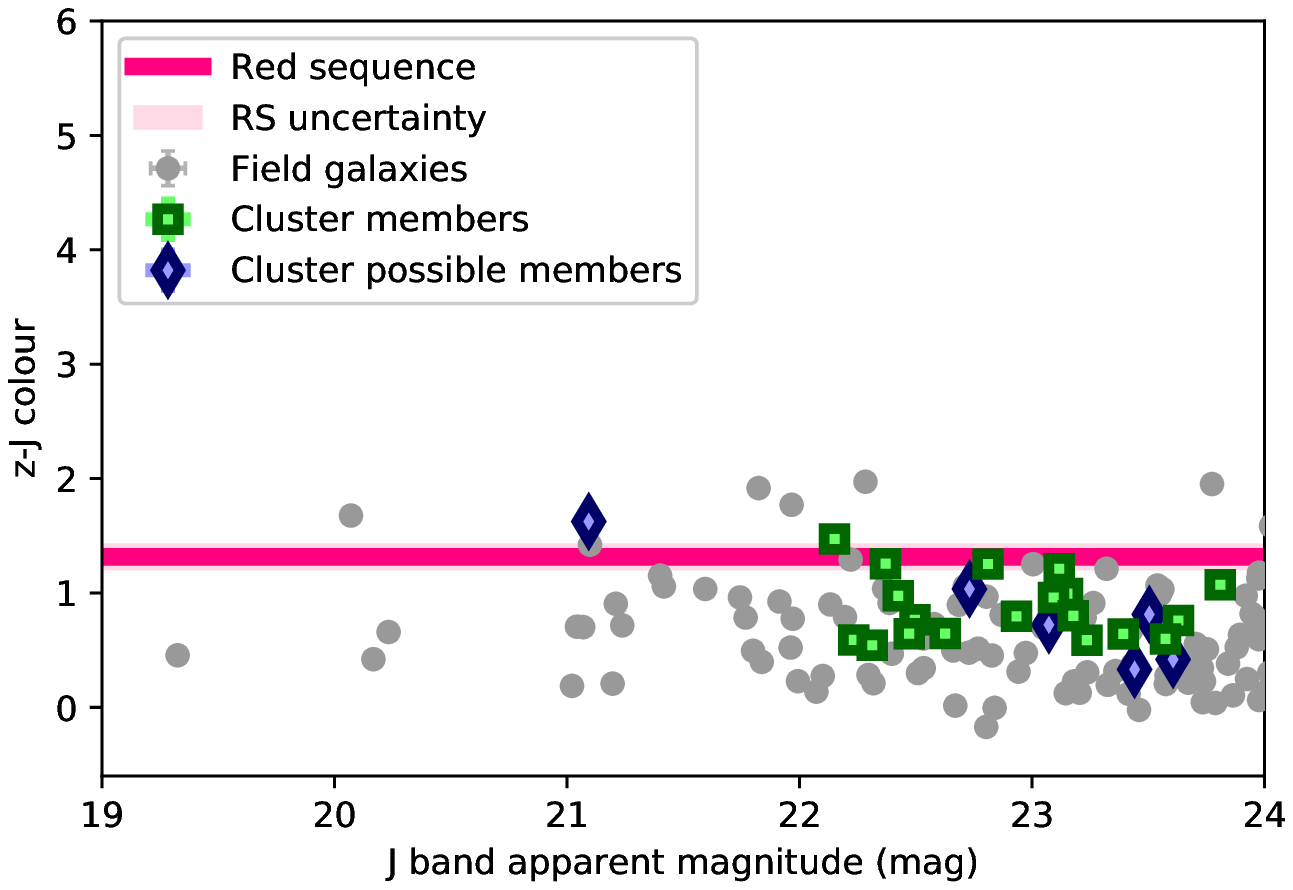}
\end{subfigure}
\caption{\textit{continued}}
\end{figure*}

\newpage

\begin{figure*}
\ContinuedFloat
\centering
\begin{subfigure}{0.3\textwidth}
\includegraphics[width=5.25cm]{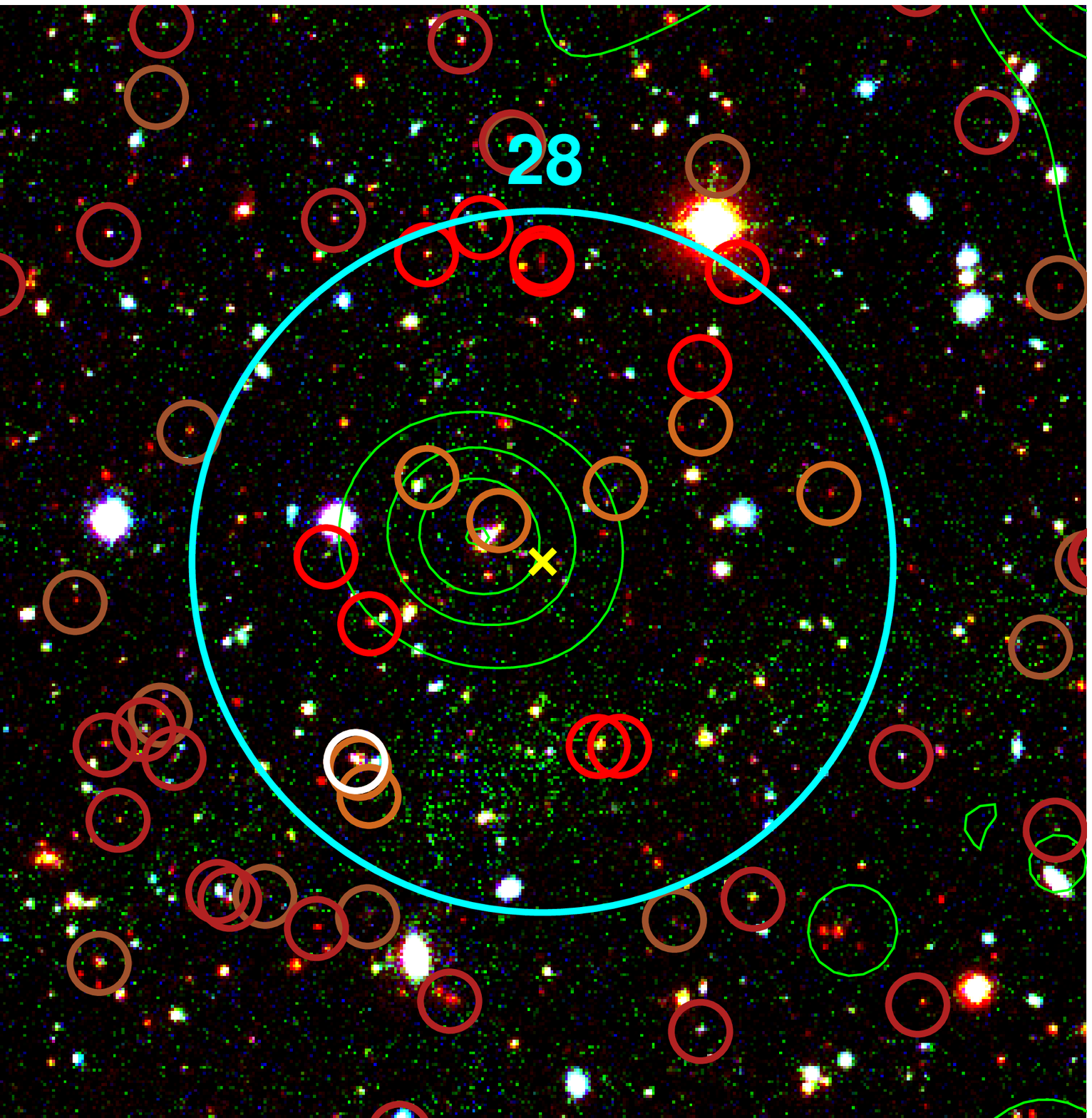}
\end{subfigure}
\hfill
\begin{subfigure}{0.3\textwidth}
\includegraphics[width=6cm]{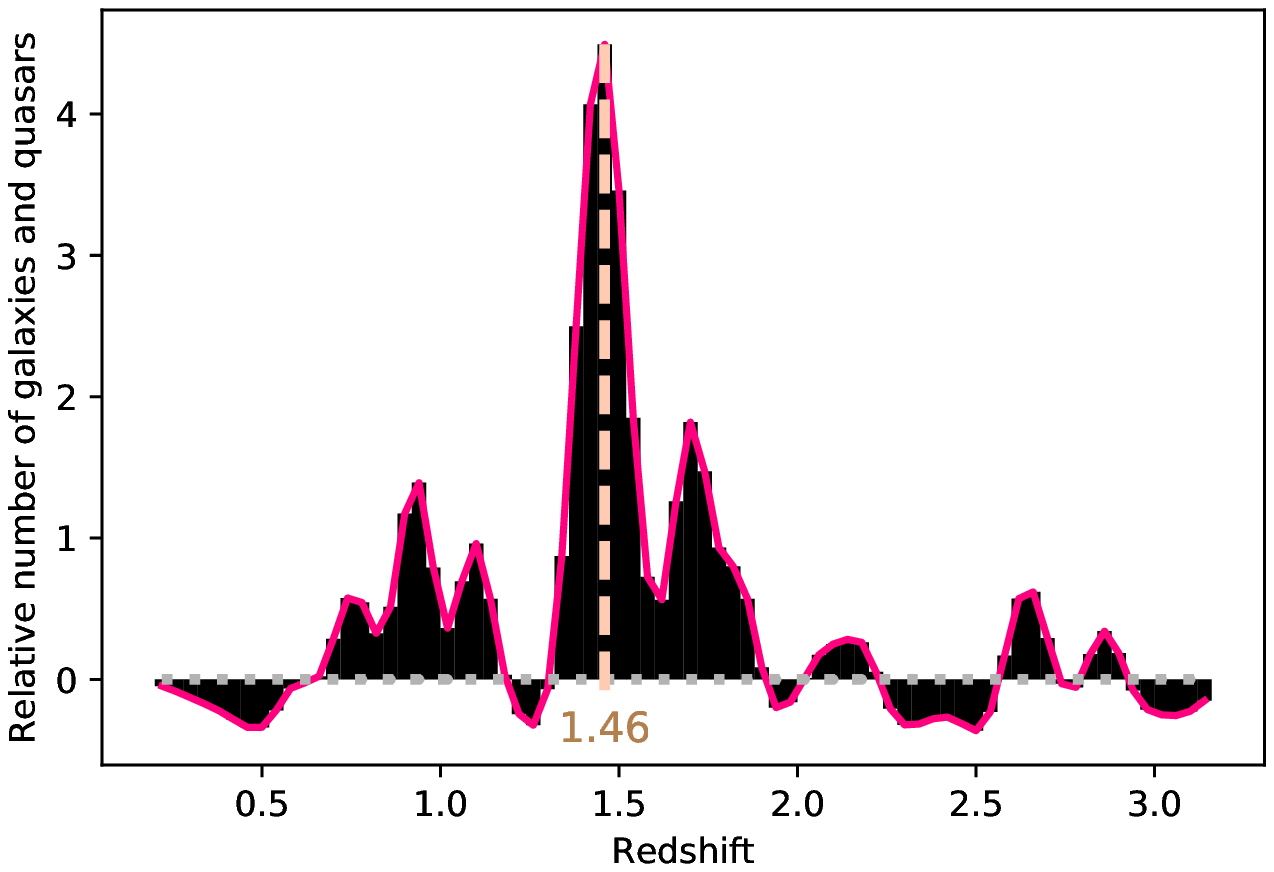}
\end{subfigure}
\hfill
\begin{subfigure}{0.3\textwidth}
\includegraphics[width=6cm]{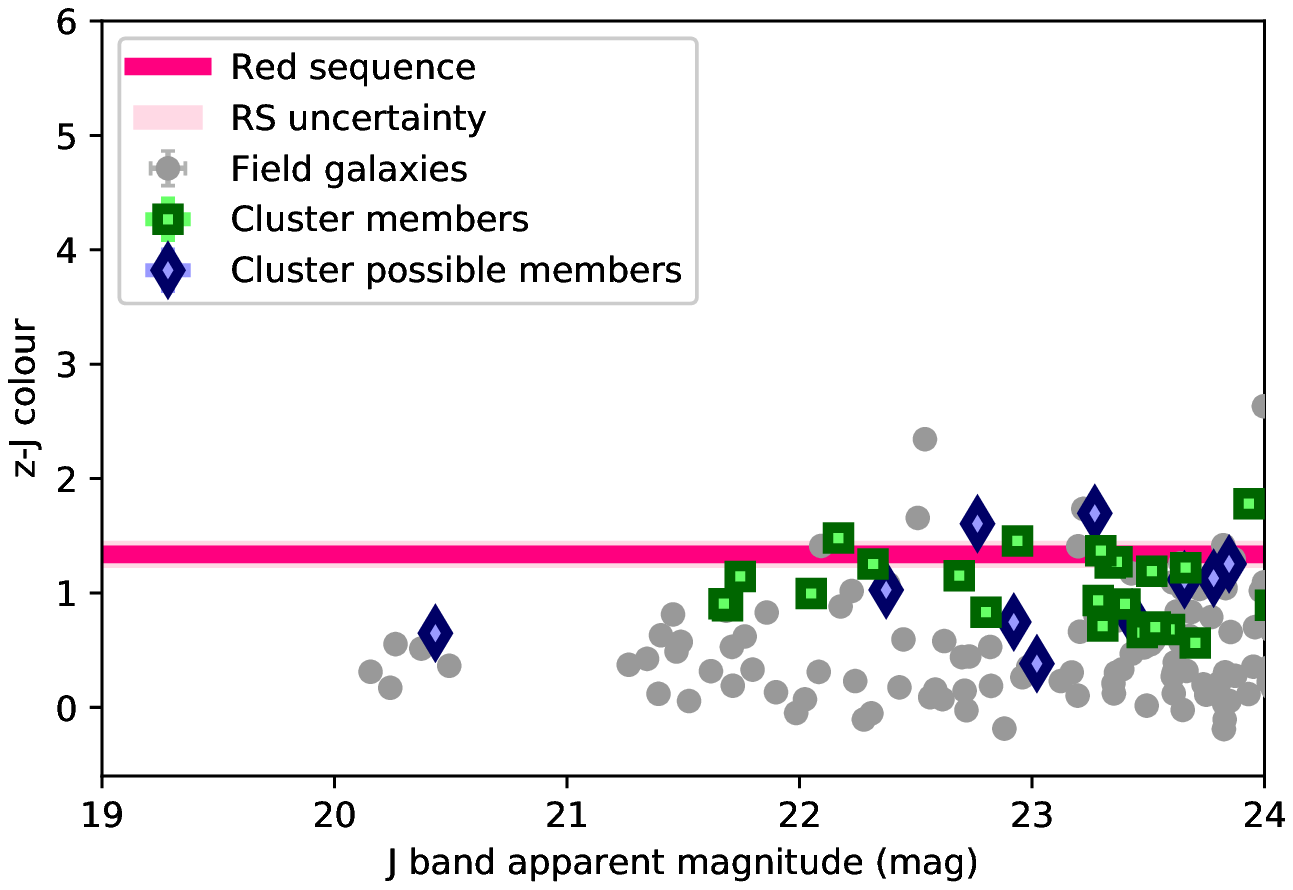}
\end{subfigure}

\centering
\begin{subfigure}{0.3\textwidth}
\includegraphics[width=5.25cm]{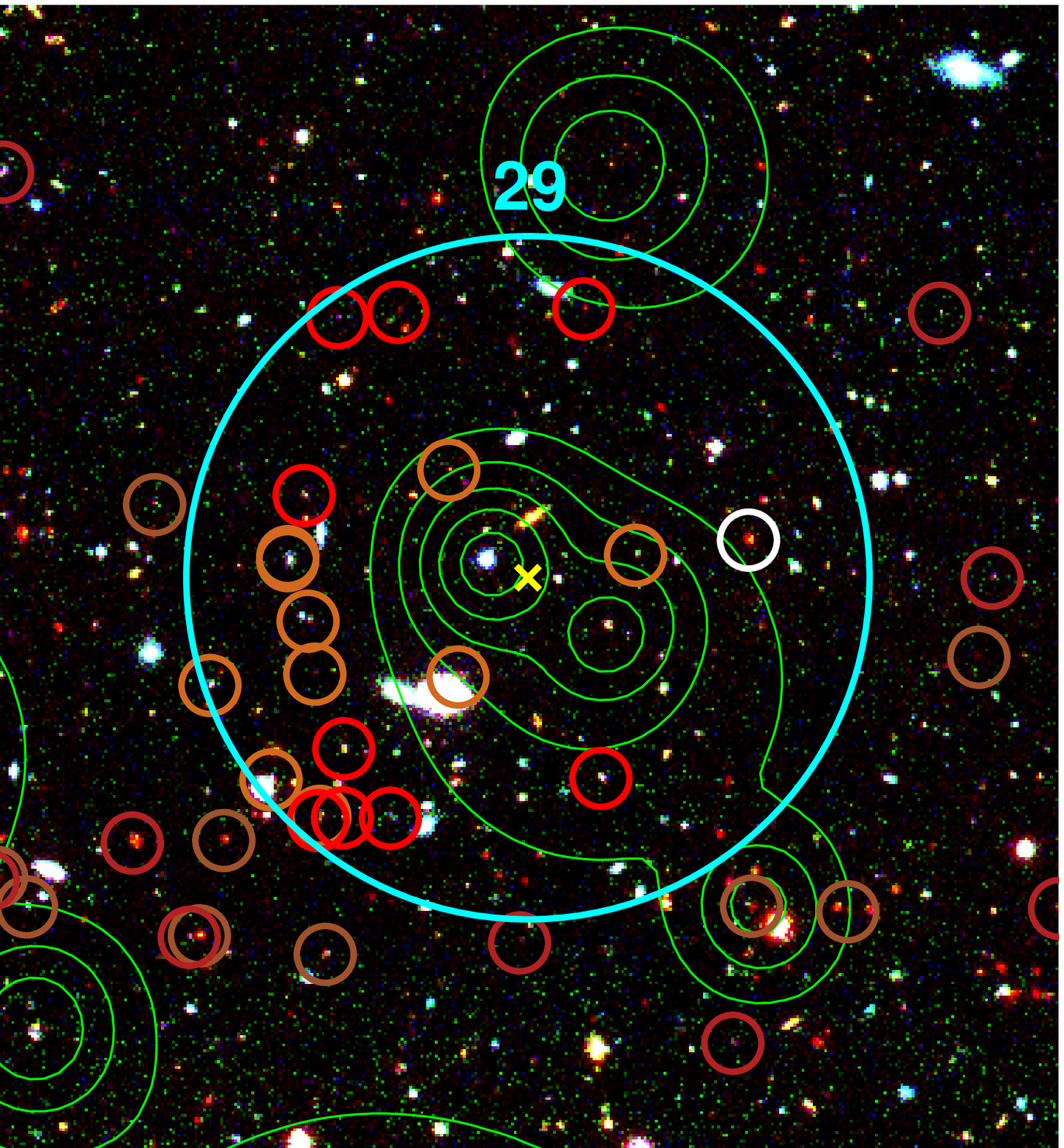}
\end{subfigure}
\hfill
\begin{subfigure}{0.3\textwidth}
\includegraphics[width=6cm]{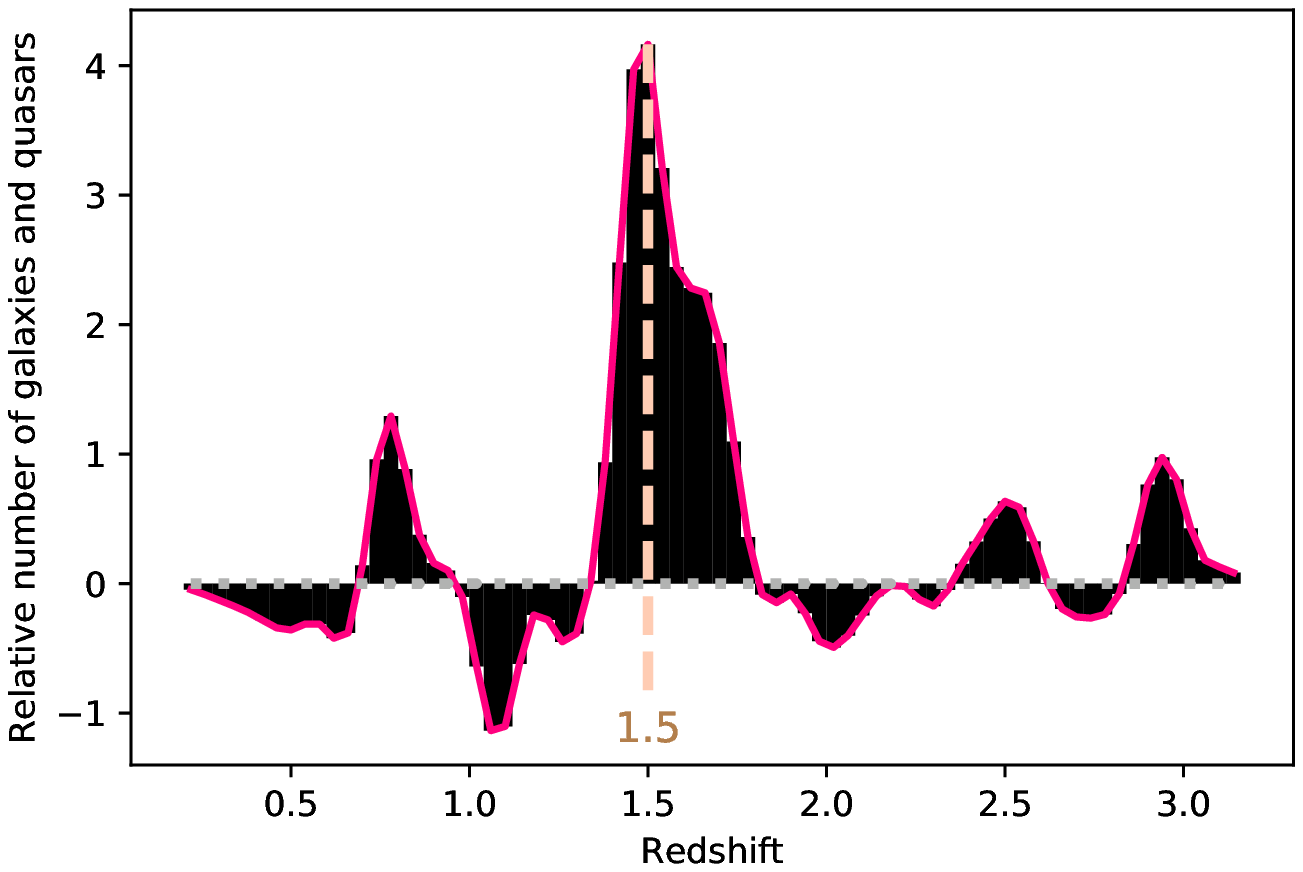}
\end{subfigure}
\hfill
\begin{subfigure}{0.3\textwidth}
\includegraphics[width=6cm]{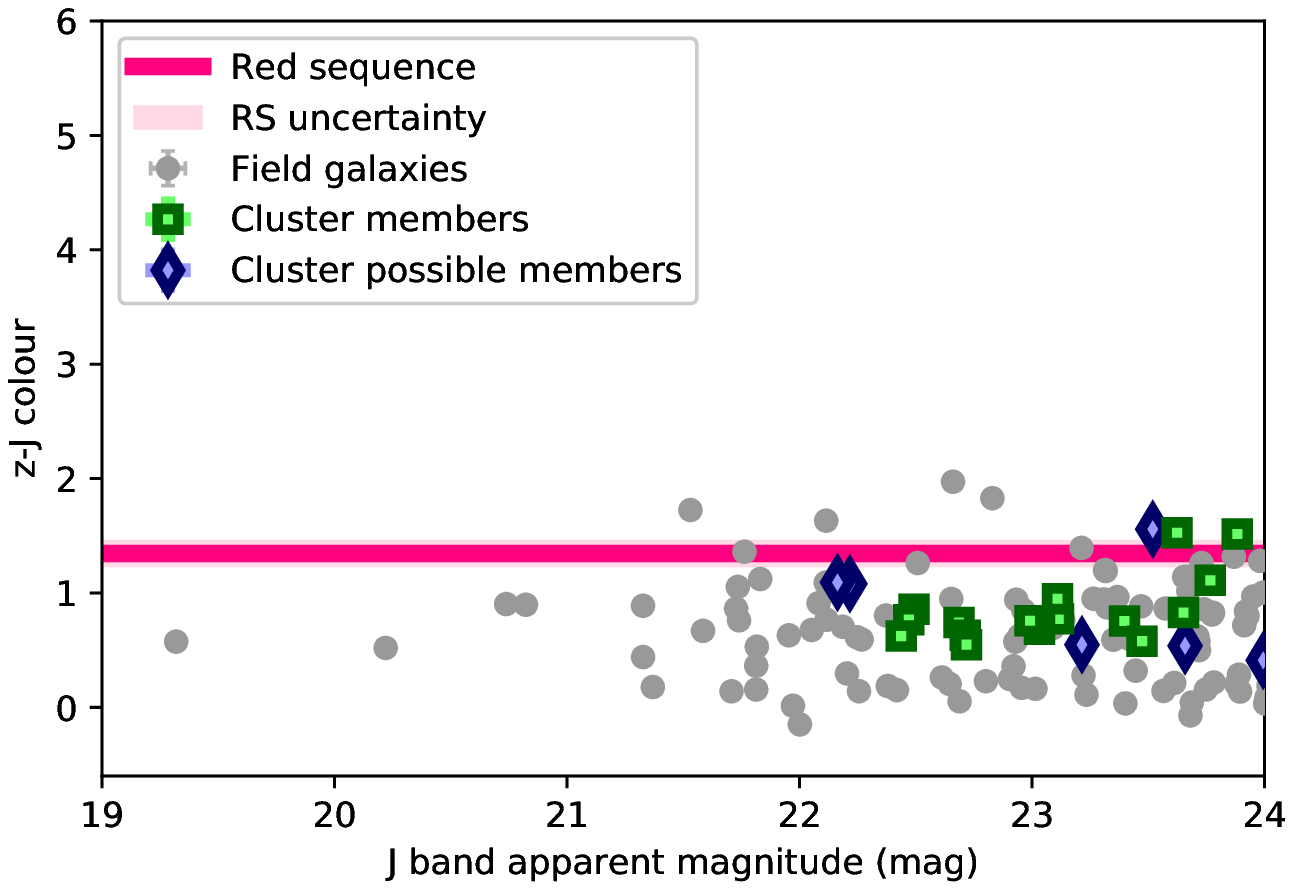}
\end{subfigure}

\centering
\begin{subfigure}{0.3\textwidth}
\includegraphics[width=5.25cm]{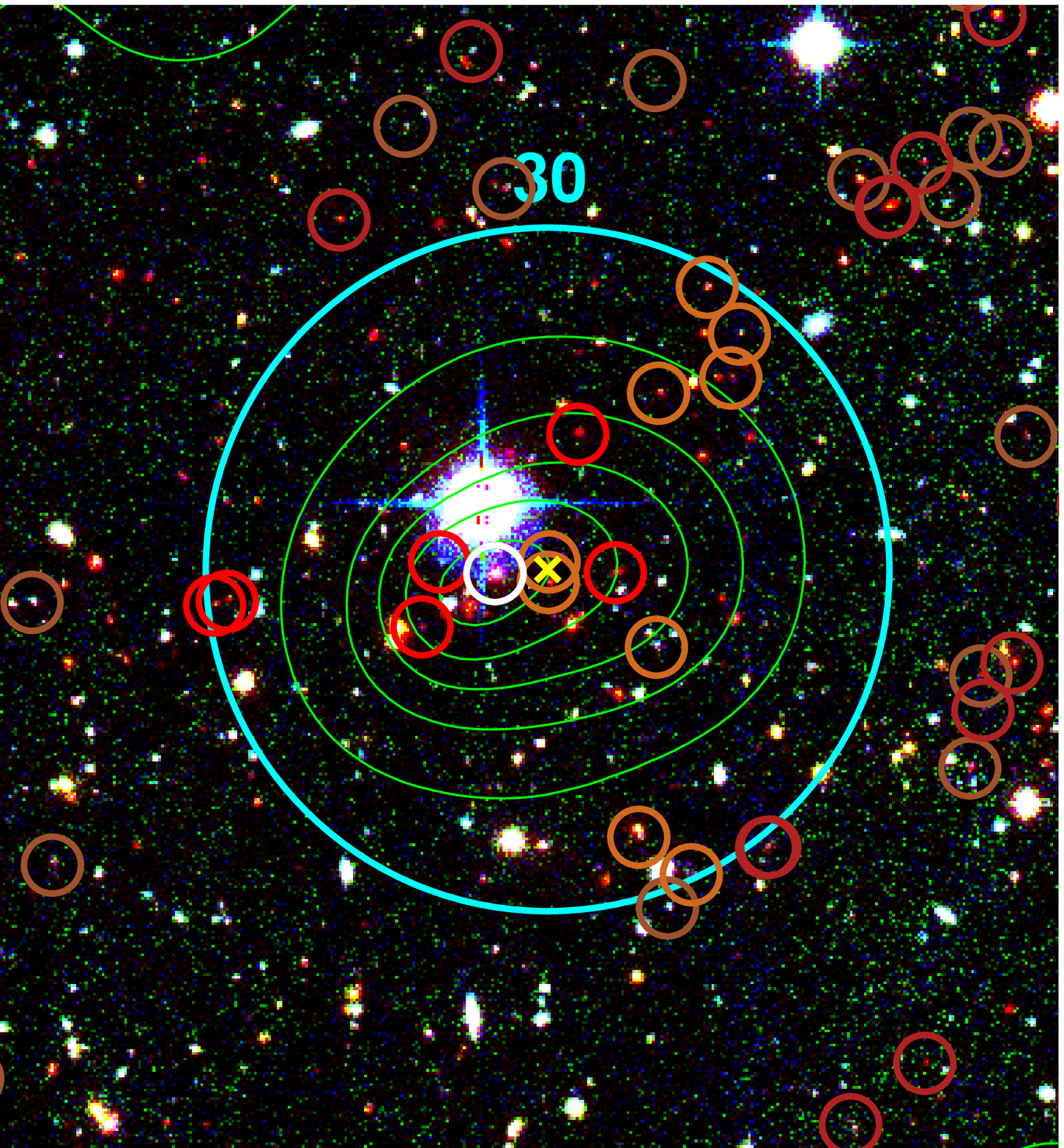}
\end{subfigure}
\hfill
\begin{subfigure}{0.3\textwidth}
\includegraphics[width=6cm]{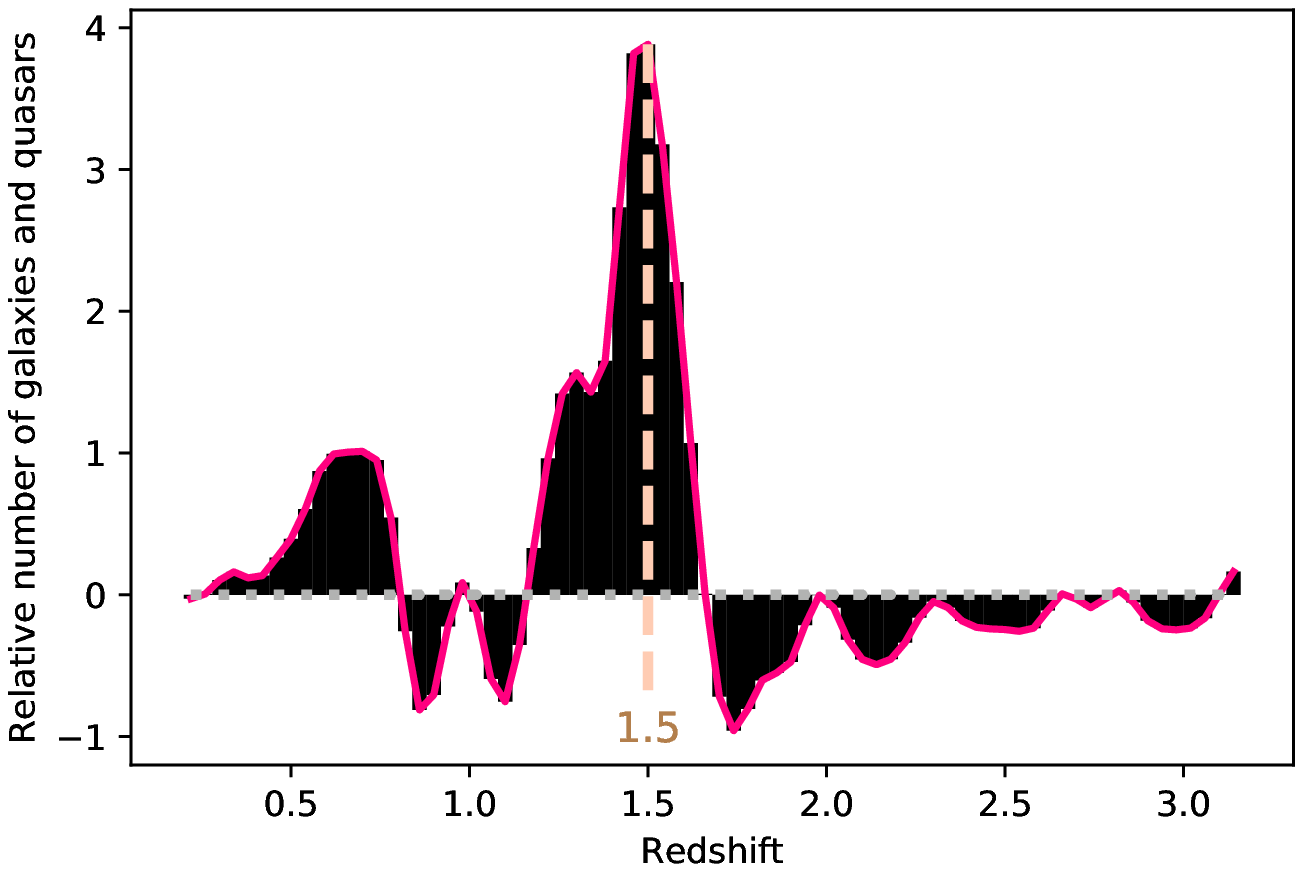}
\end{subfigure}
\hfill
\begin{subfigure}{0.3\textwidth}
\includegraphics[width=6cm]{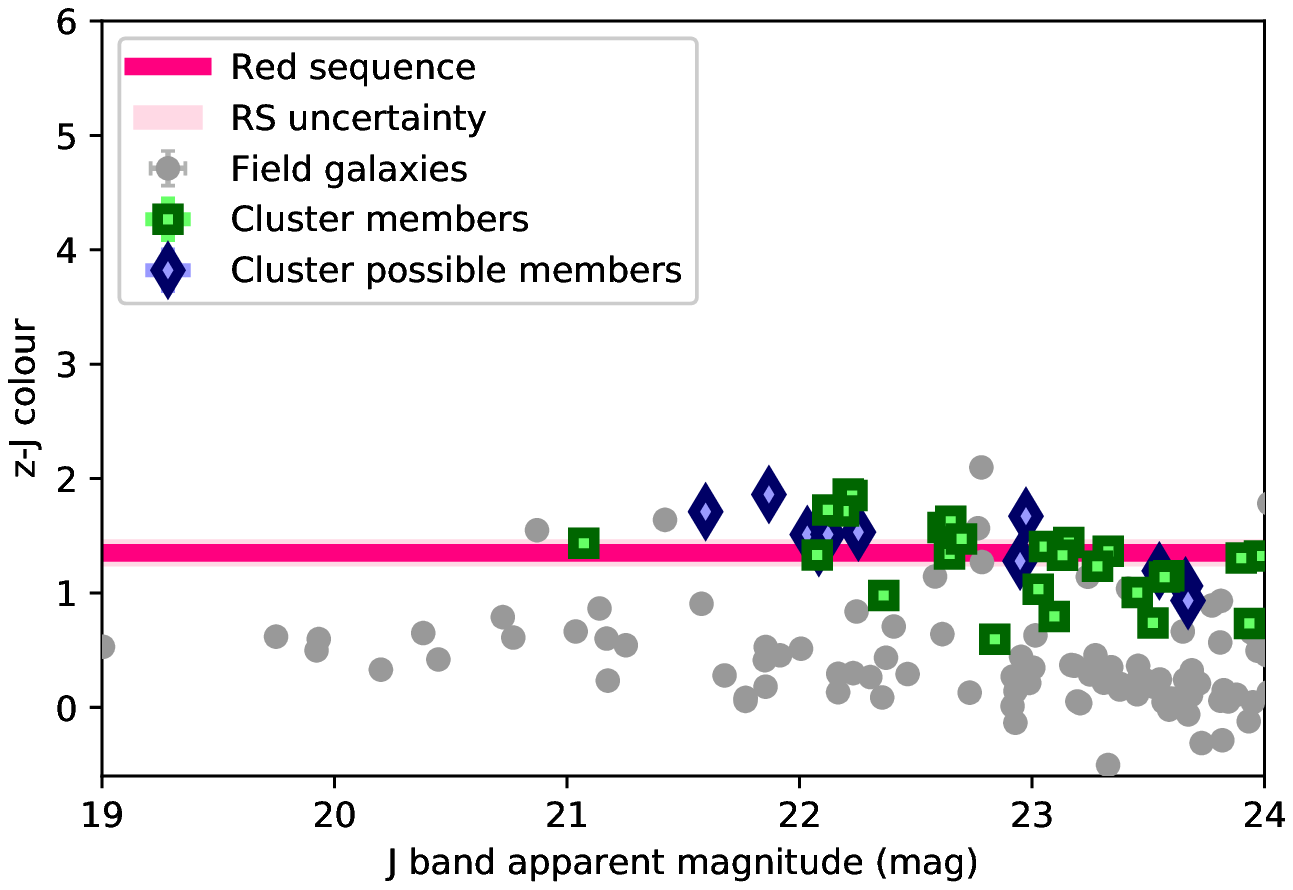}
\end{subfigure}
%
%
%
\centering
\begin{subfigure}{0.3\textwidth}
\includegraphics[width=5.25cm]{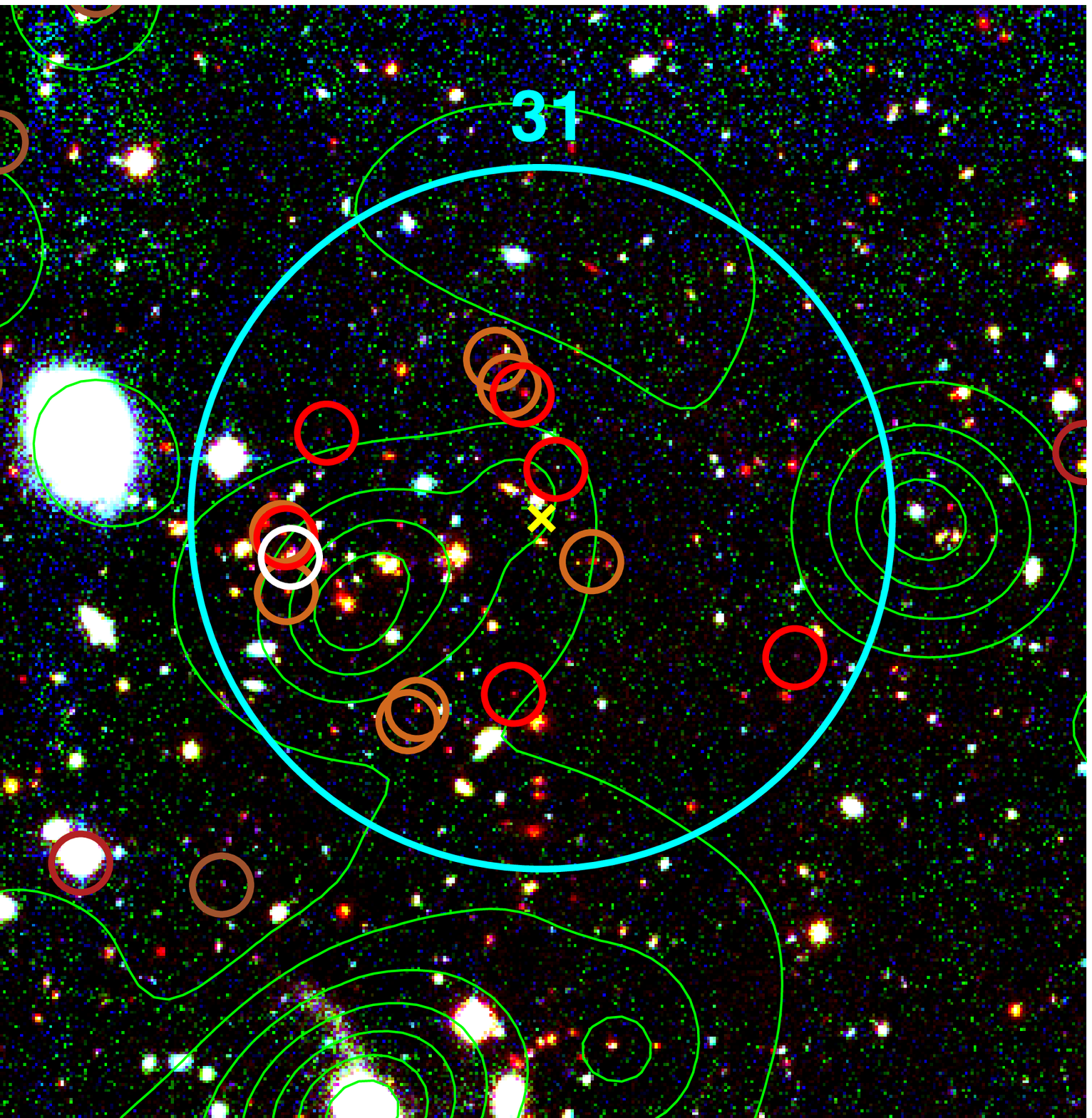}
\end{subfigure}
\hfill
\begin{subfigure}{0.3\textwidth}
\includegraphics[width=6cm]{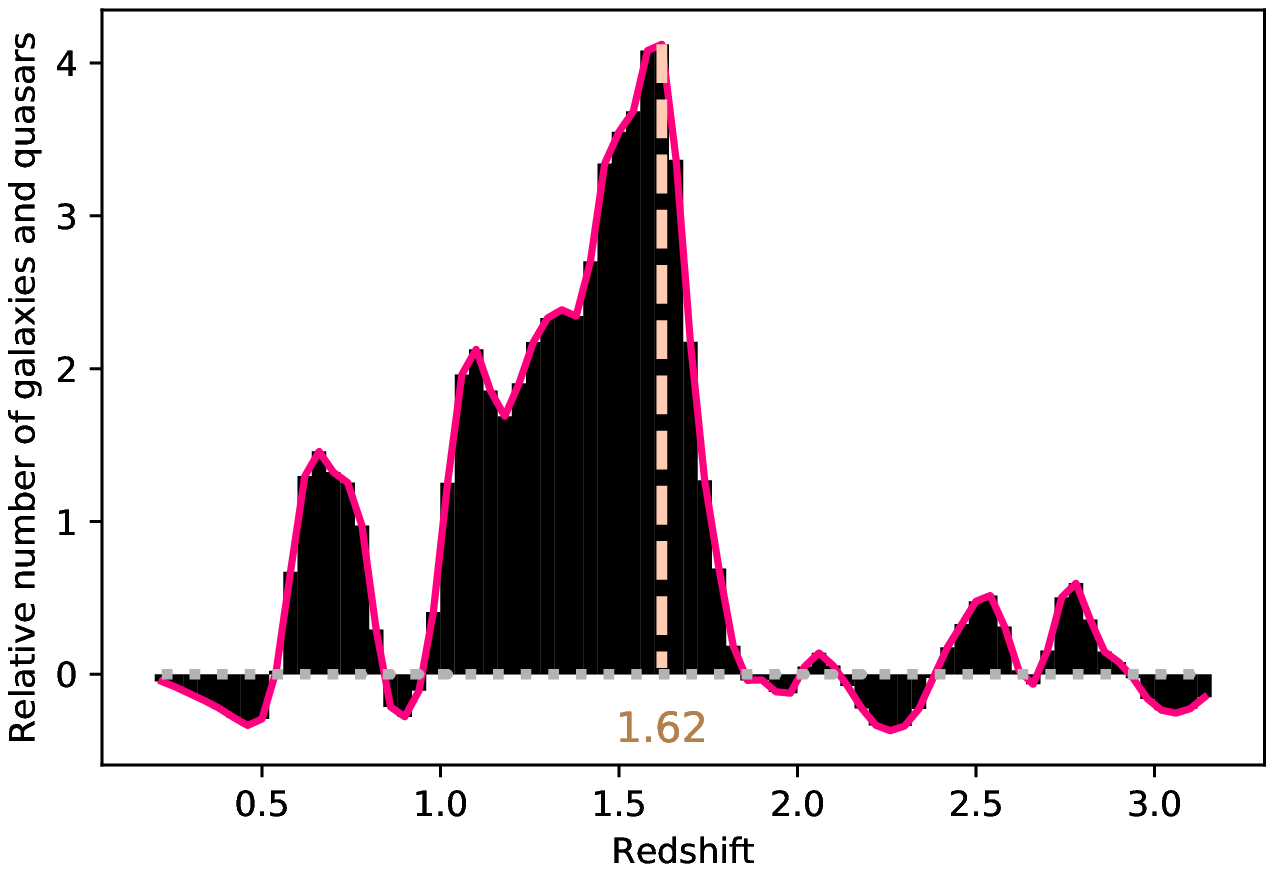}
\end{subfigure}
\hfill
\begin{subfigure}{0.3\textwidth}
\includegraphics[width=6cm]{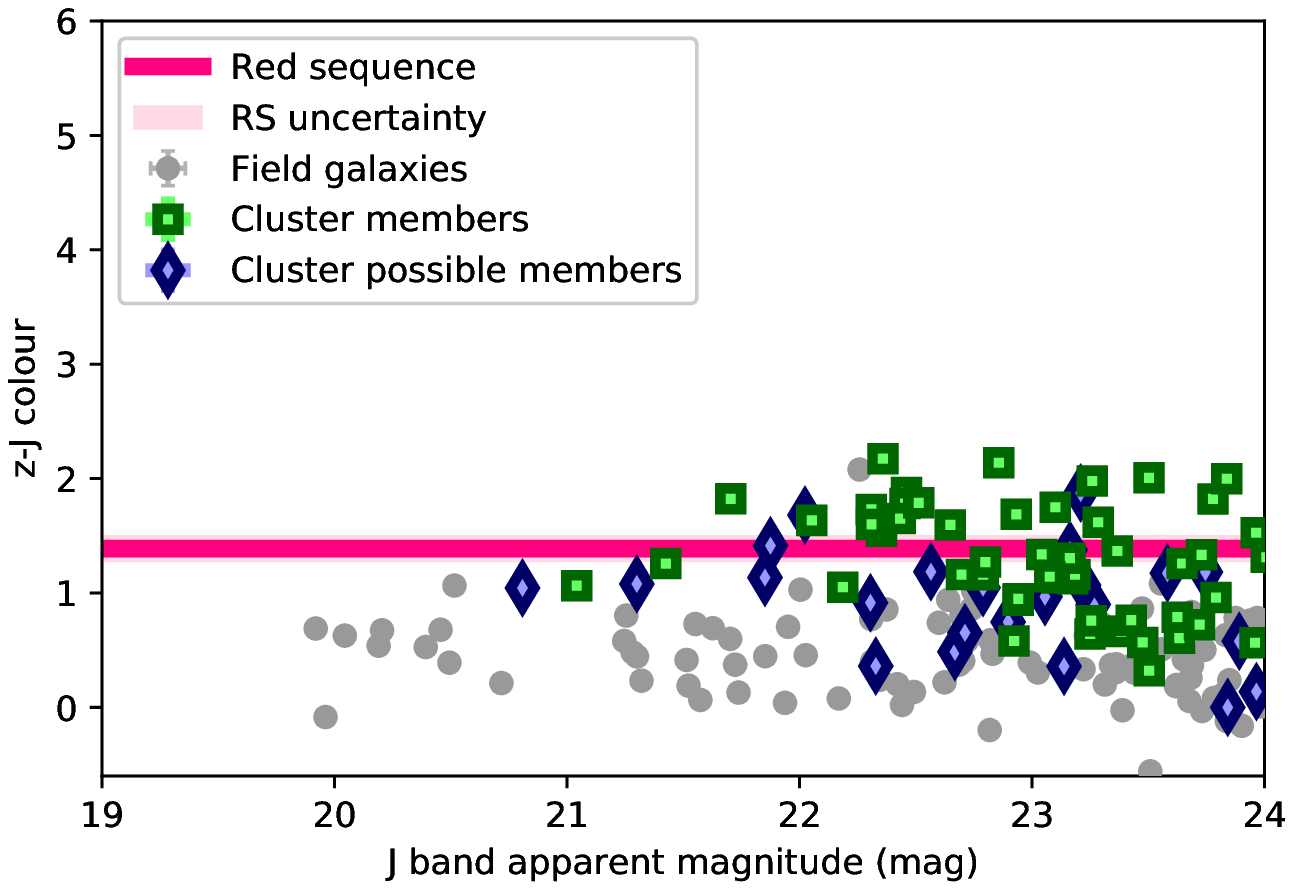}
\end{subfigure}
\caption{\textit{continued}}
\end{figure*}

\newpage

\begin{figure*}
\ContinuedFloat
\centering
\begin{subfigure}{0.3\textwidth}
\includegraphics[width=5.25cm]{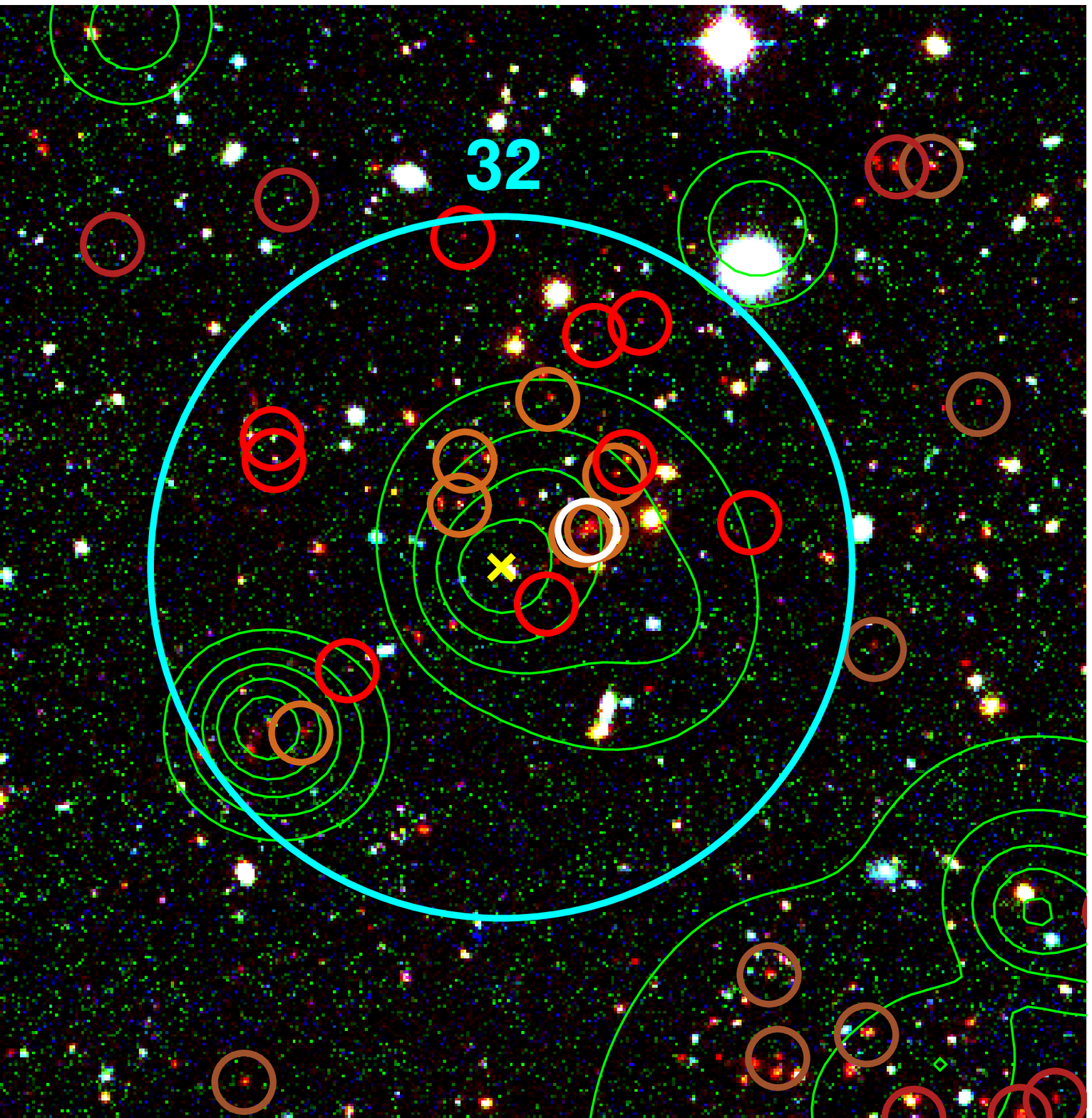}
\end{subfigure}
\hfill
\begin{subfigure}{0.3\textwidth}
\includegraphics[width=6cm]{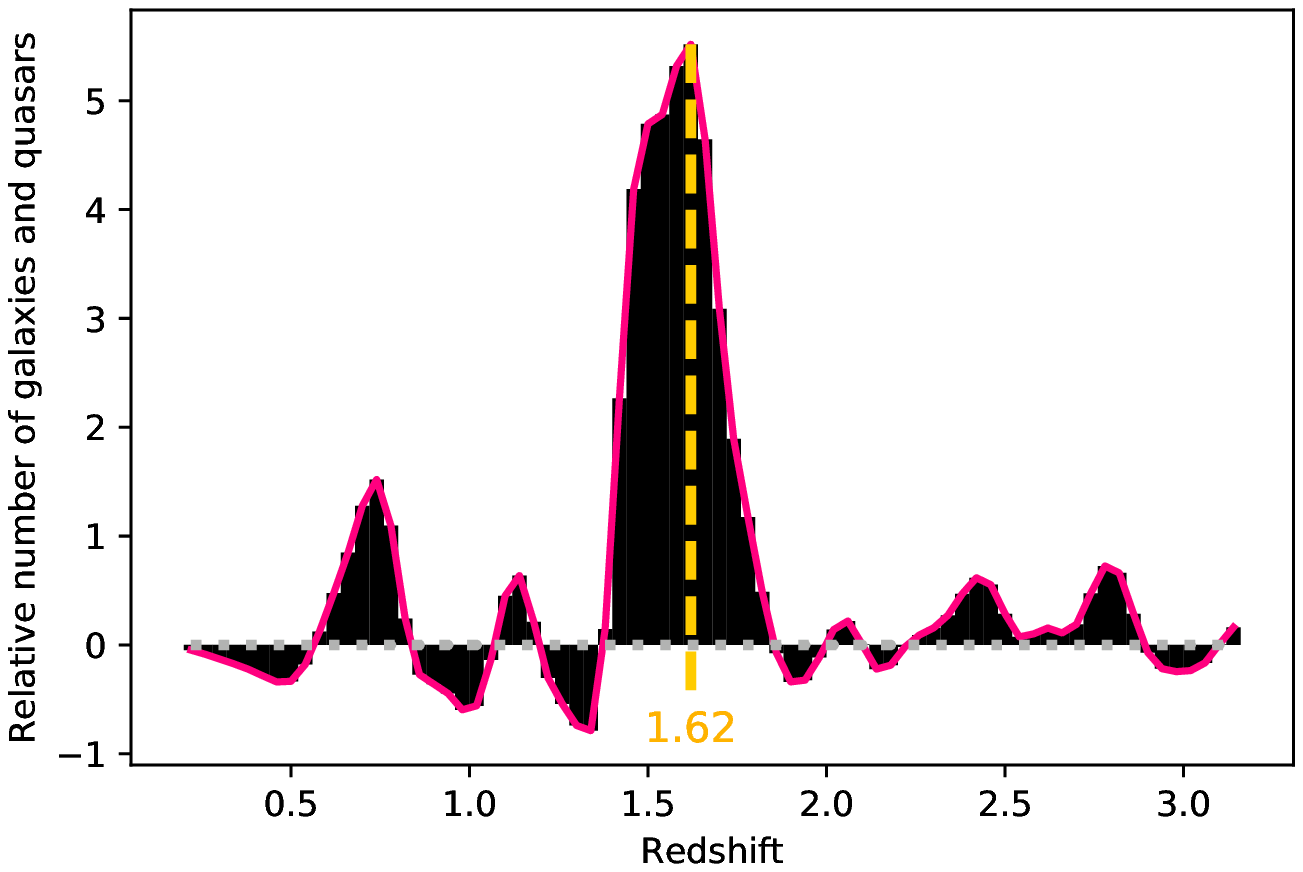}
\end{subfigure}
\hfill
\begin{subfigure}{0.3\textwidth}
\includegraphics[width=6cm]{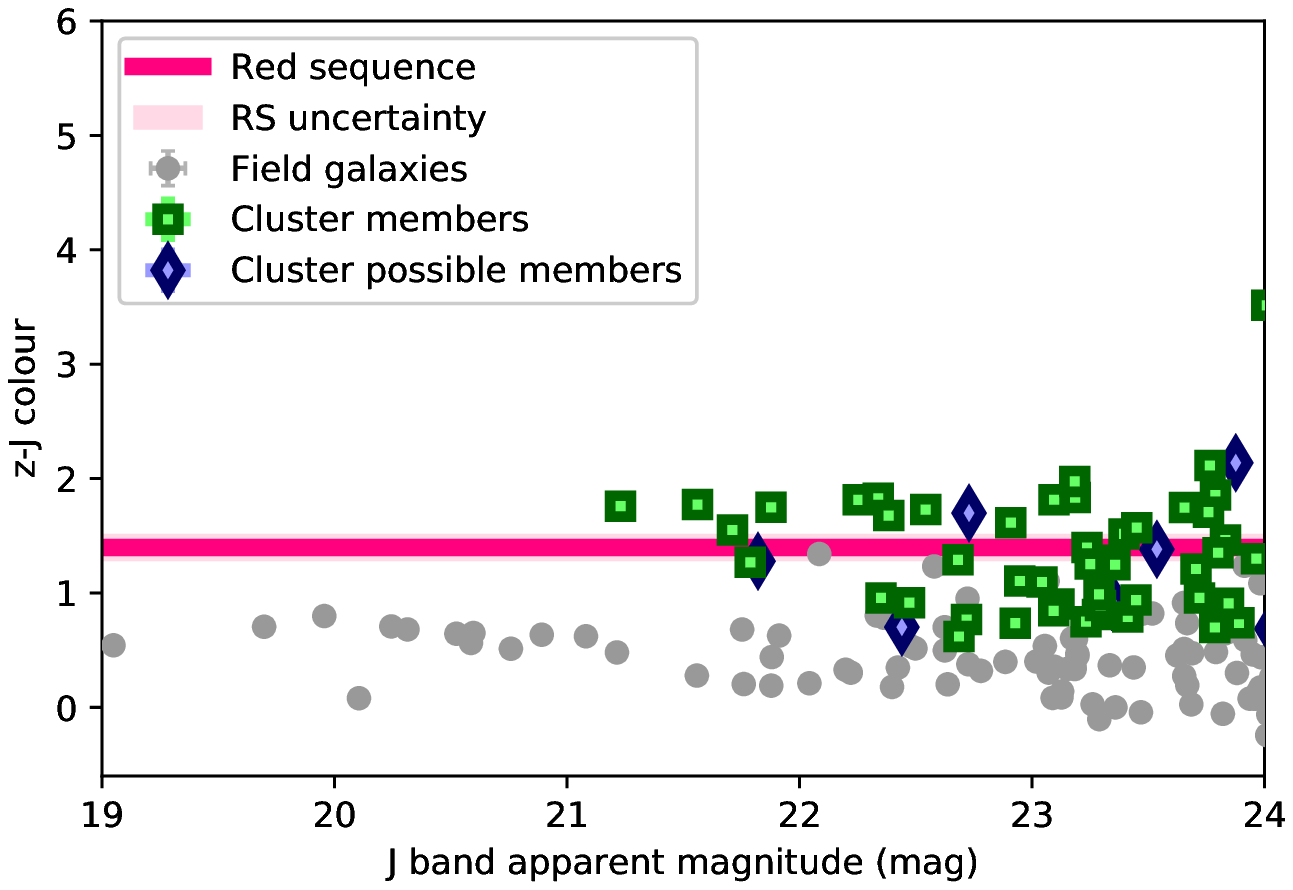}
\end{subfigure}

\centering
\begin{subfigure}{0.3\textwidth}
\includegraphics[width=5.25cm]{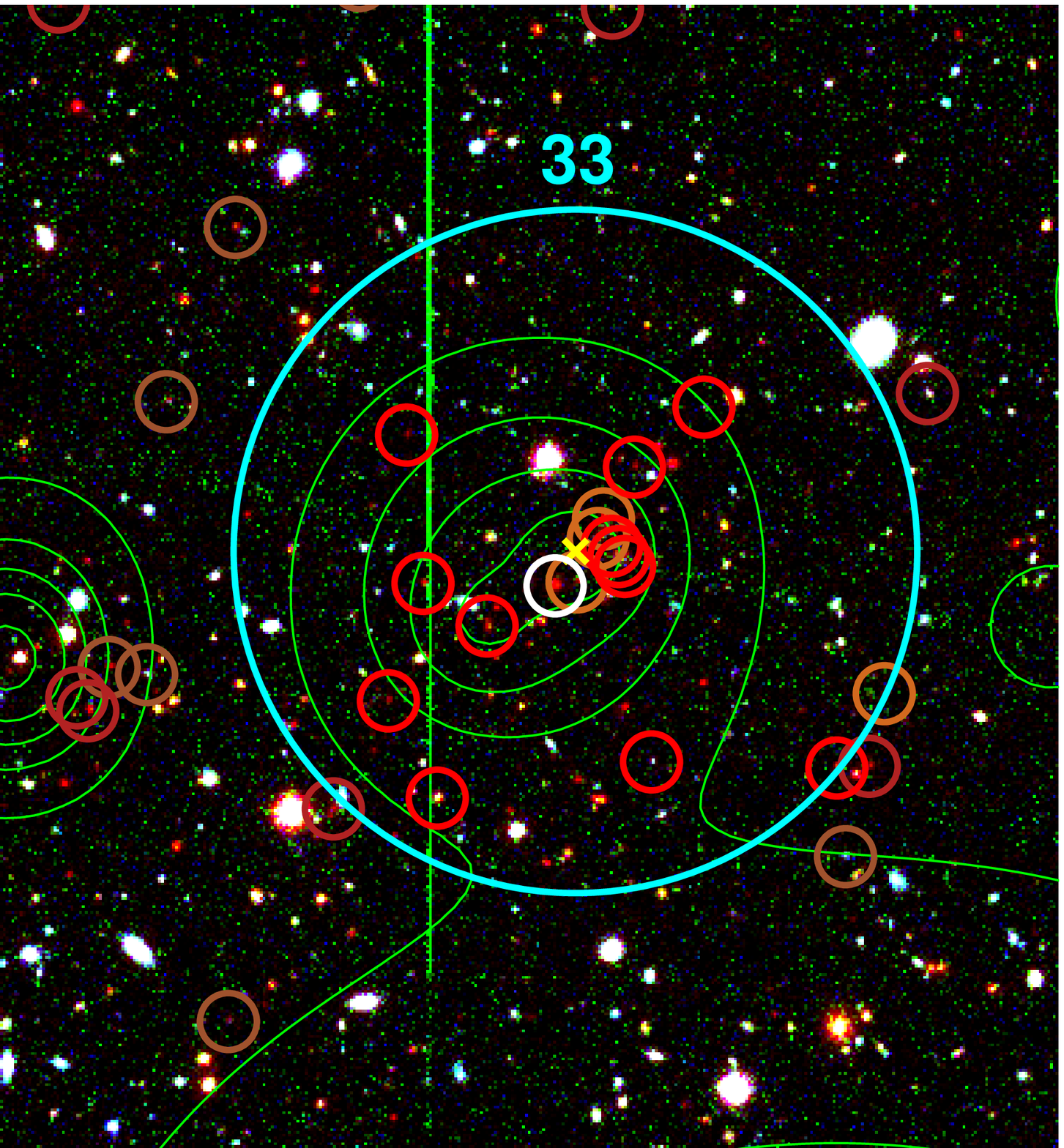}
\end{subfigure}
\hfill
\begin{subfigure}{0.3\textwidth}
\includegraphics[width=6cm]{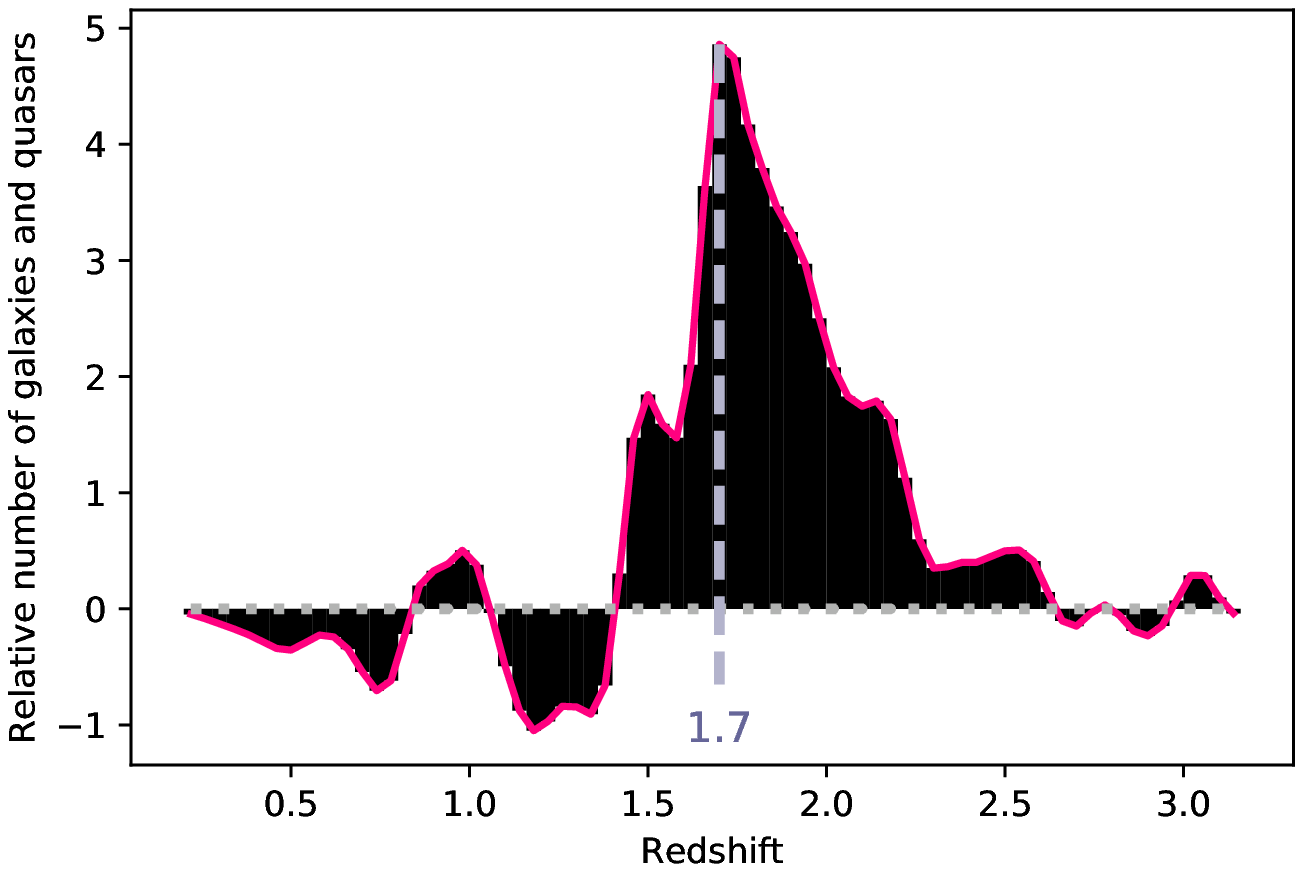}
\end{subfigure}
\hfill
\begin{subfigure}{0.3\textwidth}
\includegraphics[width=6cm]{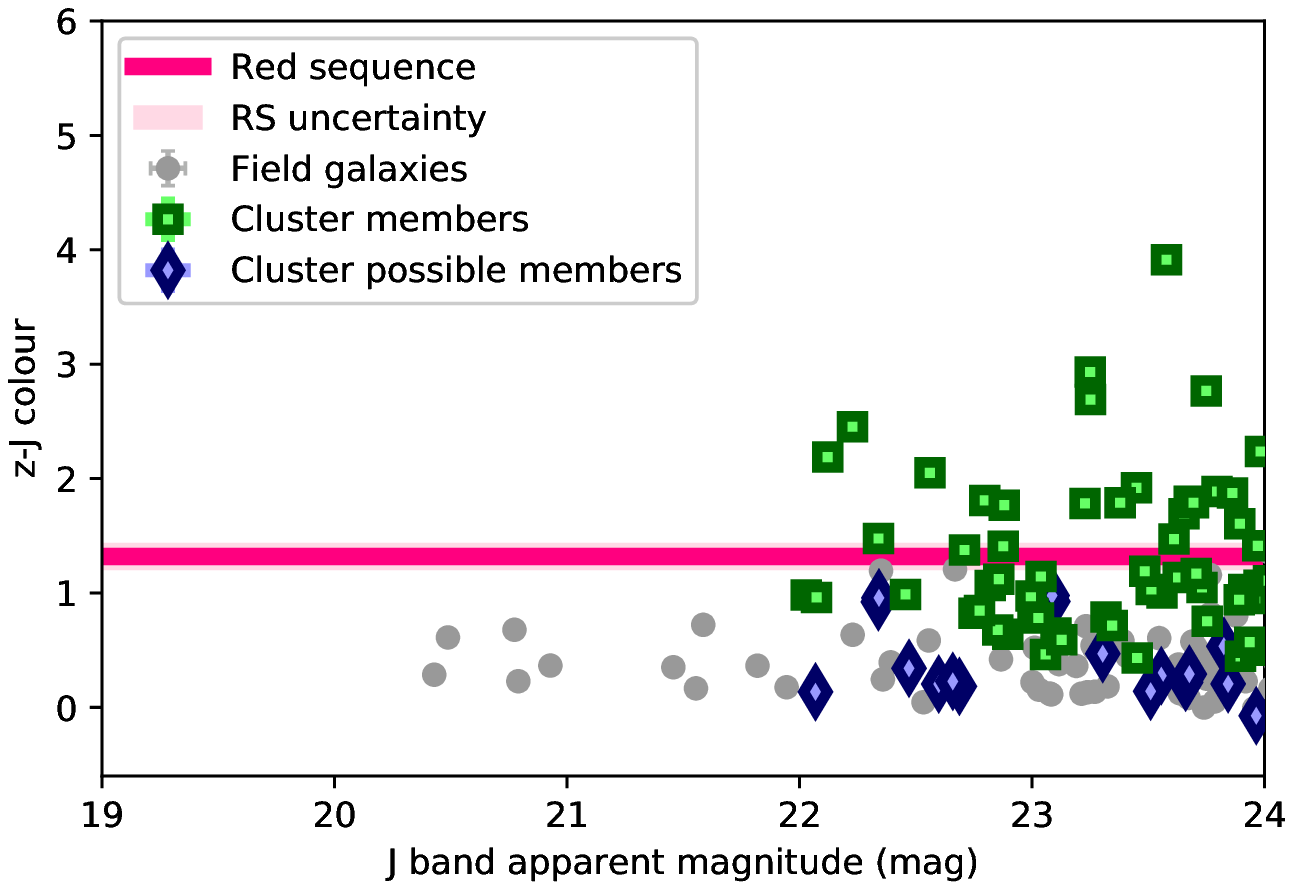}
\end{subfigure}

\centering
\begin{subfigure}{0.3\textwidth}
\includegraphics[width=5.25cm]{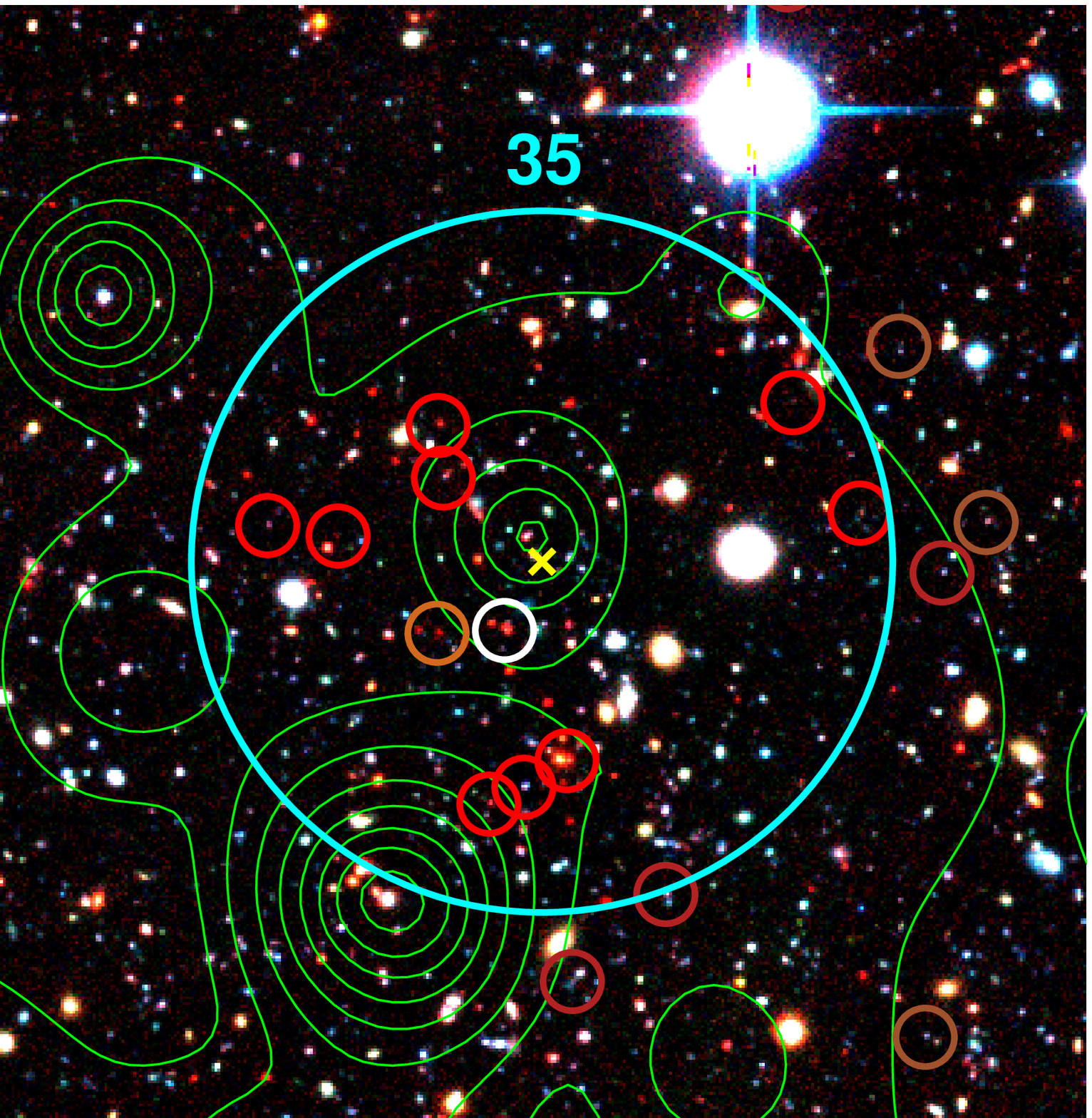}
\end{subfigure}
\hfill
\begin{subfigure}{0.3\textwidth}
\includegraphics[width=6cm]{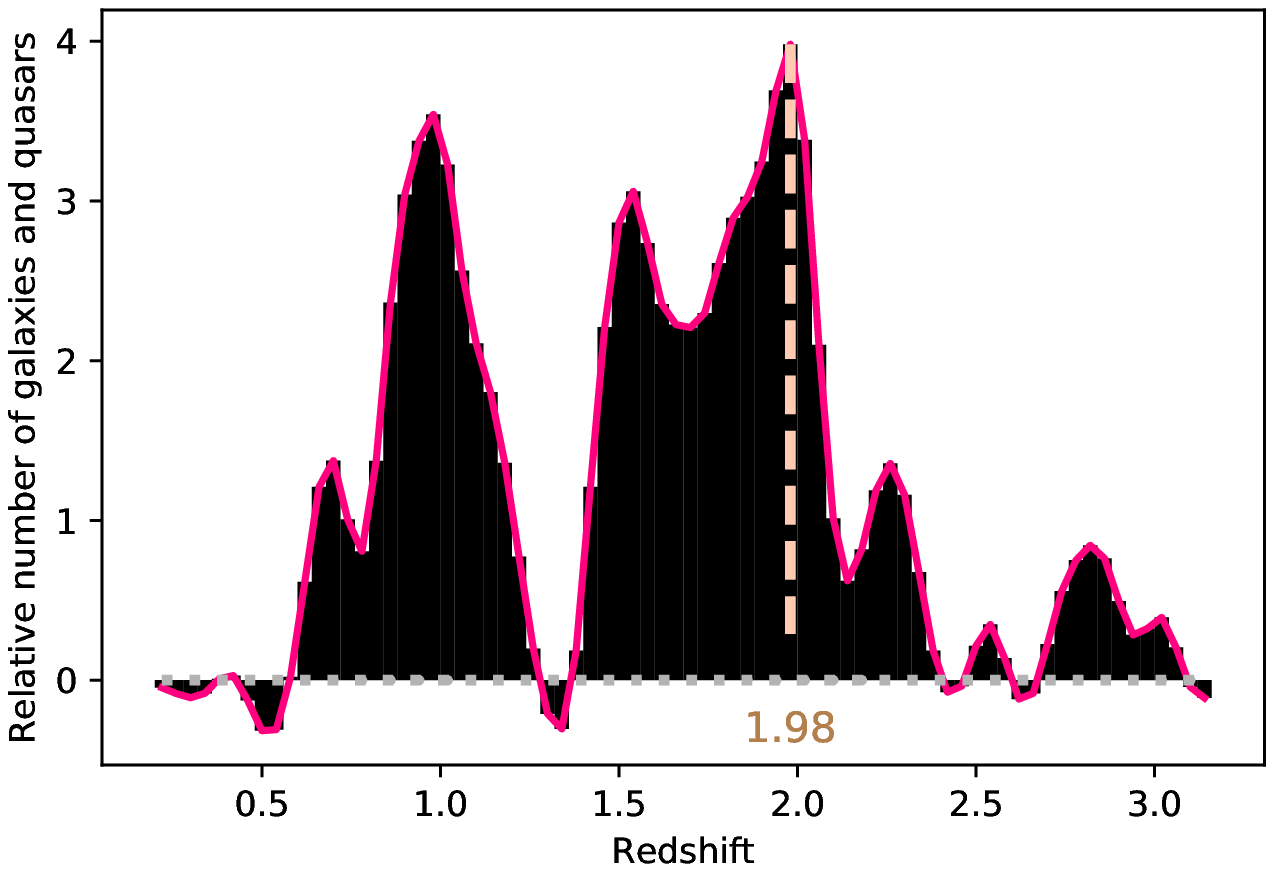}
\end{subfigure}
\hfill
\begin{subfigure}{0.3\textwidth}
\includegraphics[width=6cm]{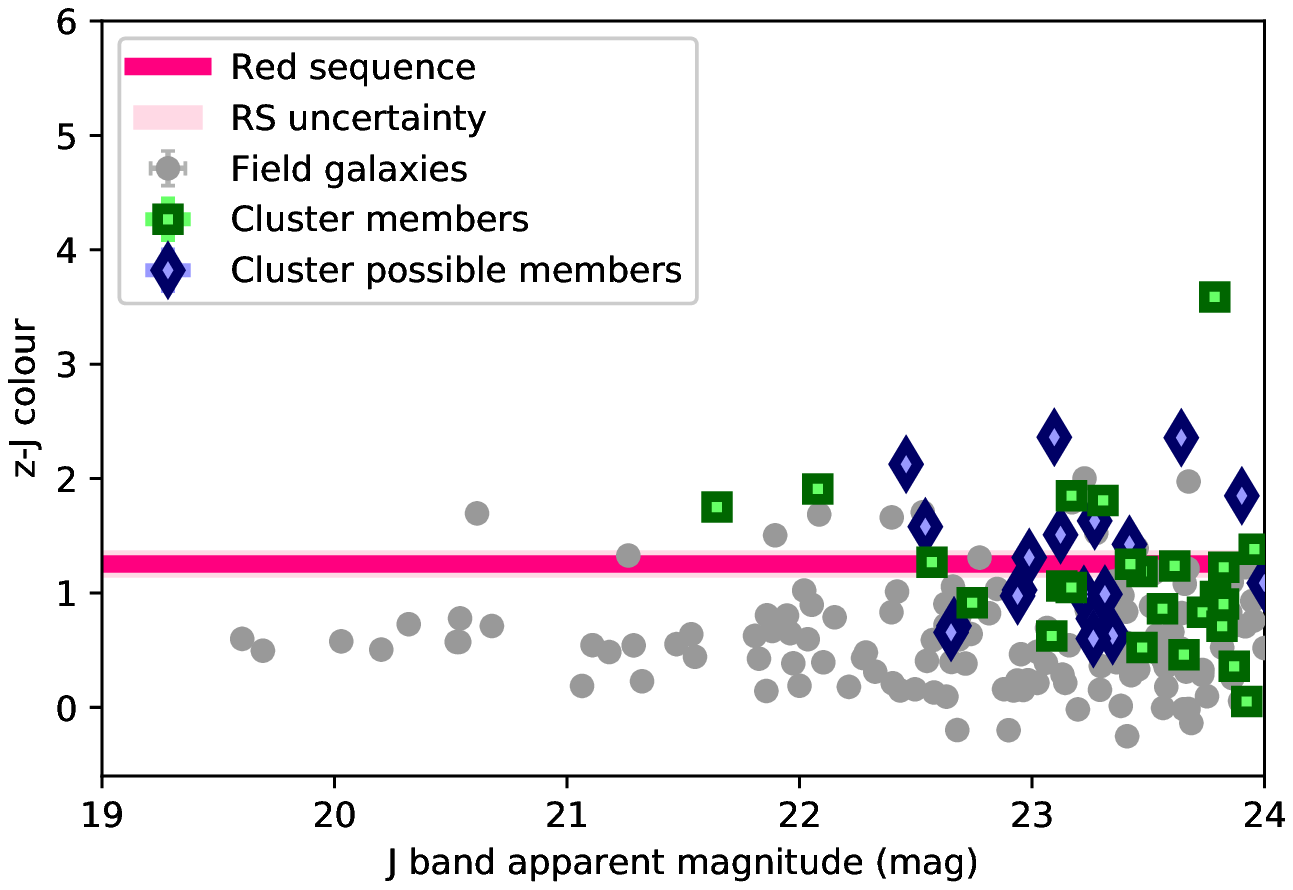}
\end{subfigure}
\raggedleft
\caption{\textit{continued}}
\label{fig_new_candidates}
\end{figure*}

\end{appendix}
\end{document}